\documentclass{ws-jai}
\usepackage[flushleft]{threeparttable}

\usepackage{graphicx}
\usepackage{natbib}
\usepackage{amssymb,amsmath}

\begin{document}

\catchline{}{}{}{}{} 

\markboth{Eric Keto}{Hierarchical configurations for cross-correlation interferometers with many elements}

\title{Hierarchical configurations for cross-correlation interferometers with many elements\\
}

\author{Eric Keto$^1$}

\address{
$^1$Harvard-Smithsonian Center for Astrophysics, 60 Garden St, Cambridge, MA 02138, keto@cfa.harvard.edu \\
}

\maketitle

\footnotetext[1]{Corresponding author.}

\begin{history}
\received{(May 29, 2012)};
\revised{(to be inserted by publisher)};
\accepted{(to be inserted by publisher)};
\end{history}

\begin{abstract}
Array configurations built on a hierarchy of simple elements
have excellent properties for cross-correlation imaging interferometers including
a smooth distribution of measured Fourier components, high angular resolution, low
side lobes, and compact array size. 
Compared to arrays with a Gaussian distribution of
antenna separations, hierarchical arrays (H-arrays) produce
beams with higher angular
resolution and a tighter concentration of the total power (encircled energy)
within a smaller
area around the main beam.
An attractive feature of
H-arrays is their
simplicity. The relationships between the Fourier coverage
and the array configuration are easy enough to understand
that they can be adjusted to achieve different design goals without the
need for numerical optimization.
H-arrays will be useful for future multi-element interferometers.
\end{abstract}


\section{Introduction}

The design or configuration of 
a cross-correlation imaging interferometer, the
placement of the antennas, presents a challenge because the properties desired
for imaging cannot all be optimized simultaneously. 
Imaging interferometers such as those used in radio astronomy construct a picture
of the sky from 
a set of measured Fourier components. Each pair of antennas in
a multi-antenna array measures one Fourier component whose scale 
or spatial wavelength is
proportional to the distance between the pair and whose direction is aligned with
their orientation. The entire set of Fourier components is given by all
the possible pairs, and the point source response or the
beam of the telescope is given by their Fourier transform. 
Thus the relative locations of the antennas determines the quality of the images
obtained by the telescope \citep{Bracewell1958}.

The impossibility of obtaining an infinite set of Fourier components requires
choices as
trade-offs among different qualities desired for imaging.
For example, the extent or size of the array should be small because
the atmospheric phase coherence decreases with distance and also because the
construction is easier. At the same time, 
the width of the beam or point-source
response should also be small to achieve high angular resolution. 
Because the sizes of the array and the beam are related through the Fourier
transform of the antenna separations, these two goals amount to minimizing
both the extent of a function and the extent of its Fourier transform.
The challenge is that the two generally have an inverse relationship, one big the other small.
Another challenge lies in the trade-off between
the angular resolution and the level of the side lobes around the main beam. 
The highest angular resolution requires
as many long spatial wavelength 
Fourier components (antenna separations) as possible. This 
requires a sharp cut-off in the number of components
at the maximum spatial wavelength (antenna separation)
because a smooth transition to the maximum
necessarily implies a decreasing number of components nearing
the limit.
The sharper the transition, the larger the side lobes so that
generally, the higher the angular resolution, the
higher the side lobe levels.
A further design goal is complete sampling of the Fourier plane.
Although imaging requirements might push the distribution to emphasize
shorter or longer wavelength spatial frequencies, gaps in the 
distribution represent missing information and allow unfaithful imaging. 
The assumption is that 
the likelihood of a significant error
in the estimate of a Fourier component increases with the distance in Fourier space
from the component to
the measurement. 
Therefore, the
best estimate of the strength of a Fourier component
is a nearby measurement,
and the distribution of separations should minimize
the maximum distance from any point in the Fourier plane to the
nearest measurement.

In summary, four design goals:

\begin{enumerate}
\item High angular resolution
\item Concentrated power, low side lobes
\item Compact array size
\item Uniform or smooth distribution of Fourier components
\end{enumerate}

How do we design arrays to fulfill these goals? How do we 
measure success? Do the trade-offs 
define a continuous space with the different goals at the vertices of a polyhedral
boundary? Can we position our design within this space to best meet the
imaging requirements of a particular application? Must we rely on numerical
optimization that might provide a configuration but no explanation leaving doubt
that we have found the best possible design?

This paper introduces array configurations built on a hierarchy made by repeating one
simple array configuration on different scales.
In this construction,
the relationships between the antenna locations, their
separations, and the beam are simple enough to provide answers to these questions.
For example, hierarchical or H-arrays can be designed to
achieve either higher angular resolution or a more concentrated beam with
lower side lobes by simply changing the scaling between the levels of the hierarchy
to emphasize shorter or longer spatial wavelength Fourier components.  Because H-arrays
are nested and scaled,
the choice can be made in the design stage by scaling the 
separation between the hierarchies, or for arrays designed with
more possible locations than antennas, the beam can be varied by
populating different levels of the hierarchy. This makes H-arrays particularly
well suited for the strategy
 employed by most
radio astronomy interferometers of periodically rearranging a smaller number of
antennas among a larger number of locations to observe different
angular scales.
H-arrays will be useful for future interferometers with many antennas.

\section{Classic pairs} \label{FTpairs}

Two
well-known Fourier transform pairs \citep{Bracewell1999} illustrate the trade-offs 
and suggest that there is a functional form for 
a continuously varying aperture distribution resulting in a beam pattern that can be
varied smoothly between the design goals. For the moment, suppose that 
we have
complete control over the aperture distribution,
and can assume an ideal continuous distribution. 
We will come back to the
question of how an array can be configured to provide the required aperture.

\begin{figure}[t]
$
\begin{array}{cc}
\includegraphics[trim=0.10in 0.3in 0.8in 0.2in, clip, width=2.75in]{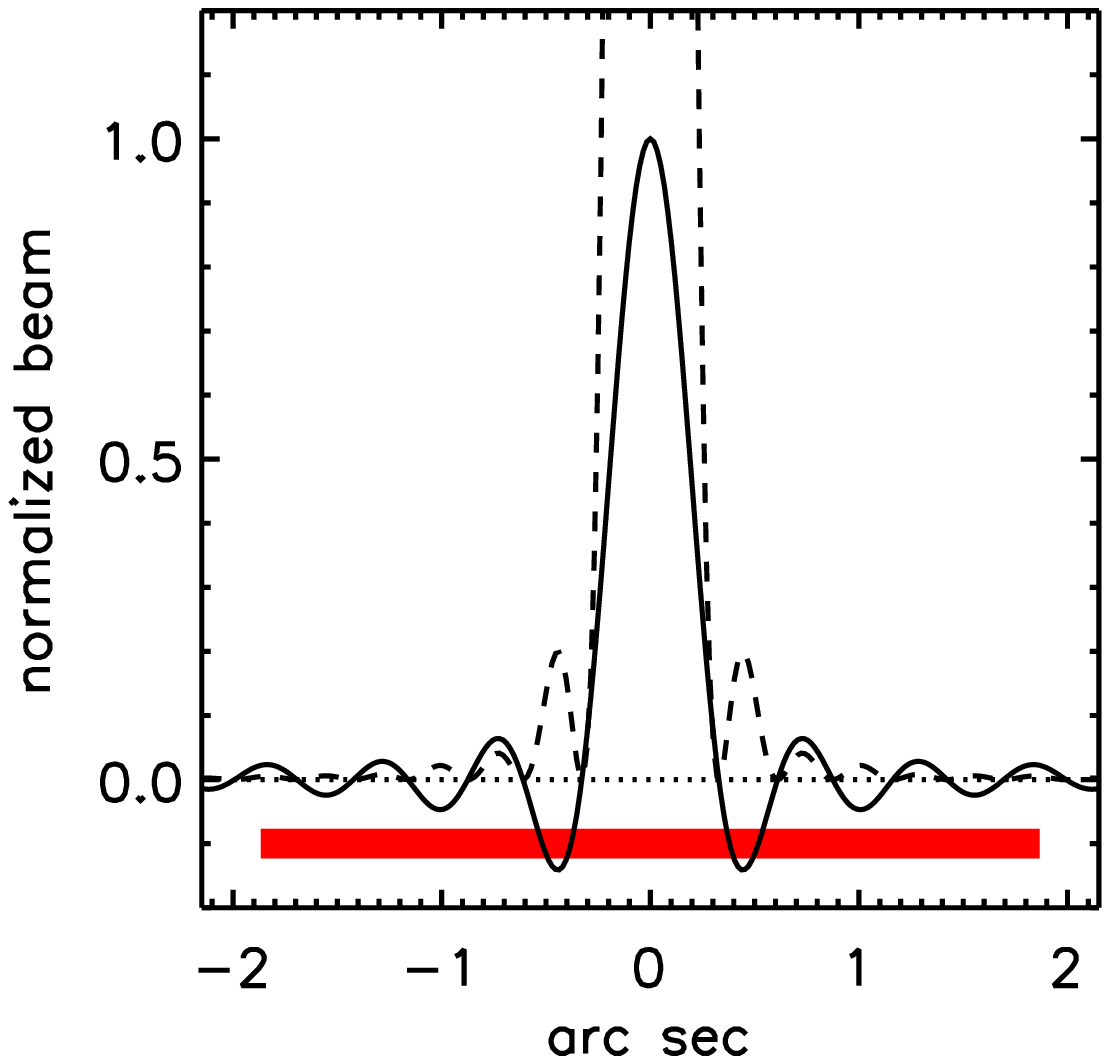} 
\includegraphics[trim=0.10in 0.3in 0.8in 0.2in, clip, width=2.75in]{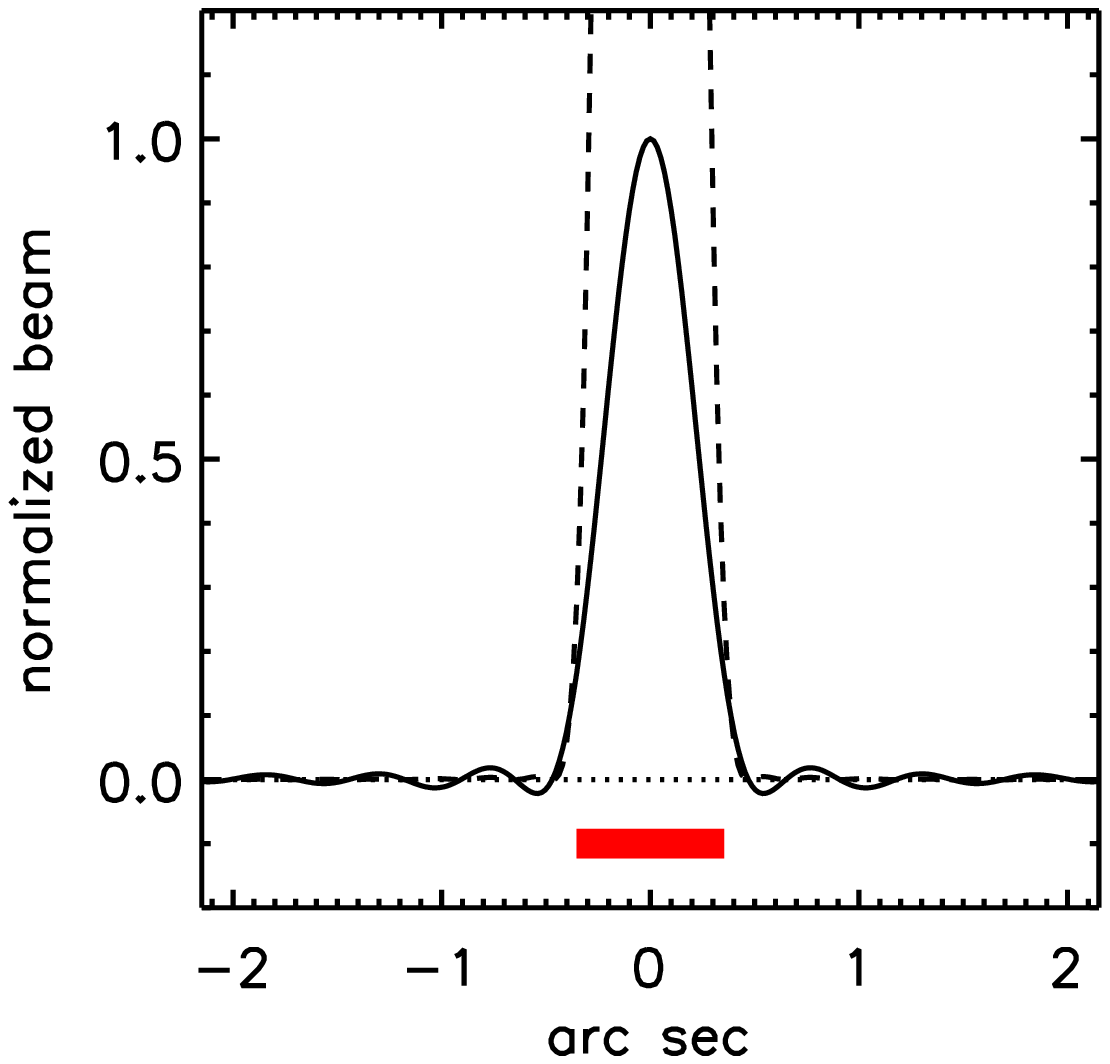} \\
\end{array}
$
\caption{{\it Left:} One dimension of the beam (solid line) and the beam power (dashed line)
of a uniform circular aperture. The beam pattern is normalized to one. The beam power
is multiplied by 10 so that the scale shows the percentage of the peak power.
The scale in arc seconds assumes
a hypothetical uniform aperture of 1000m diameter and an 
observing frequency of 230 GHz.
The main beam is relatively narrow, 0.27 arc seconds FWHM. Because of the
high side lobes, the radius that contains 98\% of the total power is
relatively wide, 1.53 arc second, indicated by the red horizontal bar. 
{\it Right:} One dimension of the beam and power pattern of a 2D Gaussian aperture
with $\sigma = 250{\rm m}$ ($2.35\sigma = {\rm FWHM}$) truncated at 1000m.
The main beam is relatively wide, 0.40 arc seconds FWHM.
Because the side lobes, caused by the truncation, are relatively 
low, 98\% of the energy is encircled within a relatively
narrow radius of 0.34 arc second.
The figures of merit are listed in table \ref{table:merit}.
}
\label{fig:FTpair}
\end{figure}

First, consider a uniform distribution across the aperture. In one-dimension
this would be 
a boxcar function or in two dimensions a uniform disk. The beam
pattern is the Fourier transform of the aperture distribution,
which for one and two dimensions
respectively, is the sinc function, $\sin{x}/x$, and the function $J_1(r)/r$
where $J_1$ is a Bessel function. In the two-dimensional case,
the power pattern, or the square of the beam, is the Airy function.
The beam and power patterns are shown in figure \ref{fig:FTpair}.
The angular resolution of the beam power is $\lambda / D$, the wavelength of the
observing frequency divided by the diameter of the aperture. 
For example, if
the aperture is 1000m in diameter and the observing frequency is 230 GHz, then
the angular resolution, measured by the width of the power in the main beam, 
is 0.27 arc seconds, full-width at half-maximum (FWHM). 
However, a significant amount of the power is outside the 
main beam in the side lobes.
The encircled energy, or the percentage of the total beam power 
as a function of radius
is a measure of the concentration of the power in the main beam.
The beam of a 1000m uniform aperture at 230 GHz contains 98\% of the power 
within a radius of 1.53 arc seconds. 


In contrast, consider a Gaussian distribution whose Fourier transform is also
a Gaussian with an inverse relation in widths, $\sigma$ and $\lambda/\sigma$.
Because the Gaussian has no side lobes other
than its own extended wings, all the power is concentrated in the main beam.
However, the inverse relation between the widths
in the real and Fourier domains means that a large array size is required to achieve
a small beam size. 
Furthermore a Gaussian aperture must be truncated at a finite radius. 
A Gaussian aperture of 1000m diameter with 
a width of one-half the radius or 
$\sigma = 250{\rm m}$ has an
angular resolution of $\lambda / \sigma = 0.40 $ arc seconds.
This is significantly broader than the beam of a uniform aperture,
but the encircled energy of the Gaussian
is more concentrated. 
The radius encircling 98\% of
the total power in 2.15 arc sec is
0.35 arc seconds. 
The beam
and power patterns are shown in figure \ref{fig:FTpair}. 
The figures of merit are listed in table \ref{table:merit}.

These two Fourier transform pairs illustrate two different trade-offs: 
first, between the angular resolution and the size 
of the array; and second between the angular resolution 
and the encircled energy which is sensitive to the amount of power
in side lobes. In general, the best apertures are 
a compromise between these two extremes,
a uniform distribution with a smoothed or apodized boundary. 

\begin{figure}[t]
$
\begin{array}{cc}
\includegraphics[trim=0.10in 0.3in 0.8in 0.2in, clip, width=2.75in]{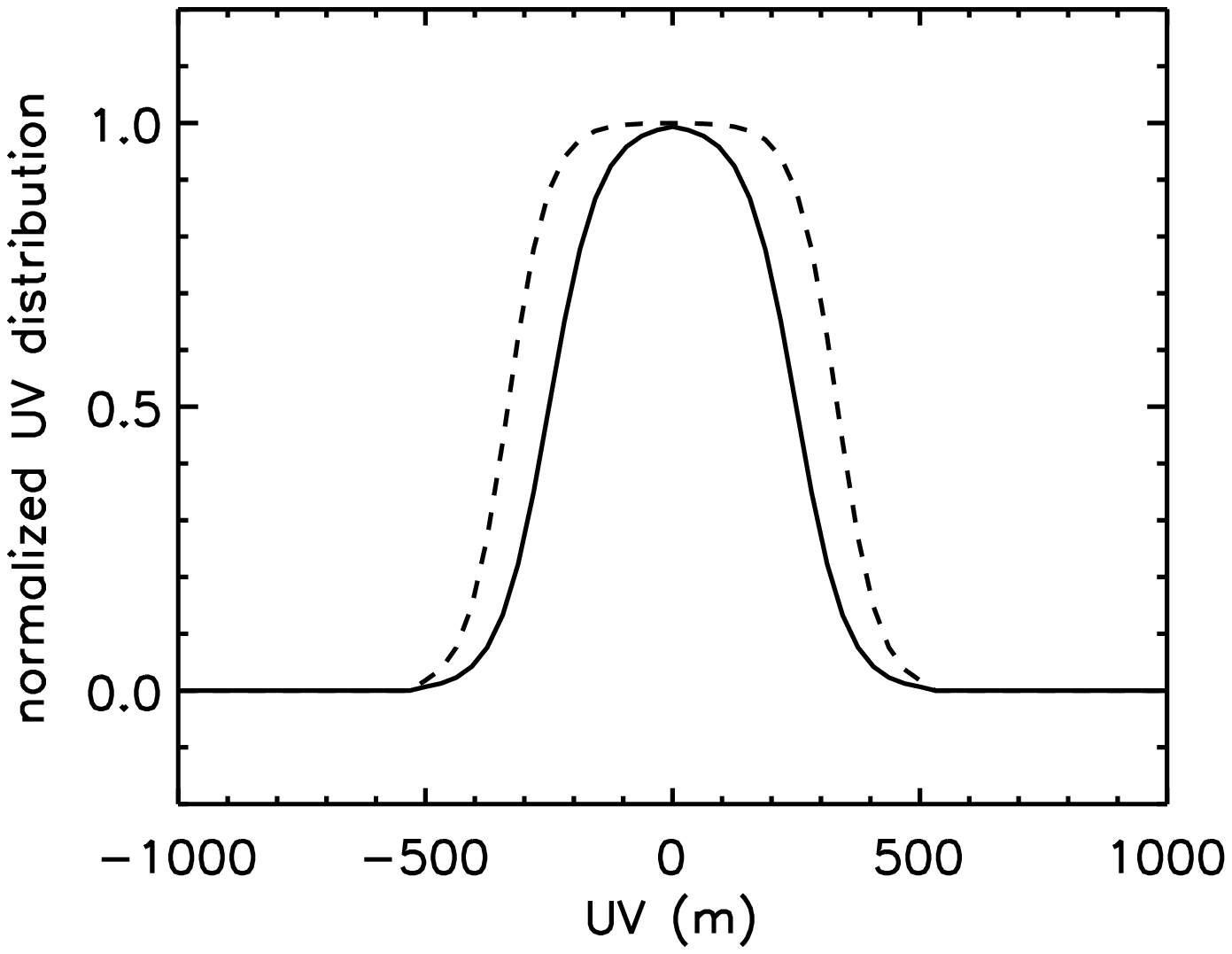} 
\includegraphics[trim=0.10in 0.3in 0.8in 0.2in, clip, width=2.75in]{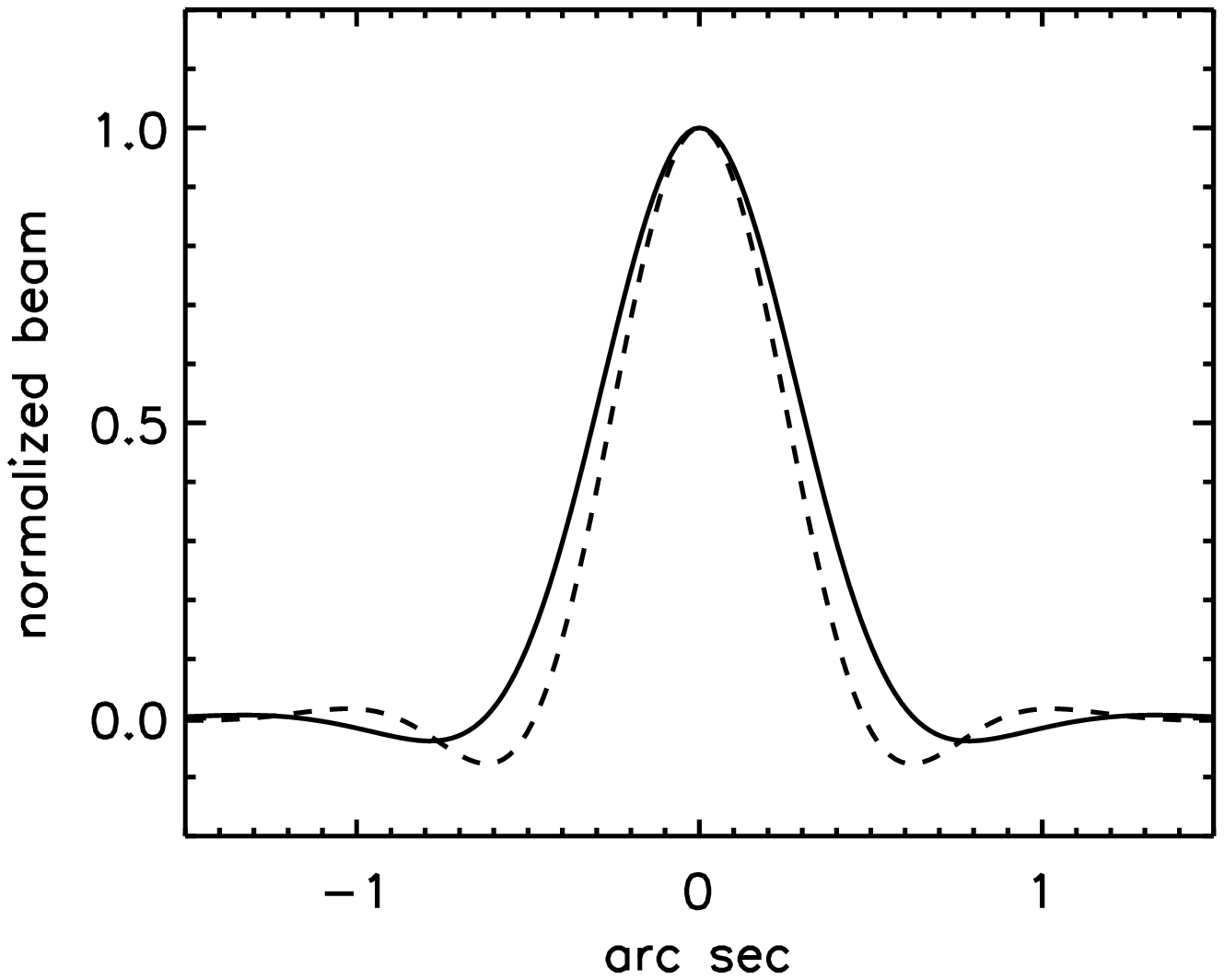} \\
\end{array}
$
\caption{{\it Left:} Two examples of a uniform distribution with a smoothed boundary
using the logistic function with different constants to create two different
boundary widths.
{\it Right:} Beam patterns corresponding to these two aperture distributions.
The comparison shows that the side lobe levels decrease with the
smoothness of the distribution while the beam width increases and the angular
resolution decreases
}
\label{fig:logistic}
\end{figure}

For example,
a  sigmoid such as the logistic function $y = 1 / (1 + \exp{(-(x-a)/b}))$
defines a curve which is uniform, equal to one in its interior, and has a
smooth boundary at $a$ whose transition width, $b$, 
can be continuously varied.
Figure \ref{fig:logistic} ({\it left}) shows two aperture distributions 
made with the logistic function
with different values for
the width of the boundary, $b$ = 50 and 42 m, and its location, 
$a$ = 250 and 333 m. Both apertures are truncated at a radius of  500 m. 
Both beams (figure \ref{fig:logistic} {\it right}) 
have broader FWHM and lower side lobes than the beam of the idealized
uniform aperture with a sharp boundary (figure \ref{fig:FTpair}).
The comparison here shows additionally that 
the side lobes decrease with the increasing width or smoothness of the boundary
while the beam width increases, and the angular resolution decreases.

Finally, we must come back to the constraint that
the aperture distribution 
is not arbitrary but given by the separations of the individual antennas.
The relationship between the pattern of antenna locations and the pattern of
their separations is not always obvious.
Particularly with Earth rotation synthesis (ERS) \citep{Ryle1962}, it is a challenge to arrange
the individual antennas to create a desired distribution from the sum
of their separations. The fourth design goal indicates how well the antenna
separations approximate a continuous distribution.

\section{Previous array designs}\label{previous}

One of the first arrays for radio astronomy, the
Mills cross  \citep{MillsLittle1953}
had regularly-spaced dipole antennas aligned in the shape of a cross which
provided a square aperture with a regularly-spaced, two-dimenensional 
grid of Fourier components.  
\citet{RyleHewish1960}
developed an interferometer with more sampling efficiency 
by eliminating
one of the arms, which changed the cross to a T-shape, and by 
locating the antennas with non-uniform spacing in such a way as to minimize
the number of separations of the same length.
Minimum redundancy is an advantange particularly for interferometers that use expensive
parabolic-dish antennas rather than dipoles. The theory 
of minimum redundancy was further explored
by \citet{Moffett1968} and extended to a large number of antennas by
\citet{Ishiguro1980}. 
Interferometers such as 
the Hat Creek Observatory/BIMA \citep{Welch1996},
the Owens Valley Radio Observatory Millimeter Array \citep{Padin1991},
and the Plateau de Bure Interferometer \citep{Guilloteau1991}, among others, 
are T-shaped arrays designed for Fourier sampling with minimum redundancy.

The Very Large Array (VLA) \citep{Thompson1980} 
improves on the T-shape by arranging the 3 arms in a Y-shape which 
provides better sampling when used in ERS. A typical astronomical observation
is about 8 hours, or one-third of a complete rotation.
The antennas are distributed along the arms according to a power-law
spacing.  This creates a non-uniform sampling that
allows a higher dynamic range between the smallest and the largest measured
Fourier component. 
In contrast, the Submillimeter Array interferometer is designed to 
to provide uniform and non-redundant 
Fourier sampling for high image fidelity \citep{Keto1997}. 
Uniform coverage is discussed more immediately 
below (\S\ref{CW}) 
and power-law spacings in \S\ref{spiralArrays}. 

Yet another design goal, 
a Gaussian-shaped beam,
was used to design the configurations for the Combined Array for Research in
Millimeter Astronomy (CARMA) \citep{Helfer2004}. 
The large number of antennas of the recently built
Atacama Large Millimeter Array (ALMA) \citep{WootenThompson2009}
allows more flexibility to shape the beam, and there are three
different design goals
for configurations of different scale. The most compact configuration
seeks to minimize
side lobe levels, the intermediate is a spiral pattern with a Gaussian beam in mind,
and the most extended is a "Y" shape modified to reduce the side lobes of the beam
\citep{Conway2006,Holdaway2007}. Examples of the three configurations are shown
in \S\ref{alma}. 

The Long Wavelength Array (LWA) distributes 256 antennas in
a uniform random pattern 
that is modified to reduce the side lobe levels \citep{Kogan2000}. Random
arrays are discussed in \S\ref{random}. 

One proposed design for the Square Kilometer
Array (SKA) puts the antennas along logarithmic spirals and seeks to minimize large
gaps in the Fourier sampling \citep{Millenaar2011}. 
Spiral arrays are discussed in \S\ref{spiralArrays} and
the minimum gap criterion in \S\ref{minimax}.

The study presented here in this article does not discuss the merits of different
design goals. These depend on the particular scientific aims of the observatories
and their particular constraints, for example, the number of antennas.
The different designs of the observatories listed above
represent a significant diversity of
goals, and the list does not even include all radio astronomy
arrays. 
Rather, this paper shows how array design can be understood in terms of
trade-offs between different goals and how the design process can be simplified by a
hierarchical strategy that allows the designer to place the array at 
a desired point between the trade-offs.

\section{Uniform sampling by curves of constant width}\label{CW}

The hierarchical arrays introduced in this paper are constructed by repeating
a simple configuration, a subarray, on multiple scales. A good choice for the
repeating pattern is one of the configurations that approximates
uniform sampling in the Fourier plane. 
This property ensures that for each scale
the gaps in the Fourier coverage, or more precisely the
separations between sampling points are all about the same size. 
Since they are a basis for
the H-arrays, it is worth reviewing
the properties of the configurations that provide the best uniform coverage.

A previous paper showed that arrays designed on the figures known as curves of 
constant width (CW) come closest to approximating a uniform 
aperture distribution \citep{Keto1997}. These figures are in the shape of closed rings.
The name, constant-width, refers to the diameter. These curves have the property
that from any point on the curve, the maximum distance to the opposite side, 
effectively the diameter, is always the same. This guarantees that the boundary of the
aperture and its Fourier transform, the beam, are both circular.
For example, the circle, which by definition has a constant diameter,
is a curve of constant width. The circle can be thought of
as a CW-curve with an infinite number of sides. The
Reuleaux triangle is the limiting case of a CW-curve with 
the least (three) number of sides. Of the two,
the Reuleaux triangle is preferred for array design and was adopted for
the design of the Submillimeter Array. 

The uniform aperture distribution
with its $J_1(r)/r$ beam approximated by a CW-array lies
toward one end of our space of design trade-offs accomplishing three of the
design goals, high angular resolution, compact array size, and uniform sampling but 
at the expense of rather high side lobes. The uniform sampling has another attractive feature.
In imaging applications, 
the uniform sampling allows the measured Fourier components
to be combined with equal weights ("natural weighting" in radio astronomy)
and therefore achieves the highest sensitivity simultaneously with
uniform sampling. 
CW-arrays are a good choice 
for interferometers that have relatively few antennas and therefore
deeply cherish each individual
Fourier component and regret weighting down any of them. 
Furthermore, 
a small number of antennas limits the flexibility 
to shape the beam,  and
the small number of
measured Fourier components necessarily results in high
side lobes
which, unavoidable, are therefore less of a design concern.
Imaging with high side lobes relies on numerical techniques to 
deconvolve the beam pattern from the image \citep{Hogbom1974}. 

CW-arrays are also the best choice for dithering patterns for flat-fielding multi-pixel
imaging arrays \citep{Arendt2000}. In this application the goal is to measure the
relative sensitivity of all the pixels in a sensor array such as a CCD. 
One possibility is to point the telescope repeatedly
until each pixel has viewed the same patch of sky. This requires many observations, 
one for each pixel in the array, but each comparison to determine the relative
sensitivity of two pixels requires the use
of only two measurements. Another possibility is to point the telescope only twice
in such a way that the view is shifted by one pixel.
Each pixel  then
views the patch of sky previously viewed by its neighbor, and the sensitivity of
neighboring pixels may be measured relative to each other. 
The relative sensitivity of all the pixels is found by working
across the array comparing neighboring pixels.
This requires only two observations, but the number of comparisons needed to determine the
sensitivity of any two pixels increases with their separation.  The noise in each intermediate 
measurement and comparison reduces the 
accuracy of the flat-fielding. The optimal strategy uses as few observations
and as few intermediate comparisons as possible. This is achieved by
distributing the pointing shifts uniformly across the space of all possible shifts, exactly
analagous to the problem of distributing the separations of antennas 
uniformly across their separation space.  
In their study, \citet{Arendt2000} objectively compared a number of different array designs
and verified that the CW-arrays provide the best approximation to uniform sampling.
The Spitzer Space Telescope uses an observing pattern based on the Reuleaux 
triangle for imaging with its infrared cameras.

\subsection{A 6-element CW-array}

\begin{figure}[t]
$
\begin{array}{cc}
\hskip 0.65in \includegraphics[trim=0.10in 0.3in 0.8in 0.2in, clip, width=2.75in]{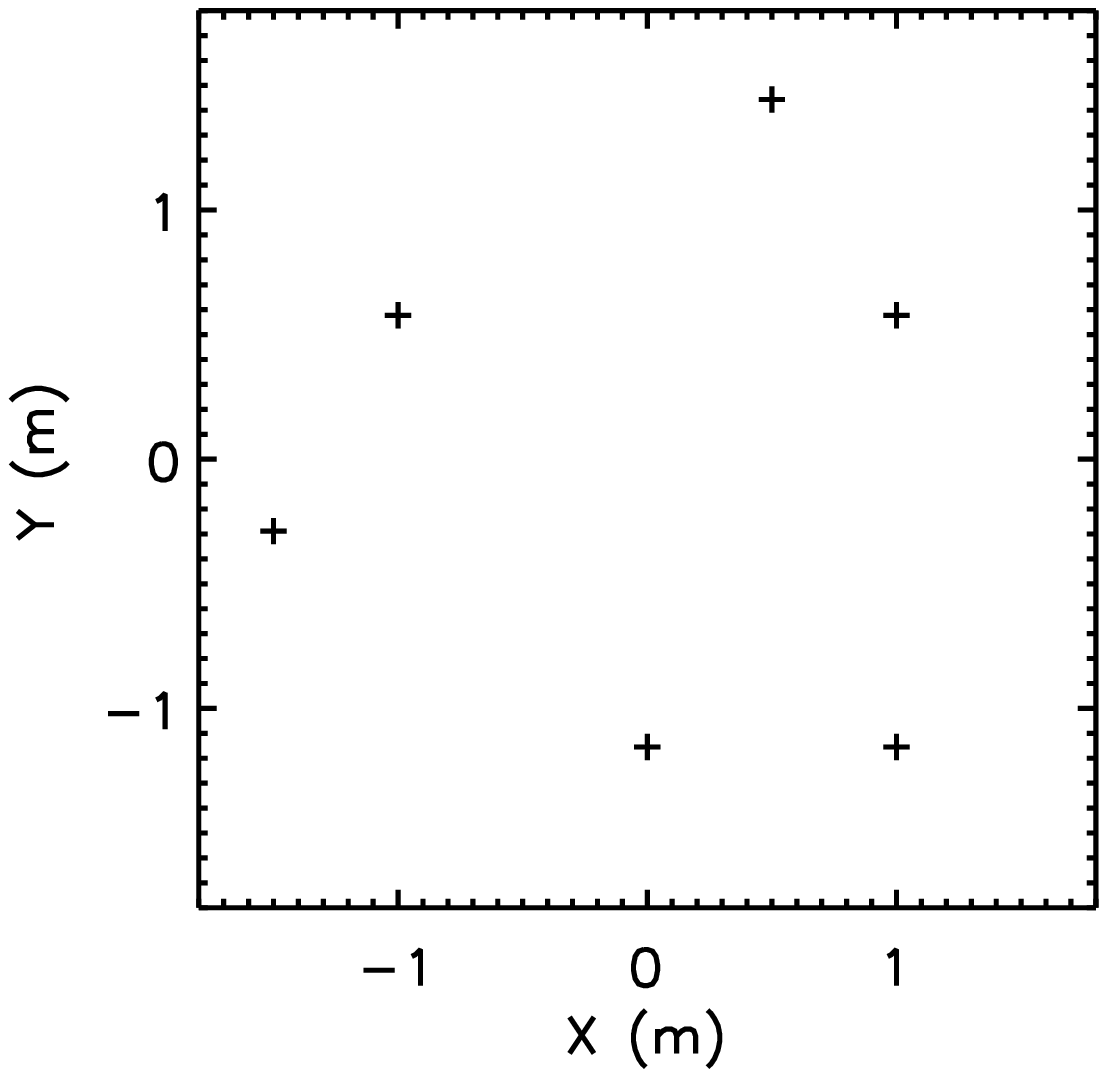} \\
\hskip 0.65in \includegraphics[trim=0.10in 0.3in 0.8in 0.2in, clip, width=2.75in]{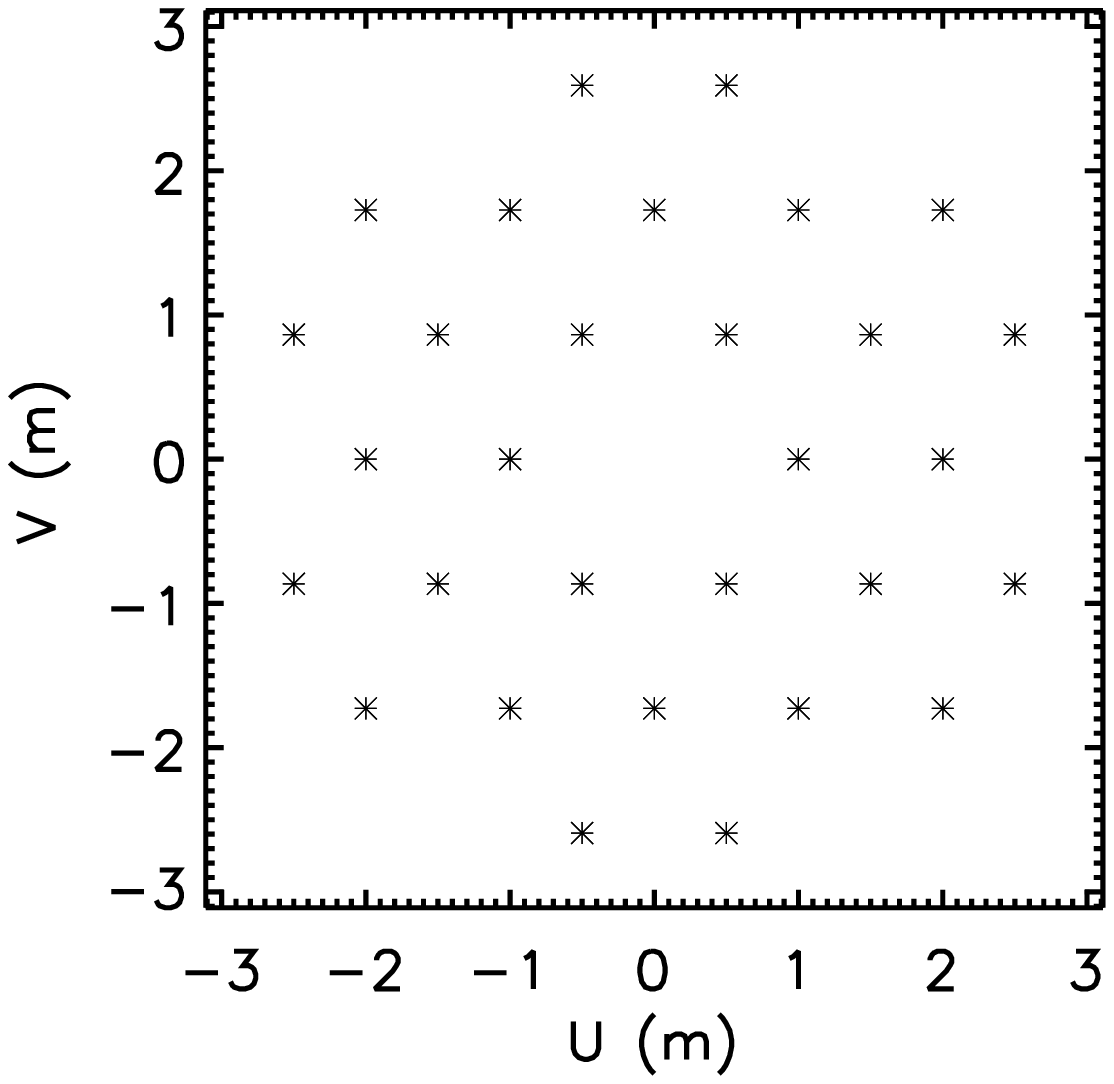} \\
\includegraphics[trim=0.10in 0.3in 0.8in 0.2in, clip, width=2.75in]{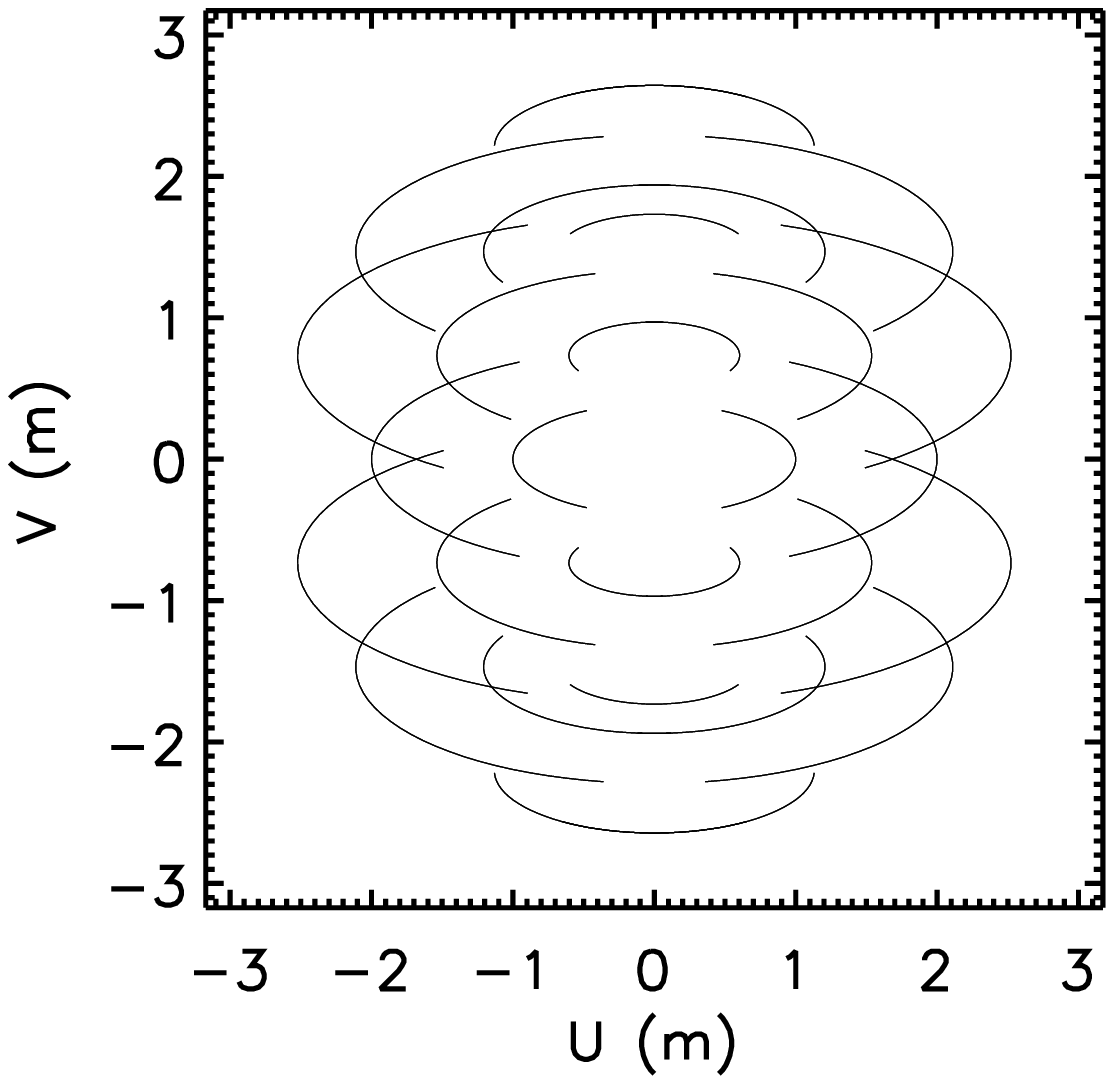} \\
\end{array}
$
\caption{{\it Top:} Antenna locations of a simple 6-element array whose separations 
(UV coverage) 
({\it middle}) 
are uniformly spaced on a hexagonal grid.
{\it Bottom: } Tracks of the antenna separations 
(baselines) in an 8.2 hr Earth rotation
synthesis (ERS). The uniform distribution of separations ({\it middle}) produces
a reasonably uniform distribution of baseline tracks.}
\label{fig:s6}
\end{figure}

The simplest example of CW-array is the six-element pattern in
figure \ref{fig:s6} whose separations 
are uniformly distributed on an hexagonal grid. 
The antenna coordinates are given in table \ref{table:s6positions}.
Visually there appears to be a hole in the center that is missing
a separation. This is a separation of zero, or equivalently the Fourier
component of zero spatial wavelength corresponding to the total power.
All interferometers are missing this measurement. Figure \ref{fig:s6}
shows that
the distance between all the separations, including
the distance from the smallest separations to the zero point, is exactly one.
In this sense, the configuration is not missing
short spacings. The shortest spacing is set by the distance between the
pairs of antennas on each side of the Reuleaux triangle which of course
can be made arbitrarily small. The reason the
central hole in the coverage looks larger in this array than in other arrays
with non-uniform coverage is because the uniform coverage limits
the size of the largest separation. For example, arrays with a power-law
distribution of antenna separations achieve a higher ratio of the largest
to smallest separations, higher spatial dynamic range, but only by leaving
larger holes in the coverage at larger spatial wavelengths. However, in these
non-uniform arrays,
the central hole looks smaller with respect to the total extent of the
non-uniform coverage. The next section, \S\ref{example36}, shows that
one advantage of hierarchical arrays is that 
this central hole can be filled by a smaller scale in the hierarchy.

While the UV coverage in ERS is no longer strictly uniform, the mid-points of the ERS
tracks are still on the original uniform grid. 
This means that arrays that
have good snapshot coverage generally have good coverage in ERS even though
the coverage in snapshot and ERS is not the same. 
Figure \ref{fig:s6} shows the tracks of
the separations (baselines in radio astronomy) in an 8.2 hr ERS assuming the array
is located at a latitude of 23$^\circ$ and the target transits through the zenith.
As shown in the figure,
ERS generates a pattern of tracks, or an effective aperture, slightly larger in
extent north-south than east-west. This is because the target is low
on the horizon at the beginning and end of the track when it is rising and
setting. At these times from the vantage point of the target, the array and its antenna
separations appear shortened by projection. This results in UV coverage
that is narrower and a beam pattern
that is wider in the east-west direction than north-south. It is trivial to make the beam
circular by adjusting the aspect ratio of the array, but 
the correction is also a function of the declination of the target. 


\begin{table}[ht]
\caption{Antenna positions for 6-element subarray}      
\vskip 0.1in
\centering                                              
\begin{tabular}{ c c}                                   
\hline\hline                                            
& \\
X & Y \\ [0.5ex]                                        
\hline                                                  
0     & 0 \\
1     & 0 \\
1     & $\sqrt{3}$ \\
1/2   & $3\sqrt{3}/2$ \\
-1     & $\sqrt{3}$ \\
-3/2   & $\sqrt{3}/2$  \\ [1ex]                          
\hline                                                   
\end{tabular}
\label{table:s6positions}                                        
\end{table}

\subsection{The minimax metric}\label{minimax}

\begin{figure}[t]
$
\begin{array}{cc}
\includegraphics[trim=0.10in 0.3in 0.8in 0.2in, clip, width=2.75in]{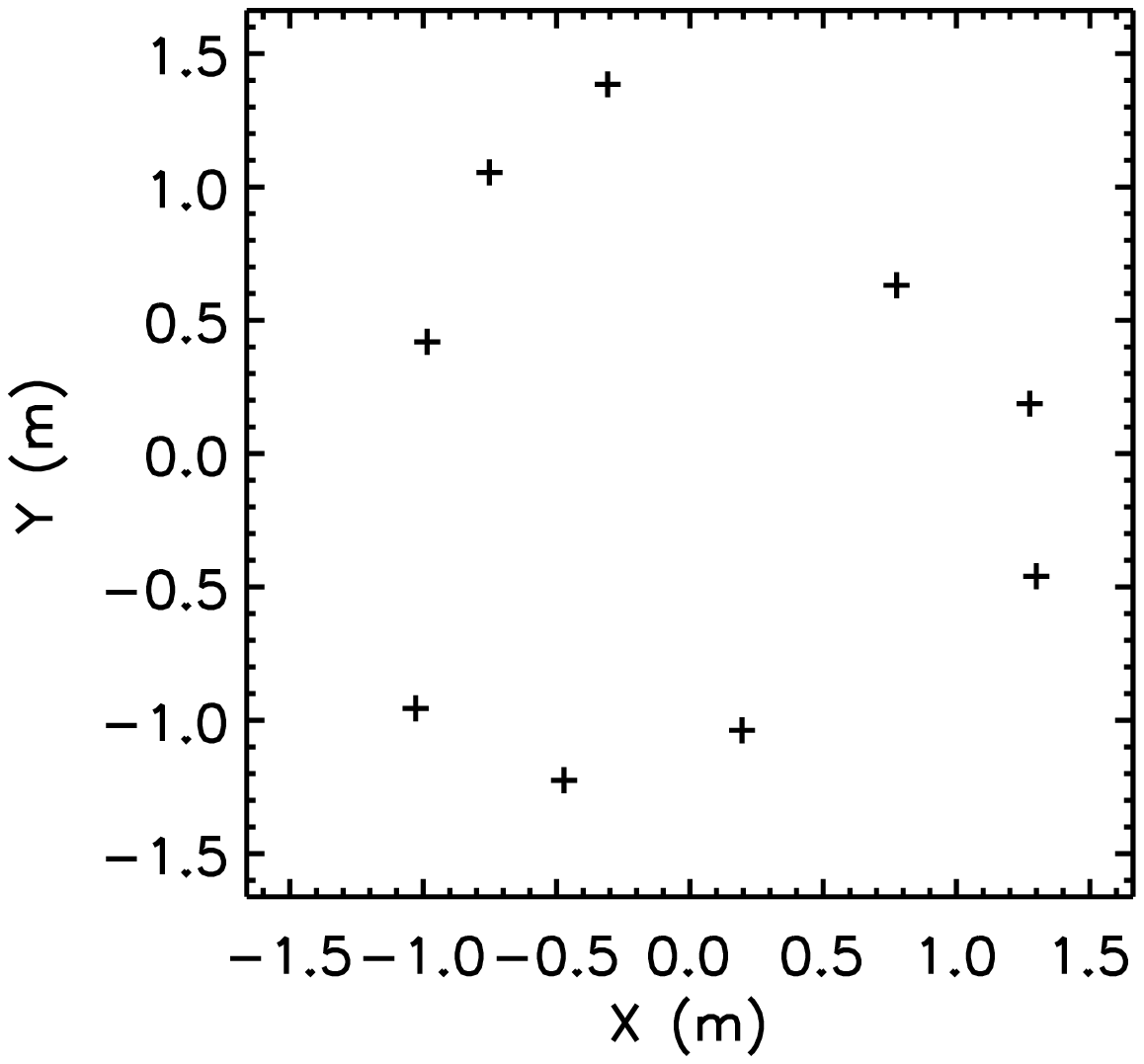} 
\includegraphics[trim=0.10in 0.3in 0.8in 0.2in, clip, width=2.75in]{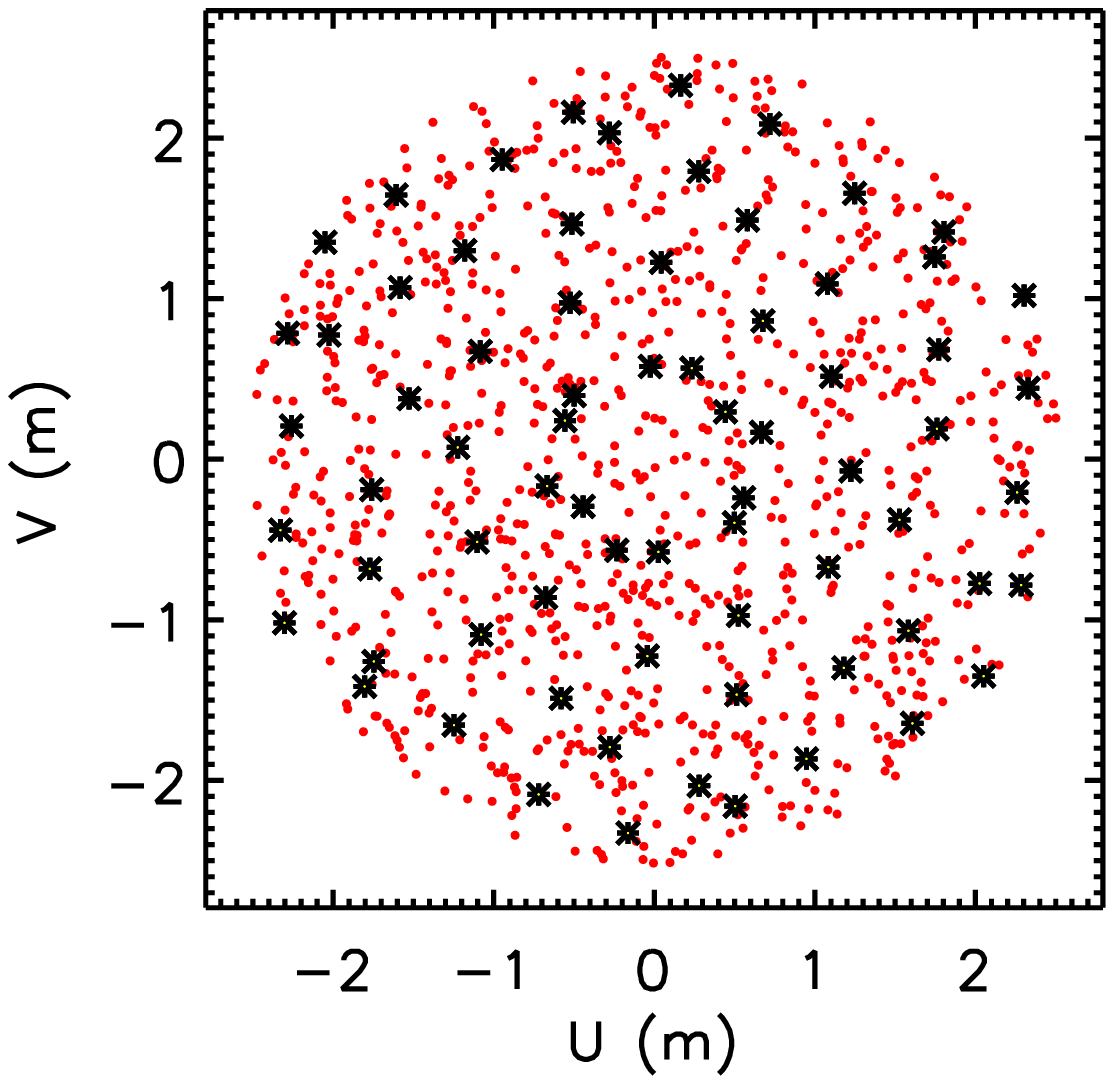} \\
\end{array}
$
\caption{{\it Left:} Antenna locations of a 9-antenna array whose separations
approximate a uniform distribution.
{\it Right: } Minimax test. Crosses show the antenna separations and red dots show
the random UV points generated by a Monte Carlo algorithm. 
}
\label{fig:s9}
\end{figure}

CW-arrays produce the most uniform sampling across a circular aperture, 
but aside from the particular 6-element pattern in snapshot observing
(discussed above), 
their separations do not lie exactly on a uniform grid. How do we measure
uniformity? For imaging arrays, the most important feature of
uniform coverage 
is that no part of the relevant Fourier space is very far from a
measurement. 
This decreases the possibility of error which 
increases with the distance in Fourier space between the component and a
nearby measurement, 
assuming that there is some smoothness or coherence in the image itself.
With this in mind,
uniformity can be defined in a statistical sense
with a particular figure of merit, a minimax that 
measures the maximum separation from
every point in the Fourier plane to the closest measured Fourier component.
With the usual distance as the metric, the best approximation is the one that
minimizes the maximum distance, $\min \sum ( \max_i [(U - u_i)^2 + (V - v_i)^2])$
where $U$ and $V$ are probability densities that represent
the desired distribution across the entire Fourier plane within the 
circular boundary allowed by the
maximum antenna separation, and $u_i$ and $v_i$ are the coordinates of each 
antenna separation.
A particularly simple way to compute this figure of 
merit is with a Monte Carlo algorithm.
To measure how closely the separations of an array
approximate a uniform distribution, generate a set of random points $U,V$,
uniformly distributed inside the boundary of separations
and compute the distance between all pairs $U,V$ and $u_i,v_i$. 
The best approximation minimizes the maximum
distance or the average maximum distance.
Figure \ref{fig:s9} illustrates the minimax for a 9-antenna array. The antenna
locations for this array are listed in table \ref{table:s9positions}.
In the right
panel the antenna separations are shown as asterisks and one particular
realization of a set of random UV points
is shown as red dots.  An actual evaluation would use many more random
points than the ones shown here for illustration.
The number of UV points necessary for an accurate measure is about 10 to 100
times the number of separations. It is easy to determine the accuracy 
of the algorithm by
running a few trials with different numbers of points and different
realizations of the random distribution.
This figure of merit was used in \citep{Keto1997}
to show that the arrays that best approximate a uniform
distribution of separations are
based on the Reuleaux triangle.
The minimax figure of merit is different from other
figures of merit that consider
the antenna separations only with respect to one another.

\begin{table}[ht]
\caption{Antenna positions for 9-element array}	
\vskip 0.1in
\centering 						
\begin{tabular}{ c c} 					
\hline\hline 						
& \\
X & Y \\ [0.5ex] 					
\hline							
       -1.02847        &	-0.955366       \\
       -0.471921       &	-1.22493        \\
       0.195772        &	-1.03746        \\
       1.29924         &	-0.459679       \\
       1.27441         &	0.187321        \\
       0.775329        &	0.631558        \\
       -0.308142       &	1.38484         \\
       -0.751473       &	1.05465         \\
       -0.984755       &	0.419072        \\
\hline							 
\end{tabular}
\label{table:s9positions}					 
\end{table}

\section{Figures of merit}

What if we want a more concentrated beam and power pattern with lower 
side lobes and a 
smaller radius of encircled energy than provided by
a uniform aperture distribution? Hierarchical arrays are one solution. First we
need a complete set of 
figures of merit to quantitatively evaluate different array designs in terms of 
each of the four design goals.
The first two goals can be combined
into the product of the array size and the beam size, leaving three.
\begin{enumerate}
\item $K_{nn}$ = maximum diameter of the array times the diameter that
contains $nn$  percent of the encircled energy.
This measures the degree to which the design is compact in both real 
and Fourier space.
Because the example arrays all have a maximum separation of 1000m 
the $K$-product is sensitive to the distribution of 
the power between the main beam
and the side lobes. For radio frequency arrays
of 1000 m with arc second resolution, the units are meter arc seconds. 
This measurement depends on the angular size scale used to define 100\% of the total power.
Since there are a limited number of Fourier components, they never completely cancel, and
the side lobes are infinite in extent.  A good practice is to integrate out
until the decrease in side lobe level slows.
For our example arrays of 1000m diameter and observing frequency of 230 GHz, 
we use a radius of 2.15 arc seconds.

\item The angular resolution, the full width at half maximum (FWHM) of the main beam.
We use the root of the product of the widths in the north-south and east-west directions.

\item A smooth distribution of UV points. There are several ways 
to measure the smoothness.
Some experimentation shows that a simple and satisfactory measure is 
the variance off a polynomial fit to the density of
antenna separations
as a function of separation. 
We use a third order
polynomial and the reduced $\chi^2$ of this fit. 
\end{enumerate}

To generate the figures of merit, 
the hypothetical observatory is located at a latitude of 23 degrees
tracking a source through zenith in Earth rotation synthesis. 
The UV tracks are calculated
for 8.2 hours between Hour Angles $\pm 4.1$ with points recorded every 0.25 hours.
The UV tracks and the beams are calculated assuming that the arrays
have a maximum diameter of 1000m, and that the observing frequency is 230 GHz.

The maps of the beam pattern are made with a particularly punishing color scale that
emphasizes the low-level side lobes.
The beam is normalized by its peak, then subject to an asymmetric sigmoid function, 
the Gompertz function $y = \exp{(-\exp{(-ax)})}$ with constant, $a=50$, large
enough to saturate the beam at a few percent revealing the side lobes.
The color table, inspired by Tang sancai (three colors) pottery, shows
the negative and positive side lobes in blue and red. The asymmetry of the
Gompertz function compensates in part for the visual bias that red appears
slightly brighter than blue.

\section{Hierarchical configurations}\label{hier}

\begin{figure}[t]
$
\begin{array}{cc}
\includegraphics[trim=0.10in 0.3in 0.8in 0.2in, clip, width=2.75in]{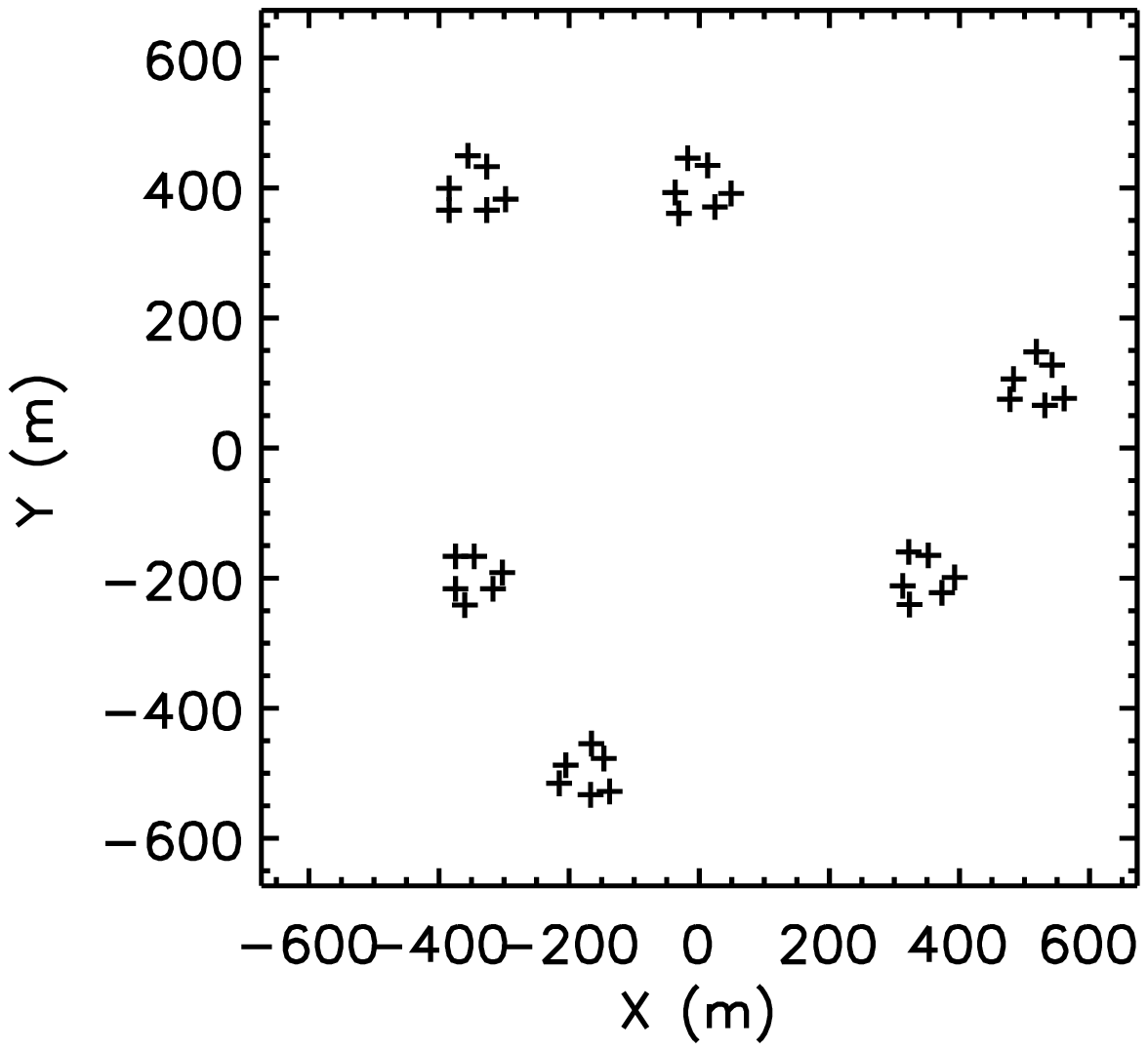} 
\includegraphics[trim=0.10in 0.3in 0.8in 0.2in, clip, width=2.75in]{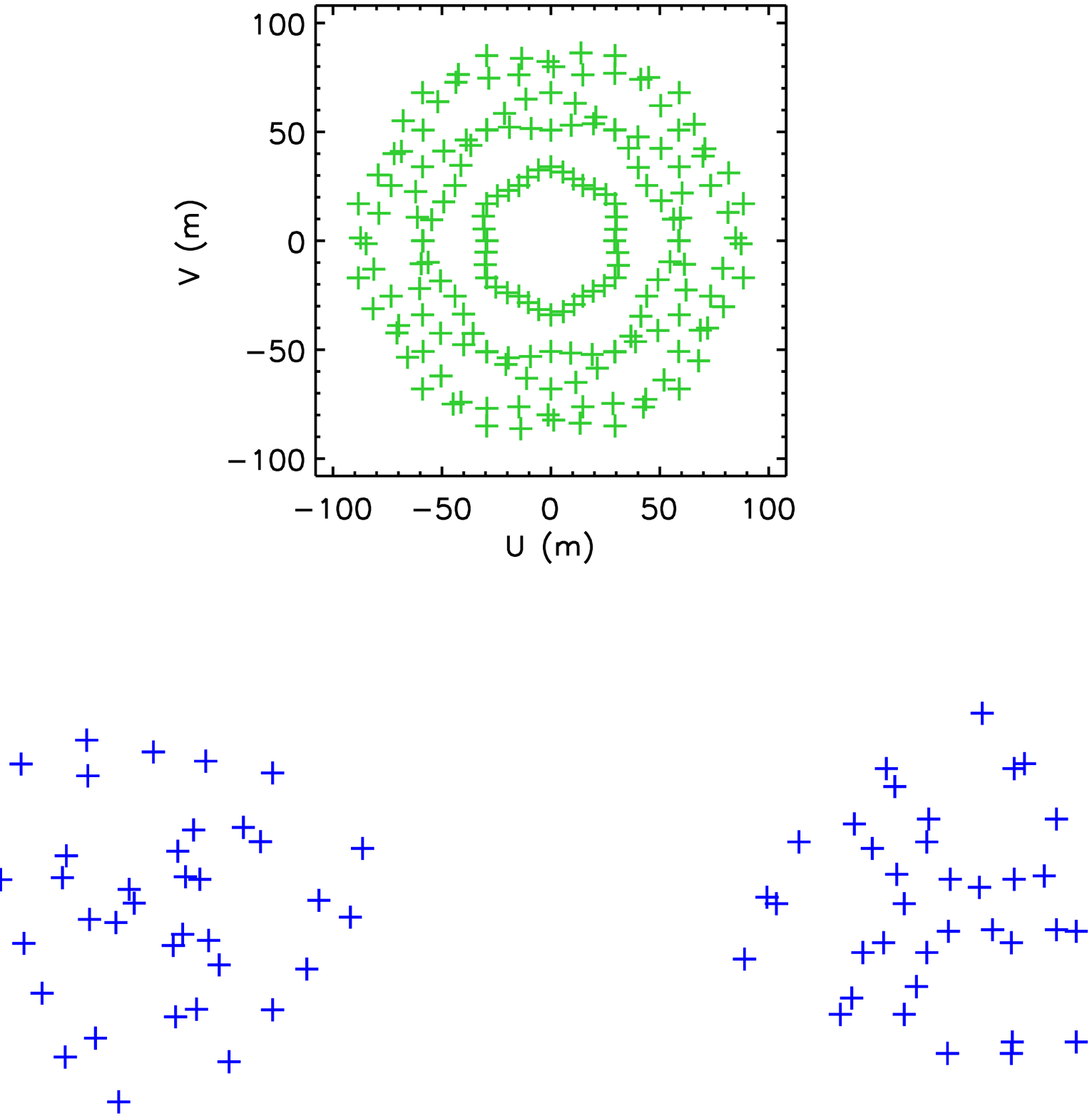} \\
\includegraphics[trim=0.10in 0.3in 0.8in 0.2in, clip, width=2.75in]{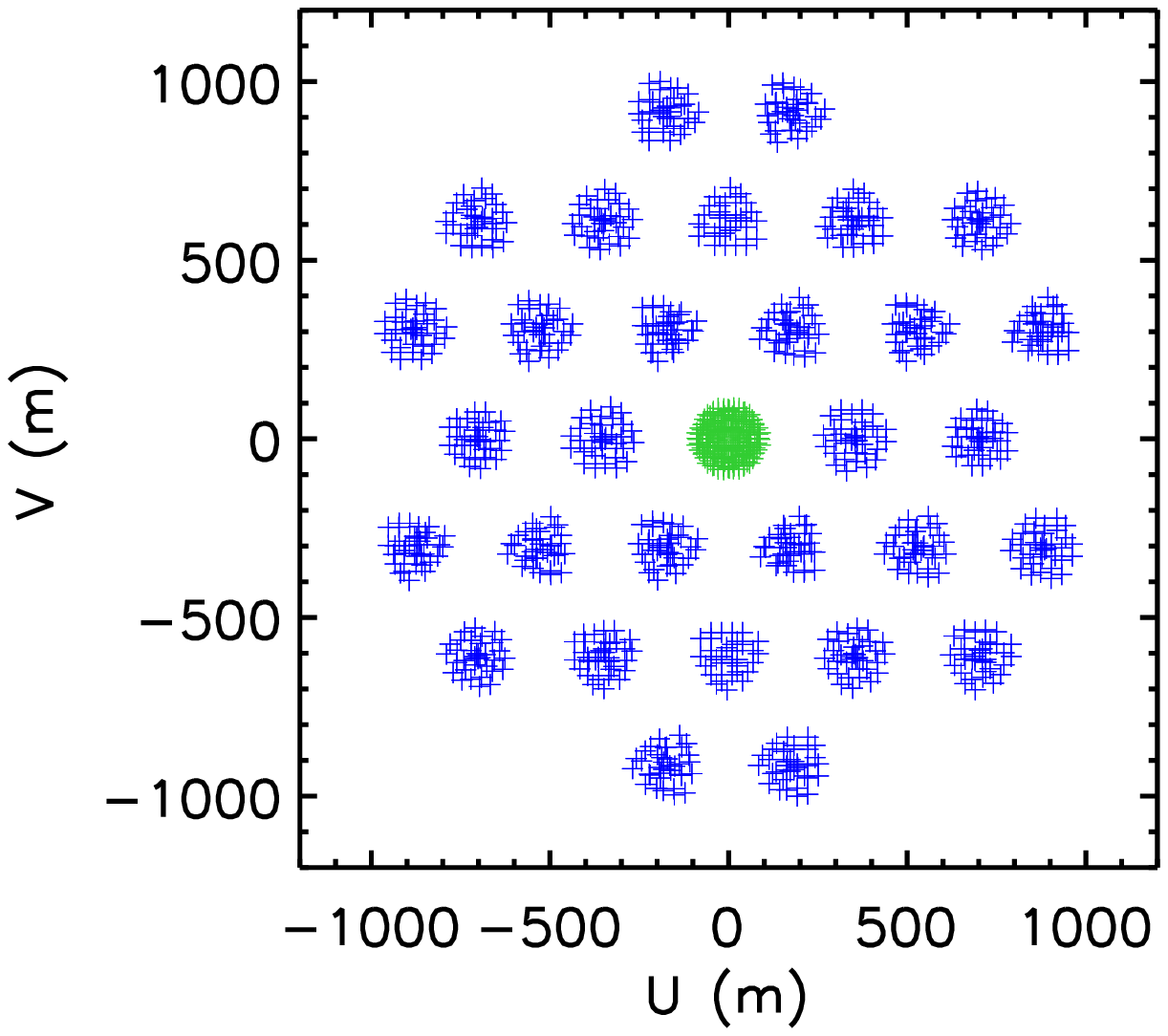} 
\includegraphics[trim=0.10in 0.3in 0.8in 0.2in, clip, width=2.75in]{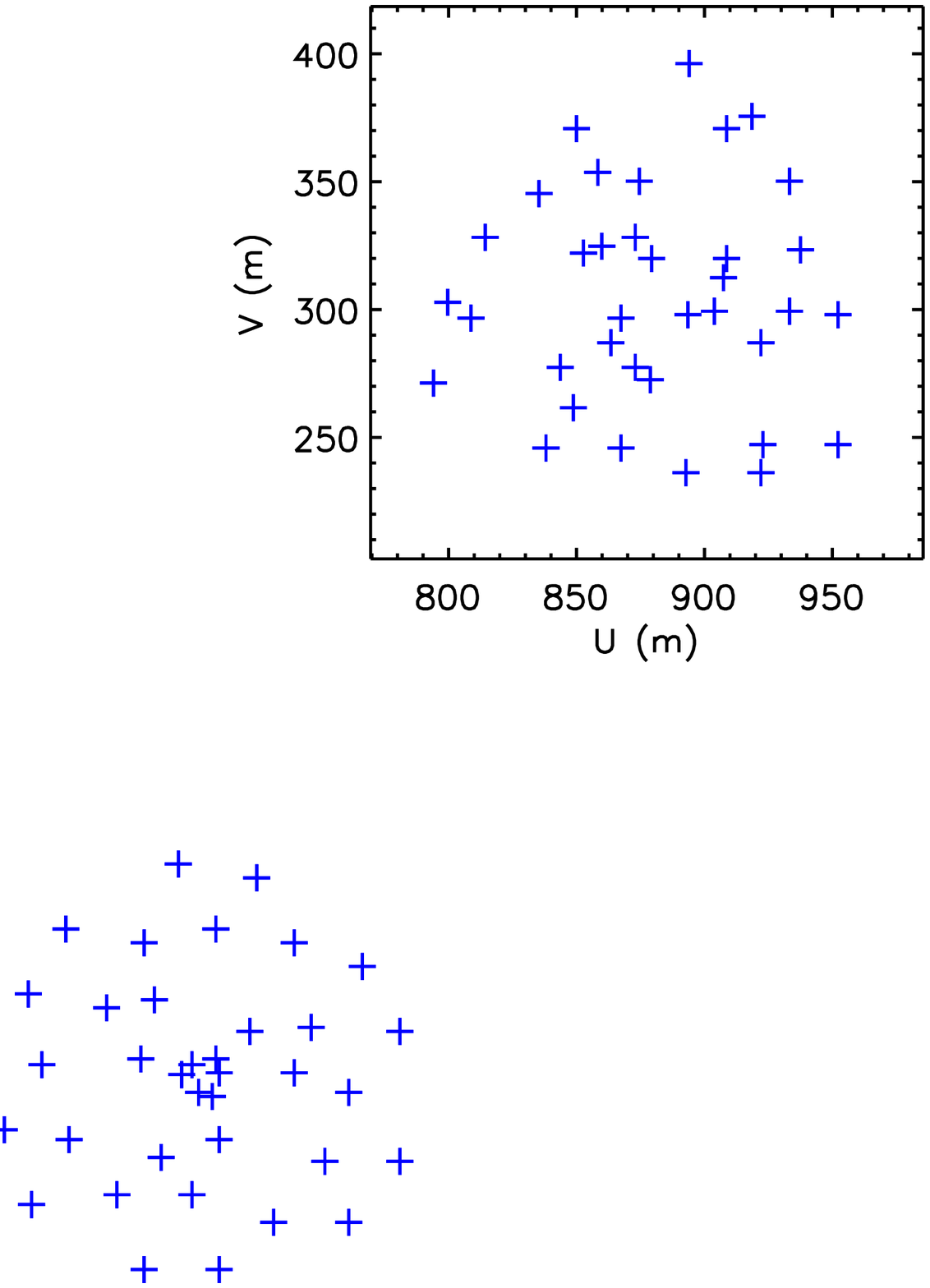} \\
\includegraphics[trim=0.10in 0.3in 0.8in 0.2in, clip, width=2.75in]{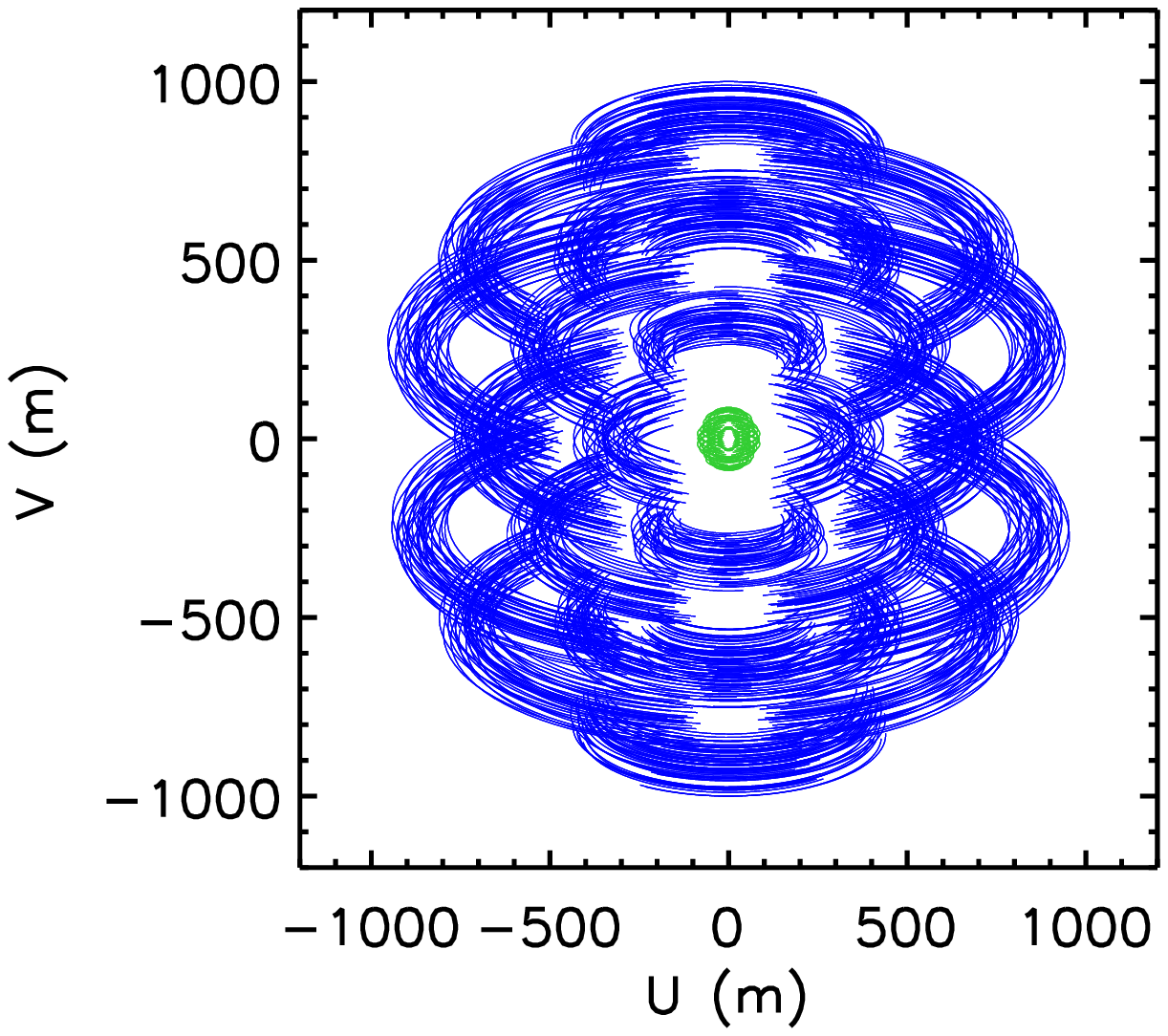} \\
\end{array}
$
\caption{{\it Top left:} Antenna locations in a  2-level s6p6
array. The subarrays are rotated relative to the subarray at the
bottom of the figure by 50, 10, 40, 20, and 30 degrees moving
counterclockwise around the pattern. 
The subarrays are also scaled by factors of 1.03,  or from the
bottom, by
1.03,      1.06,   1.09,   1.13,    1.16.
The entire pattern is
rotated 60 degrees with respect to the 6 positions in table \ref{table:s6positions}.
{\it Middle left:} Antenna separations.
The $180 = 30 \times 6$ short separations between antennas within 
each of the 6 subarrays
occupy the center of the UV plane in green and in
detail {\it top right}. 
The 30 other patches around the center
are the separations between the subarrays. The set of patches replicate the
uniform spacing of the basic 6-element array (figure \ref{fig:s6}). 
Detail of one patch
{\it middle right}. The exact pattern is different in each
patch.
({\it bottom}) UV tracks in an 8.2 hr ERS. Each of the 30 patches
moves across the UV plane like a paint brush.}
\label{fig:s6p6}
\end{figure}


To construct a hierarchical array, repeat 
the subarray pattern on a larger scale with
copies of the subarray distributed on a pattern equal to itself.
If we call the first level the s-level, the second the p-level, 
the third the d-level, and the fourth the f-level, a two-level 
array with 6 elements in the first
level and 6 in the second is indicated by s6p6. Figure \ref{fig:s6p6}
shows this configuration with the second level scaled to 12 times the
first. 
The two hierarchies of antennas create the two
hierarchies of separations. 
In this example, the scaling between the hierarchical levels is large enough
that the separations in different levels do not overlap.

In the very center marked in green are the 
$30 \times 6$ shortest separations within each of the 6 subarrays. 
To eliminate the redundancy in the separations 
that would occur if the subarrays were identical,
the subarrays are rotated and scaled very slightly
with respect to each other. 
Relative to the subarray at the
bottom of the figure, the subarrays are rotated 
by 50, 10, 40, 20, and 30 degrees moving
counterclockwise around the pattern,
and scaled by factors of 1.03,  or from the
bottom, by
1.03,      1.06,   1.09,   1.13,    and 1.16.
This 
produces 180 unique short separations
rather than 6 copies of the same 30 separations
(figure \ref{fig:s6p6} {\it top right}). 
The entire pattern is
rotated 60 degrees with respect to the 6 positions in table \ref{table:s6positions}.

The second level of the 
hierarchy of separations is made up of all possible separations of the
antennas of different subarrays. These larger scale separations
show the same uniform pattern as
in figure 3, but with patches made of the 30 possible pairs of the 6 subarrays
instead of the 30 UV points from the 6 individual antennas.  
In each of the 30 patches there are 36 points  
from the 36 possible pairs of antennas between 2 subarrays.
Figure \ref{fig:s6p6} ({\it middle right}) shows 
one set of these 36 separations. Each patch has a slightly different
pattern.  Finally, figure \ref{fig:s6p6} ({\it bottom}) shows how ERS drags each of the
30 patches across the UV plane like a wide paint brush across a white canvas.
Because the separations 
of the 6-element subarray are uniformly distributed, 
so are the midpoints of the tracks
of the second level of the hierarchy.

The extension to further hierarchical levels is obvious. The entire two-level array is
repeated on the same basic pattern of the subarray. To keep the figures
from becoming too crowded, the following examples refer to just two levels or three levels.

\section{The flexibility of a 2-level H-array}\label{example36}

The simple 36-element 2-level hierarchy pattern can be scaled to select a point source response
along the continuum of choices in the trade-off between high
angular resolution and low side lobes. First, to design a high angular resolution
beam, scale the second level just large enough to place the 30 patches of
separations side-by-side. Next, to design a beam with low side lobes,
scale the second level so that the patches overlap and create a
distribution that tapers gracefully to the boundary.

\subsection{A high angular resolution beam}

\begin{figure}[t]
$
\begin{array}{cc}
\includegraphics[trim=0.10in 0.3in 0.8in 0.2in, clip, width=2.75in]
{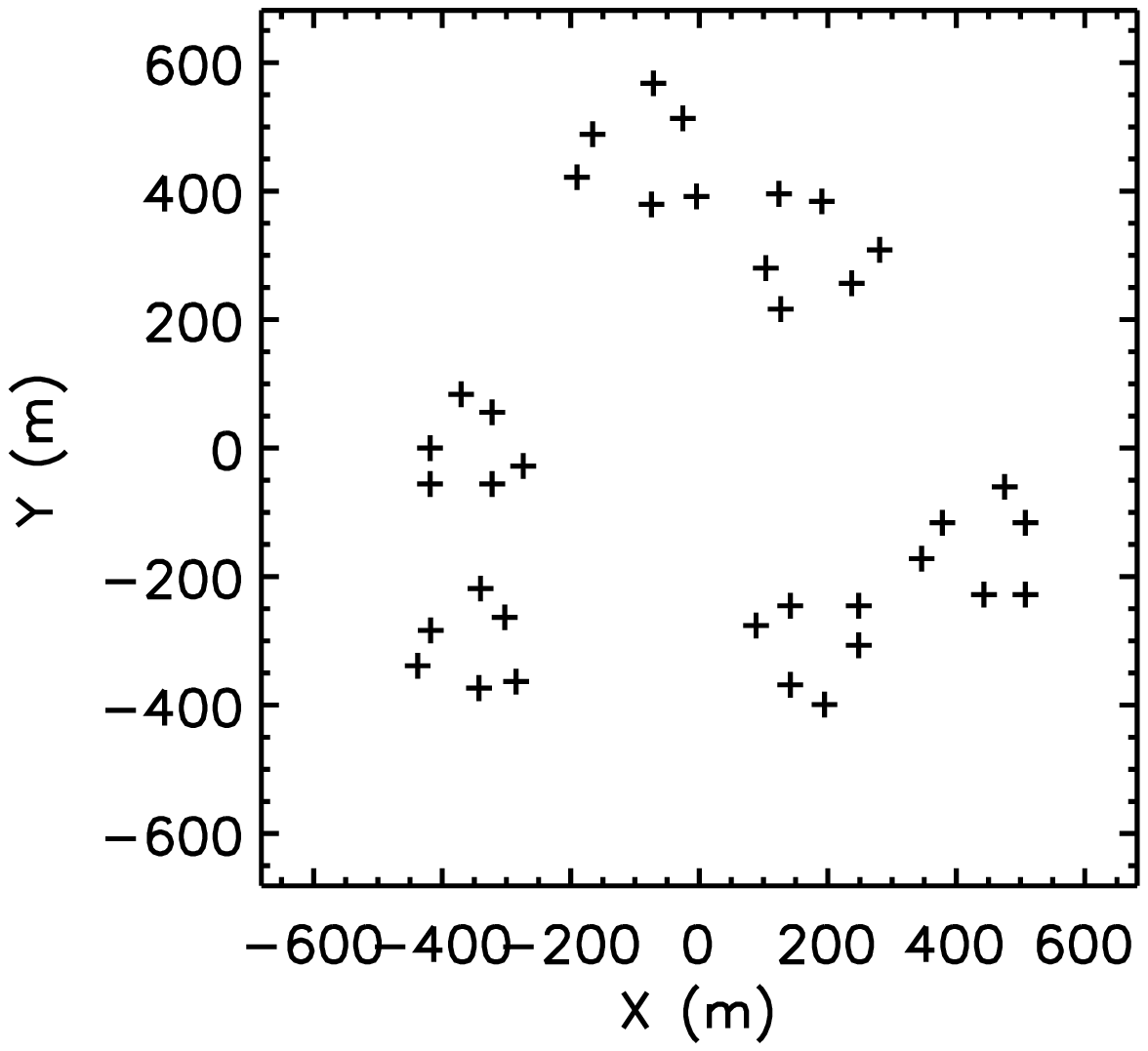}
\includegraphics[trim=0.10in 0.3in 0.8in 0.2in, clip, width=2.75in]
{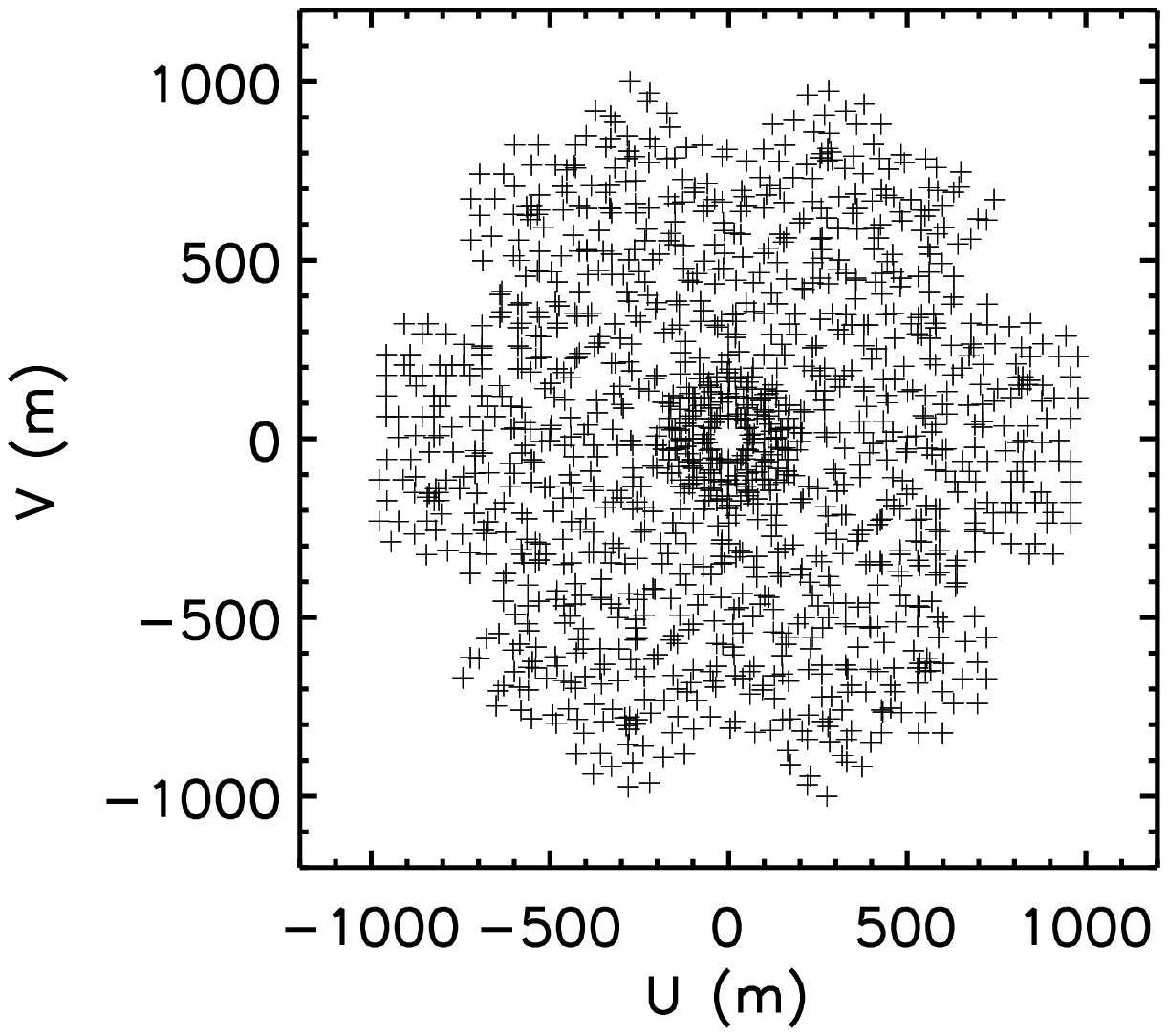}  \\
\hskip 0.65in \includegraphics[trim=0.10in 0.3in 0.8in 0.2in, clip, width=2.75in]
{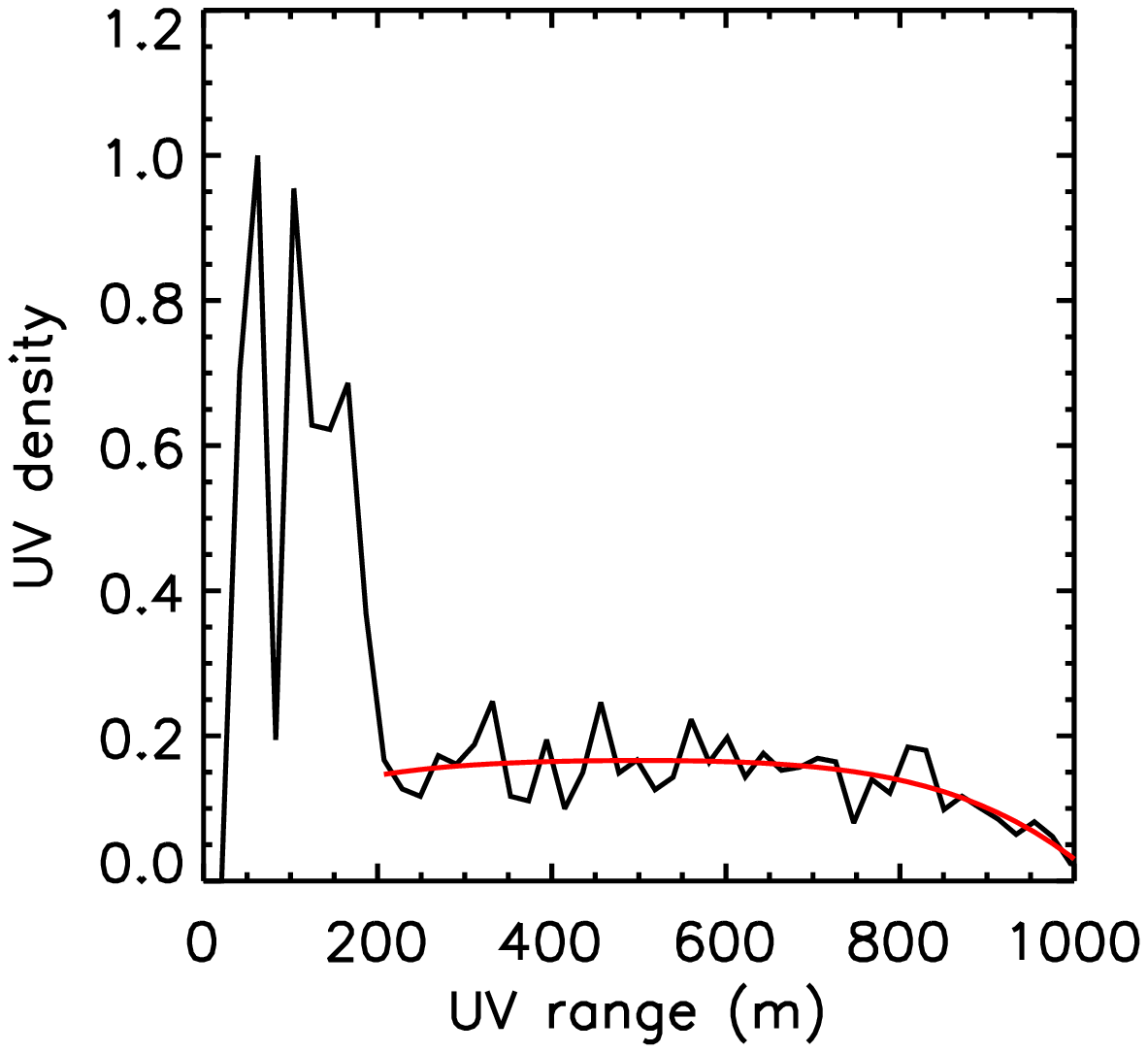}  
\includegraphics[trim=0.10in 0.3in 0.8in 0.2in, clip, width=2.75in]
{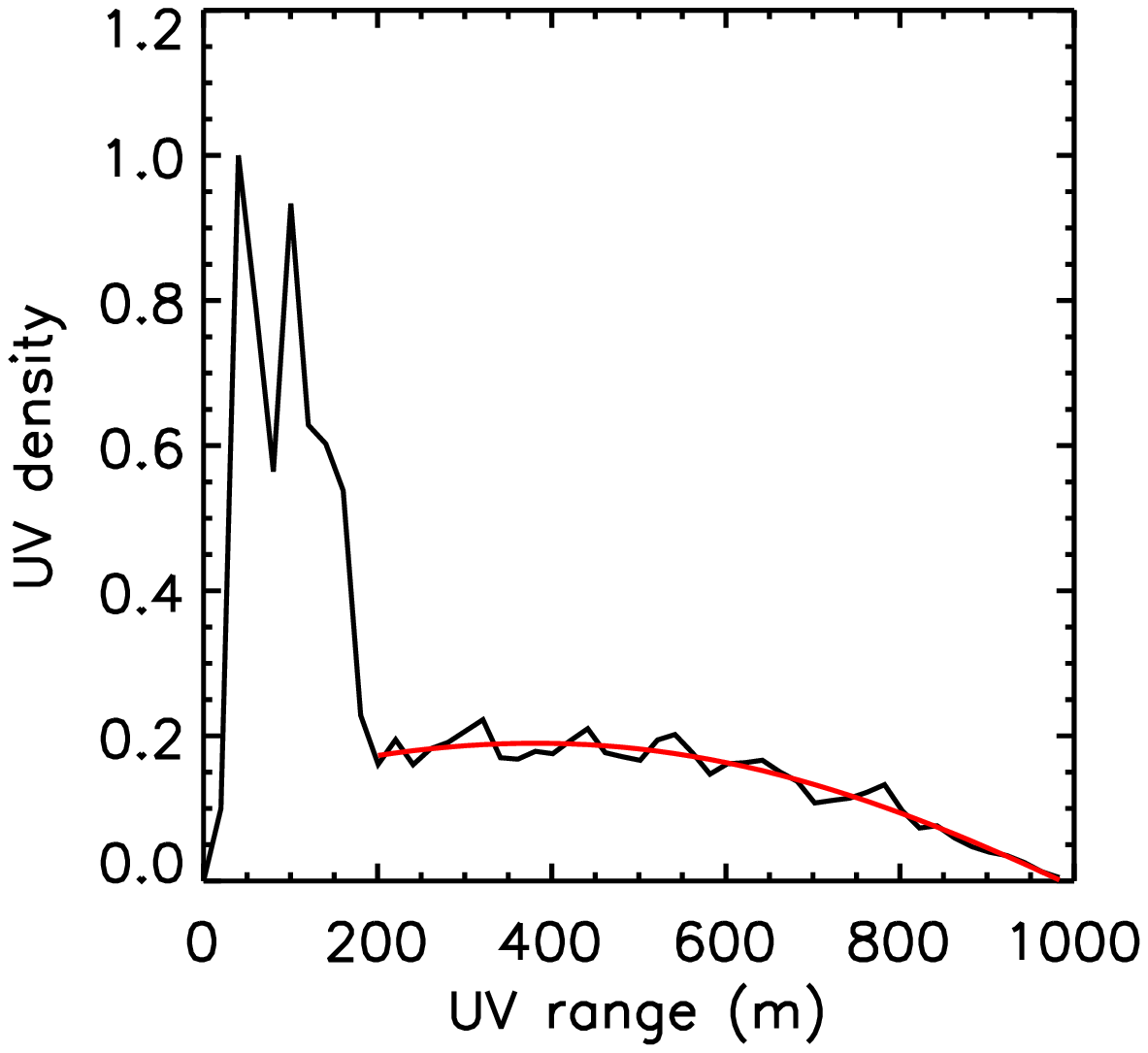}  \\
{\hskip 0.35in \includegraphics[trim=0.10in 0.3in 0.8in 0.2in, clip, width=2.75in]
{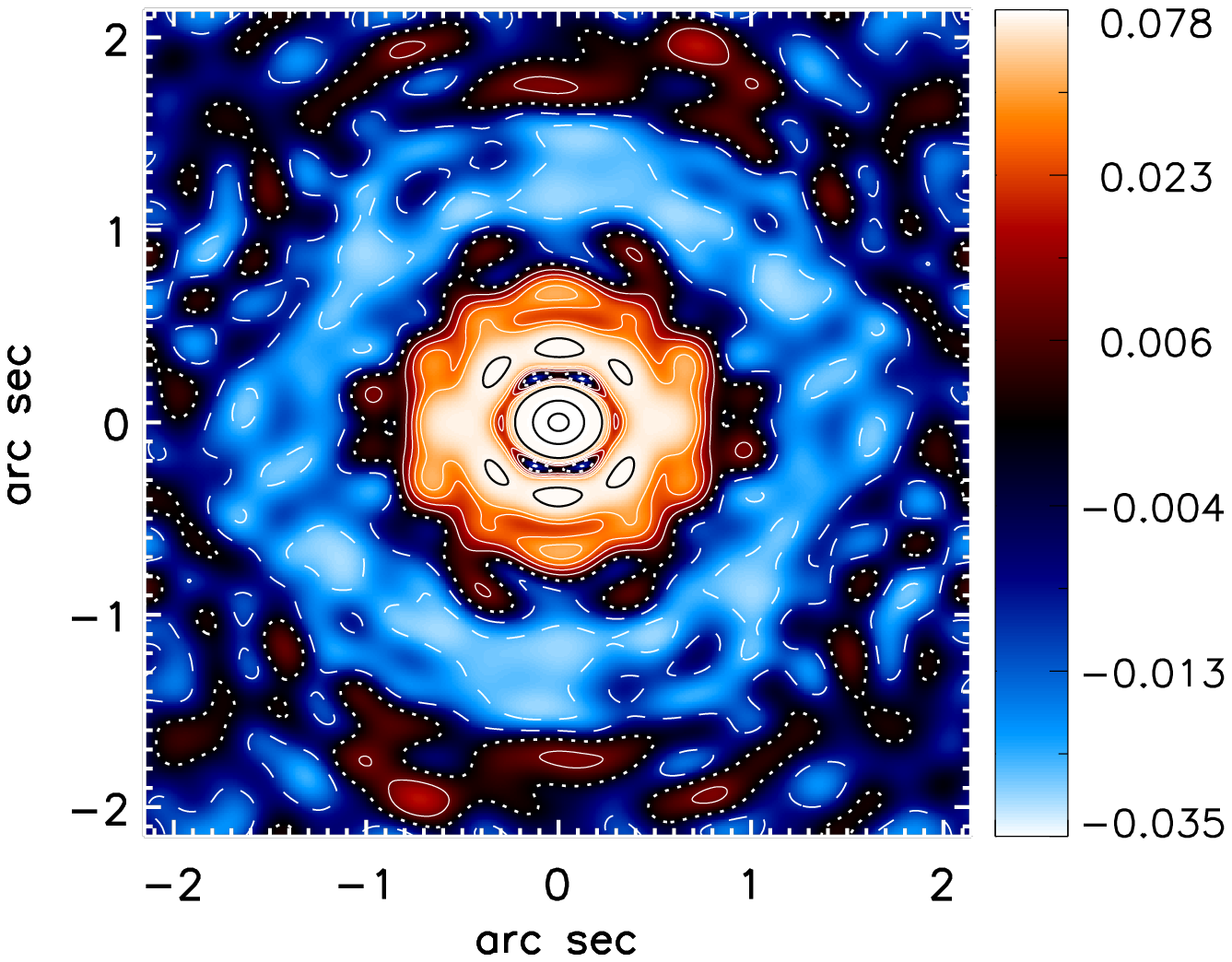}  }
{\hskip -0.30in \includegraphics[trim=0.10in 0.3in 0.8in 0.2in, clip, width=2.75in]
{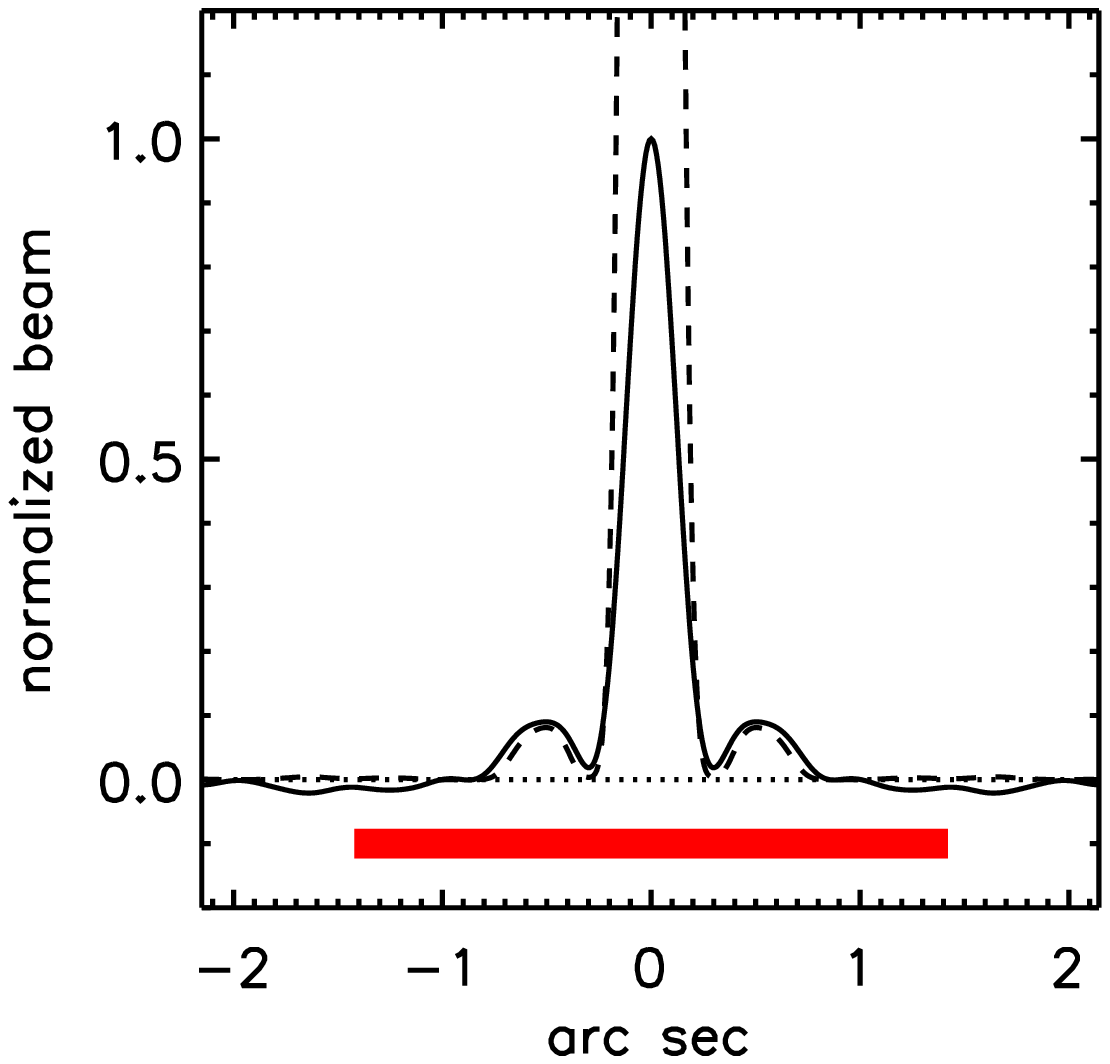}}  \\
\end{array}
$
\caption{A 36-element s6p6 array designed for high angular resolution. 
{\it Top left:} Antenna positions. 
Moving counterclockwise around the figure  
the subarrays are rotated with
respect to the left subarray of the two on the bottom by 20, 60, 30,
100, and 20 degrees.  
The subarrays are also scaled by factors of 1.05 with respect
to each other,  or counterclockwise from the
left bottom, by
1.05,    1.10,    1.16,    1.22, and    1.28. 
{\it Top right:} Separations and UV coverage in snapshot at zenith.
{\it Middle left:} Radial distribution of the density of separations in snapshot at zenith.
The smooth red line shows a polynomial fit to the outer part of the distribution. 
The variance from this fit is a measure of the smoothness of the coverage in the outer zone.
{\it Middle right:} Radial distribution of the density of separations in ERS.
{\it Bottom left:} Beam pattern in ERS.
The black contours are at 10, 50, and 90\% of
the peak. From the peak, the
second black contour shows the FWHM. The white contours are at
1, 2, and 3\% of the peak. The negative contours are dashed. The zero contour is white and dotted.
{\it Bottom right:} Beam and power pattern in the same format as figure \ref{fig:FTpair}. The
beam power (dashed line) is multiplied by 10.
The figures of merit are listed in table \ref{table:merit}.
}
\label{fig:ska36_sinc_revB}
\end{figure}

Scaling the second level by a factor of 5.5 relative to the first
(figure \ref{fig:ska36_sinc_revB} {\it top left}) produces a tiered
distribution of separations with a nearly uniform density in two
regions. Figure \ref{fig:ska36_sinc_revB} ({\it top right}) shows 
the two-dimensional pattern of separations and the UV coverage 
obtained in snapshot imaging at zenith. 

Figure
\ref{fig:ska36_sinc_revB} ({\it middle left}) shows the corresponding
density of separations
as a function of radius averaged over angle. In ERS, the 
elliptical arcs across the UV plane 
improve the UV
coverage, particularly as a function of angle.
Because the UV points of the shorter separations move more slowly
across the UV plane,
ERS also affects the radial
distribution by increasing the density of the shorter separations.
Nonetheless, the radial distribution in ERS still maintains its essential character
(figure \ref{fig:ska36_sinc_revB} {\it middle right}).
The smooth red line shows the fit of a third order polynomial to the
distribution in the outer zone. In snapshot the reduced $\chi ^2$ is 0.00094.
The better UV coverage in ERS fills in gaps and improves
the smoothness to 0.00017.

The two-dimensional beam pattern in ERS 
(figure \ref{fig:ska36_sinc_revB} {\it bottom})
shows the high side lobes characteristic of arrays 
with uniform UV coverage, but the angular resolution is exquisite,
0.17 arc seconds FWHM. The radius encircling 98\% of the beam power
is 1.38 arc seconds and the K-product, $K_{98}=1378$
(figure \ref{fig:ska36_sinc_revB} {\it bottom right}).
The figures of merit are listed in table \ref{table:merit}.

\subsection{A beam with low side lobes}

\begin{figure}[t]
$
\begin{array}{cc}
\includegraphics[trim=0.10in 0.3in 0.8in 0.2in, clip, width=2.75in]
{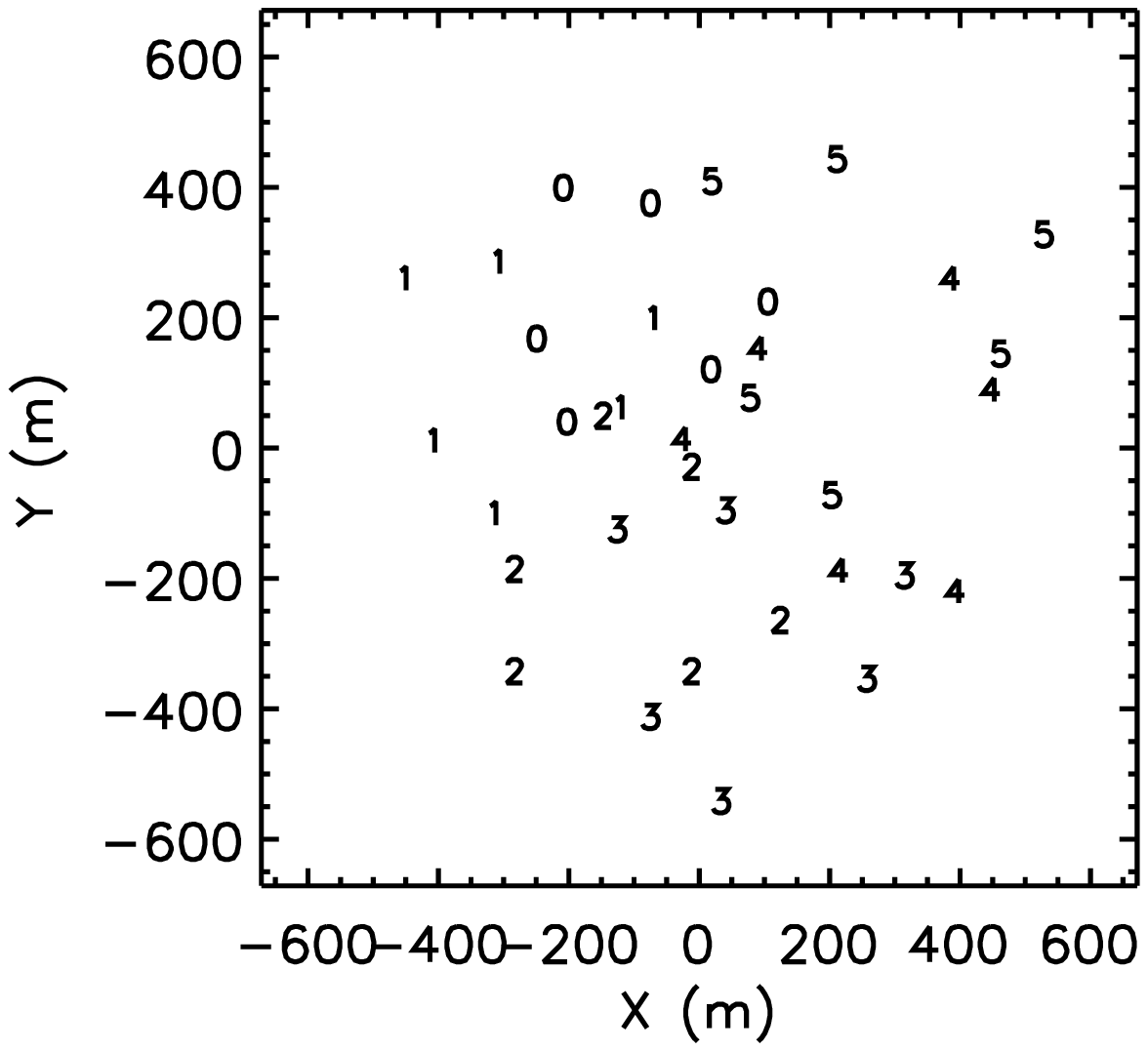}
\includegraphics[trim=0.10in 0.3in 0.8in 0.2in, clip, width=2.75in]
{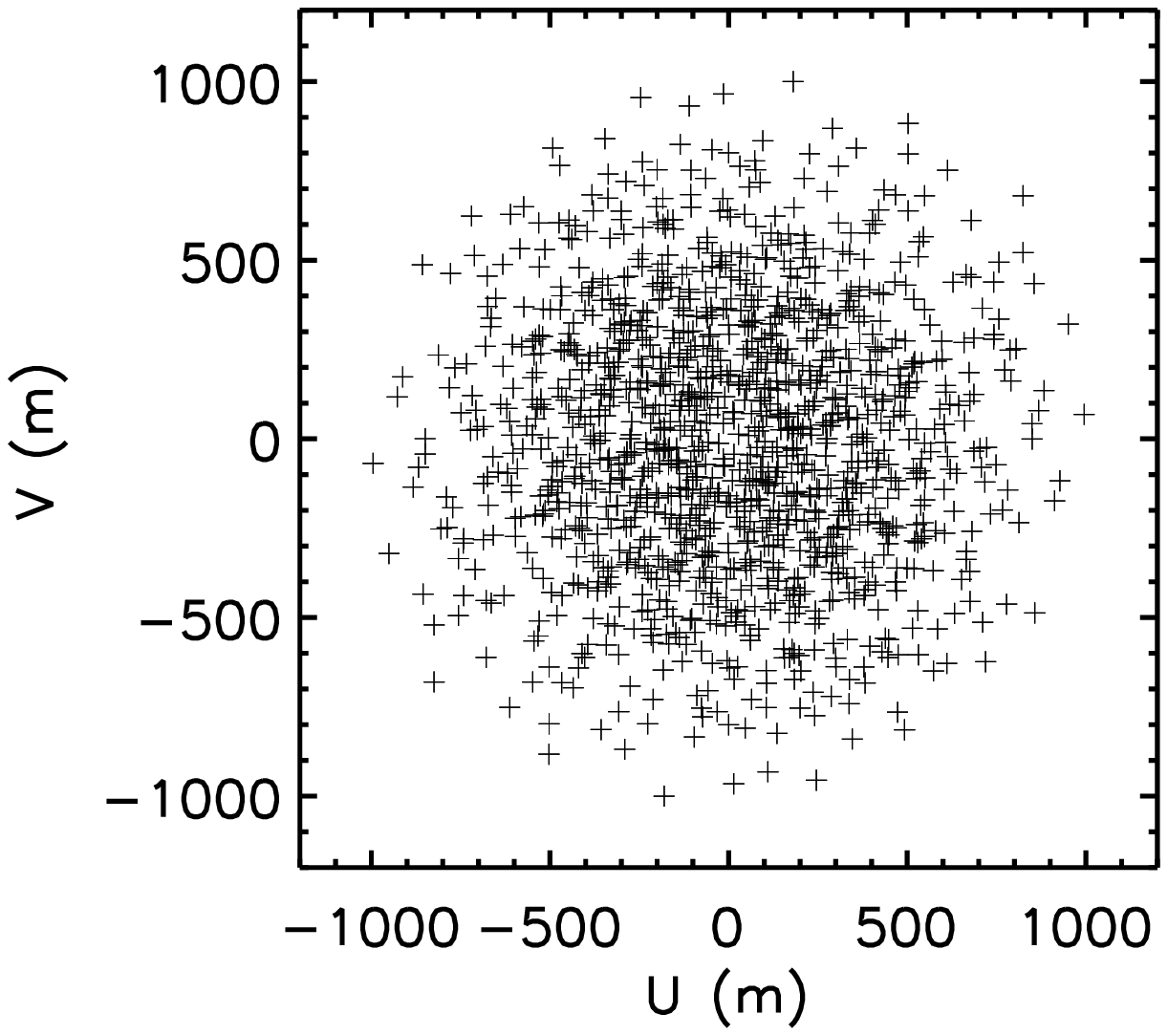}  \\
\hskip 0.65in \includegraphics[trim=0.10in 0.3in 0.8in 0.2in, clip, width=2.75in]
{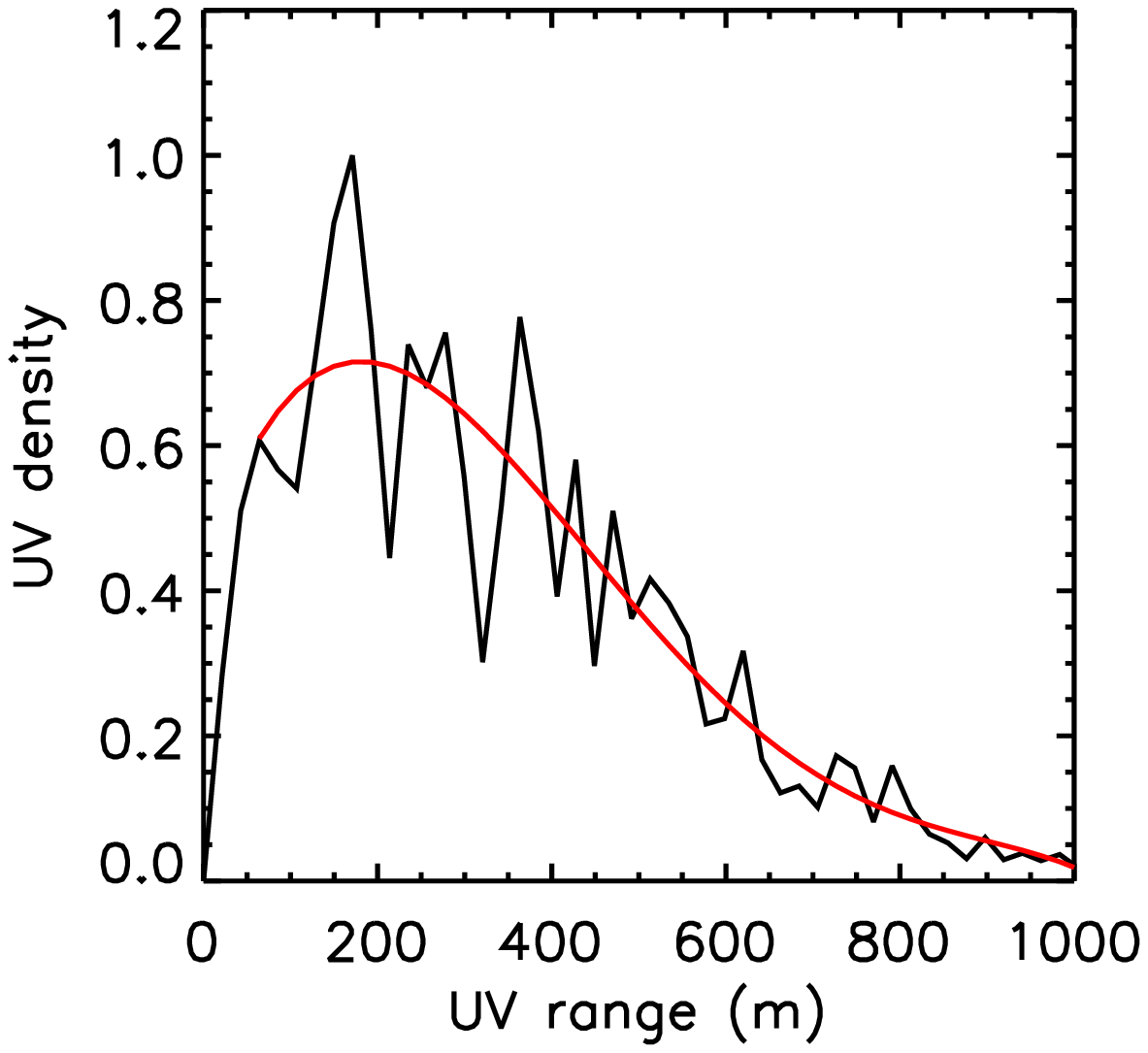}  
\includegraphics[trim=0.10in 0.3in 0.8in 0.2in, clip, width=2.75in]
{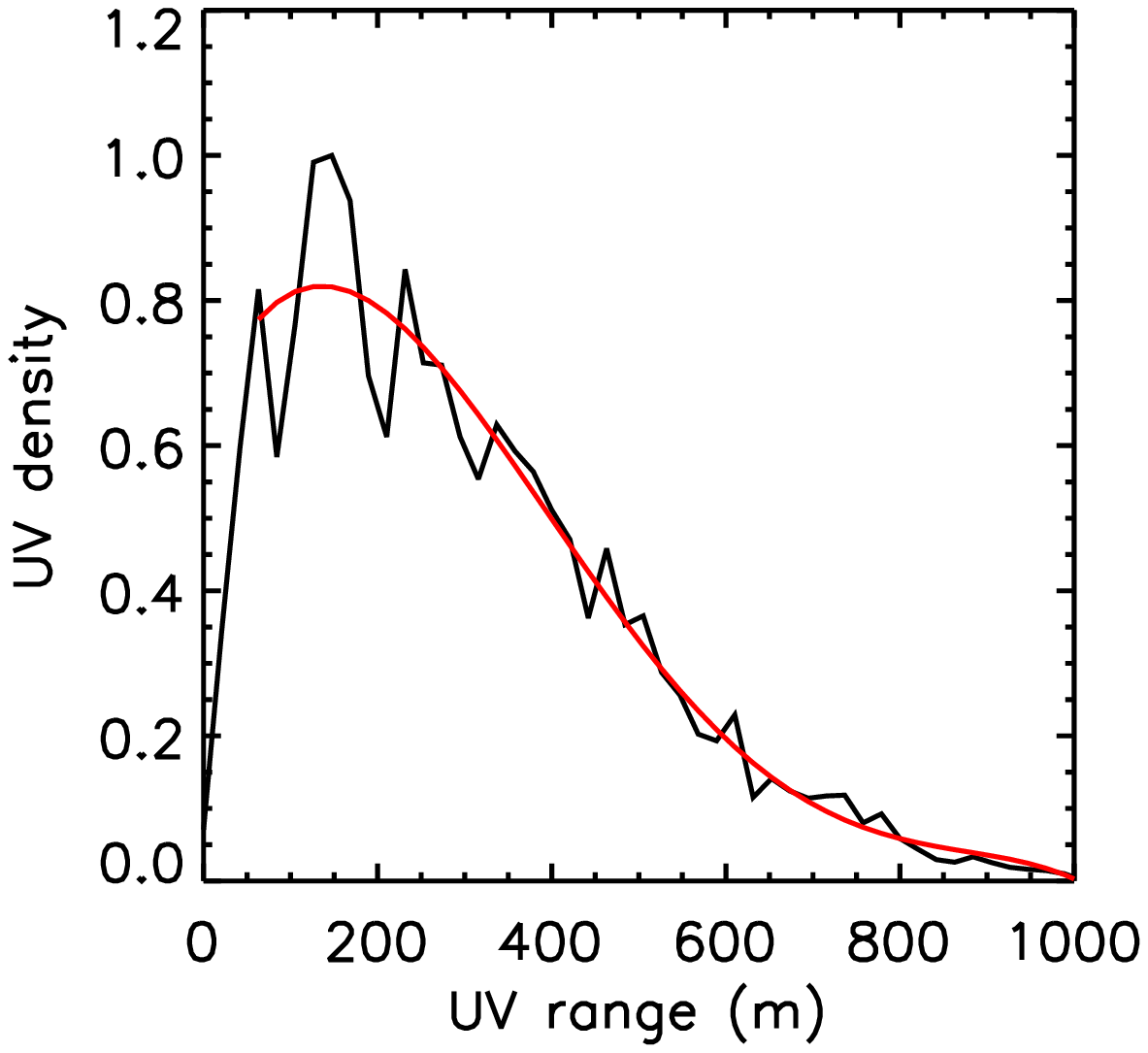}  \\
\hskip 0.35in \includegraphics[trim=0.10in 0.3in 0.8in 0.2in, clip, width=2.75in]
{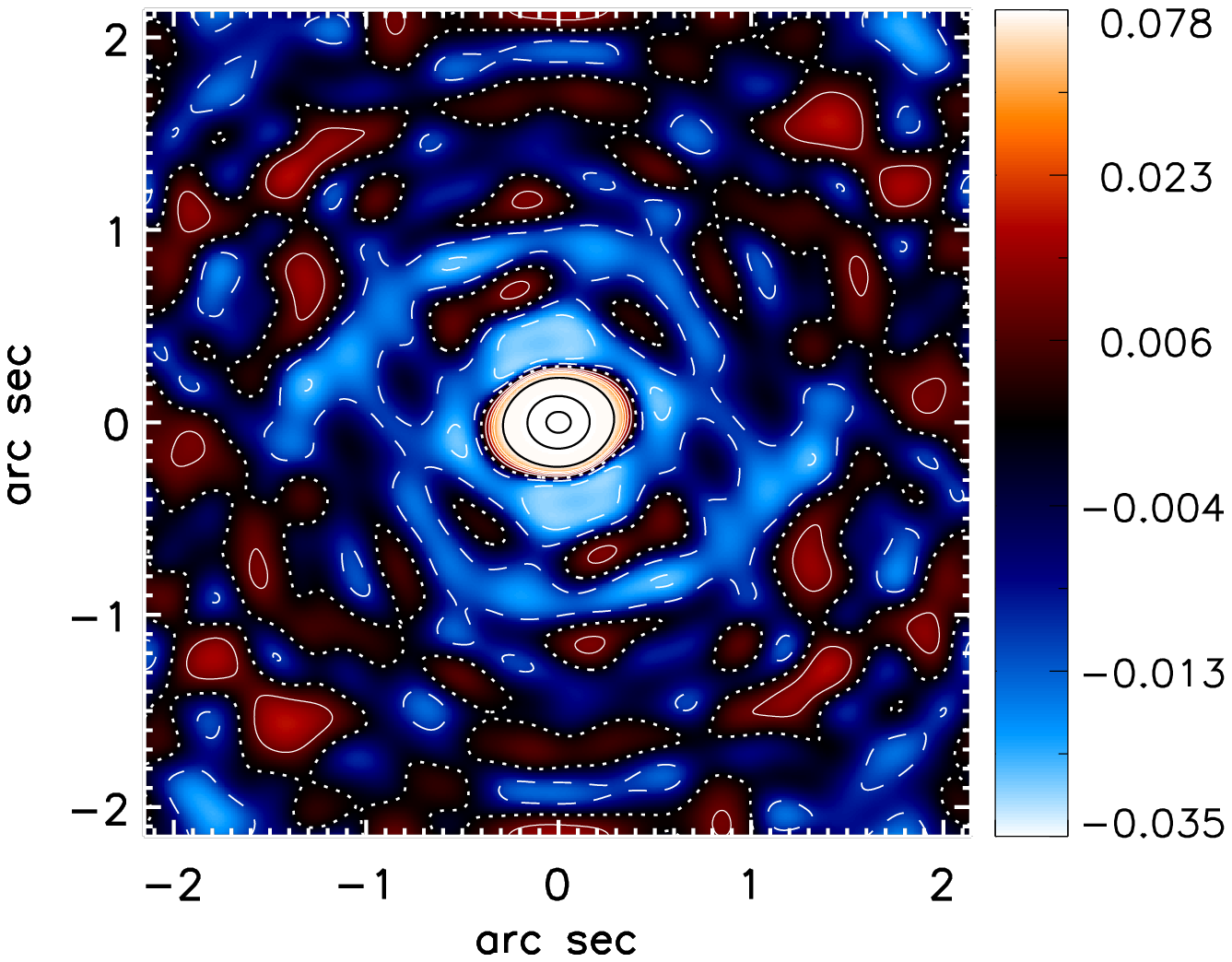}  
\hskip -0.30in \includegraphics[trim=0.10in 0.3in 0.8in 0.2in, clip, width=2.75in]
{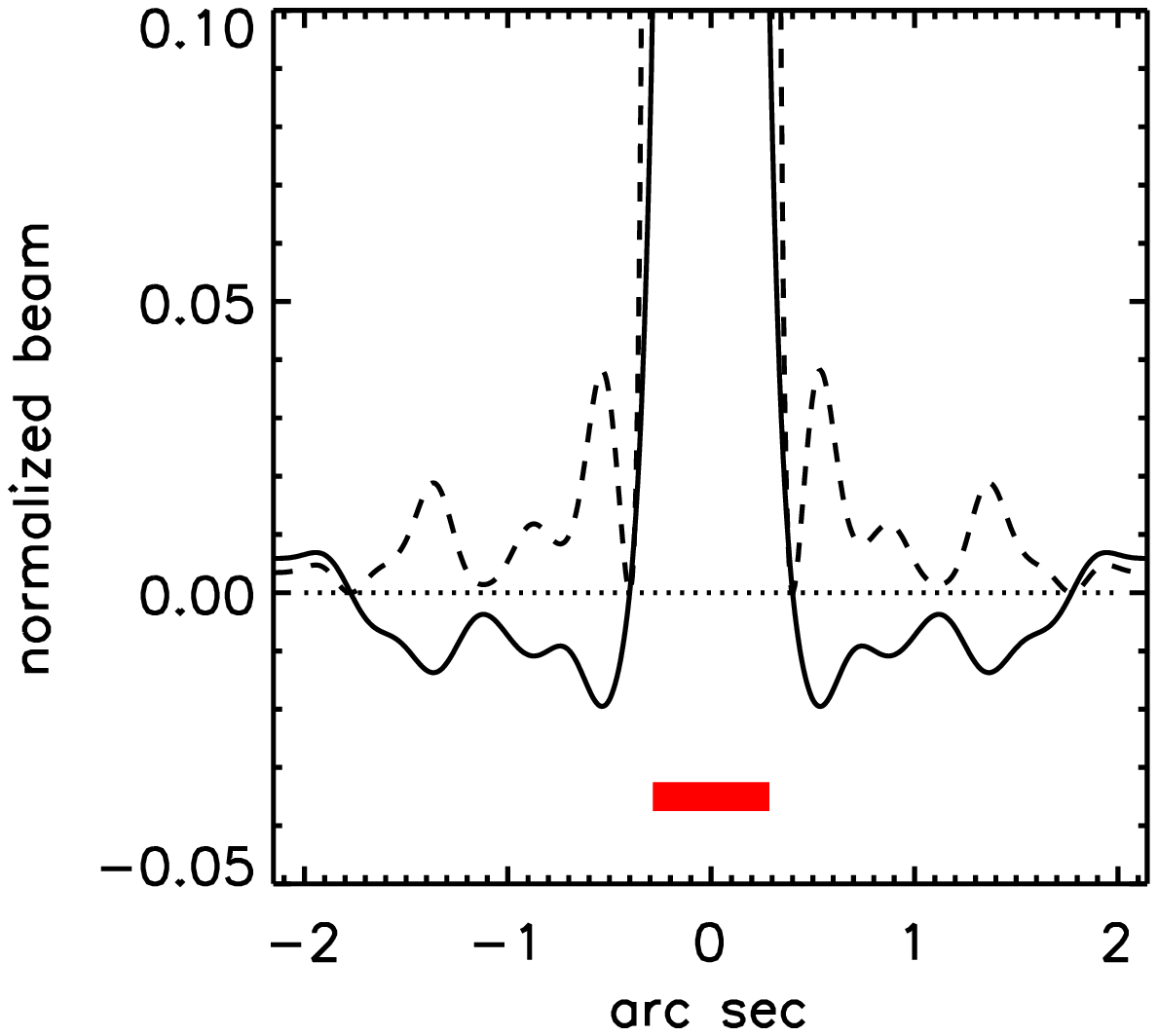}  \\
\end{array}
$
\caption{A 36-element s6p6 array designed for a high concentration of 
power in the main beam and low side lobes. 
{\it Top left:} Antenna positions. 
The 6 subarrays overlap and each is 
drawn with a different number.  The 6 positions marked with the
number 0 belong to the same subarray.
Moving counterclockwise
around the figure,
relative to the subarray marked by zeroes,
the subarrays
are rotated by 40, 20, 60, 20, 100, and 20 degrees.  
The subarrays are also scaled by factors of 1.075 with
respect to each other,
or counterclockwise from the upper left by
1.075,   1.163,   1.24,   1.34,   1.44.
The entire pattern
is rotated 30 degrees with respect to the locations in table
\ref{table:s6positions}.
Same format
as figure \ref{fig:ska36_sinc_revB} except for the panel in the lower
right. Here the solid line shows the bottom of the beam which is normalized to one.
The dashed line shows the beam power multiplied by 100. The scale then shows
the power as a percentage of the peak power.
The figures of merit are listed in table \ref{table:merit}.
}
\label{fig:ska36fix}
\end{figure}

To reduce the side lobes, reduce the scaling of the second level
of the hierarchy 
to pull the 30 patches of separations inward until they are overlapping
and create a smoothly tapered distribution. Figure  \ref{fig:ska36fix} 
shows the same plots as in the previous example but for an array with
the second level only 1.5 times larger than the first. 
The radial distribution of the separation density is now shifted to
shorter separations and gradually tapers to the maximum. The beam
has lower angular resolution, 0.21 arc seconds, but also lower
side lobes and more concentrated beam power. The radius encircling
98\% of the energy is 0.29 arc seconds, and the K-product is
$K_{98} = 285$. 
This beam has better angular resolution than the 0.35 arc sec
Gaussian beam of our idealized example discussed in the introduction,
and the smaller $K$-product indicates that
the side lobes of this beam contain less energy than the
extended wings of the Gaussian beams of either the truncated or idealized 
Gaussian apertures.

The concentrated UV distribution leaves fewer points to cover the
outer zone resulting in wider gaps than in the more uniform distribution
of the array in figure \ref{fig:ska36_sinc_revB}. The $\chi ^2$ figure of
merit measures the 
local smoothness and is accordingly larger even though the distribution
is globally tapered to the boundary. 
For snapshot and ERS, $\chi ^2$ is 0.0098 and 0.0041, respectively.

\subsection{Design principles}

These examples show how H-arrays allow control over different scales
in the UV plane. A change in the first-level subarrays affects the
UV distribution of the smallest separations in the center of the 
UV plane and also the distribution inside the 30 patches or
brush strokes in ERS of the
second level but does not much affect the distribution of these
patches across the UV plane. This is controlled by
the placement of the subarrays in the second level. 
In this way we have simple and understandable control over
the aperture distribution.
We can shape the aperture distribution by scaling
the levels to emphasize 
Fourier components of different scales.

The conceptual separation of scales possible with H-arrays
can be thought of as accomplishing more than one
observation at the same time. Arrays for astronomical imaging
are often built with multiple configurations of different
sizes to measure Fourier components on different angular
scales. The configurations are nested so that the separations
of each configuration are just larger than the next smaller
size, usually with some overlap. The full range of Fourier
components is measured in multiple observations by moving the
antennas to each of the configurations between observations.
The multiple scales of the H-arrays accomplish this in one
observation.

\section{Sparse H-arrays}  \label{sparse}

Is the uniform pattern of separations provided by the 
6-element array of figure \ref{fig:s6} essential? 
What if the number of available antennas is not a power of six?
Two examples with a minimal subarray of 3 elements
show that it is possible 
to obtain good coverage with sparse arrays built 
with a subset (half)
of the basic 6-element subarray. 
The lower density of UV coverage leaves some gaps,
but there is a positive trade-off as well.
Sparse
arrays can achieve a larger dynamic range, the difference between 
the smallest and largest separations,
with fewer antennas.

\subsection{A sparse array for high angular resolution}

\begin{figure}[t]
$
\begin{array}{cc}
\includegraphics[trim=0.10in 0.3in 0.8in 0.2in, clip, width=2.75in]{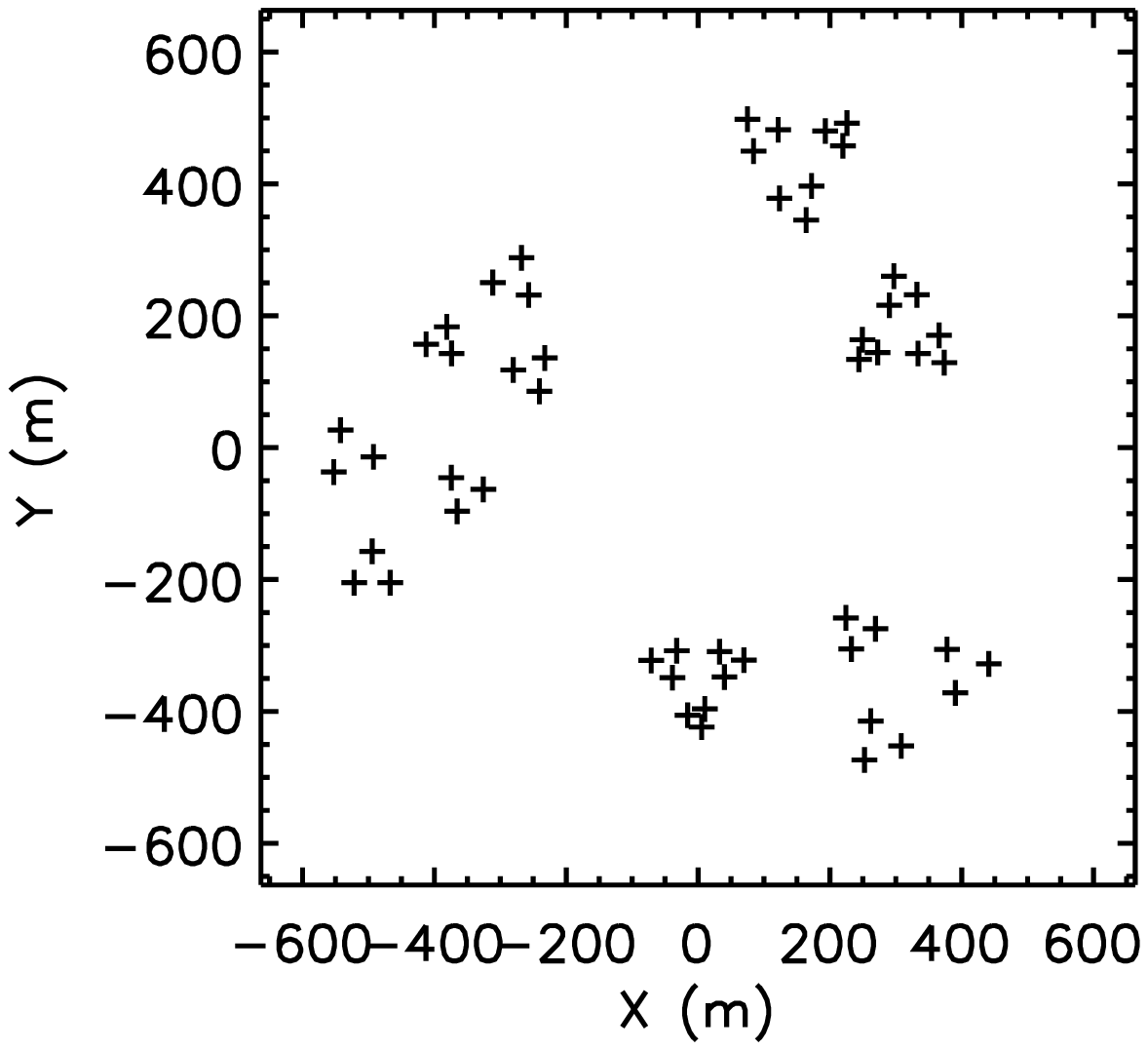} 
\includegraphics[trim=0.10in 0.3in 0.8in 0.2in, clip, width=2.75in]{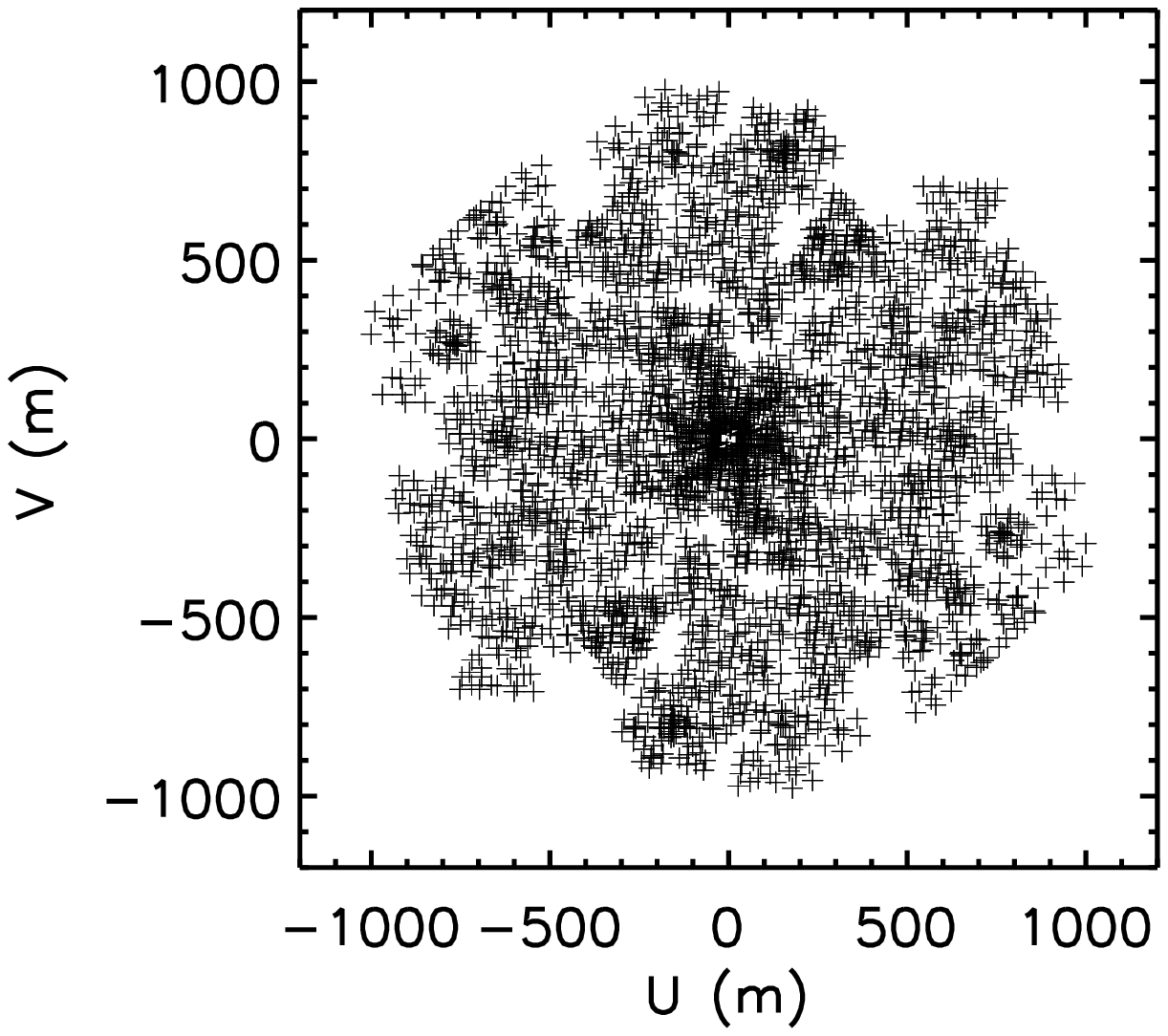} \\
\hskip 0.65in \includegraphics[trim=0.10in 0.3in 0.8in 0.2in, clip, width=2.75in]
{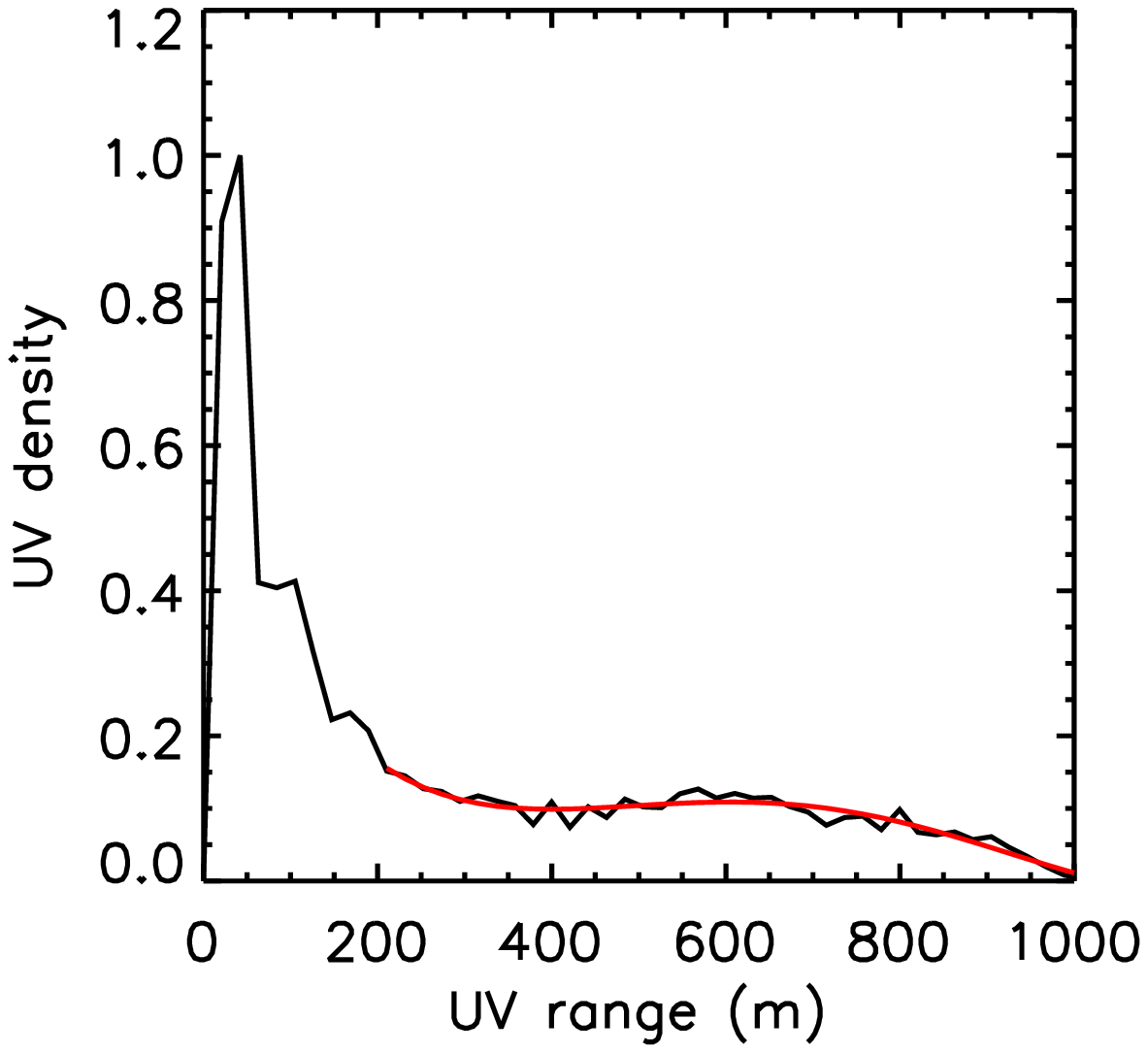} 
\includegraphics[trim=0.10in 0.3in 0.8in 0.2in, clip, width=2.75in]{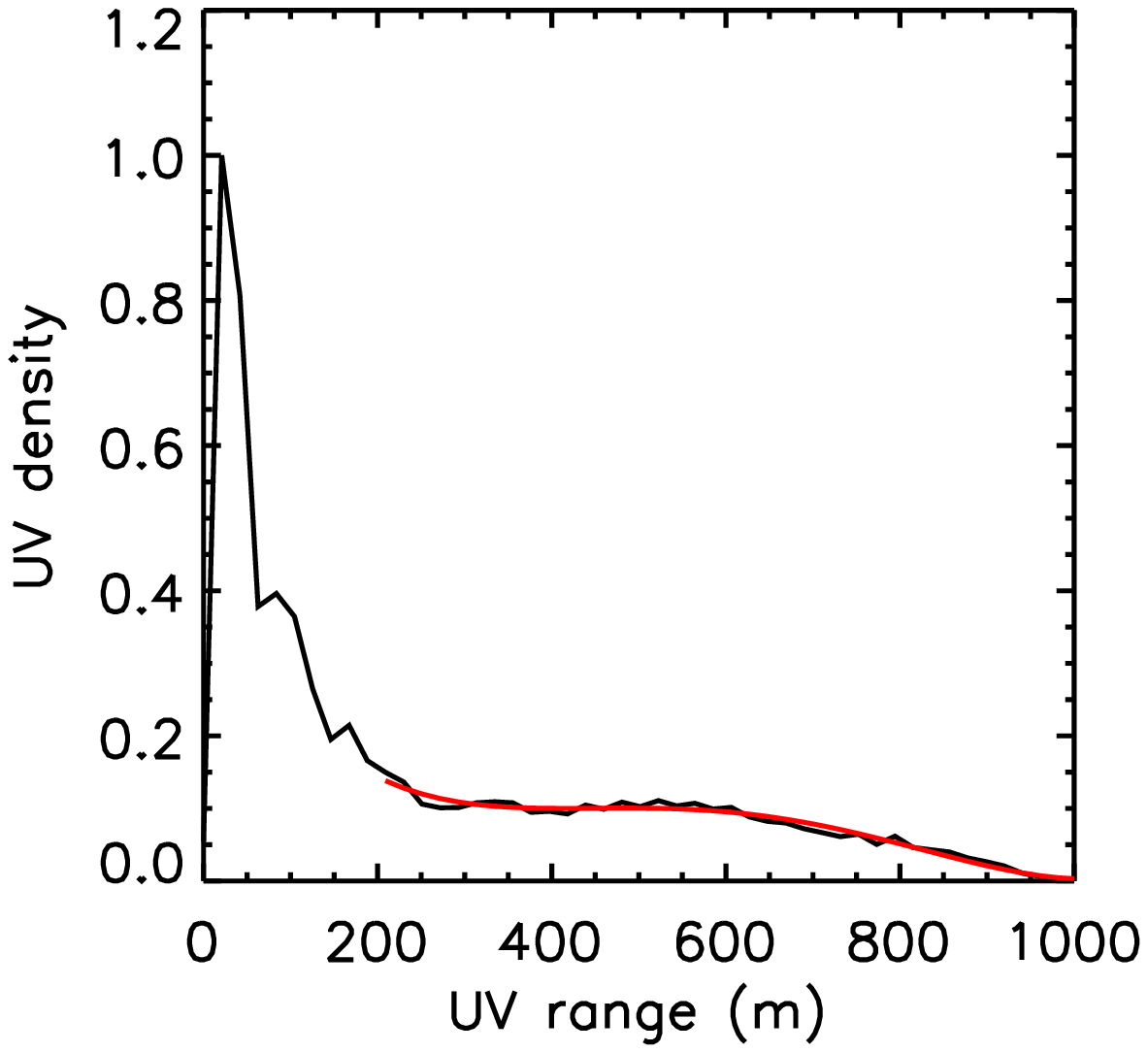} \\
\hskip 0.35in \includegraphics[trim=0.10in 0.3in 0.8in 0.2in, clip, width=2.75in]
{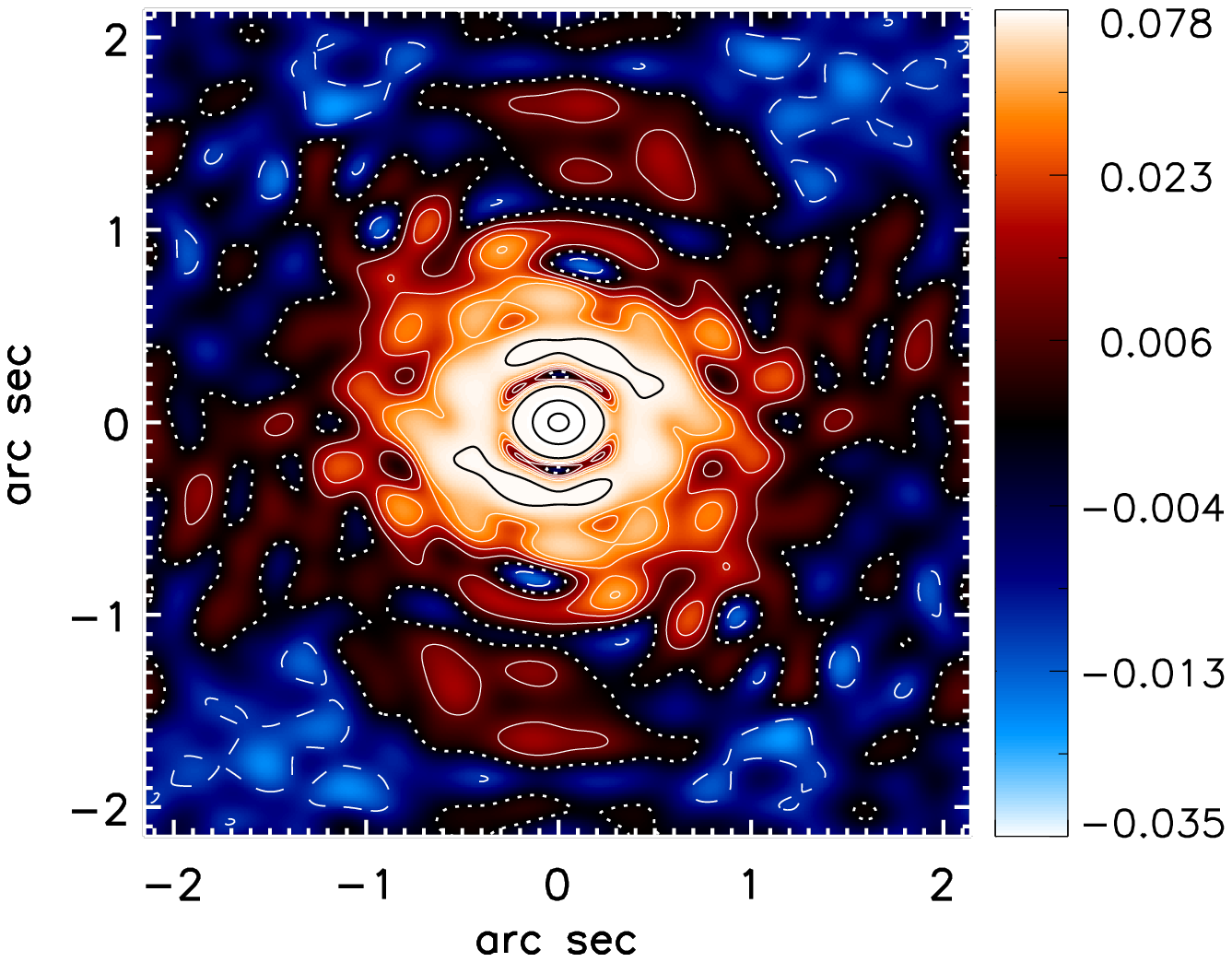} 
\hskip -0.30in \includegraphics[trim=0.10in 0.3in 0.8in 0.2in, clip, width=2.75in]
{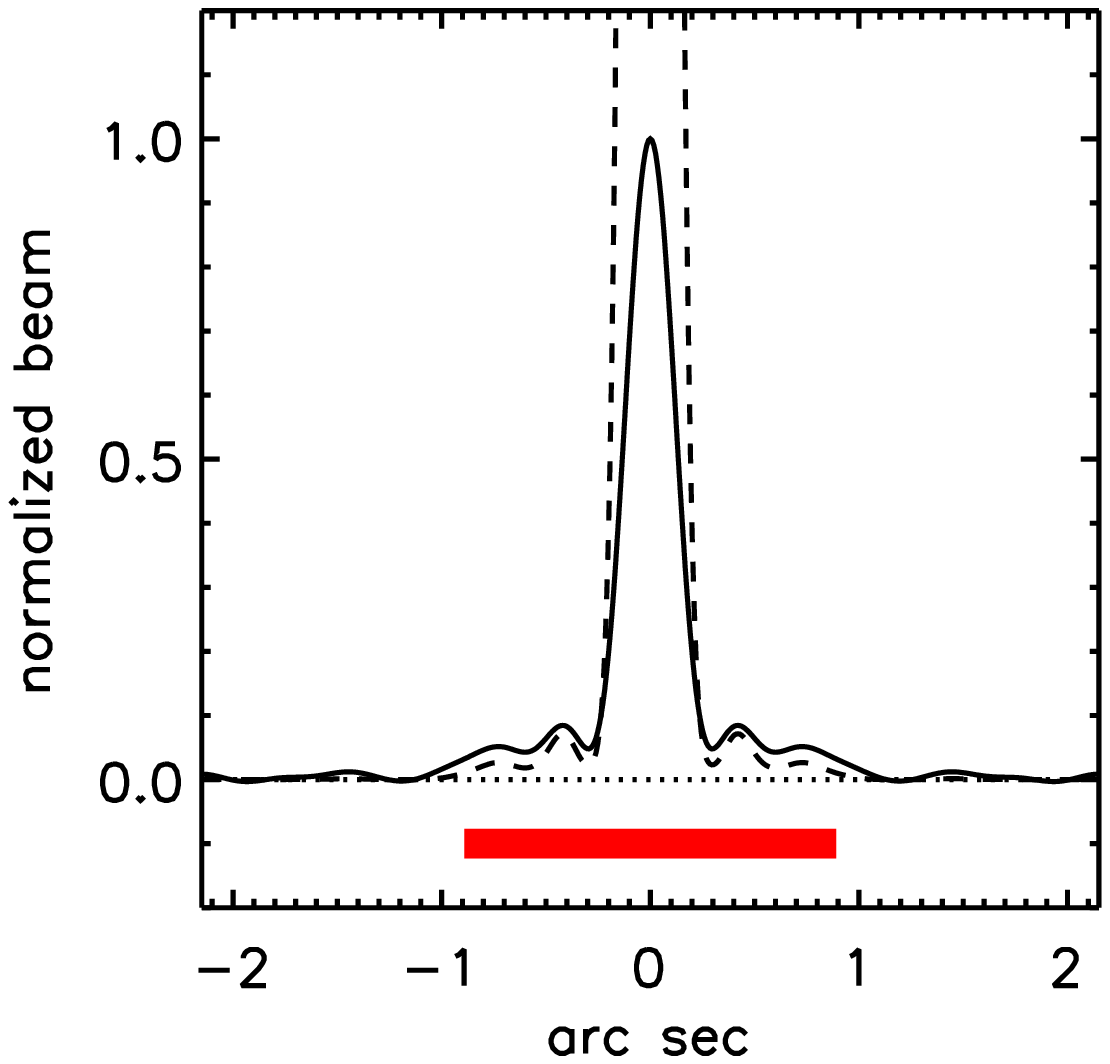} \\
\end{array}
$
\caption{{\it Top left:} Antenna locations, UV distribution, and beam for an
example configuration of 54 antennas weighted toward longer baselines. 
Same format as figure \ref{fig:ska36_sinc_revB}. The rotations and scalings of
the subarrays are listed in table \ref{table:ska216_details} The beam power (dashed
line) in the lower right panel is multiplied by 10.
The figures of merit are listed in table \ref{table:merit}.
}
\label{fig:ska54_flat}
\end{figure}

As with our earlier example, a larger scaling between 
the hierarchical levels can be used to create approximately uniform
UV coverage in separate zones. Figure \ref{fig:ska54_flat} 
shows an array with 3 hierarchical levels. The
first two levels 
use 3-element subarrays made by dropping 
every other element of the basic 6-element array.
The third level uses the full 6-element pattern. 
This array has
54 antennas in an s3p3d6 configuration (figure \ref{fig:ska54_flat}).
The second and third levels are scaled by 3.4 and 22.0
times the basic subarray. The subarrays within the two levels
are scaled by 6\% and 8\% to eliminate redundancy.  
The details describing the rotation and placement of
the subarrays are listed in table \ref{table:ska216_details}.
With a maximum separation of 1000m, 
the angular resolution is 0.18 arc seconds at 230 GHz. 
The $K$-product is $K_{98} = 891$
indicating that 98\% of the encircled energy is within a radius of 0.89 arc seconds.
In snapshot imaging, the coverage of this s3p3d6 sparse array has 
less uniform coverage than the 2-level s6p6 array but still maintains
the character of uniform coverage tiered in separate zones.
Figure \ref{fig:ska54_flat} shows the 
separations (snapshot coverage at zenith), the radial
density of the separations in snapshot and ERS, and the beam and its trace in ERS. 
With more UV points than the 36-antenna s6p6 array, the $\chi ^2$ measure of
local smoothness is lower, 0.00027 and 0.00020 in both snapshot and ERS, respectively.
The figures of merit are listed in table \ref{table:merit}.

\begin{table}[ht]
\caption{Subarray rotations and scalings for figures \ref{fig:ska54_flat},
\ref{fig:ska54-n}
}
\vskip 0.1in
\centering 			
\begin{tabular}{ c c c} 		
\hline\hline 					
& \\
Subarray & Rotation  & Exponent of the\\ [0.5ex] 
& & scaling factor\\
\hline							
& \\
s0~\footnotemark[1]	& 100 	& 0	\\
s1	& 80 	& 4	\\
s2	& 60 	& 1	\\
s3	& 20 	& 3	\\
s4	& 20 	& 2	\\
s5	& 0 	& 5	\\
p0~\footnotemark[2]	& 0 	& 0	\\
p1	& 180 	& 4	\\
p2	& 60 	& 1	\\
p3	& 240 	& 3	\\
p4	& 120 	& 2	\\
p5	& 300 	& 5	\\
& \\
\hline							 
\end{tabular}
\begin{tablenotes}
\item[1]
{\small $^1$ The s-level subarrays for the s3p3d6 configuration for high angular resolution shown in figure \ref{fig:ska54_flat} 
are scaled by 1.06 raised to the power shown in the table. For the s3p3d6 and s6p6d6
arrays shown in figures \ref{fig:ska54-n}
the scaling factor is 1.03.
} 
\item[2]
{\small $^2$ The p-level subarrays for the s3p3d6 configuration shown in figure \ref{fig:ska54_flat}    
are scaled by powers of 1.08. For the s3p3d6 and s6p6d6 
arrays shown in figures \ref{fig:ska54-n}
the scaling factor is 1.03.
}
\end{tablenotes}
\label{table:ska216_details}			 
\end{table}

\subsection{A sparse array with a concentrated beam}

\begin{figure}[t]
$
\begin{array}{cc}
\includegraphics[trim=0.10in 0.3in 0.8in 0.2in, clip, width=2.75in]{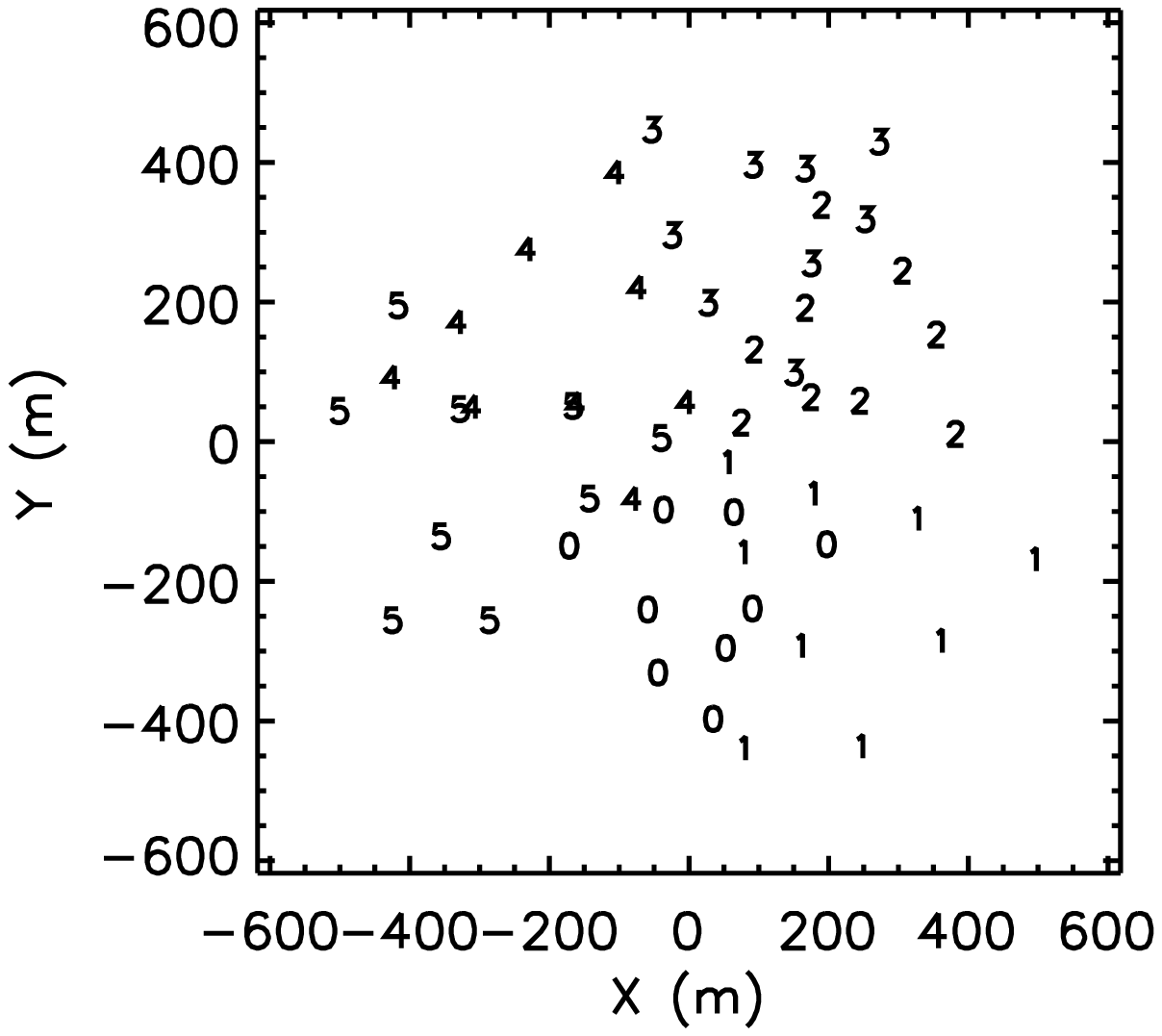} 
\includegraphics[trim=0.10in 0.3in 0.8in 0.2in, clip, width=2.75in]
{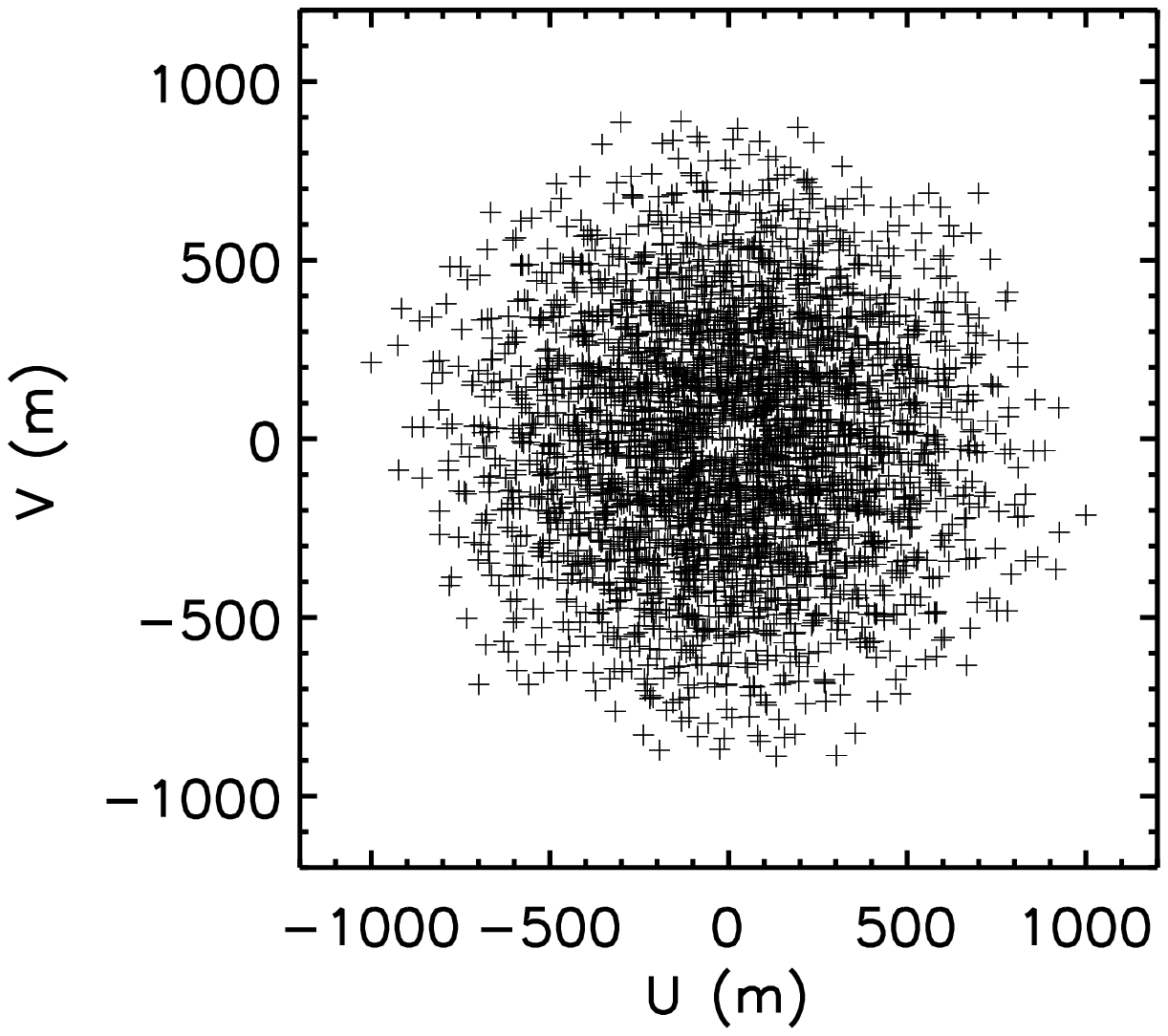}  \\
\hskip 0.65in \includegraphics[trim=0.10in 0.3in 0.8in 0.2in, clip, width=2.75in]
{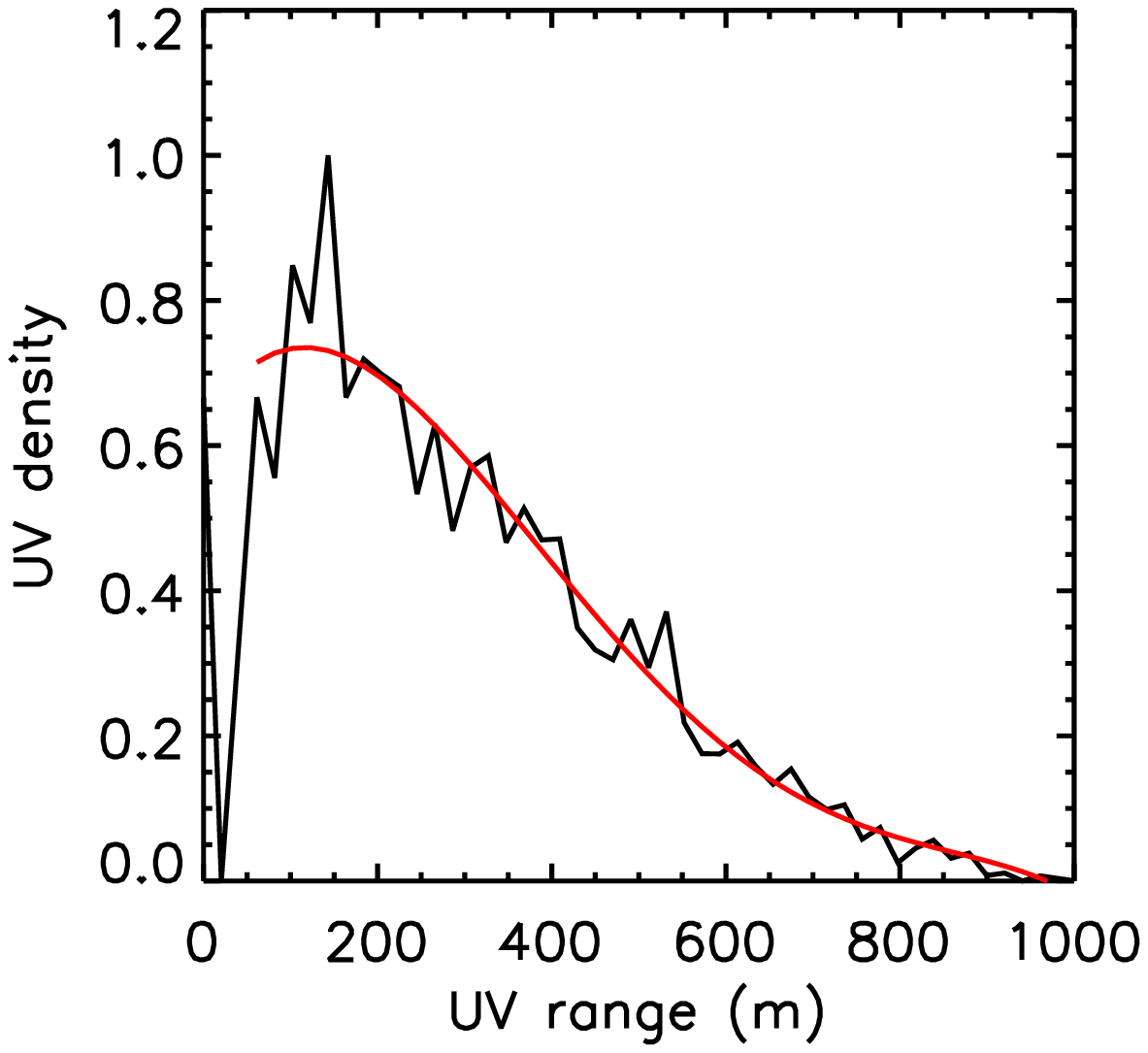} 
\includegraphics[trim=0.10in 0.3in 0.8in 0.2in, clip, width=2.75in]{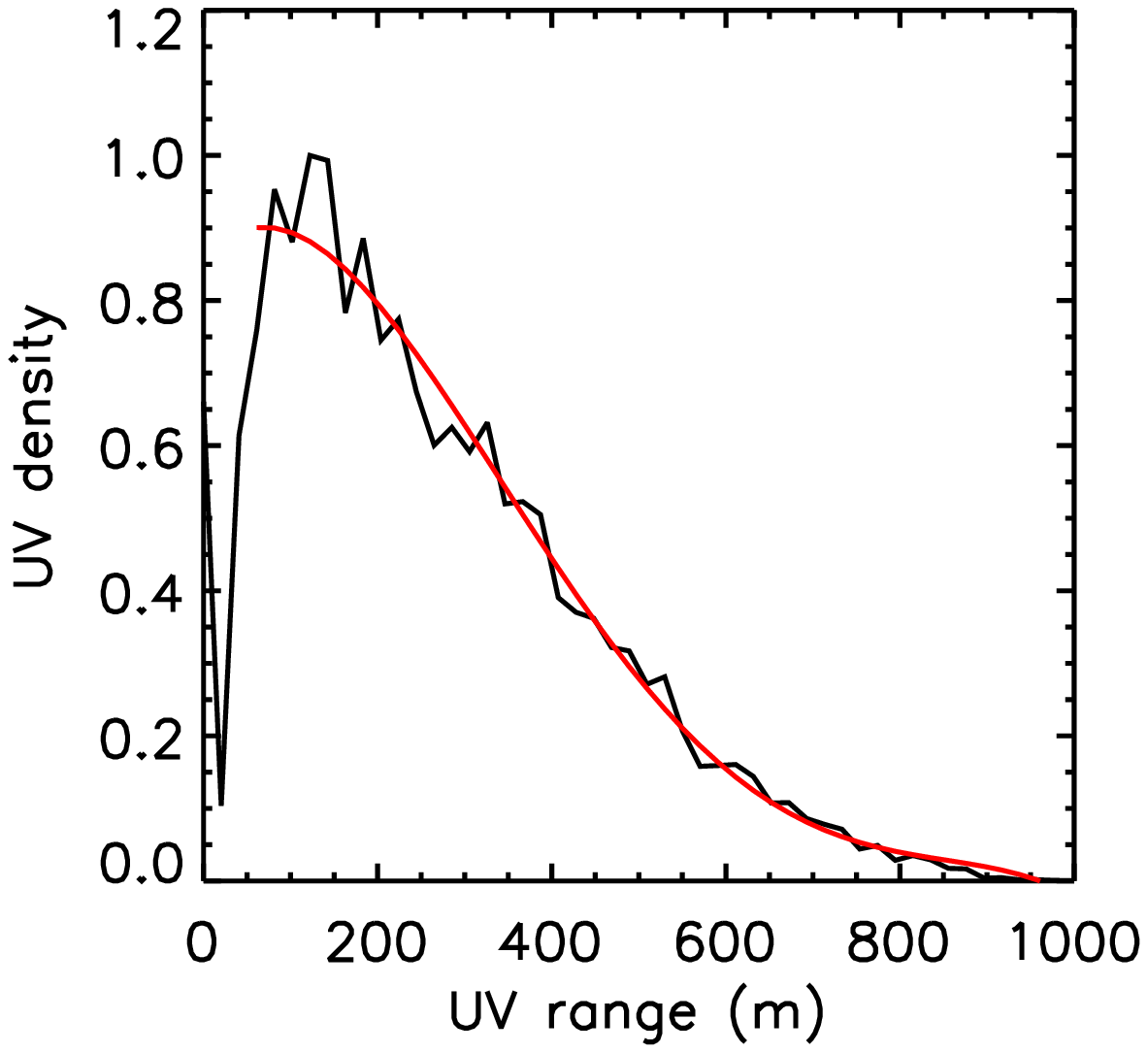}  \\
\hskip 0.35in \includegraphics[trim=0.10in 0.3in 0.8in 0.2in, clip, width=2.75in]
{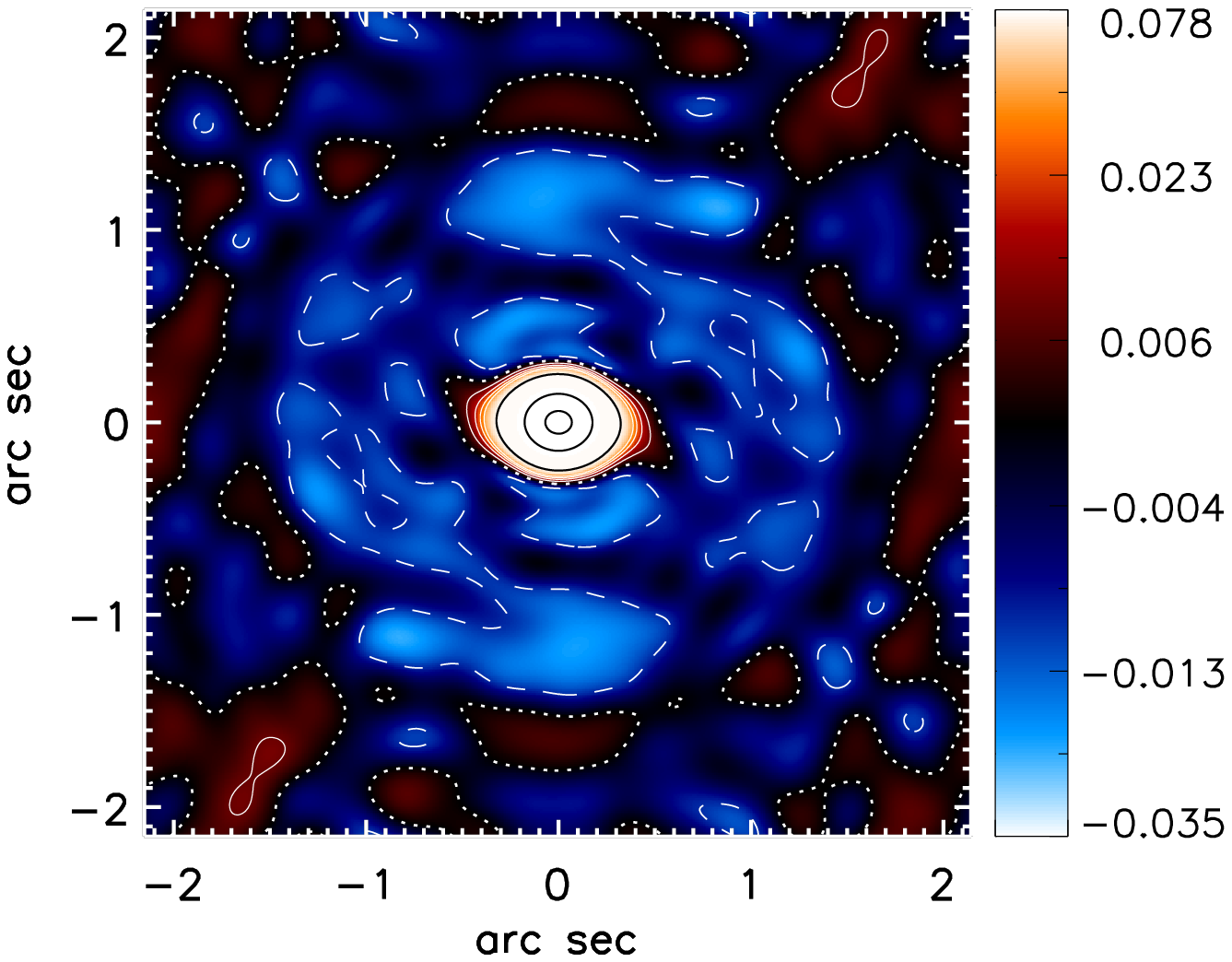} 
\hskip -0.30in \includegraphics[trim=0.10in 0.3in 0.8in 0.2in, clip, width=2.75in]
{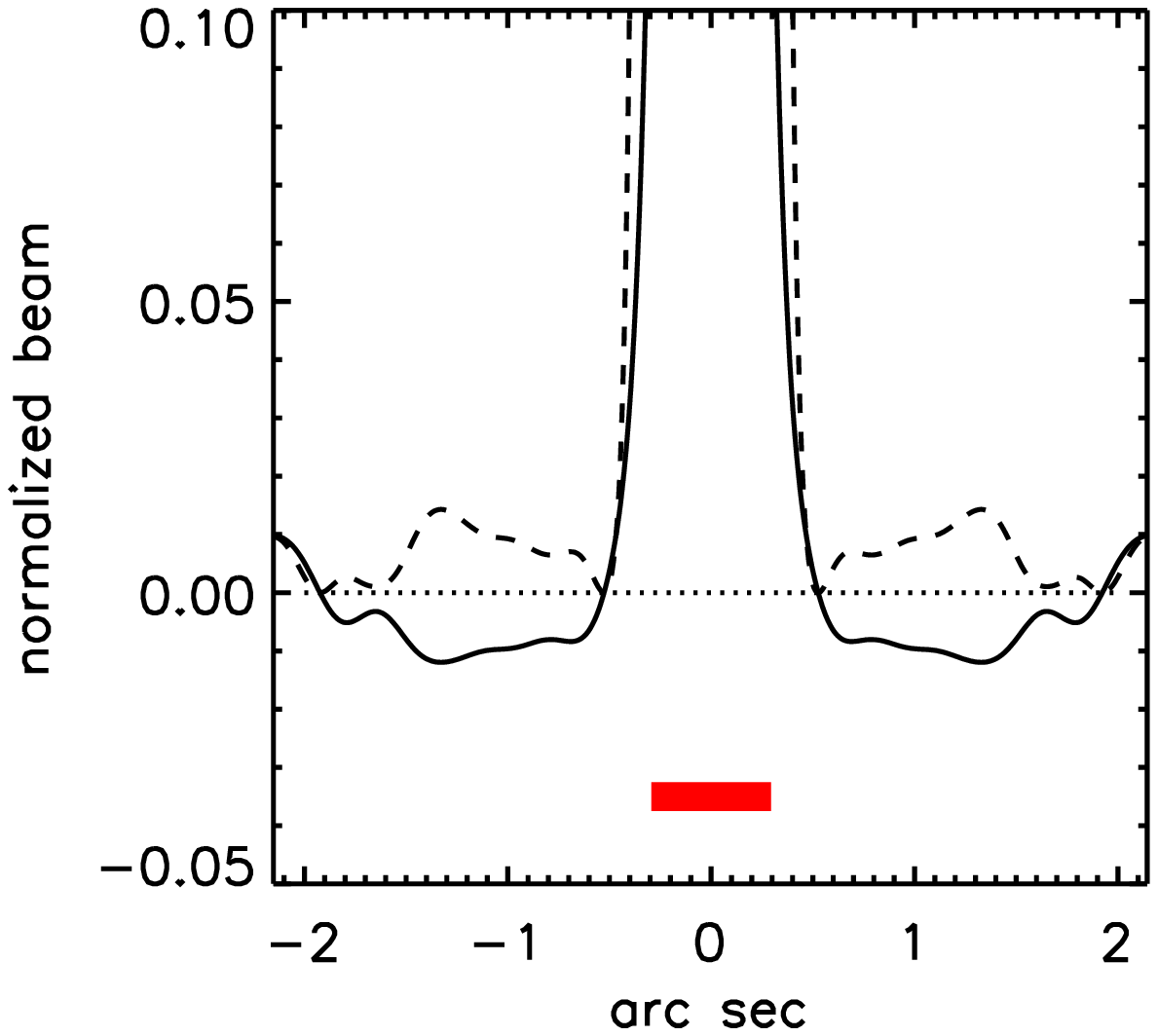} \\
\end{array}
$
\caption{{\it Top left:} Antenna locations, UV distribution, and beam for a 54-antenna s3p3d6
designed for low side lobes.  
Same format as figure \ref{fig:ska36fix}.  
The beam power (dashed line) in the lower right panel has been multiplied by 10.
The figures of merit are listed in table \ref{table:merit}.
}
\label{fig:ska54-n}
\end{figure}

Figure \ref{fig:ska54-n} shows an example with a factor of 2 scaling
between the levels and 
3\% scaling of the subarrays within each level to eliminate redundant
separations.
The details describing the rotation and placement of
the subarrays are listed in table \ref{table:ska216_details}.
The smoothly tapered distribution of separations from the overlapping subarrays 
creates a beam with an angular resolution of 0.23 arc seconds FWHM and a K-product,
$K_{98} = 294$  indicating that 98\% of the encircled energy is within 0.29 arc seconds.
Both the angular resolution and encircled energy are better than the Gaussian aperture in
\S\ref{FTpairs}.
Compared to the sparse H-array with uniform UV coverage, here the concentrated 
UV coverage results in less local smoothness and higher $\chi ^2$ measures,
0.0036 and 0.0018 in snapshot and ERS respectively.
The figures of merit are listed in table \ref{table:merit}.


\section{Hierarchical spirals, H-spirals} \label{spiralArrays}

The technique
of scaling and rotating subarrays with good uniform 
coverage can also be used to create intelligently designed high performance 
spiral arrays with many antennas. These hierarchical
spirals or H-spiral arrays are 
aesthetically  attractive and also have some practical advantages
useful in construction.
(The process of scaling and rotating produces spirals that are mathematically
equivalent to logarithmic spirals with different constants.)
Because of their concentric rather than hierarchical 
structure, spiral arrays do not 
produce the highest angular resolution beams 
with a compact array size. Nevertheless, these H-spirals share the property of
the H-arrays that
a simple change in scale again moves the design between the
trade-offs of high angular resolution and a more concentrated beam.

The Square Kilometer Array might be based on a spiral pattern. The primary design goal
is to minimize gaps in the UV coverage within the constraints of the spiral design 
\citep{Millenaar2011}.
The uniform coverage of the CW-arrays by definition provides minimal gaps (\S\ref{minimax}).
A spiral array built by scaling a CW-subarray according to a power law has larger gaps at
larger scales, but at each scale, the CW-subarray provides 
coverage with the minimum possible gaps at that scale.

\subsection{Rotation in H-spirals}

\begin{figure}[t]
$
\begin{array}{cc}
\includegraphics[trim=0.10in 0.3in 0.8in 0.2in, clip, width=2.75in]{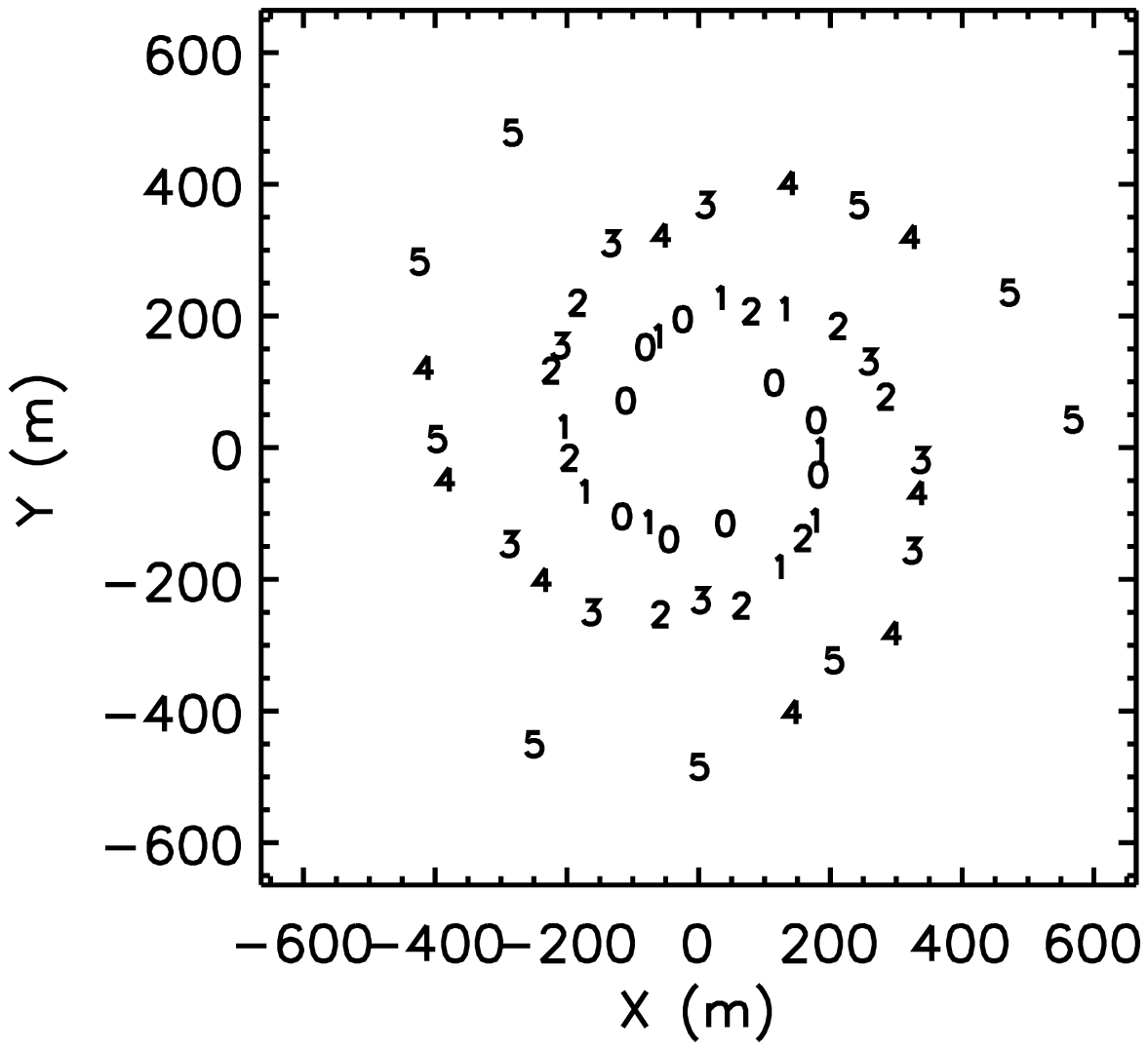} 
\includegraphics[trim=0.10in 0.3in 0.8in 0.2in, clip, width=2.75in]{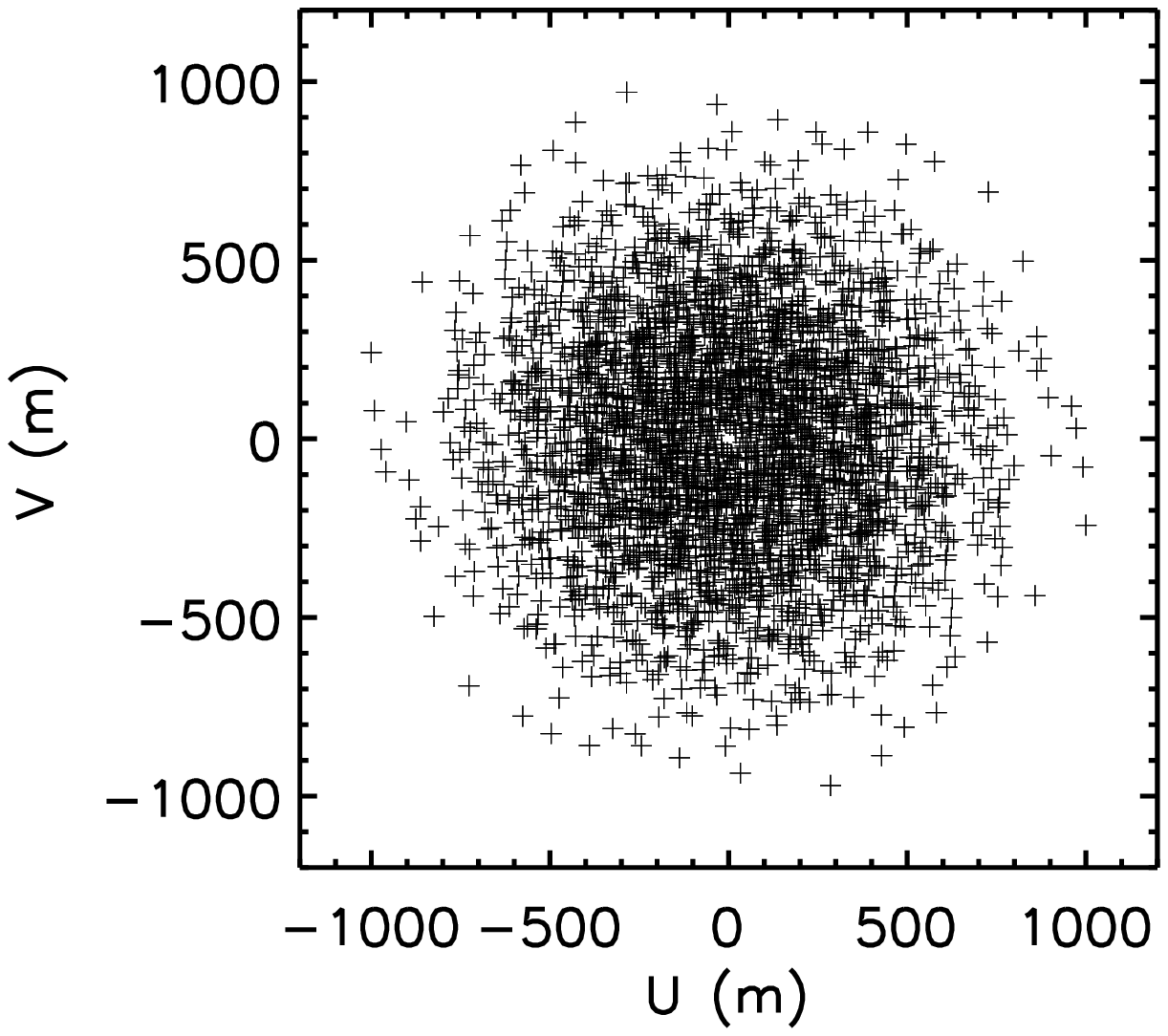} \\
\hskip 0.65in \includegraphics[trim=0.10in 0.3in 0.8in 0.2in, clip, width=2.75in]
{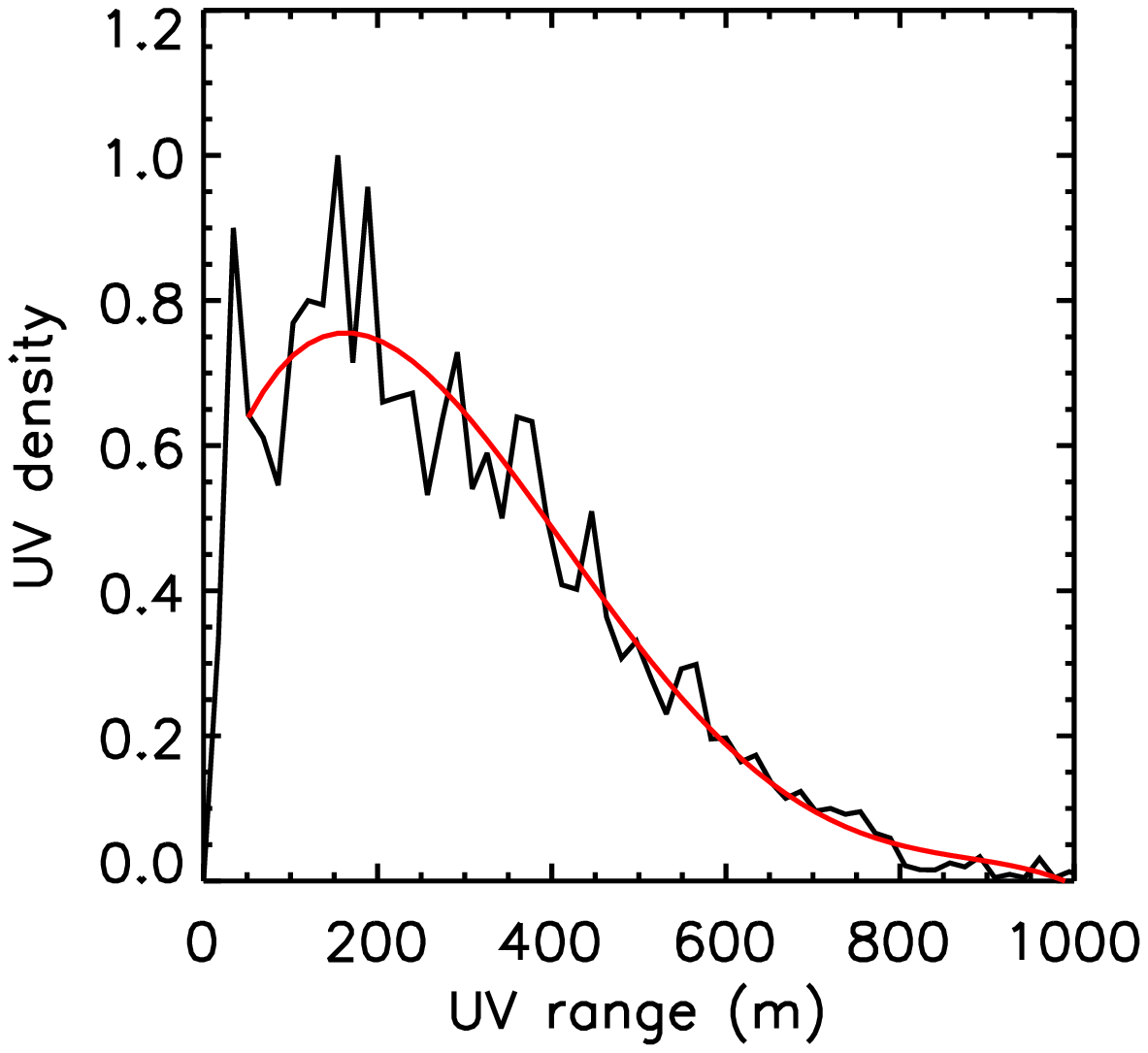} 
\includegraphics[trim=0.10in 0.3in 0.8in 0.2in, clip, width=2.75in]{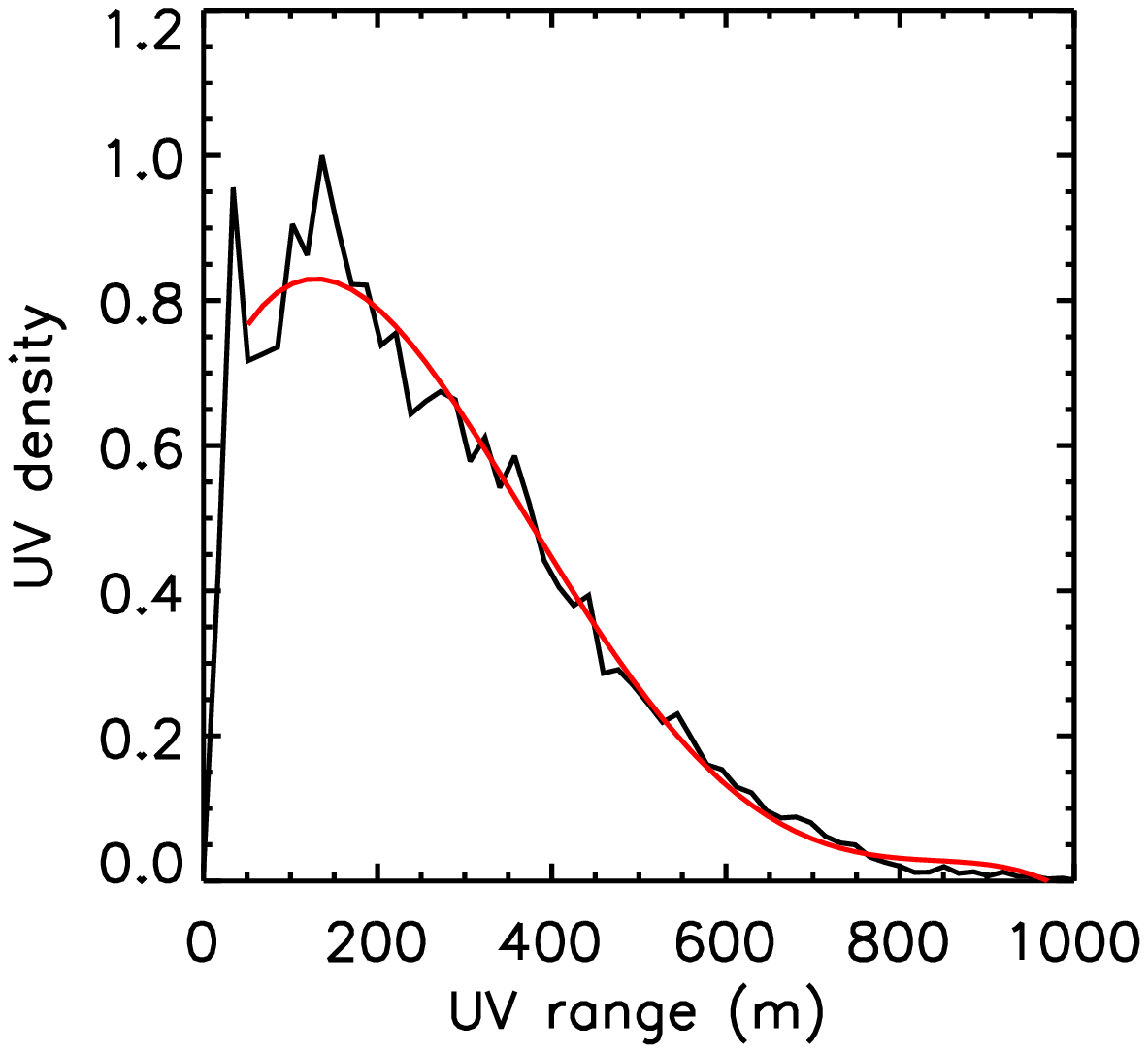} \\
\hskip 0.35in \includegraphics[trim=0.10in 0.3in 0.8in 0.2in, clip, width=2.75in]
{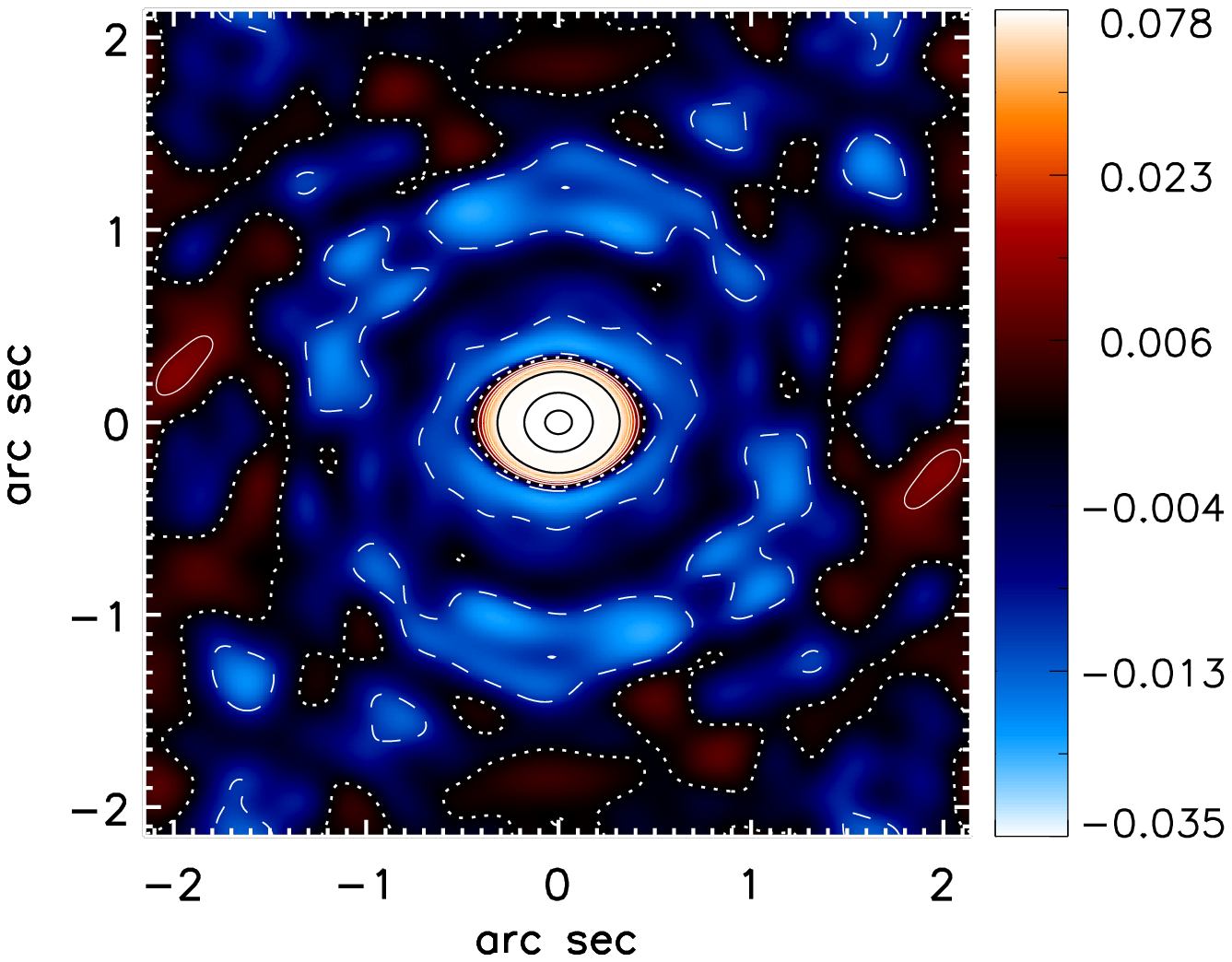} 
\hskip -0.30in \includegraphics[trim=0.10in 0.3in 0.8in 0.2in, clip, width=2.75in]
{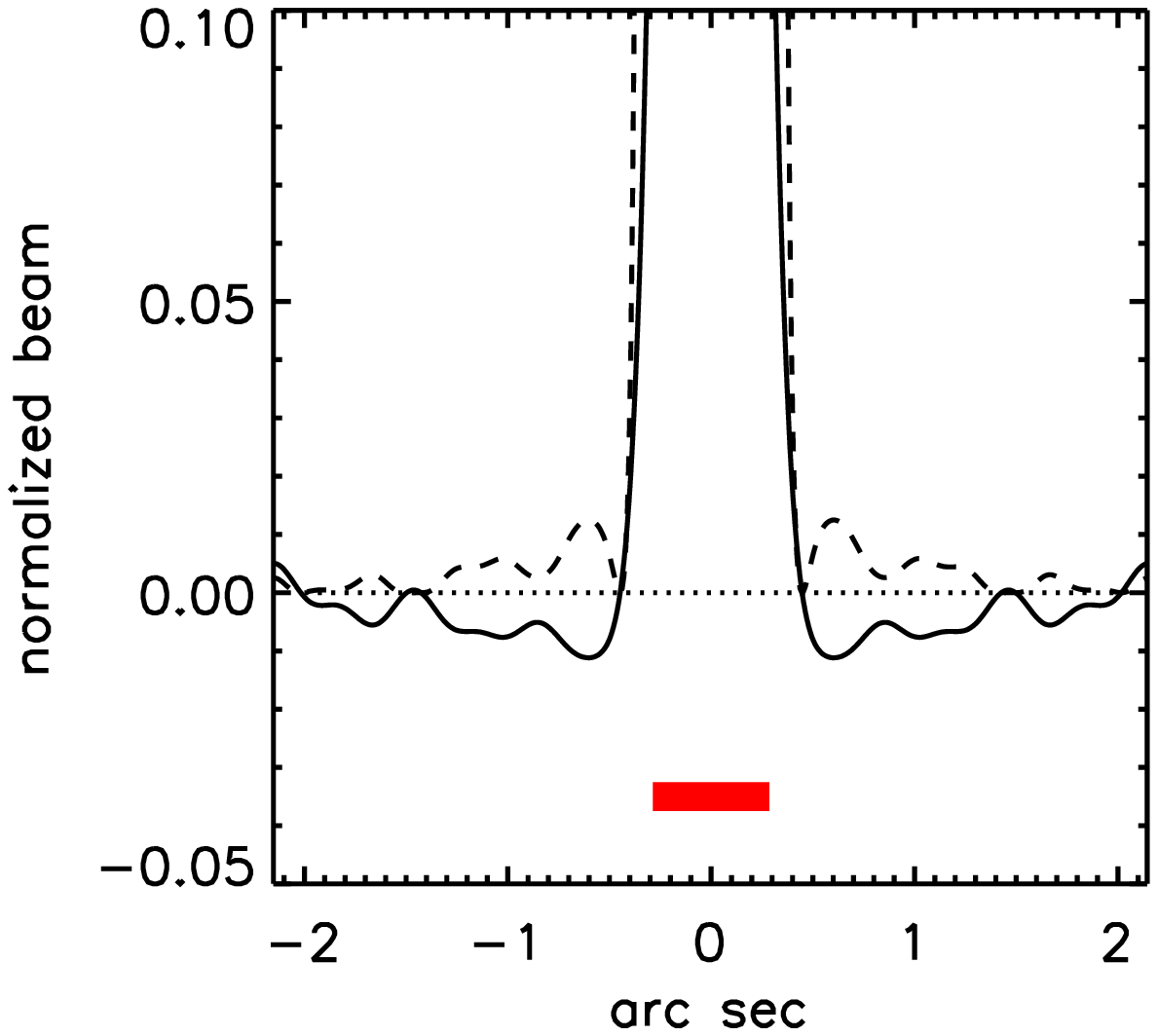}  \\
\end{array}
$
\caption{Antenna locations, UV distribution, and beam for a 54-antenna 
spiral array designed for a high concentration of power in the main beam.
The pattern uses
the 9-element CW-array in figure
\ref{fig:s9} as the simple subarray. A power law scaling by
a factor of 1.25 and 
a constant rotation of 164$^\circ$
between subarrays produces the 3-armed pattern
resembling a spiral galaxy. The difference between this array and the one
in figure \ref{fig:ispiral113} 
is the amount of rotation between the scaled subarrays.
Same format as figure \ref{fig:ska36fix}.
The beam power (dashed line) in the lower right panel has been multiplied by 100.
The figures of merit are listed in table \ref{table:merit}.
}
\label{fig:ispiral164}
\end{figure}

\begin{figure}[t]
$
\begin{array}{cc}
\includegraphics[trim=0.10in 0.3in 0.8in 0.2in, clip, width=2.75in]{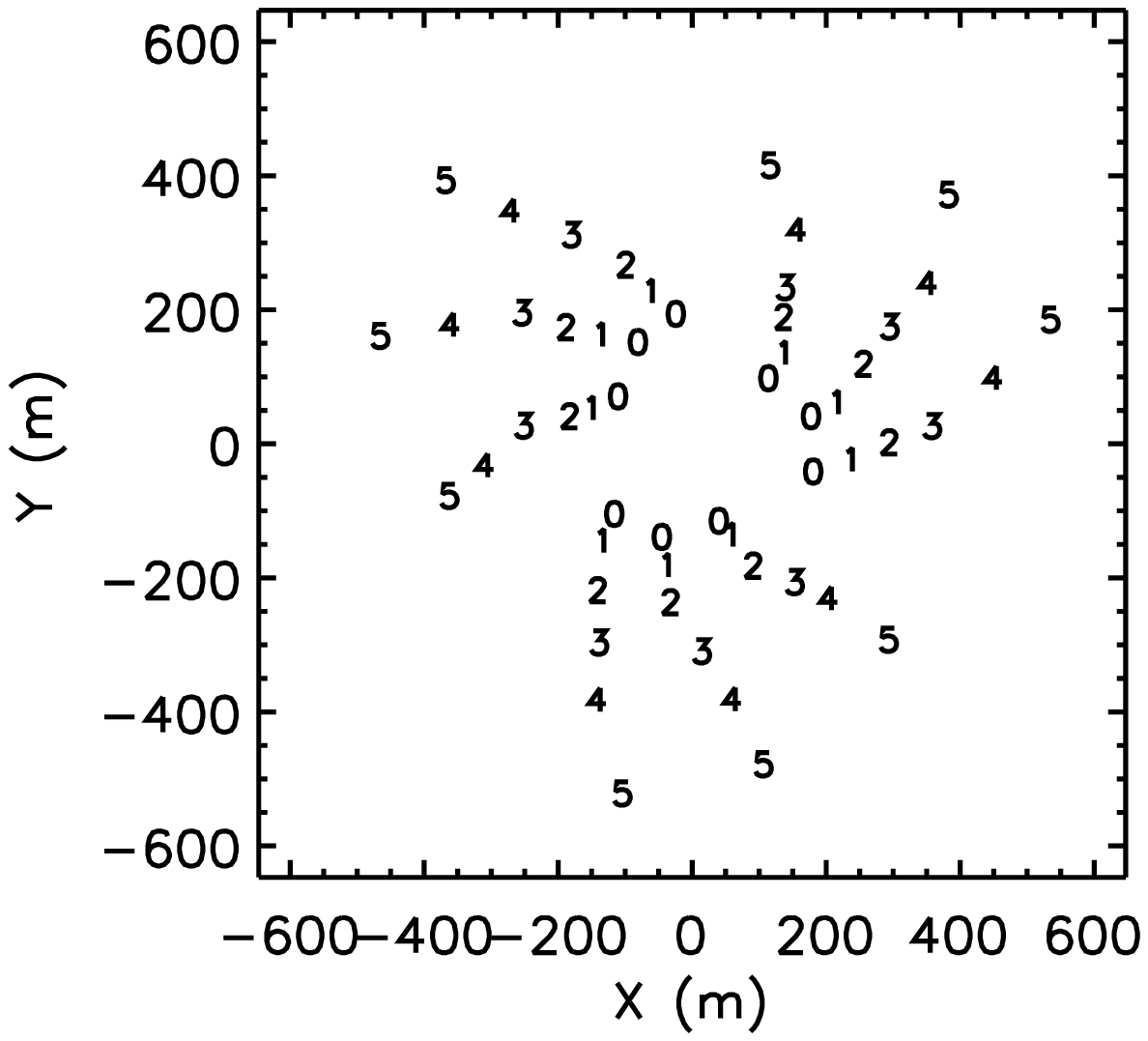} 
\includegraphics[trim=0.10in 0.3in 0.8in 0.2in, clip, width=2.75in]{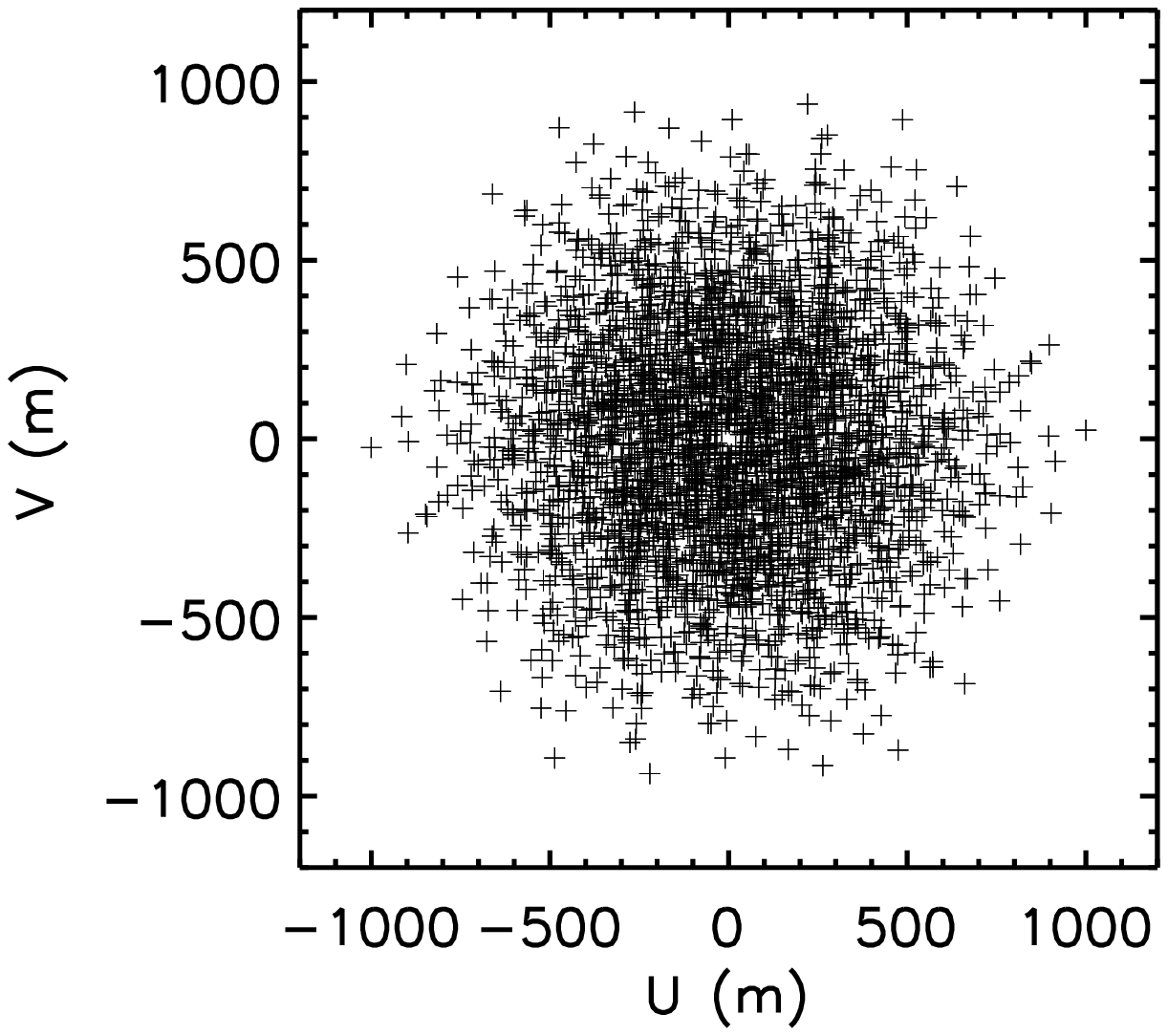} \\
\hskip 0.65in \includegraphics[trim=0.10in 0.3in 0.8in 0.2in, clip, width=2.75in]
{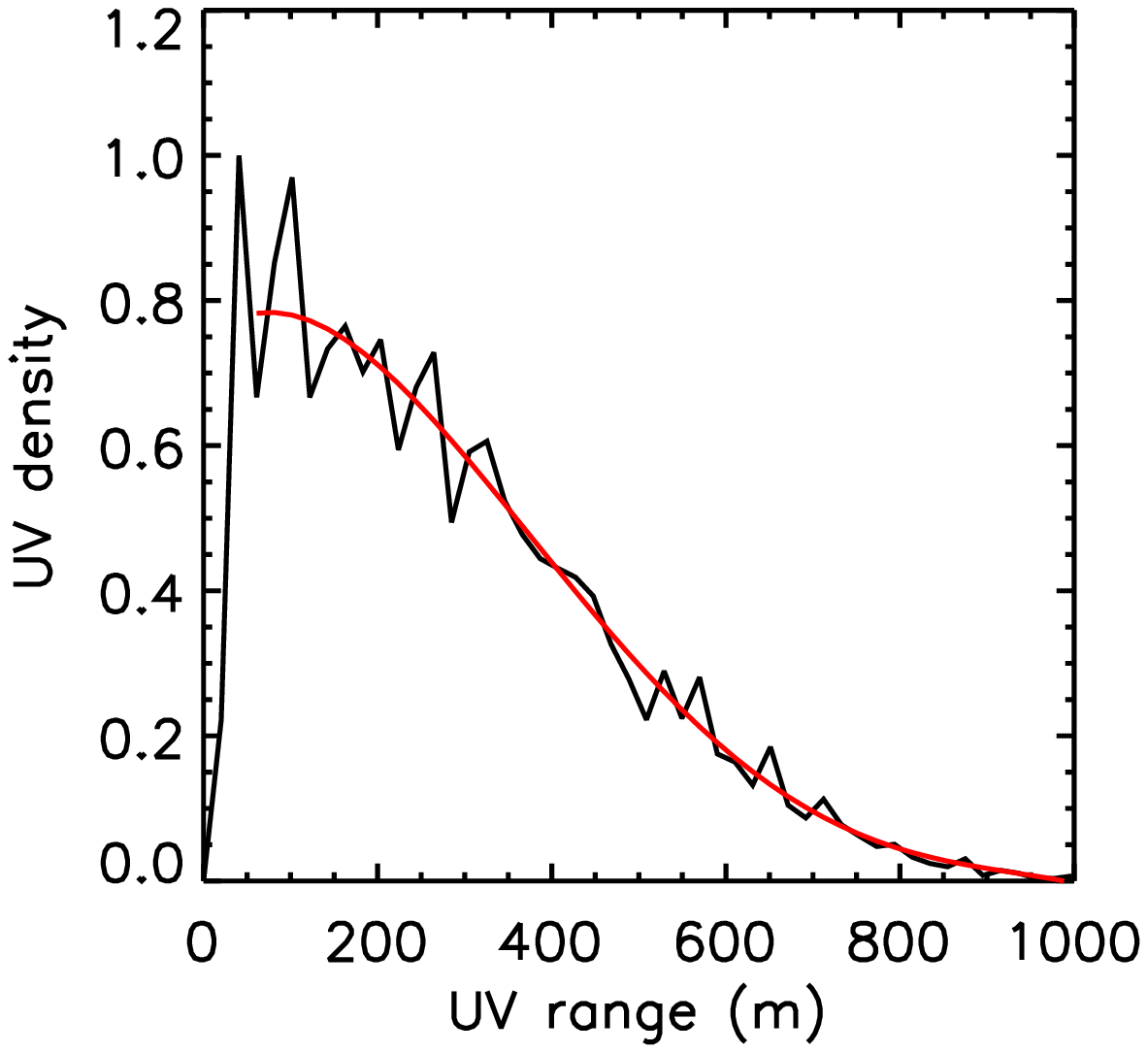} 
\includegraphics[trim=0.10in 0.3in 0.8in 0.2in, clip, width=2.75in]{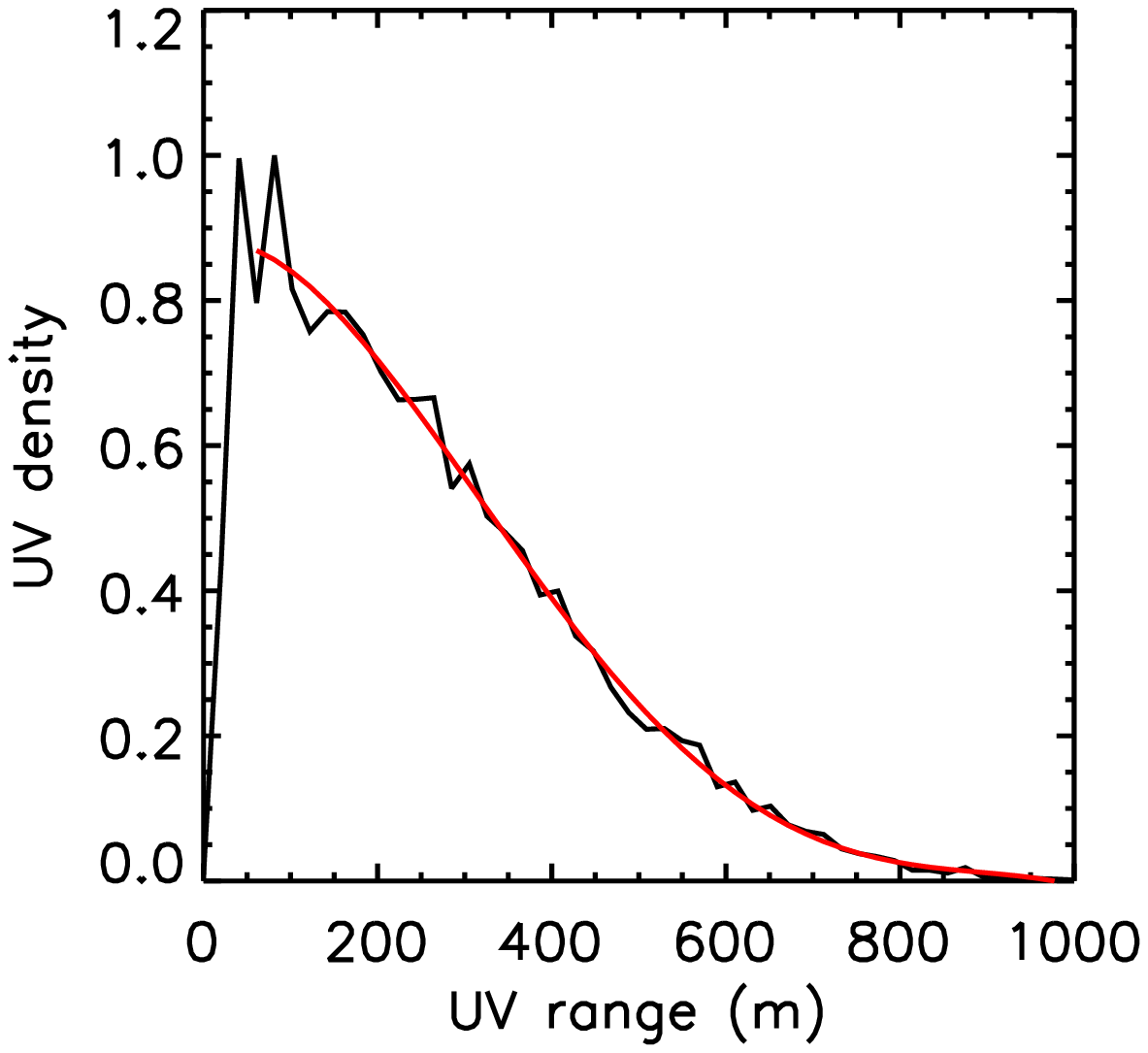} \\
\hskip 0.35in \includegraphics[trim=0.10in 0.3in 0.8in 0.2in, clip, width=2.75in]
{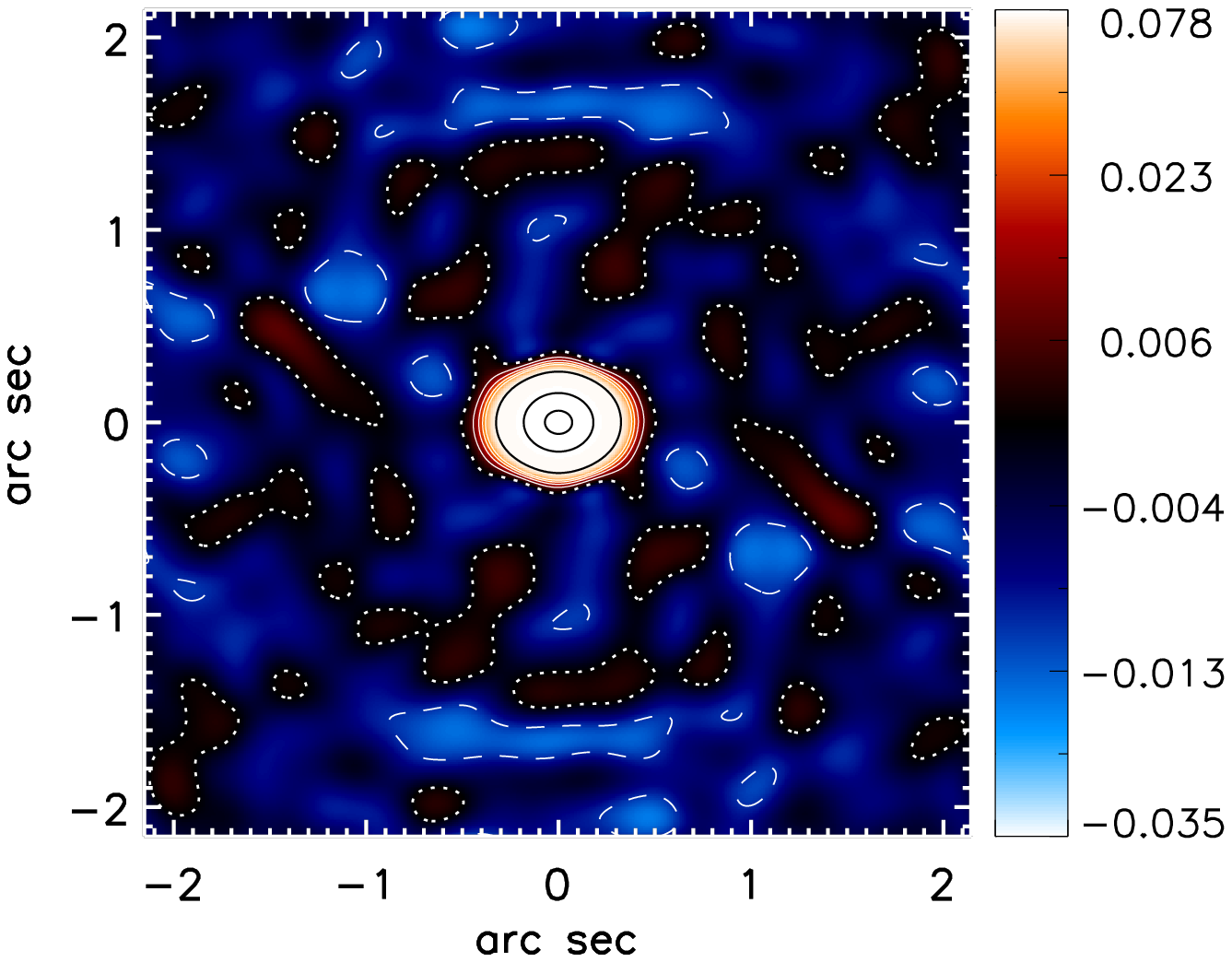} 
\hskip -0.30in \includegraphics[trim=0.10in 0.3in 0.8in 0.2in, clip, width=2.75in]
{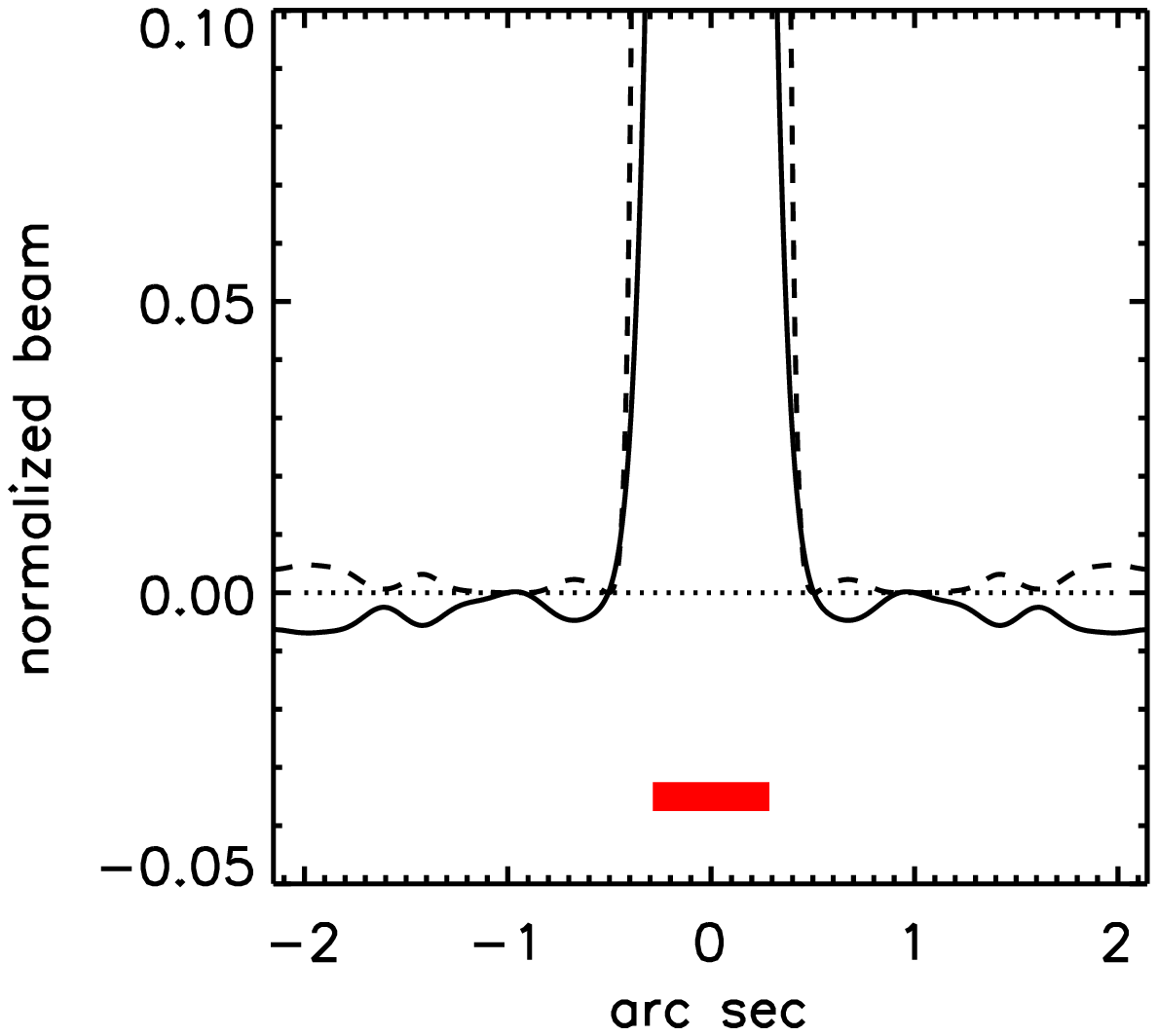}  \\
\end{array}
$
\caption{Antenna locations, UV distribution, and beam for a 54-antenna 
spiral array designed for a high concentration of power in the main beam.
The pattern uses
the 9-element CW-array in figure
\ref{fig:s9} as the simple subarray. A power law scaling by
a factor of 1.25 and 
a constant rotation of 113$^\circ$
between subarrays produces the 9-armed pattern
resembling sea stars. The difference between this array and the one
in figure \ref{fig:ispiral164} 
is the amount of rotation between the scaled subarrays.
Same format as figure \ref{fig:ska36fix}.
The beam power (dashed line) in the lower right panel has been multiplied by 100.
The figures of merit are listed in table \ref{table:meritcont}.
}
\label{fig:ispiral113}
\end{figure}

First consider the effect of rotation.
The first two examples are
H-spiral patterns 
based on the 9-element CW-array in figure \ref{fig:s9} but with
different amounts of rotation between the subarrays. The power law
scaling, 1.25,  is the same in both cases.
The first pattern
rotates the subarrays by 164$^\circ$ to create a 
tight, wrapping spiral resembling a late-type galaxy
(figure \ref{fig:ispiral164}). 
In the second, a smaller rotation
of 113$^\circ$ produces a 9-ray pattern 
resembling the nine-armed sea stars of the Florida Keys
(figure \ref{fig:ispiral113}). 
Both have essentially
identical performance metrics.
In ERS, the H-spiral galaxy has a angular resolution 
of 0.23 arc seconds and a K-product, $K_{98}=285$.
The $\chi ^2$ measure of smoothness is 0.0029. The H-spiral 
sea star has the same angular resolution,
0.23 arc seconds and a K-product, $K_{98}=285$, but the
$\chi ^2$ indicates that the distribution of separations is 
slightly smoother at 0.0026.
In snapshot, the $\chi ^2$ measures of local smoothness for the galaxy 
and sea star H-spirals are 
0.0039 and 0.0024 respectively. 
Both these arrays have better angular resolution and tighter encircled 
energy than the Gaussian beam
in \S\ref{FTpairs}.
The figures of merit are listed in table \ref{table:merit}.

\subsection{Scaling in H-spirals}

\begin{figure}[t]
$
\begin{array}{cc}
\includegraphics[trim=0.10in 0.3in 0.8in 0.2in, clip, width=2.75in]{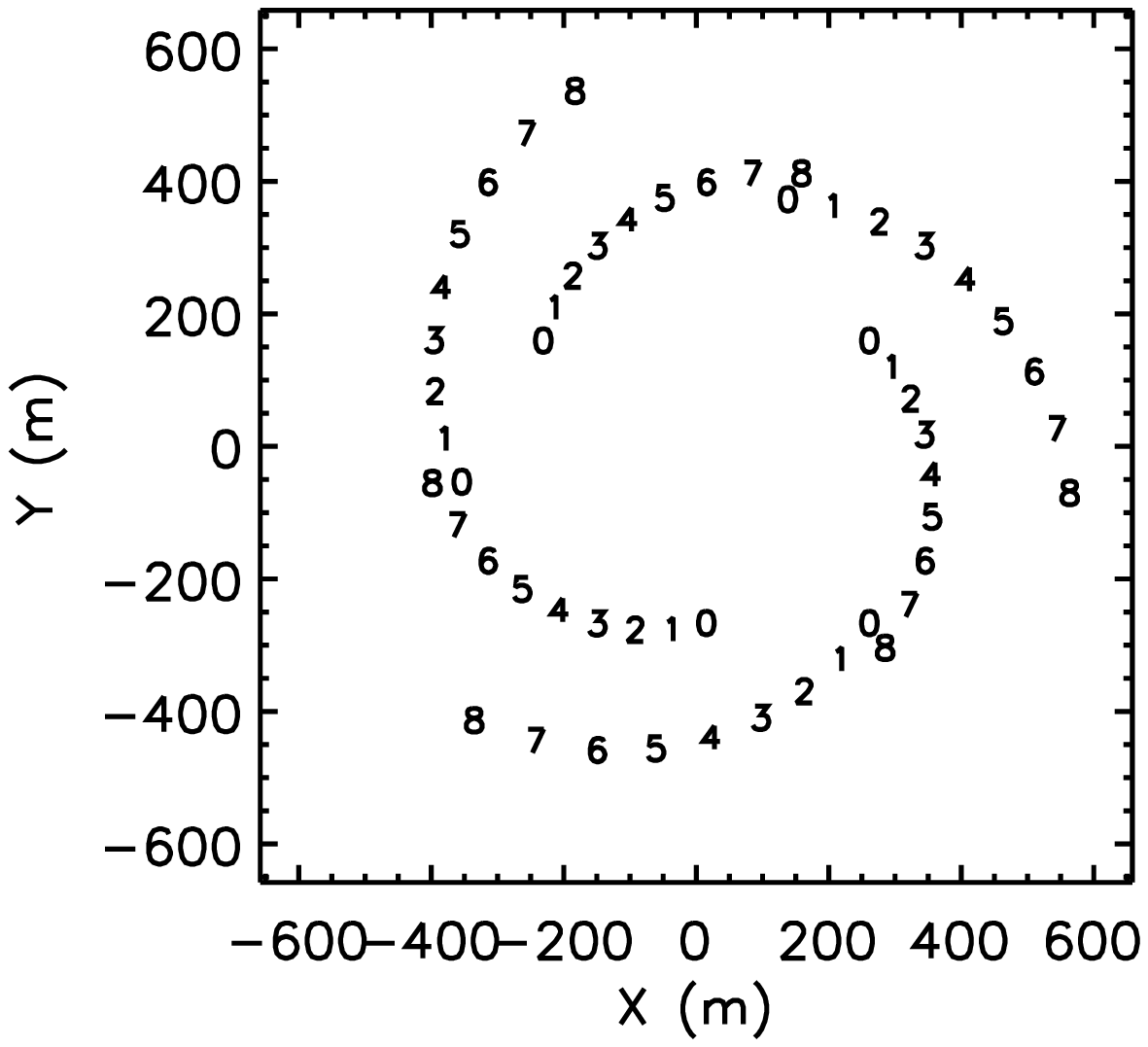} 
\includegraphics[trim=0.10in 0.3in 0.8in 0.2in, clip, width=2.75in]{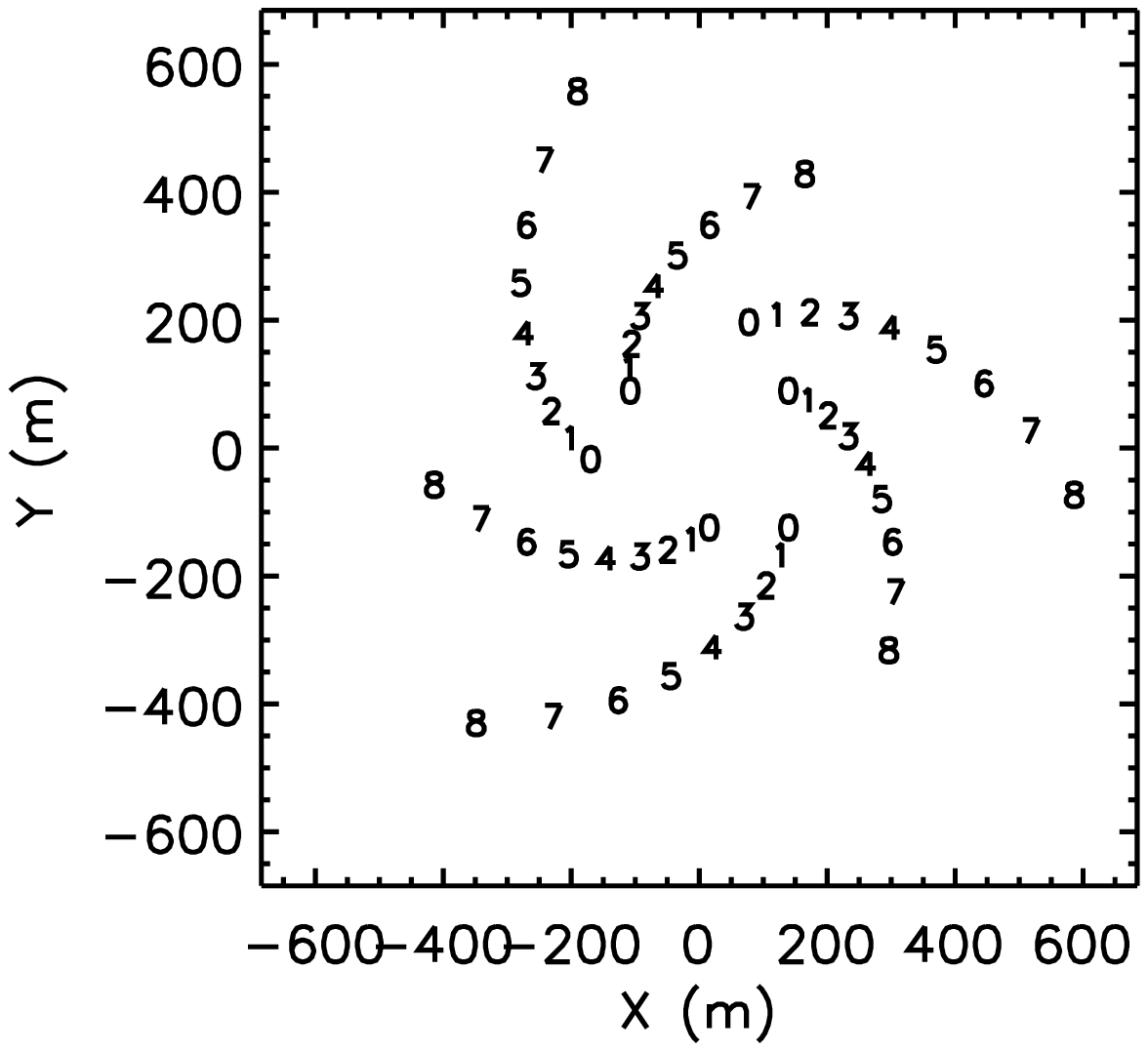} \\
\hskip 0.65in \includegraphics[trim=0.10in 0.3in 0.8in 0.2in, clip, width=2.75in]{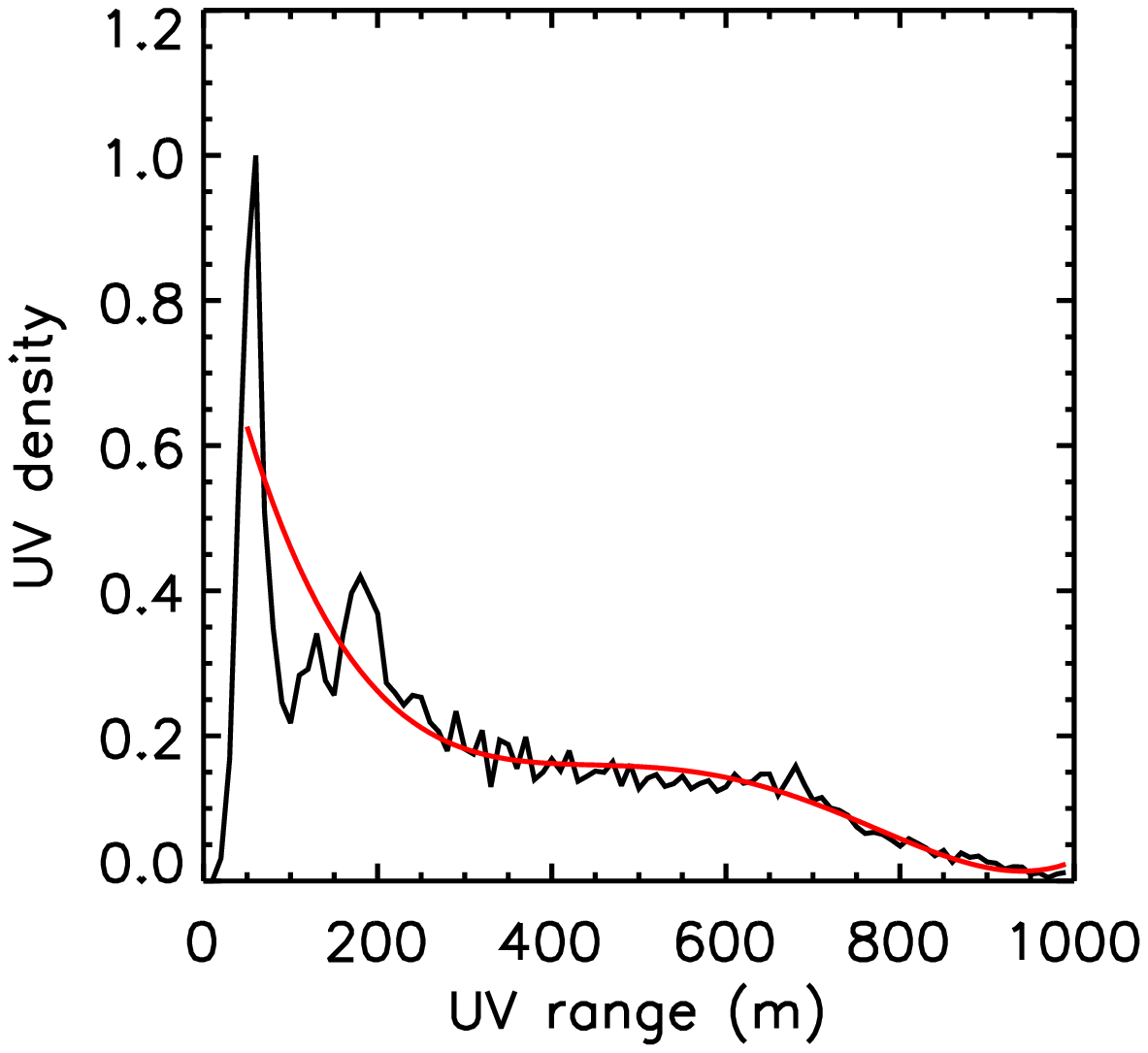} 
\includegraphics[trim=0.10in 0.3in 0.8in 0.2in, clip, width=2.75in]{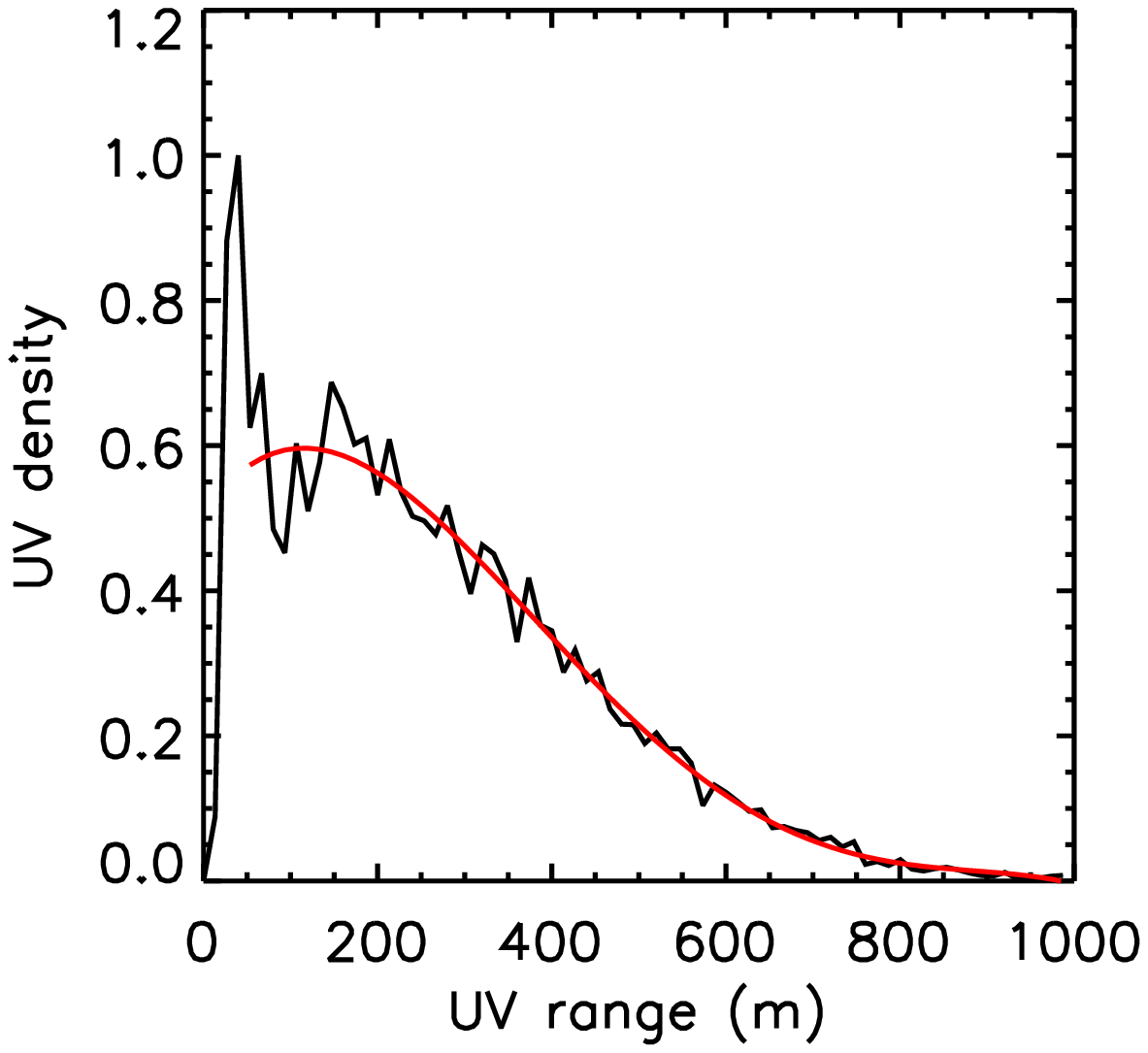} \\
\includegraphics[trim=0.10in 0.3in 0.8in 0.2in, clip, width=2.75in]{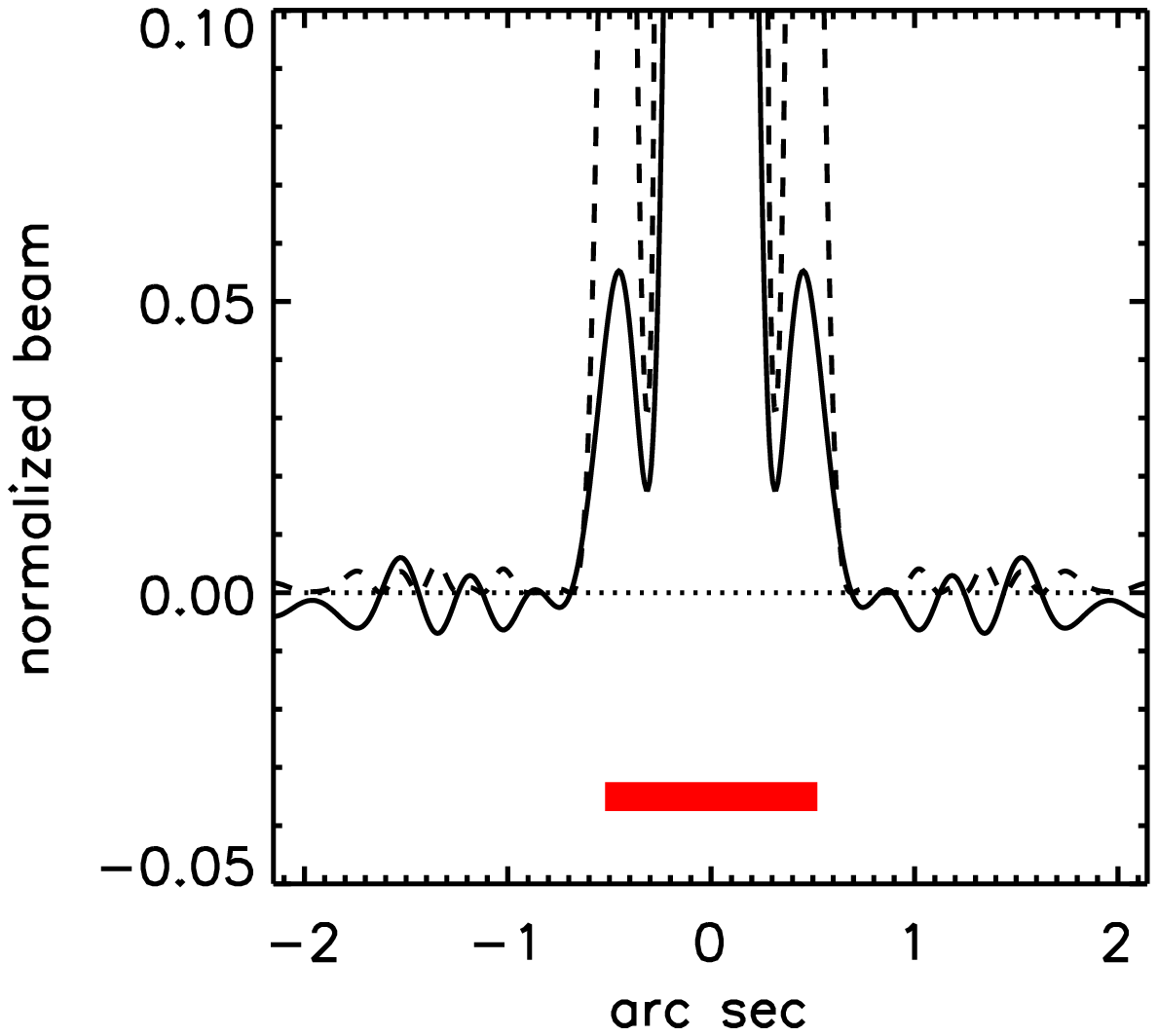} 
\includegraphics[trim=0.10in 0.3in 0.8in 0.2in, clip, width=2.75in]{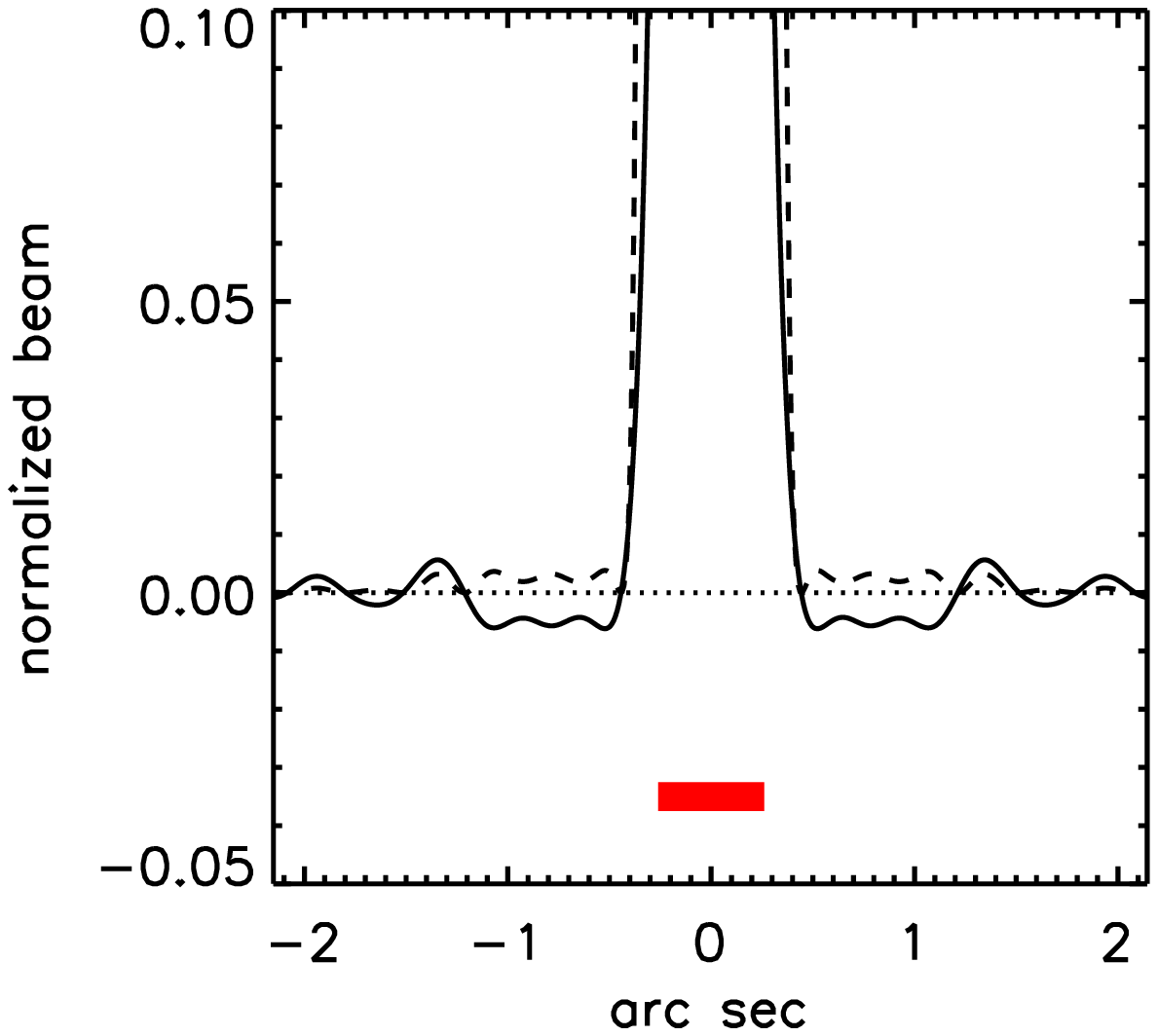}  \\
\hskip 0.65in \includegraphics[trim=0.10in 0.3in 0.8in 0.2in, clip, width=2.75in]{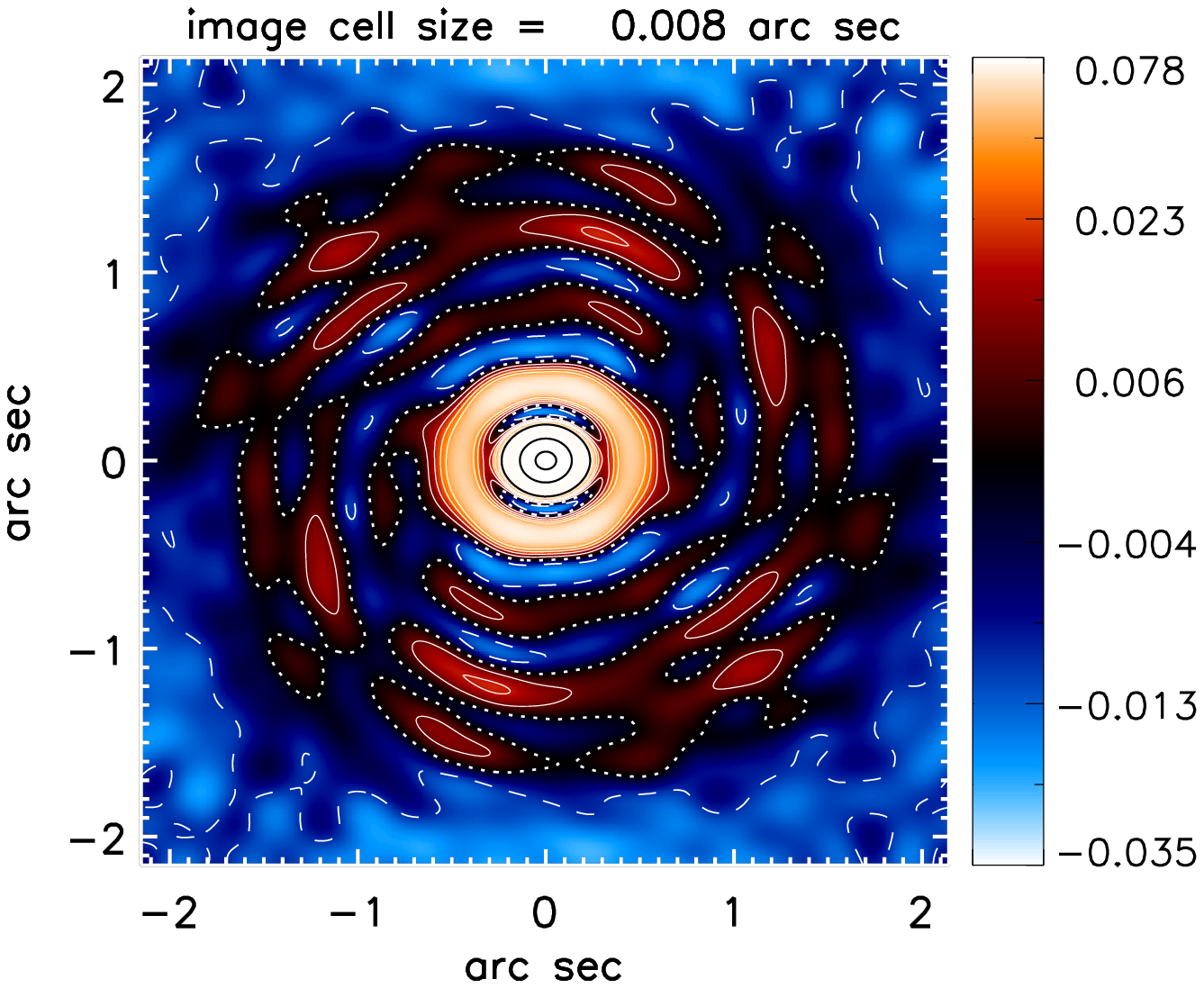} 
\includegraphics[trim=0.10in 0.3in 0.8in 0.2in, clip, width=2.75in]{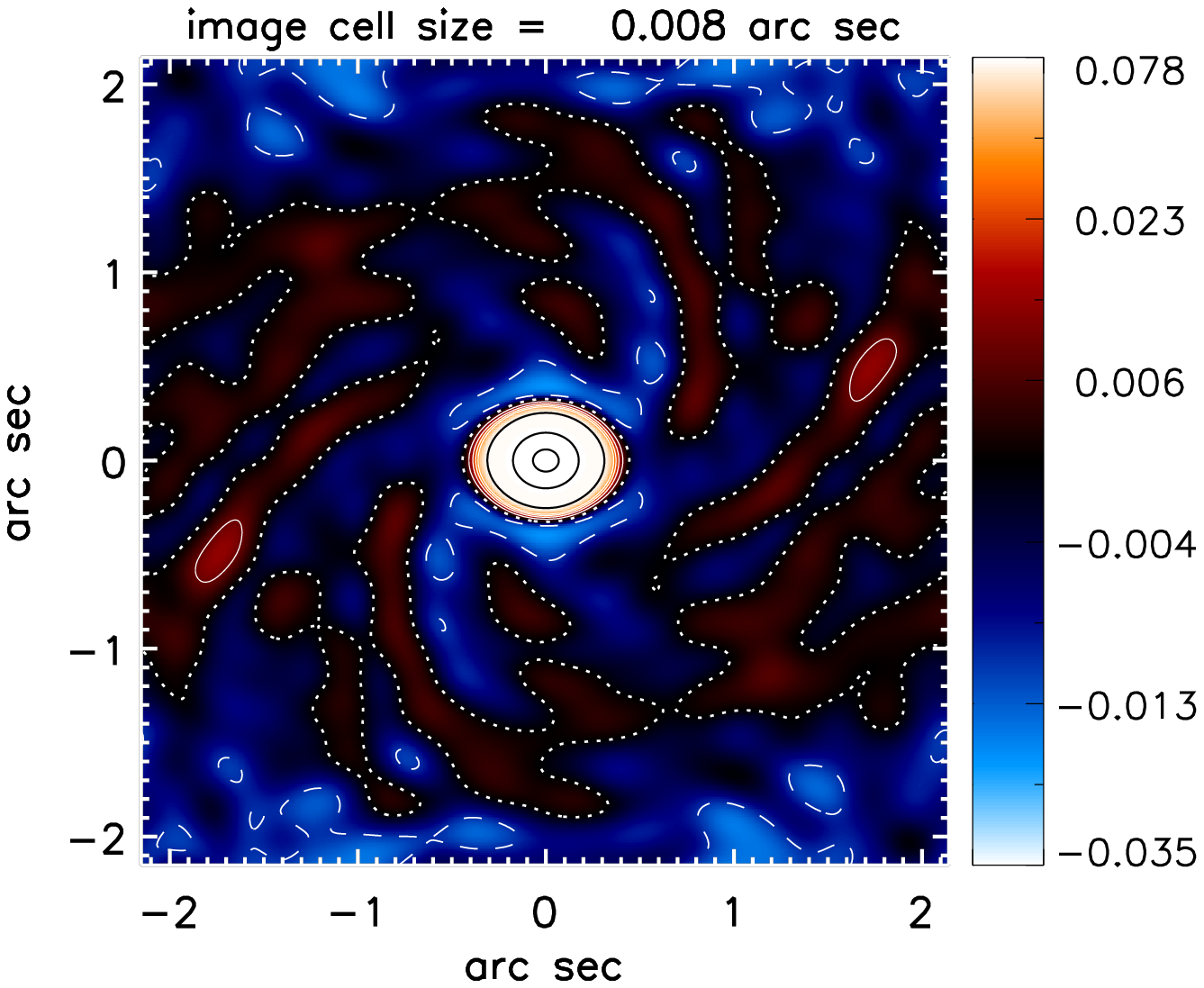}  \\
\end{array}
$
\caption{Antenna locations, UV distribution, and beam for 54-antenna 
spiral arrays showing the effect of 
the power law scaling between subarrays.
The pattern uses
the 6-element CW-array in figure
\ref{fig:s6} as the simple subarray. In the left column the configuration is
built by scaling successive 
subarrays by a factor of 1.05. The right column shows the results for a scaling of by 1.15.
The same arrays but with power law scalings of 1.25 and 1.35 are shown
in figure \ref{fig:ispiral6b}. The lower scalings here produce beams
of higher angular resolution.
The panels have the same format as figure \ref{fig:ska36fix}, except that here
the UV coverage is not shown and the UV density is shown
for ERS only.
The beam power (dashed line) in the third panel down in both columns has been multiplied by 100.
The figures of merit are listed in table \ref{table:meritcont}.
}
\label{fig:ispiral6a}
\end{figure}

\begin{figure}[t]
$
\begin{array}{cc}
\includegraphics[trim=0.10in 0.3in 0.8in 0.2in, clip, width=2.75in]{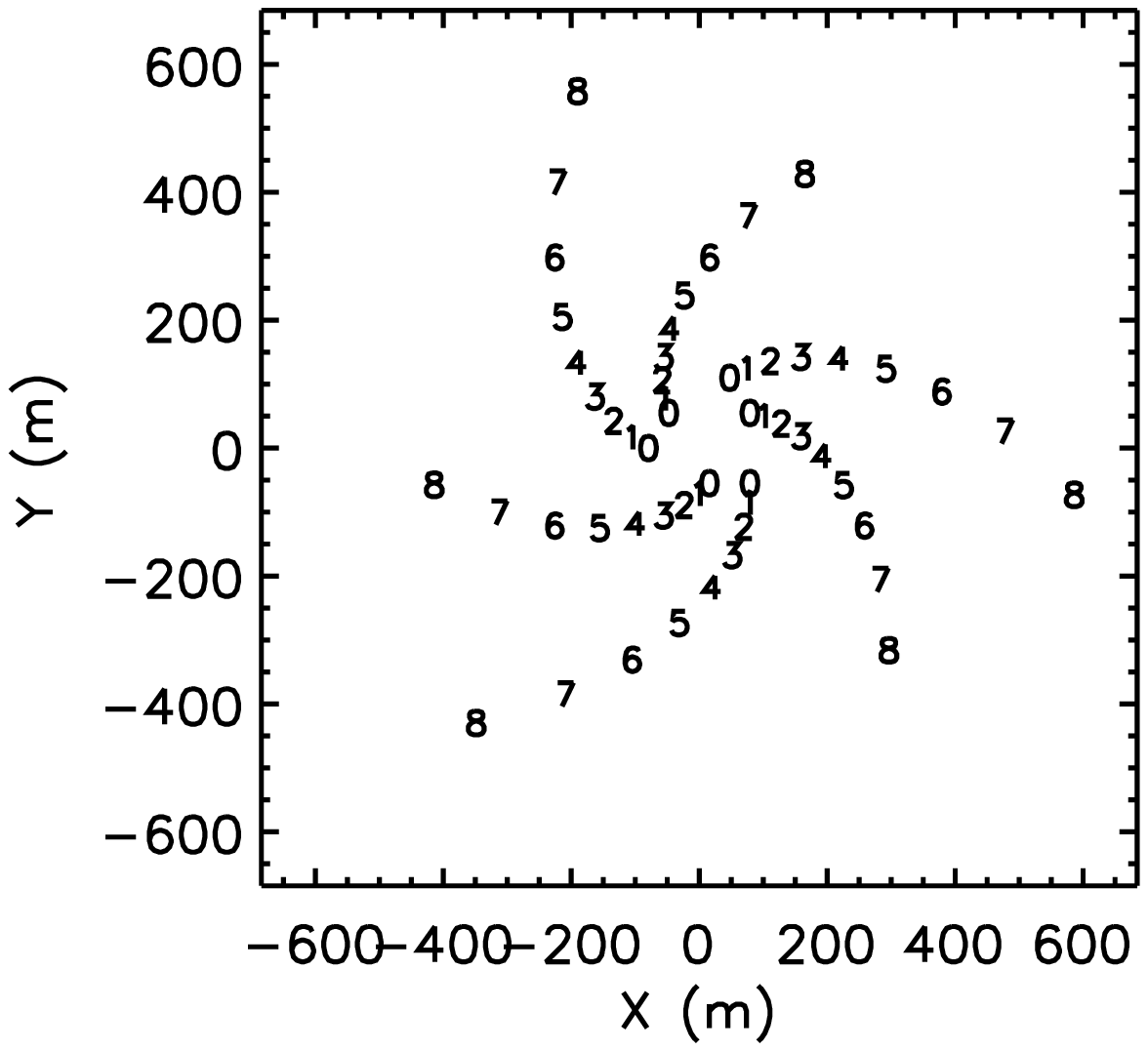} 
\includegraphics[trim=0.10in 0.3in 0.8in 0.2in, clip, width=2.75in]{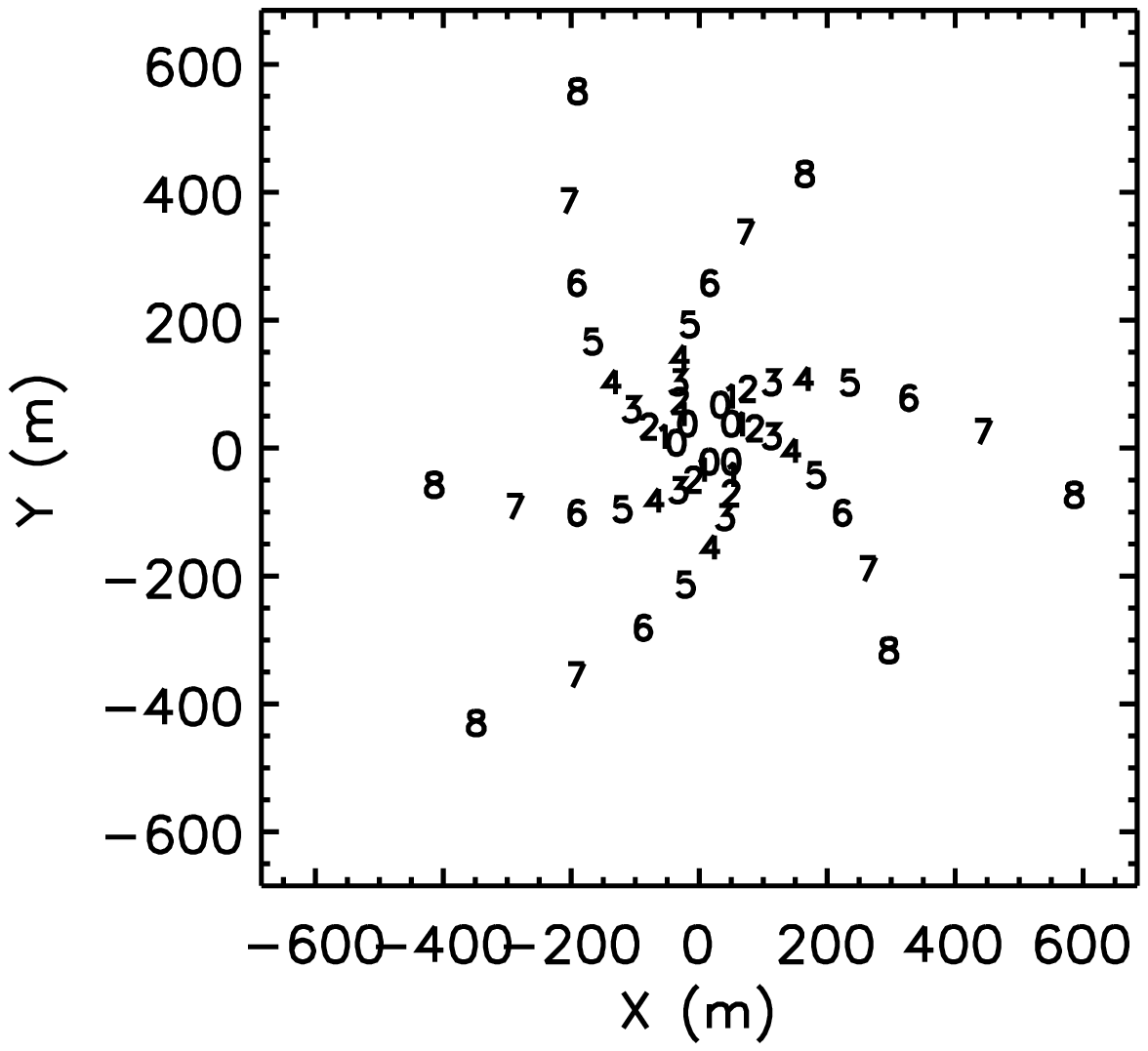} \\
\hskip 0.65in \includegraphics[trim=0.10in 0.3in 0.8in 0.2in, clip, width=2.75in]{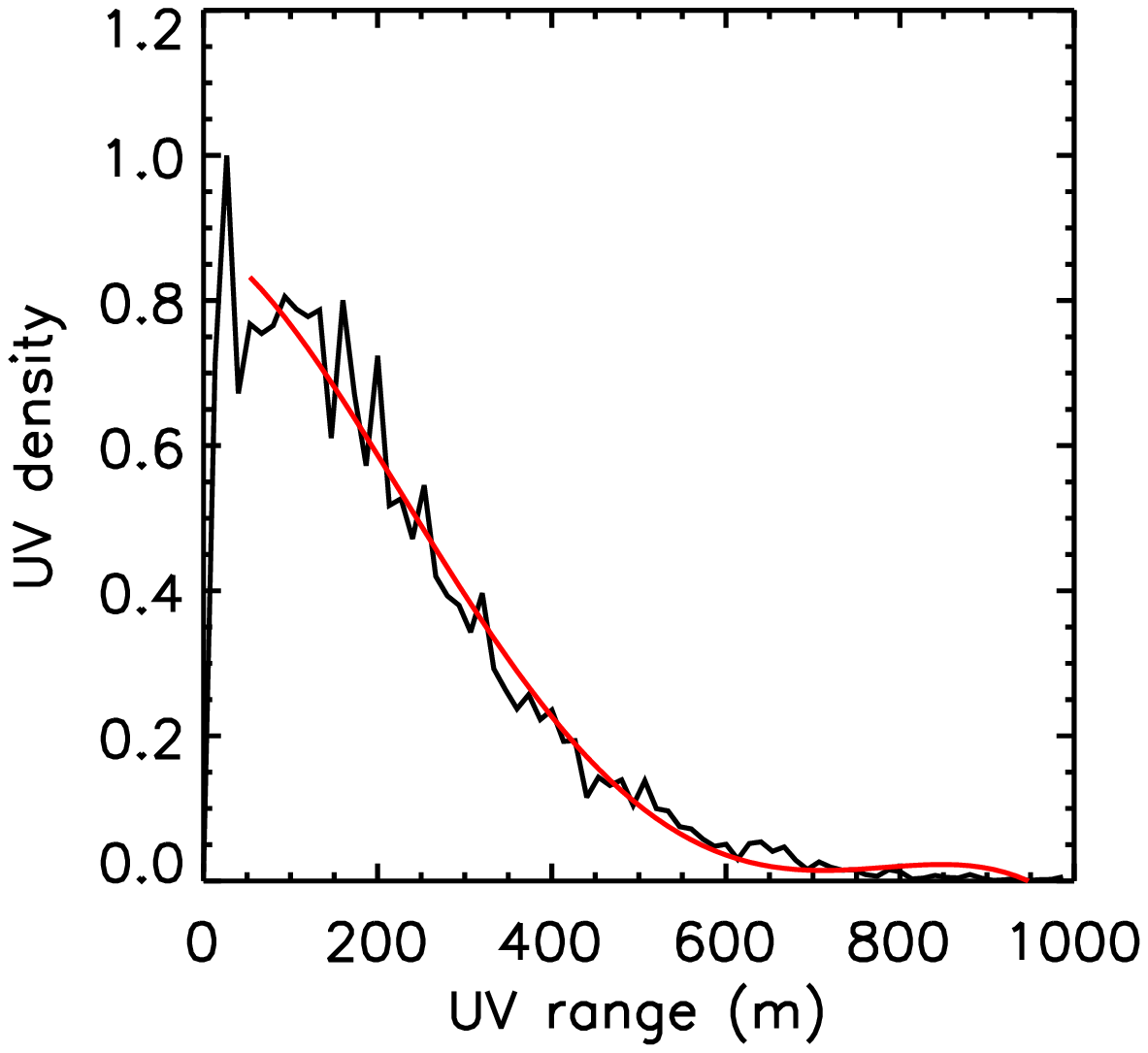} 
\includegraphics[trim=0.10in 0.3in 0.8in 0.2in, clip, width=2.75in]{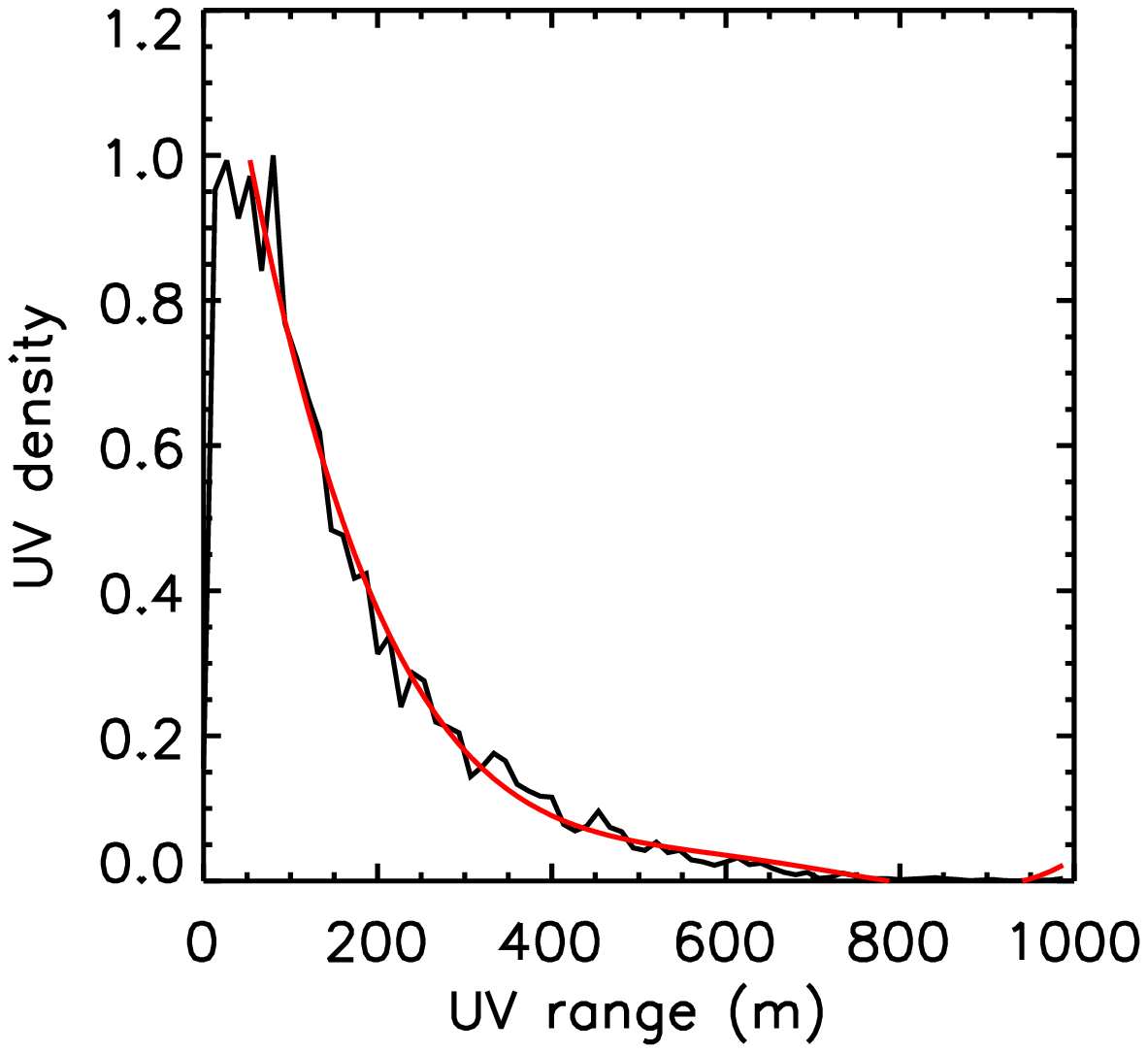} \\
\includegraphics[trim=0.10in 0.3in 0.8in 0.2in, clip, width=2.75in]{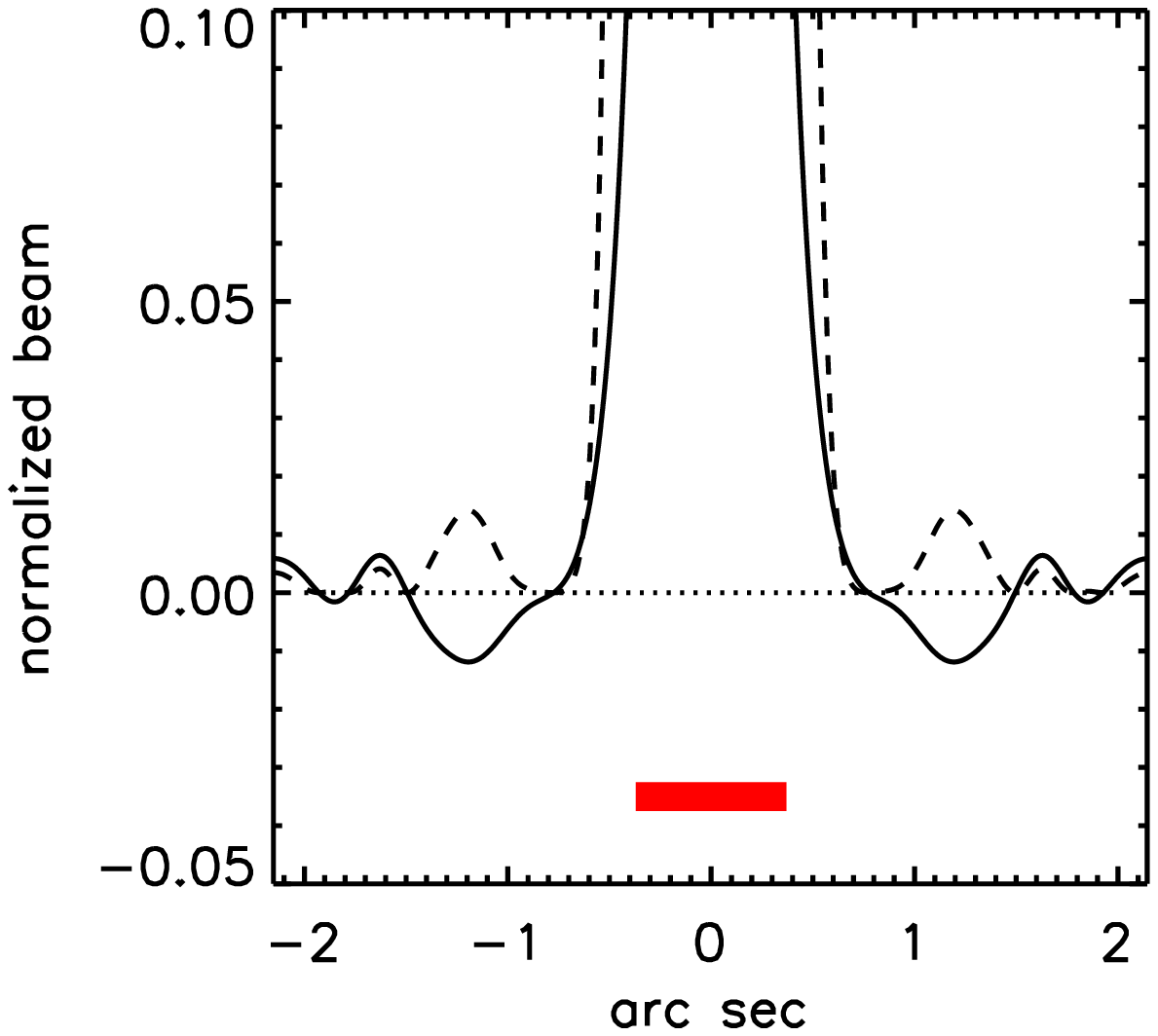} 
\includegraphics[trim=0.10in 0.3in 0.8in 0.2in, clip, width=2.75in]{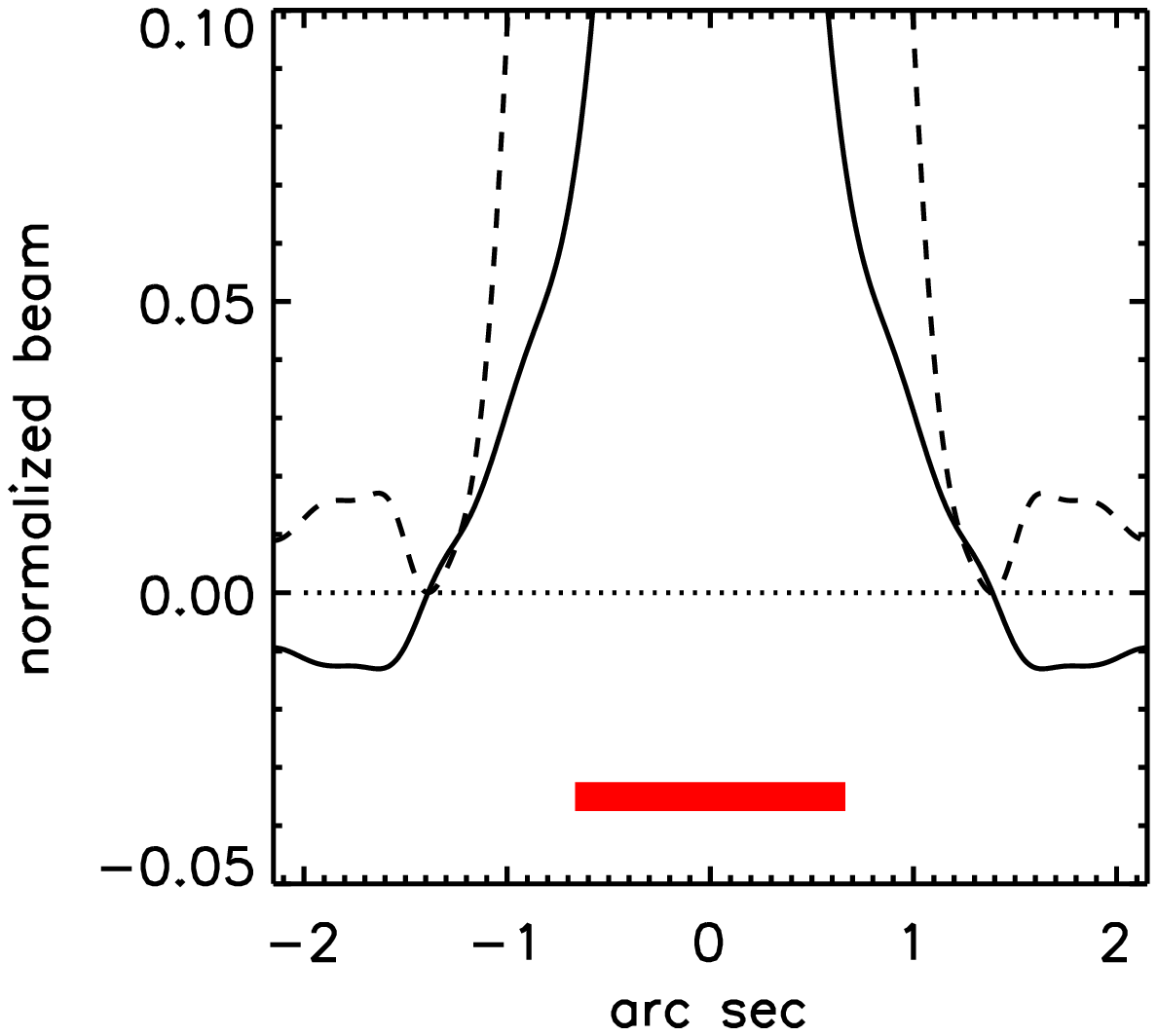}  \\
\hskip 0.65in \includegraphics[trim=0.10in 0.3in 0.8in 0.2in, clip, width=2.75in]{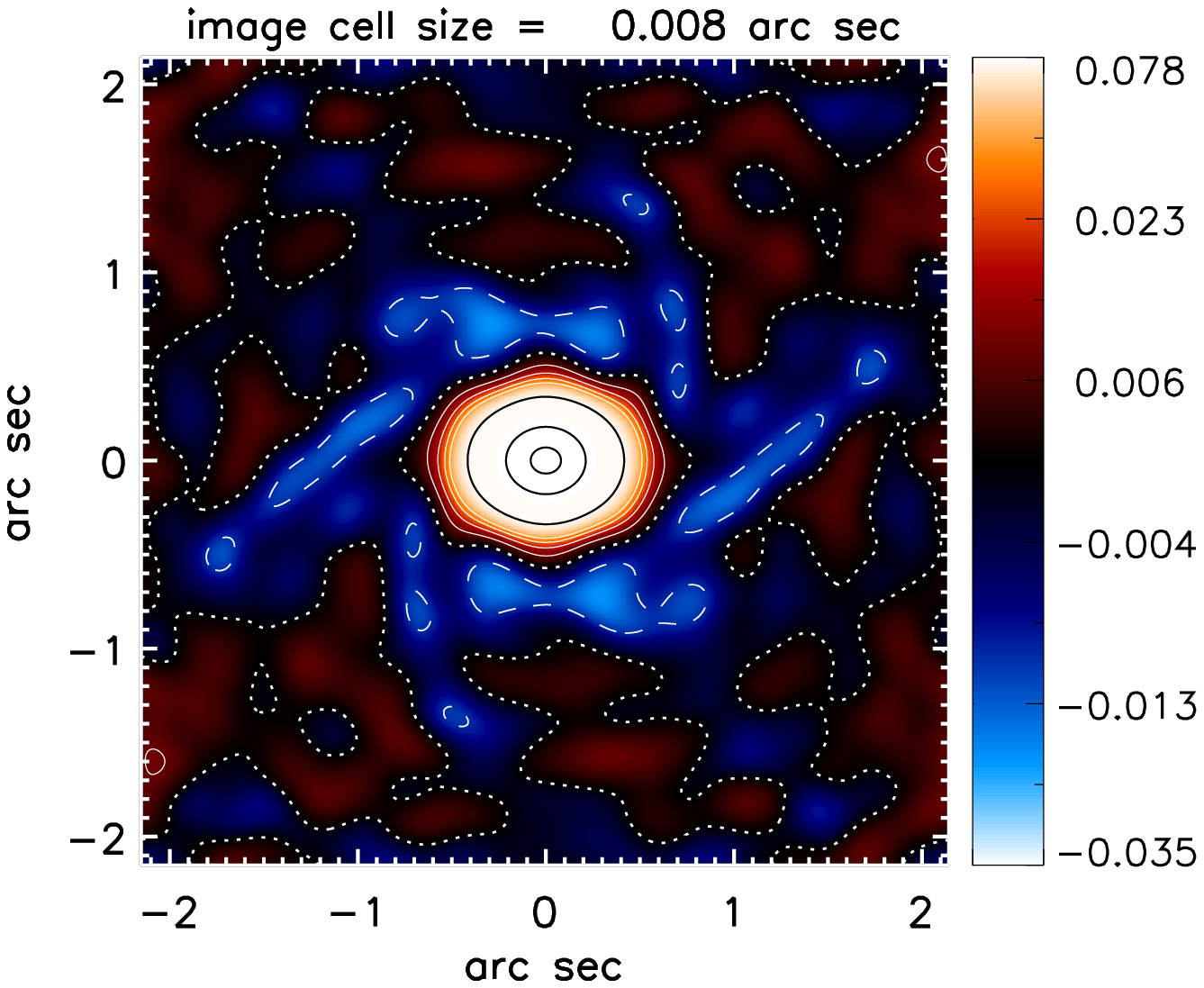} 
\includegraphics[trim=0.10in 0.3in 0.8in 0.2in, clip, width=2.75in]{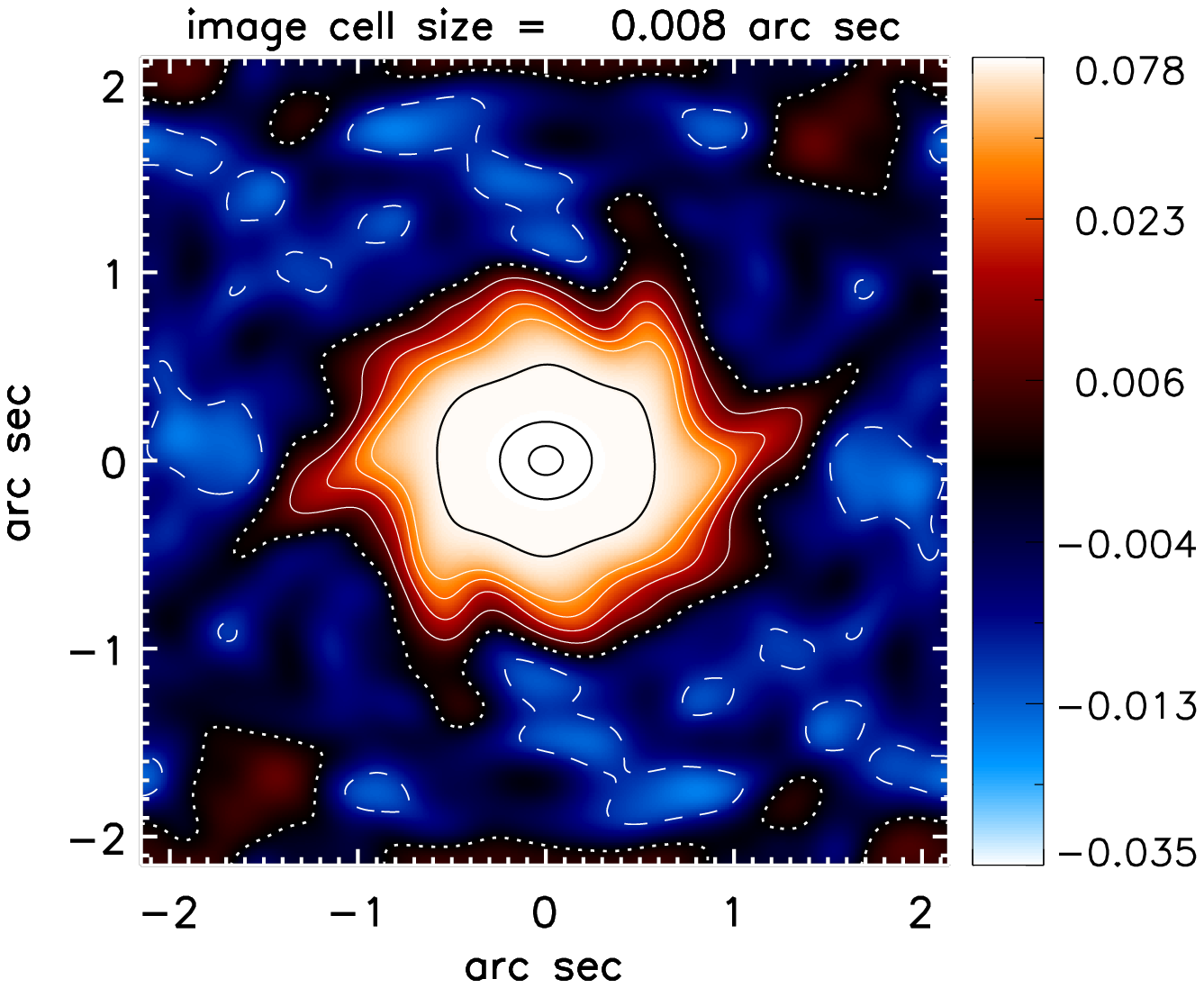}  \\
\end{array}
$
\caption{Antenna locations, UV distribution, and beam for 54-antenna 
spiral arrays showing the effect of 
the power law scaling between subarrays.
The pattern uses
the 6-element CW-array in figure
\ref{fig:s6} as the simple subarray. In the left column the successive 
subarrays are scaled by a factor of 1.25, and the right by 1.35.
The same arrays but with power law scalings of 1.05 and 1.15 are shown
in figure \ref{fig:ispiral6a}. The higher scalings here produce beams
of lower angular resolution.
Same format as figure \ref{fig:ispiral6a}. 
The beam power (dashed line) in the third panel down in both columns has been multiplied by 100.
The figures of merit are listed in table \ref{table:merit}.
}
\label{fig:ispiral6b}
\end{figure}

Differences in the amount of rotation between the levels of an H-spiral do not much affect the
properties of the beam; 
however, their relative scaling is quite important as the next four examples show (figures 
\ref{fig:ispiral6a} and \ref{fig:ispiral6b}).
These H-spirals are built from the basic 6-antenna pattern (figure \ref{fig:s6}) scaled 9 times. 
Different values are used in each example
for the power law exponent, 1.05, 1.15, 1.25, and 1.35. The figures of merit are listed
in table \ref{table:meritcont}.
At the lower value, the UV coverage and beam resemble those of a CW-array with its raptor resolution.
At the high end, the scaling produces a soft beam with wide Lorentzian wings similar to the beam of VLA
with its scaling factor of 1.716. However, the curved arms of the spiral produce a rounder beam with
less concentrated sidelobes than the straight arms of the VLA. 
Similarly, amateur astronomers sometimes use curved instead of straight
vanes in their Newtonian telescopes as supports for the secondary in order
to spread out the diffraction pattern of the supports.

\clearpage
\clearpage



\clearpage

\section{Random arrays} \label{random}

Successful array designs represent a curious blend of symmetry and randomness. 
Configurations with too much symmetry concentrate their separations in specific
patterns in the UV space
resulting in the equivalent of diffraction patterns in
the beam.  
Symmetry in circular and triangular arrays is discussed in \citet{Keto1997}
where the better performance of the triangular array is attributed to its lower degree of 
symmetry. 
Better performance for both arrays 
is obtained by slightly perturbing the antenna locations off of a regular
distribution around the circle or Reuleaux triangle. \citet{Keto1997} used a numerical
algorithm to optimize the perturbations. \citet{Arendt2000} 
simply used random perturbations. Their slightly randomized CW-arrays 
look to be at least nearly as good as those designed by optimization, 
but a direct comparison has not been done. 

In contrast, fully random patterns, rather than randomly perturbed patterns, 
generally do not have as good performance as more 
thoughtfully designed patterns.
The reason is that random patterns may by chance position antennas too close
to one another resulting in a concentration of separations and correspondingly
higher side lobes as well as larger gaps elsewhere
in the UV distribution. 

\begin{figure}[t]
$
\begin{array}{cc}
\includegraphics[trim=0.10in 0.3in 0.8in 0.2in, clip, width=2.75in]{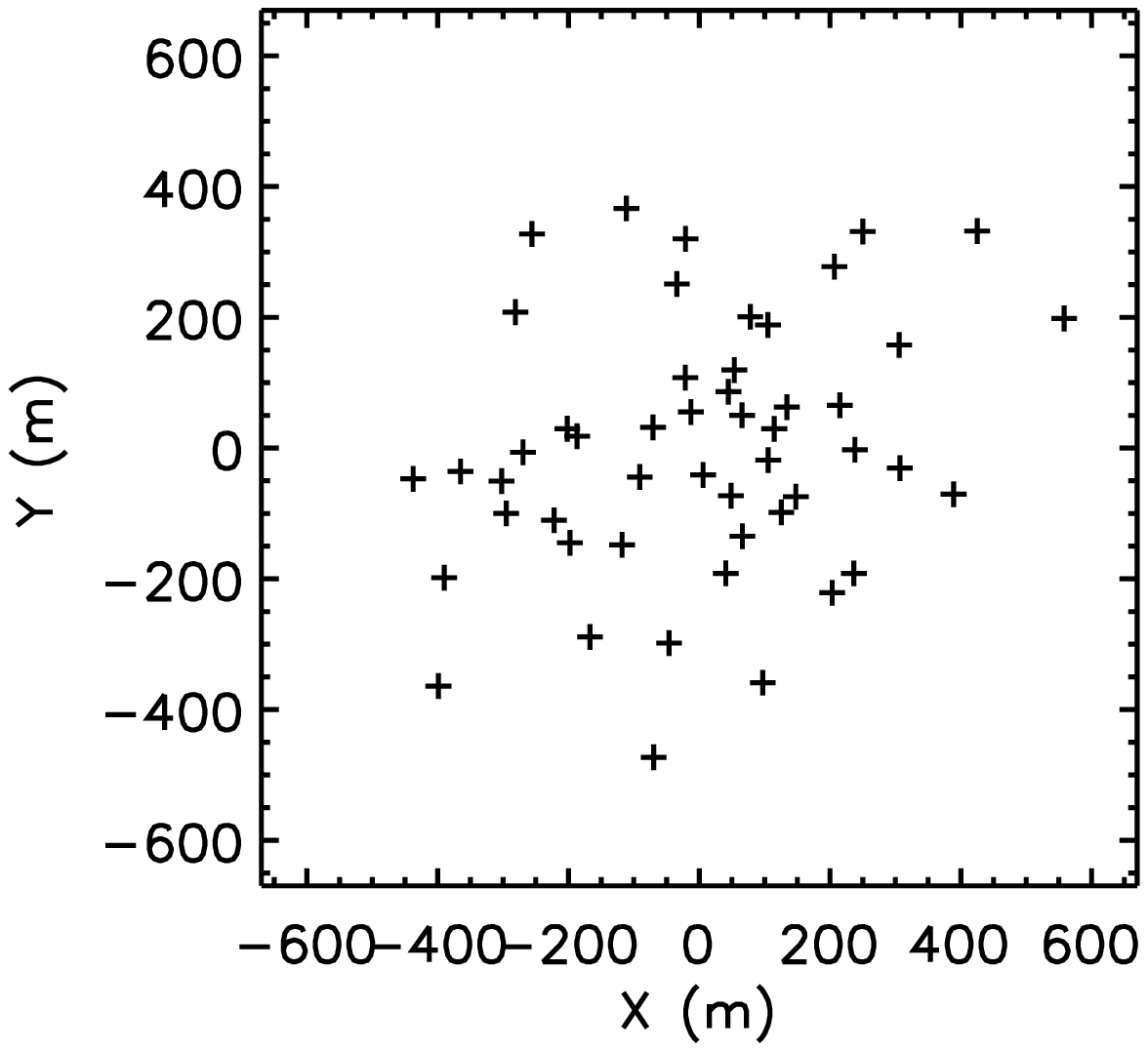} 
\includegraphics[trim=0.10in 0.3in 0.8in 0.2in, clip, width=2.75in]{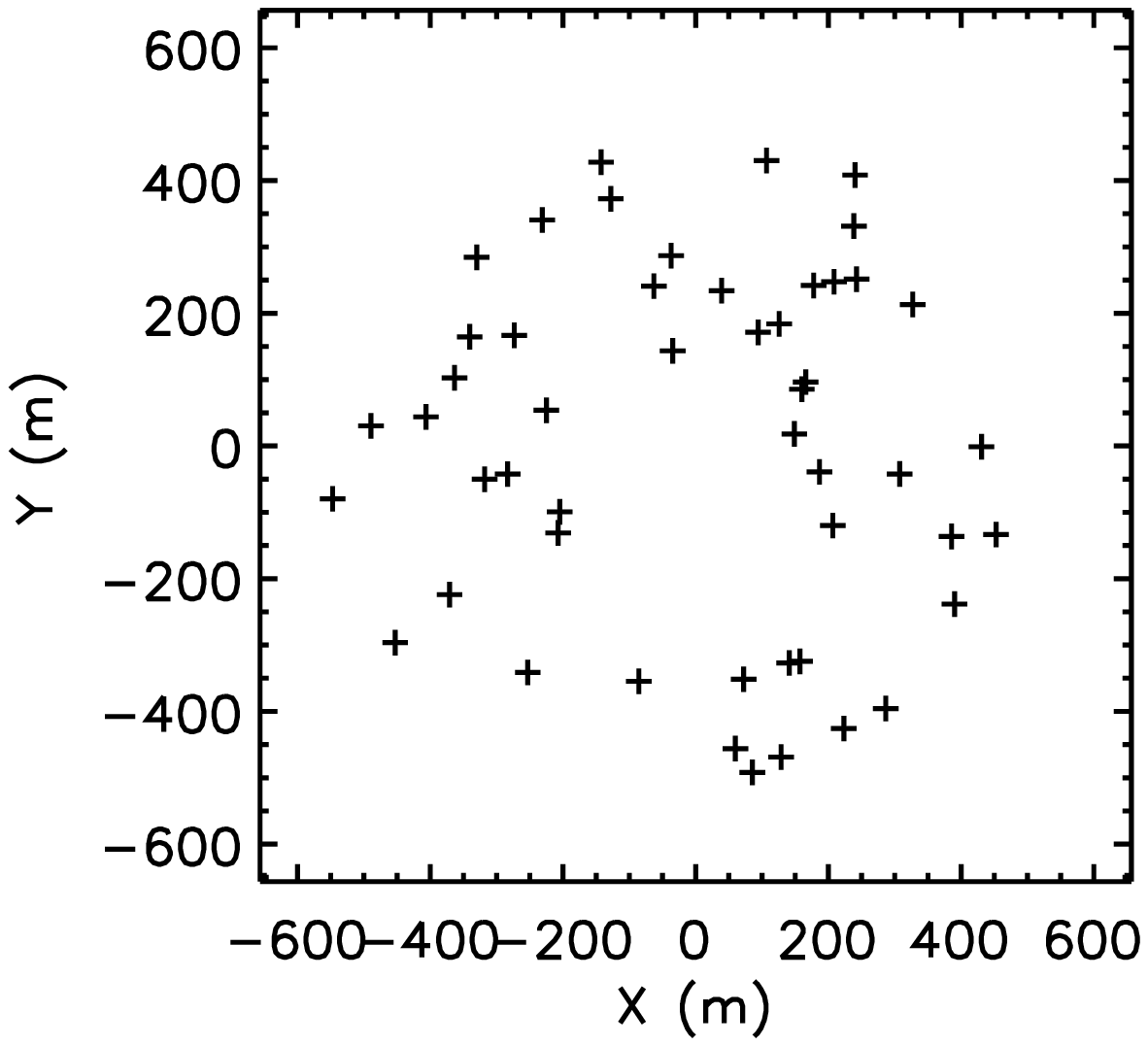} \\
\hskip 0.35in \includegraphics[trim=0.10in 0.3in 0.8in 0.2in, clip, width=2.75in]
{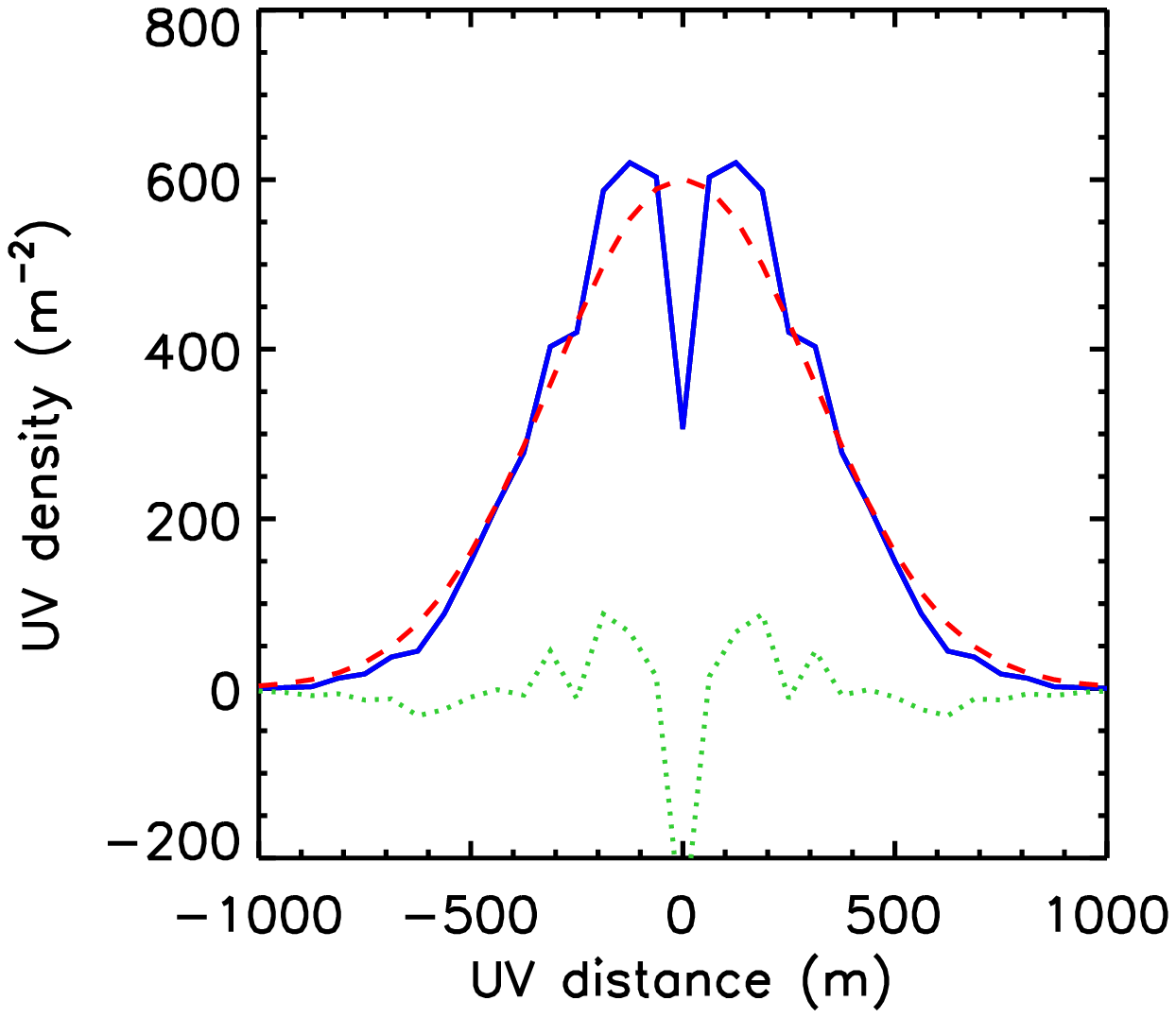}
\hskip 0.30in \includegraphics[trim=0.10in 0.3in 0.8in 0.2in, clip, width=2.75in]
{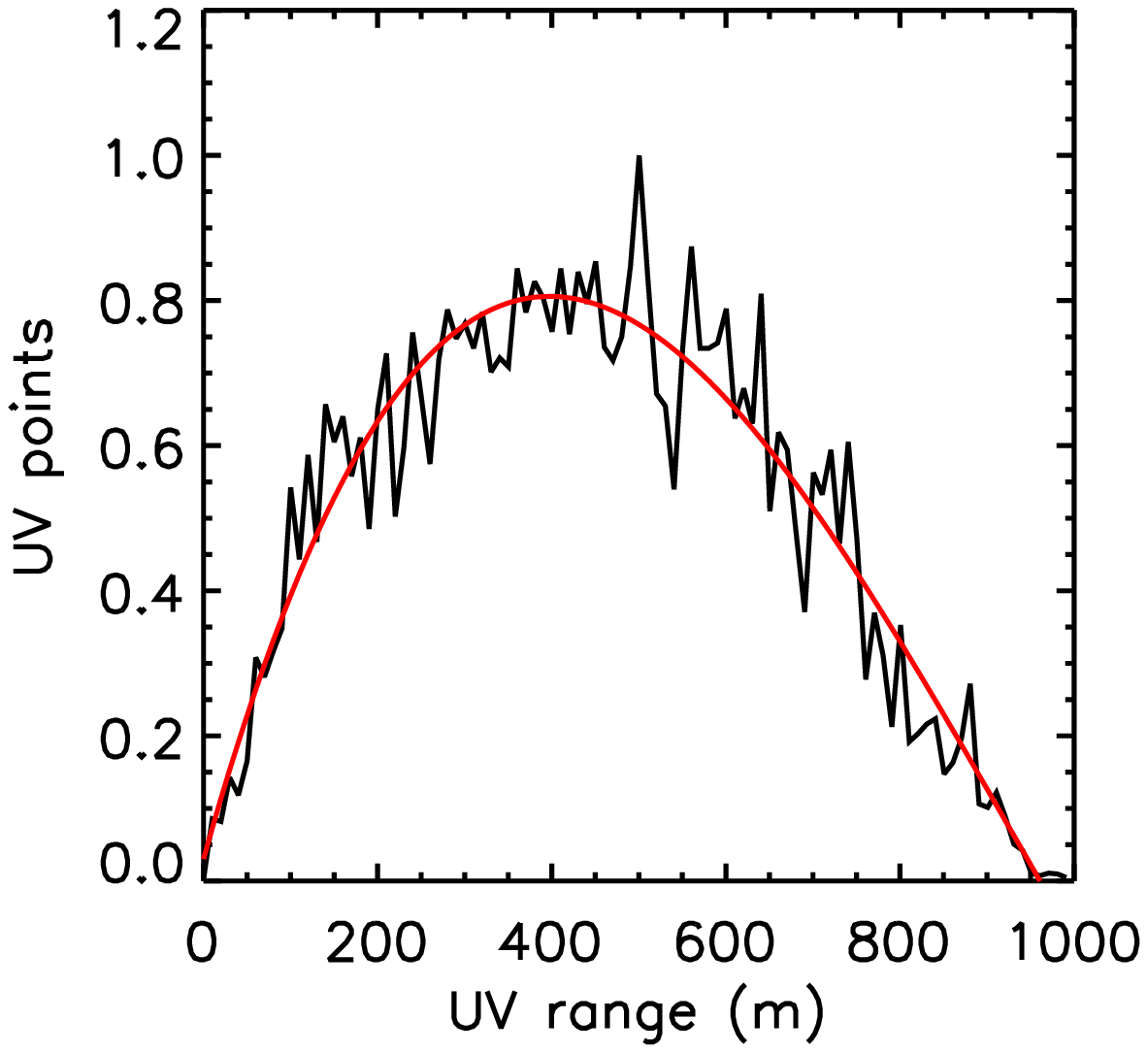} \\
\includegraphics[trim=0.10in 0.3in 0.8in 0.2in, clip, width=2.75in]{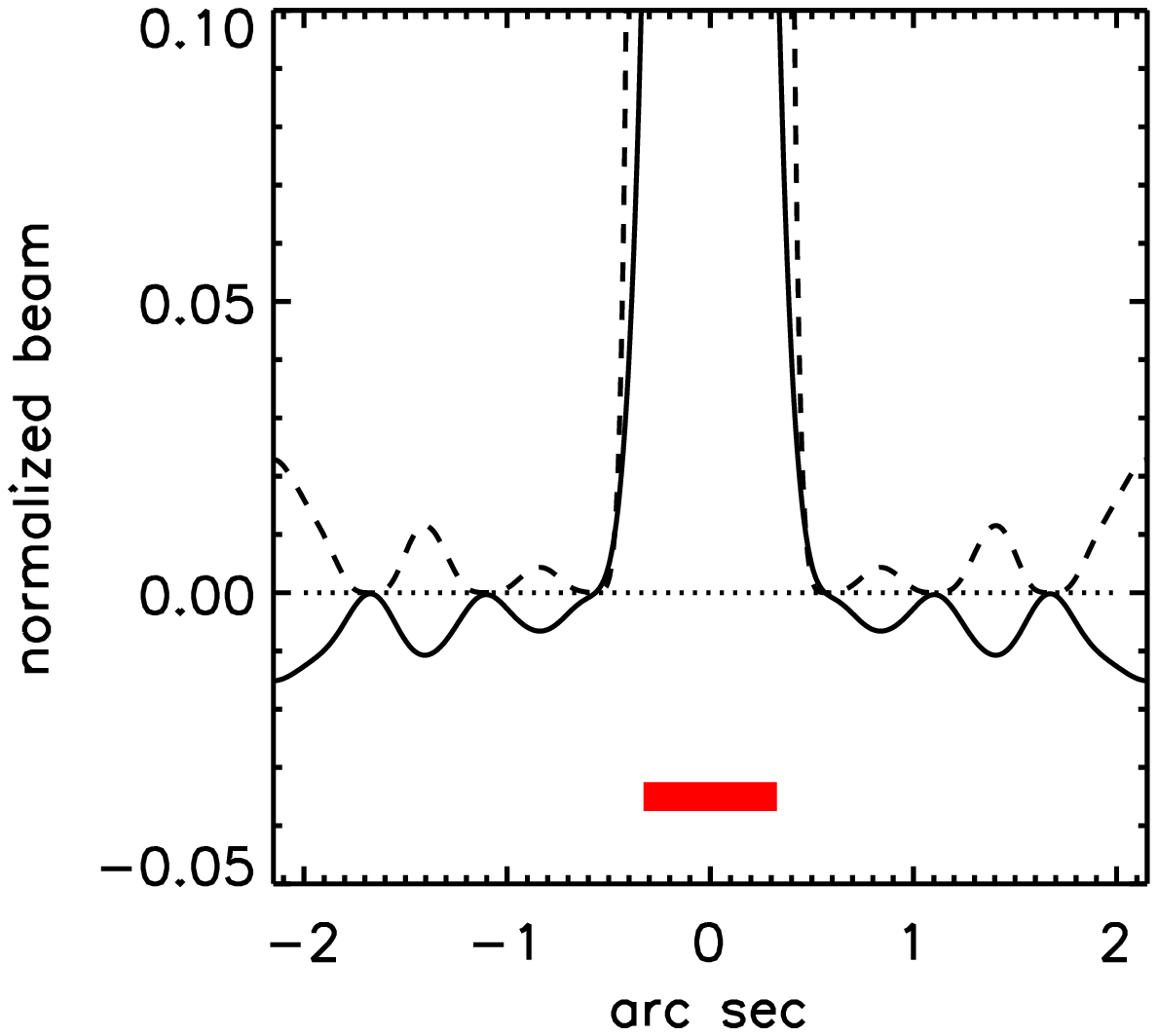} 
\includegraphics[trim=0.10in 0.3in 0.8in 0.2in, clip, width=2.75in]{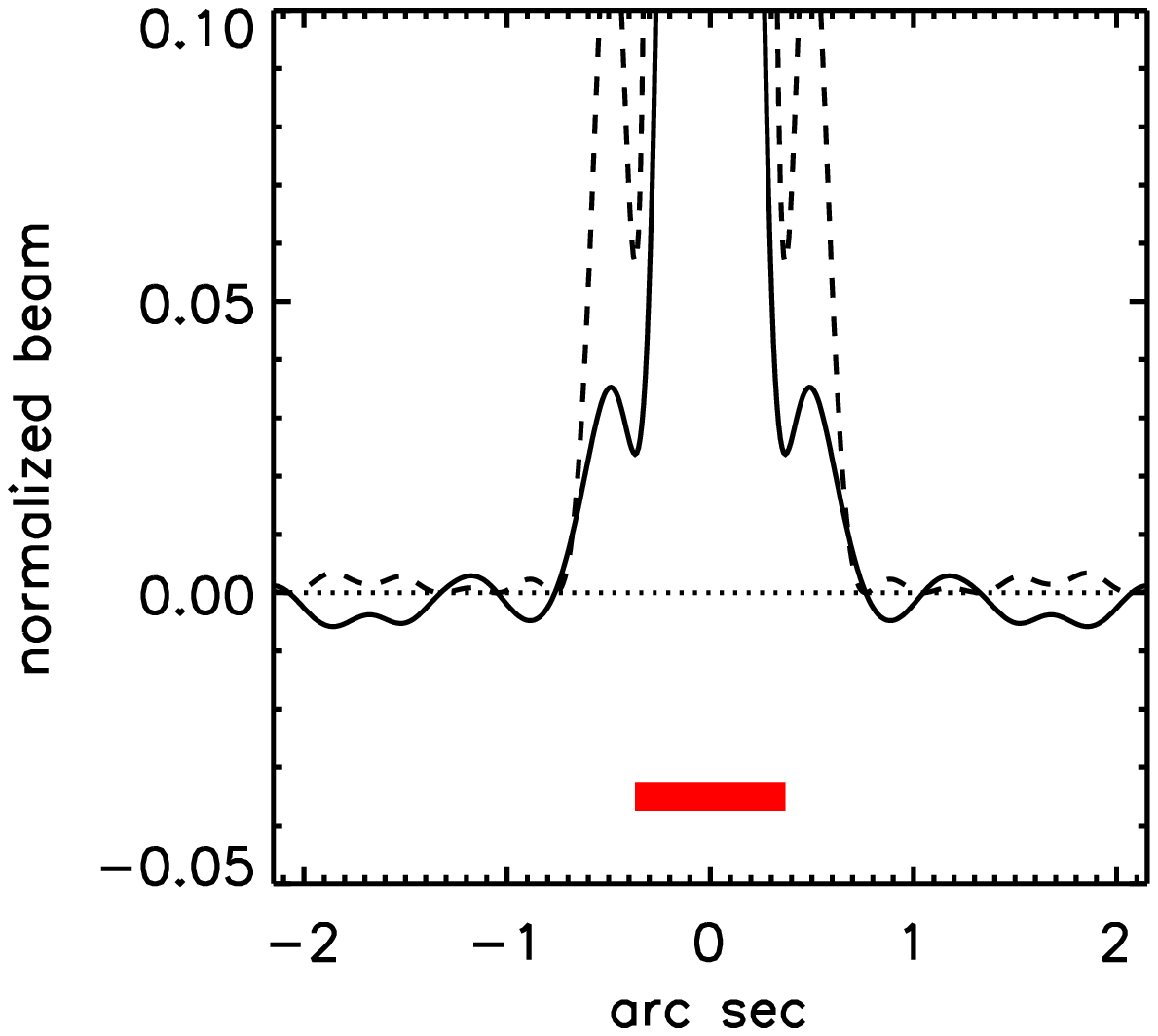} \\
\hskip 0.65in  \includegraphics[trim=0.10in 0.3in 0.8in 0.2in, clip, width=2.75in]{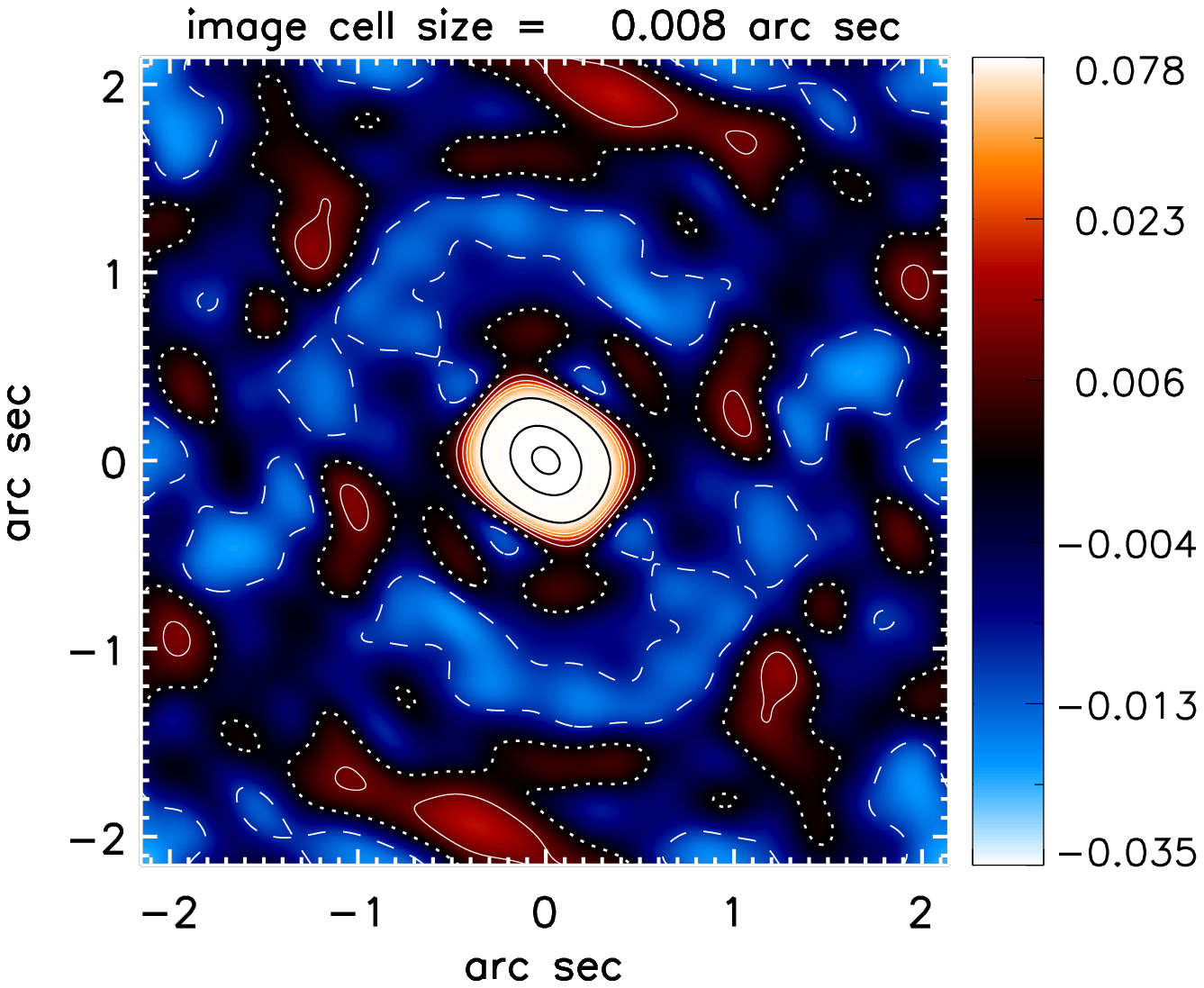} 
\hskip  0.10in \includegraphics[trim=0.10in 0.3in 0.8in 0.2in, clip, width=2.75in]{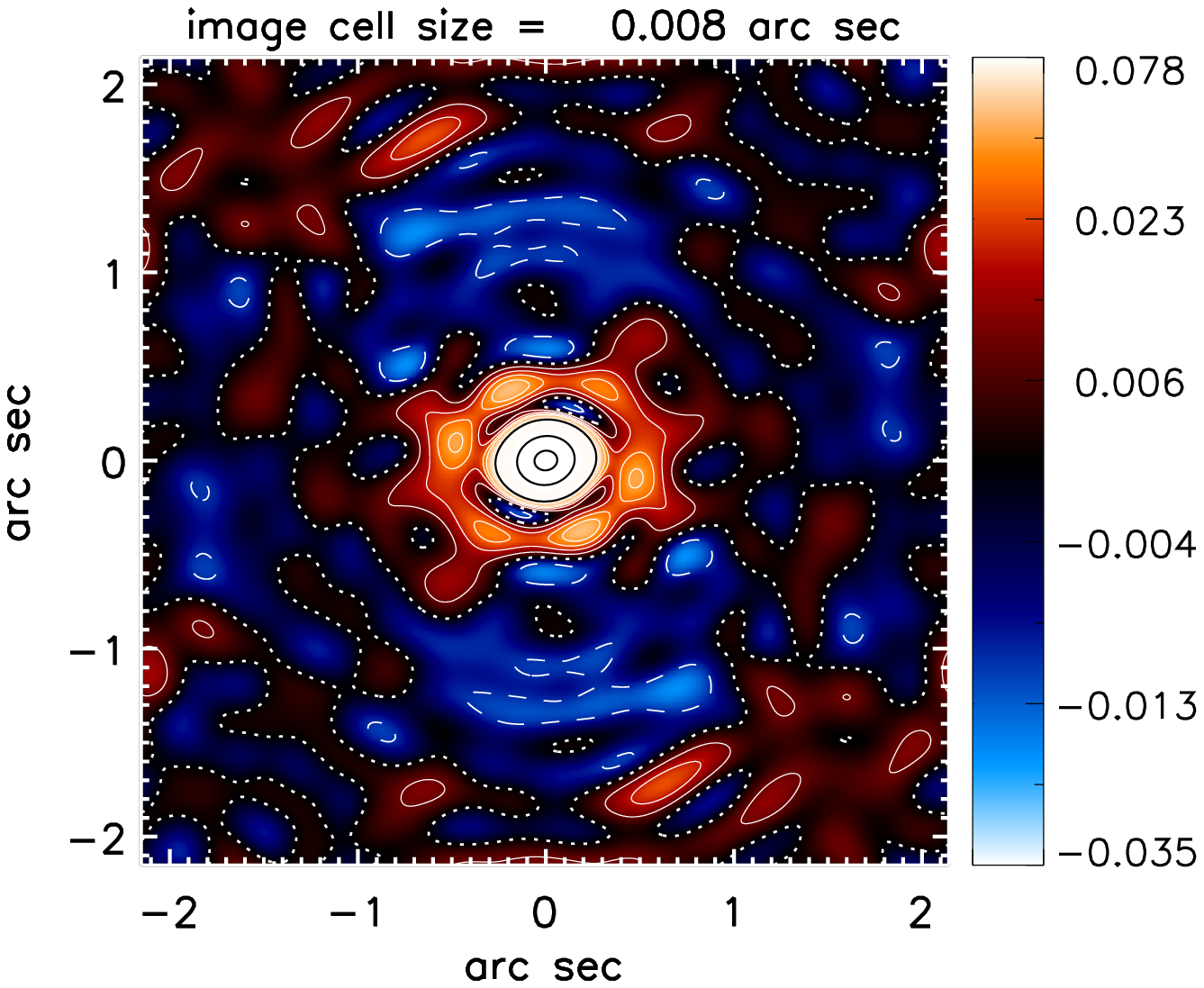}  \\
\end{array}
$
\caption{Antenna locations, UV distributions, and beams for two 
54-antenna arrays with different random 
antenna locations. The left column shows an array created from a Gaussian or
normal distribution of antenna locations. The right column shows an array
with a uniform distribution of antennas.
Same format as figure \ref{fig:ska36fix} except that the second panel in
the left column plots a trace across the UV distribution ({\it solid blue})
to show the
Gaussian distribution of antenna separations resulting from
the Gaussian distribution of locations. The dashed red line shows
a Gaussian fit and the dotted green line the residuals.
The figures of merit are listed in table \ref{table:merit}.
}
\label{fig:random}
\end{figure}

However, it is easy to create arrays with random patterns, and there is an
attractiveness to the savage simplicity of the process. For example, a Gaussian
distribution of antenna locations produces a Gaussian distribution of separations
because the auto-correlation function of two Gaussians is another Gaussian. This
in turn provides a Gaussian beam through the Fourier transform relationship. 
A Monte Carlo algorithm can generate a large number of possible arrays from
which the best can be selected according to whatever criteria. Figure \ref{fig:random}
shows an example selected for the best K-product
from a very small number of trials, specifically 100 trials.

In contrast, a uniform distribution of antenna locations results in an
array at the other end of our trade-off between high resolution and a concentrated
beam. In one dimension, the auto-correlation function of a uniform distribution
is a triangular function whose Fourier transform is the sinc function squared.
The power pattern of the beam then follows the fourth power of the sinc.
The beam of a 2-dimensional uniform distribution with a circular boundary
is conceptually the same, approximately a sinc$^2$ as a function of radius.
An example of a two-dimensional array based on a uniform random
distribution of antennas is shown in figure \ref{fig:random}. This array
had the best angular resolution out of 100 trials.
The figures of merit are listed in table \ref{table:merit}.



\clearpage

\section{Outriggers}\label{outriggers}

Interferometers are sometimes built with a few outrigger antennas around a main array
to obtain a limited sample of longer Fourier components. What do the H-arrays 
say about the placement of outriggers?
H-arrays may be designed with concentric or asymmetric outriggers. Each has different
properties. 

\begin{figure}[t]
$
\begin{array}{cc}
\includegraphics[trim=0.10in 0.3in 0.8in 0.2in, clip, width=2.75in]{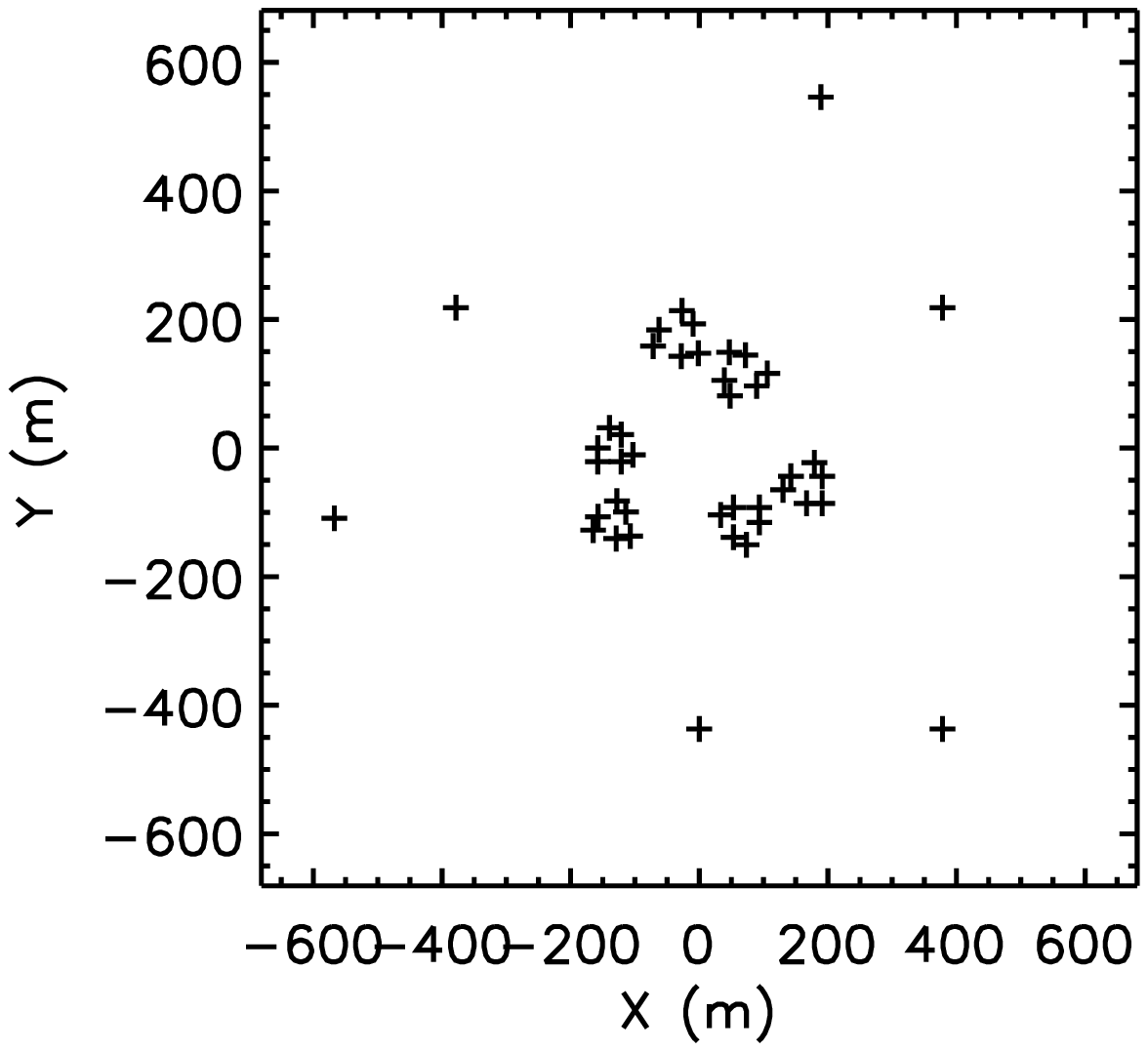} 
\includegraphics[trim=0.10in 0.3in 0.8in 0.2in, clip, width=2.75in]{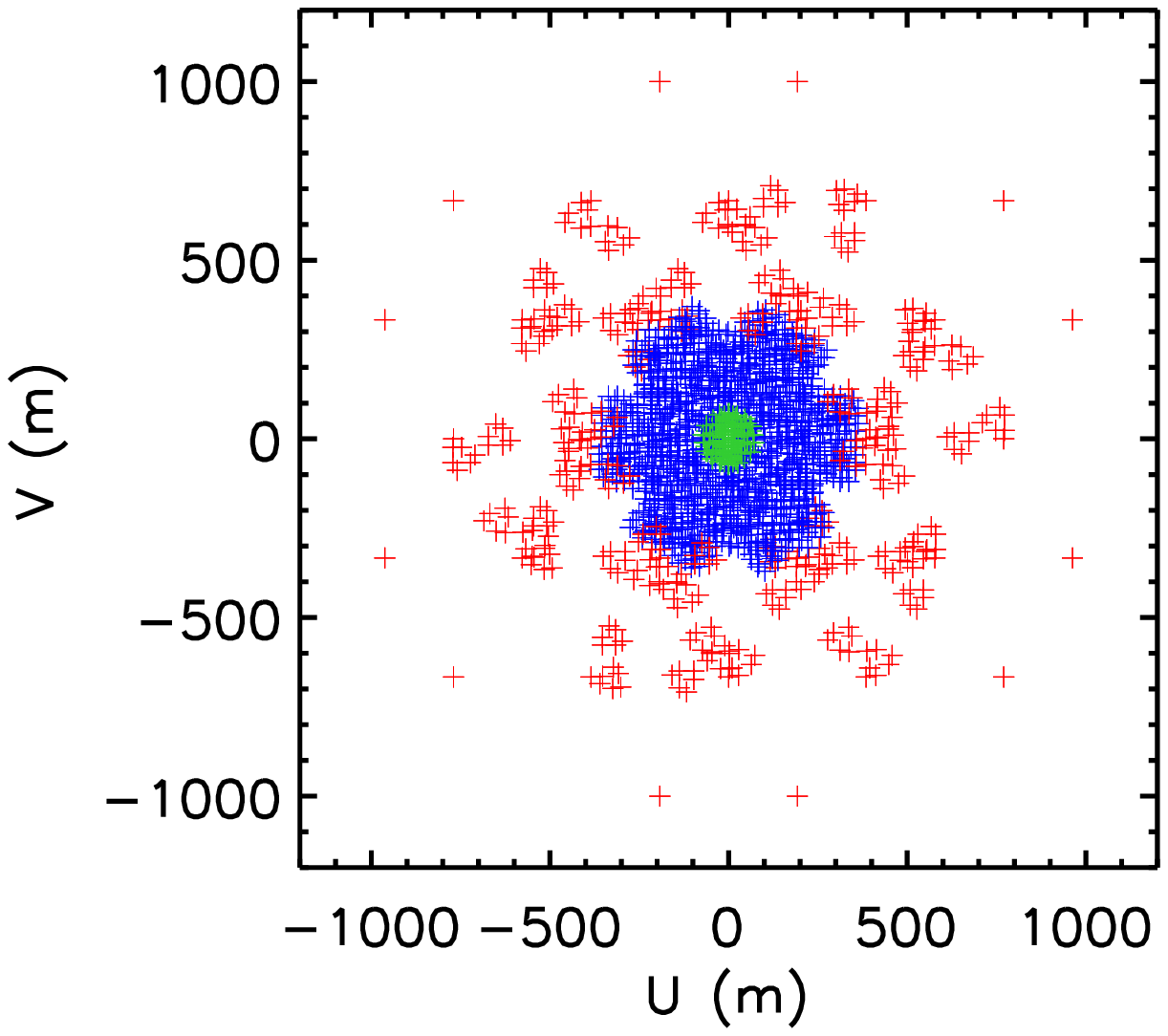} \\
\hskip 0.65in \includegraphics[trim=0.10in 0.3in 0.8in 0.2in, clip, width=2.75in]{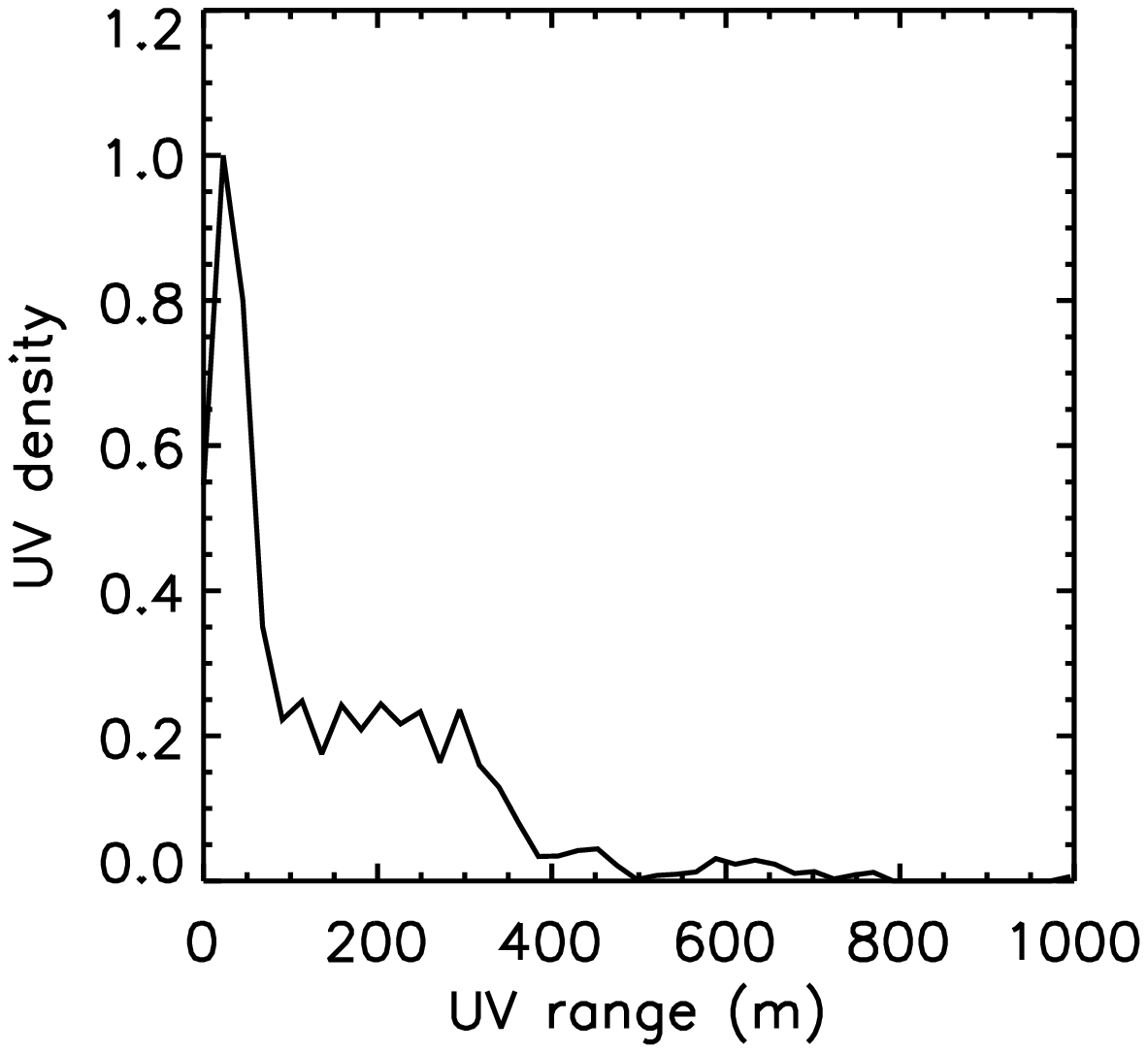} 
\includegraphics[trim=0.10in 0.3in 0.8in 0.2in, clip, width=2.75in]{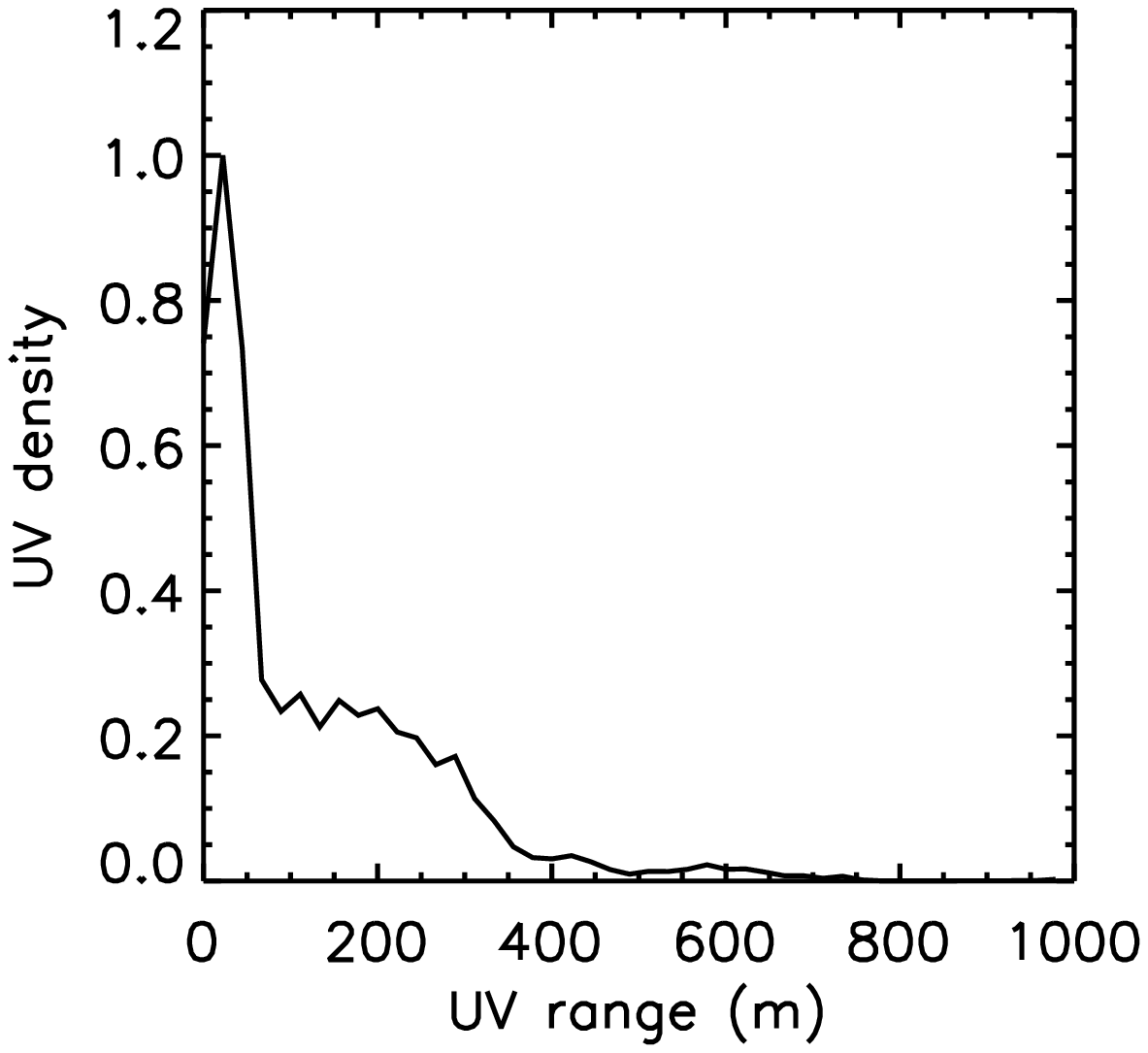} \\
\end{array}
$
\caption{An s6p6-array with 6 concentric outriggers. The s6p6-array is the same as
shown in figure \ref{fig:ska36_sinc_revB}, and the 6 outriggers take the positions
of the 6-element subarray in figure \ref{fig:s6} scaled by a factor of 5.5
times the first s-level subarray.
Same format as figure \ref{fig:ska36_sinc_revB} except the plots of the beam are not shown.
}
\label{fig:outrigger2}
\end{figure}

Figure \ref{fig:outrigger2} shows a concentric configuration with the main
array at the center of 6 outriggers. In this example, the main array is the
36-antenna s6p6 array designed for high angular resolution; the outriggers
use the locations of the 6-antenna subarray to make a 42-antenna s6p6+d6 array. 
Figure \ref{fig:outrigger2} shows four concentric zones of UV coverage. First, 
on the largest scale, the 30
separations between the outriggers uniformly cover the UV space within the
boundary. 
The next zone includes the separations between the outriggers and the main array,
also colored red in figure \ref{fig:outrigger2}. 
The two inner zones, blue and green in figure \ref{fig:outrigger2} ({\it top right}),
are the separations of the main s6p6 H-array as described 
in section \S \ref{example36}.

In most applications, the
inner UV points from the main array 
would be used in imaging and the outer UV points from the outriggers would
be used separately to
locate point sources to high angular resolution. The outer UV points would generally
not be used for imaging because they create
uncancelled sine waves that do not improve the image
fidelity. If the inner array, not including the outriggers, is scaled to 1000m and the outriggers
are not included in the Fourier transform of the beam, then the array design is the
same as shown in figure \ref{fig:ska36_sinc_revB}, 
and the beam pattern and figures of merit 
are also the same. The UV coverage of the outriggers alone 
would be as shown in figure \ref{fig:s6}.

\begin{figure}[t]
$
\begin{array}{cc}
\includegraphics[trim=0.10in 0.3in 0.8in 0.2in, clip, width=2.75in]{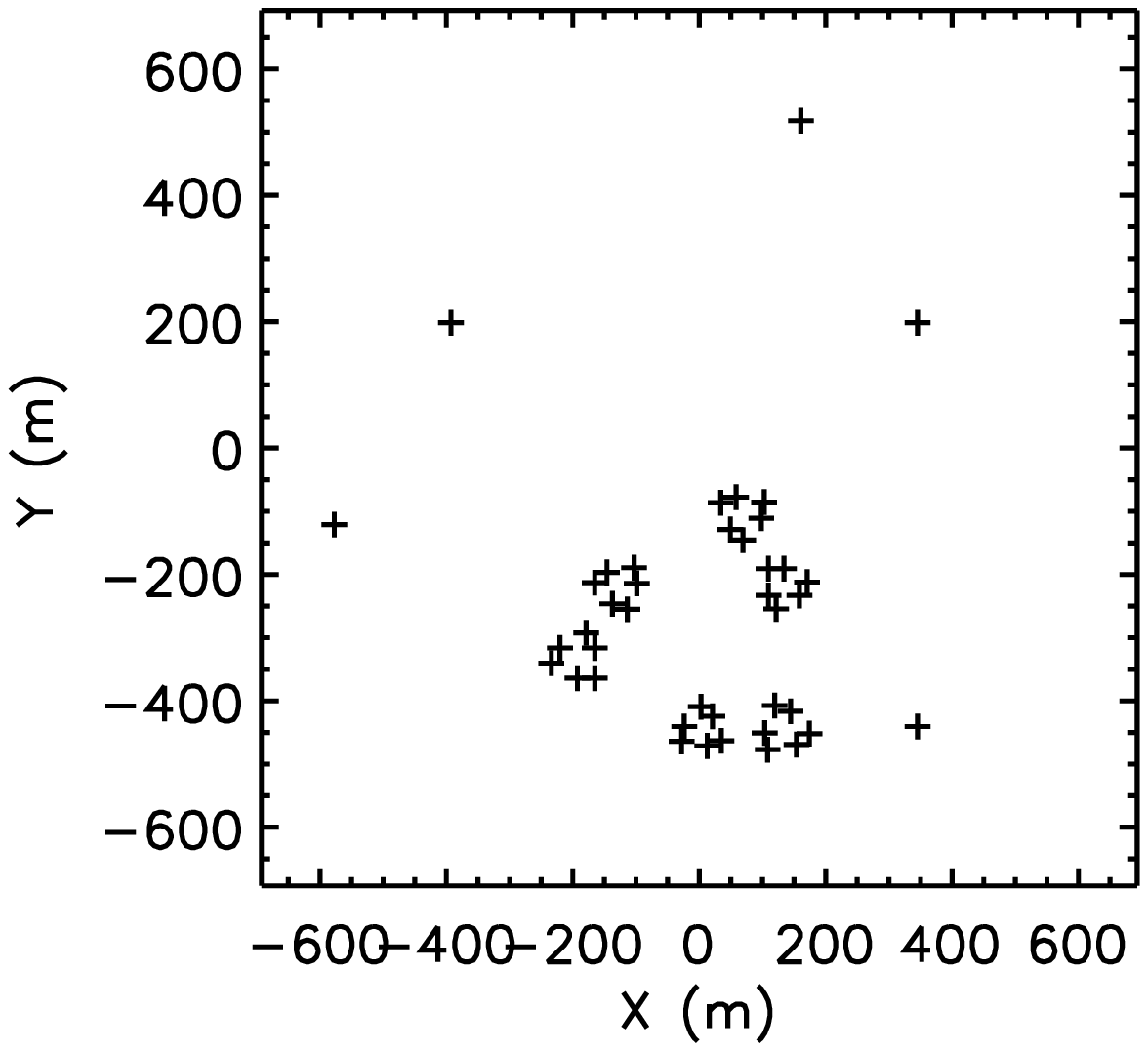} 
\includegraphics[trim=0.10in 0.3in 0.8in 0.2in, clip, width=2.75in]
{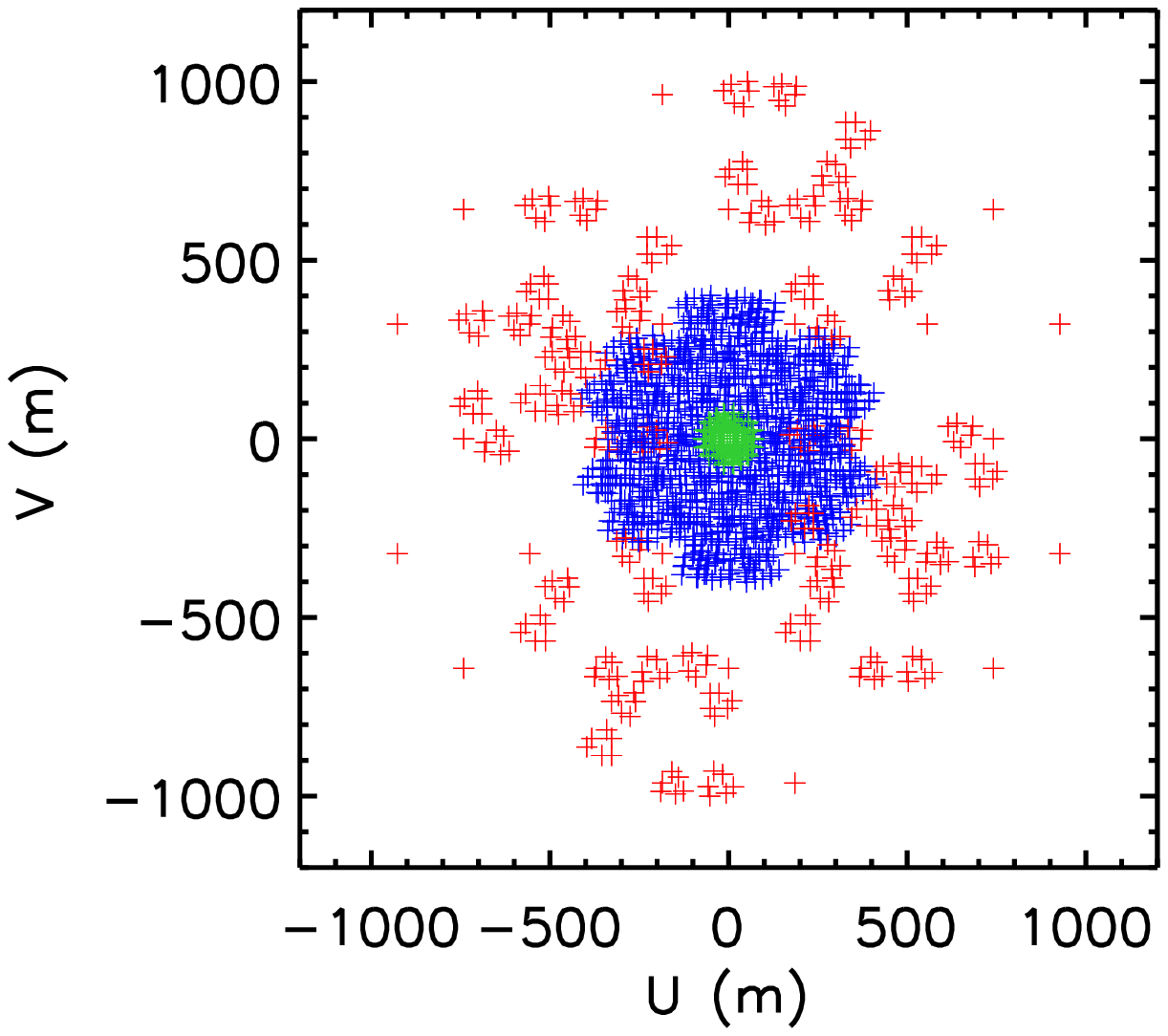} \\
\hskip 0.65in \includegraphics[trim=0.10in 0.3in 0.8in 0.2in, clip, width=2.75in]{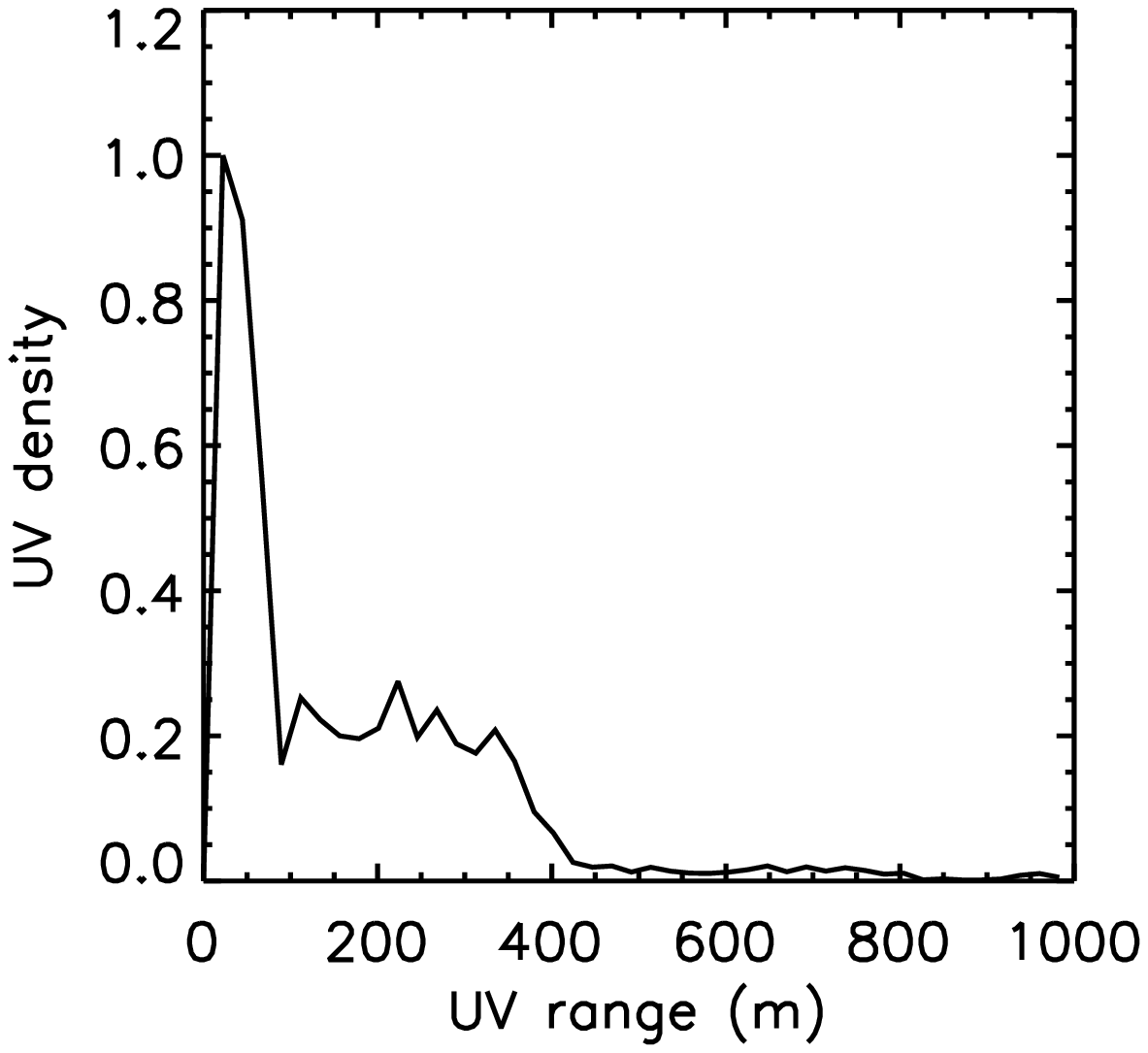} 
\includegraphics[trim=0.10in 0.3in 0.8in 0.2in, clip, width=2.75in]{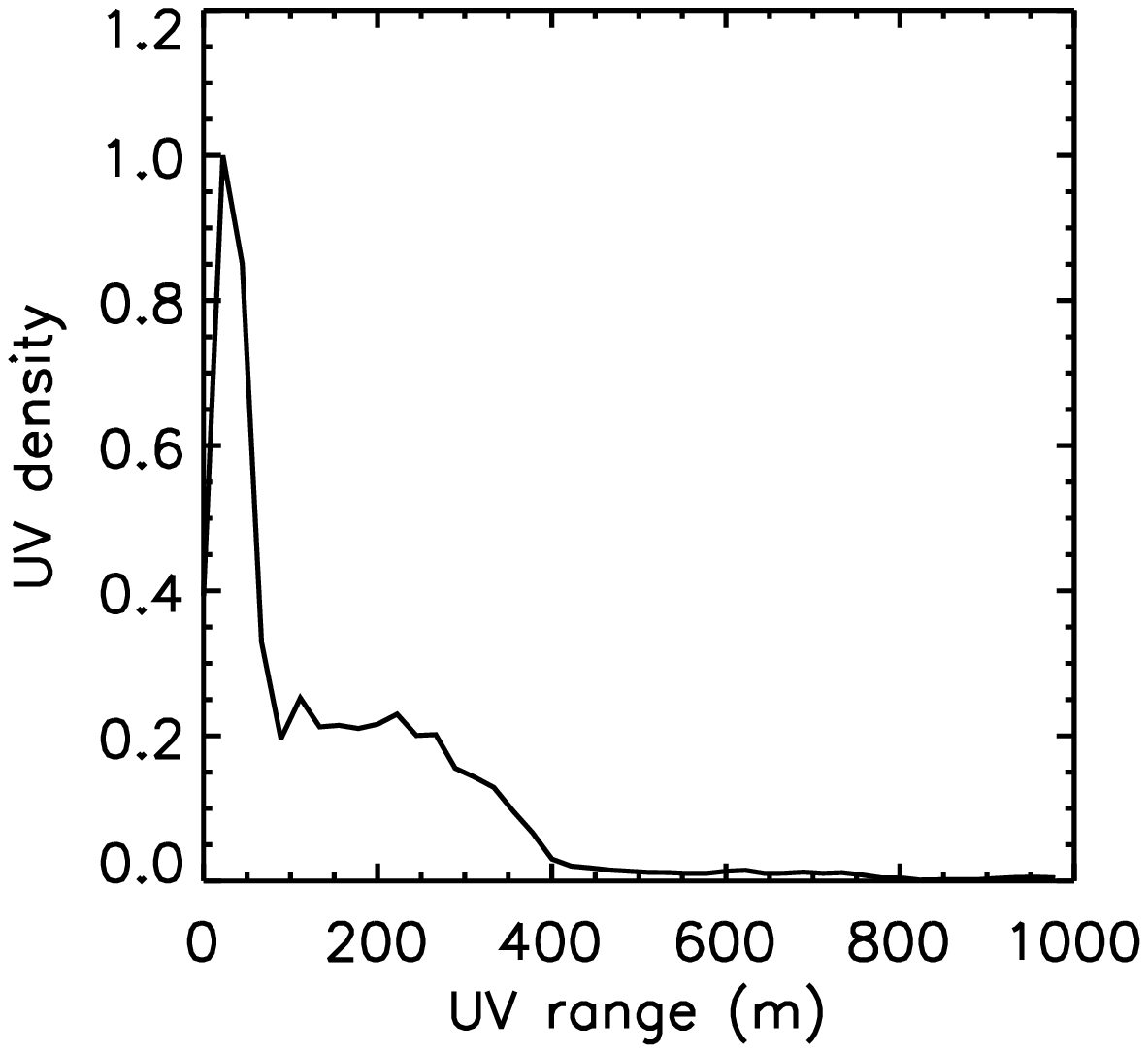} \\
\end{array}
$
\caption{An s6p6-array with 5 asymmetric outriggers. The s6p6-array is the same as
shown in figure \ref{fig:ska36_sinc_revB}, and the 5 outriggers take 5 of the positions
of the 6-element subarray in figure \ref{fig:s6} scaled by a factor of 5.5
times the first s-level subarray. The main s6p6-array takes the sixth position of the s-level
subarray.
Same format as figure \ref{fig:outrigger2}.
}
\label{fig:ska216thin}
\end{figure}

Alternatively, we can design an asymmetric outrigger array
with the main array placed at the one of the 6 locations of the 
6-element pattern as shown in figure \ref{fig:ska216thin}. This creates
a 41 antenna array, one less than the concentric design. This array again shows
the four zones of UV coverage but has higher angular resolution, FWHM = 0.30 arc
seconds because the location of the main array on the border of the 6-element
CW pattern creates more longer spacings than the concentric placement. However,
the UV coverage and the beam are asymmetric with more of the longer spacings in
the north-south direction.
The asymmetric outrigger array results in better
signal to noise on some of the longer separations whereas the concentric
design improves some of the shorter.

\section{Beam patterns with different observations}\label{snapshots}

How does the performance of an array change for different types of 
observations, for example when
tracking targets at different declinations that do not pass directly overhead? 
In astronomical imaging, the 
beams of good arrays, those that have a good distribution of antenna separations,
generally maintain their good properties
when observing sources in either instantaneous (snapshot) imaging or earth rotation
synthesis and also when observing at different astronomical declinations. 

\begin{figure}[t]
$
\begin{array}{cc}
\includegraphics[trim=0.10in 0.3in 0.8in 0.2in, clip, width=2.75in]{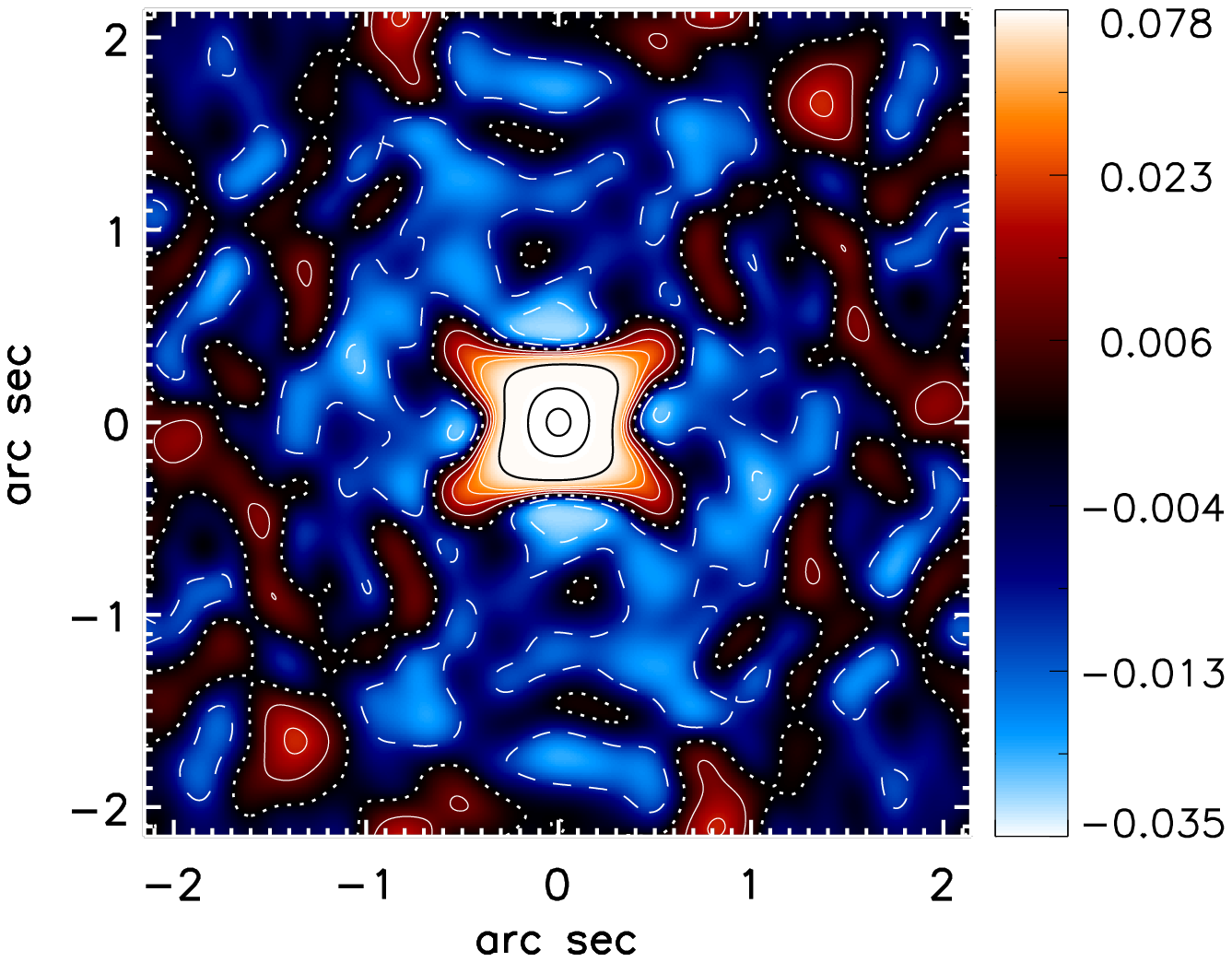} 
\includegraphics[trim=0.10in 0.3in 0.8in 0.2in, clip, width=2.75in]{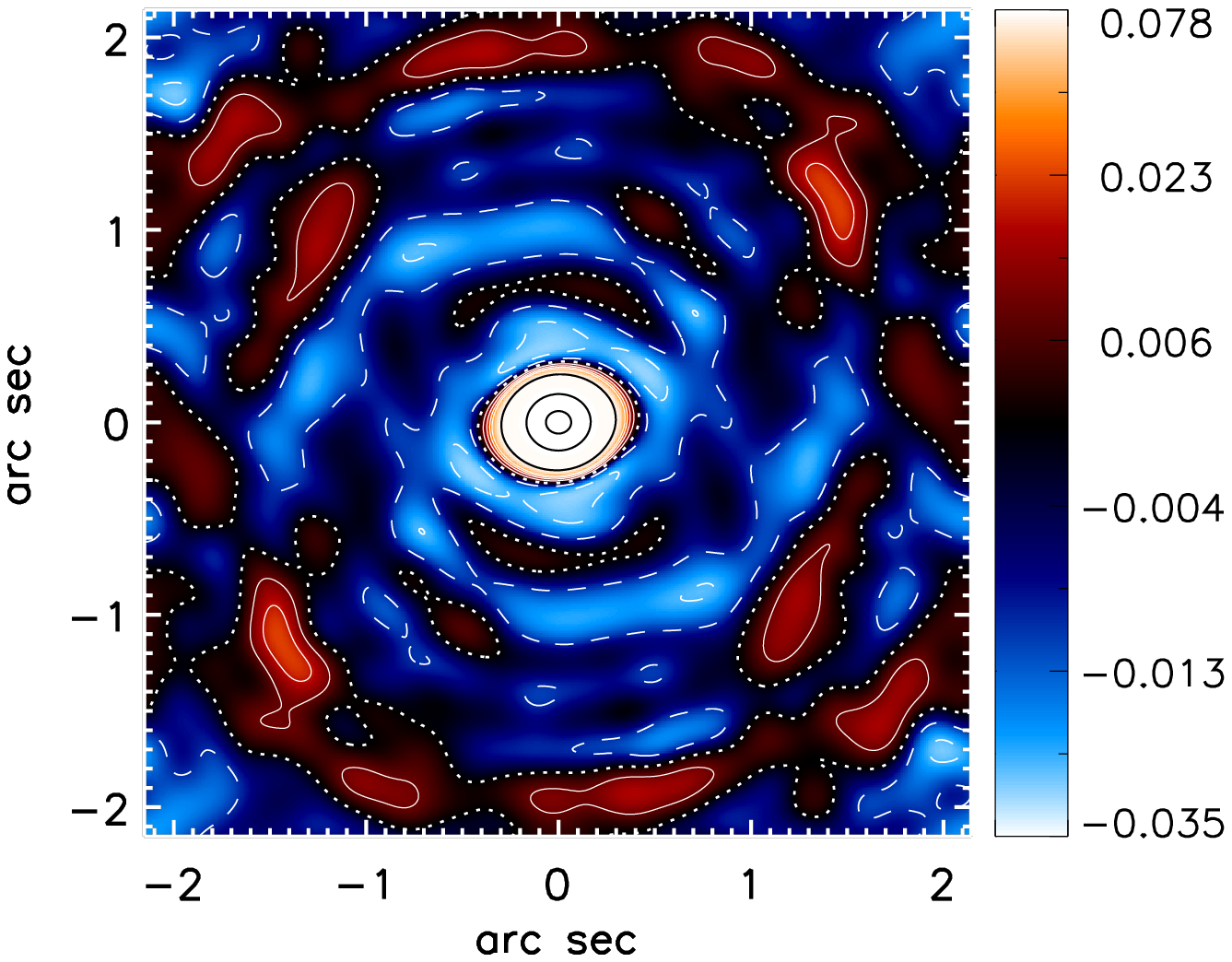} \\
\includegraphics[trim=0.10in 0.3in 0.8in 0.2in, clip, width=2.75in]{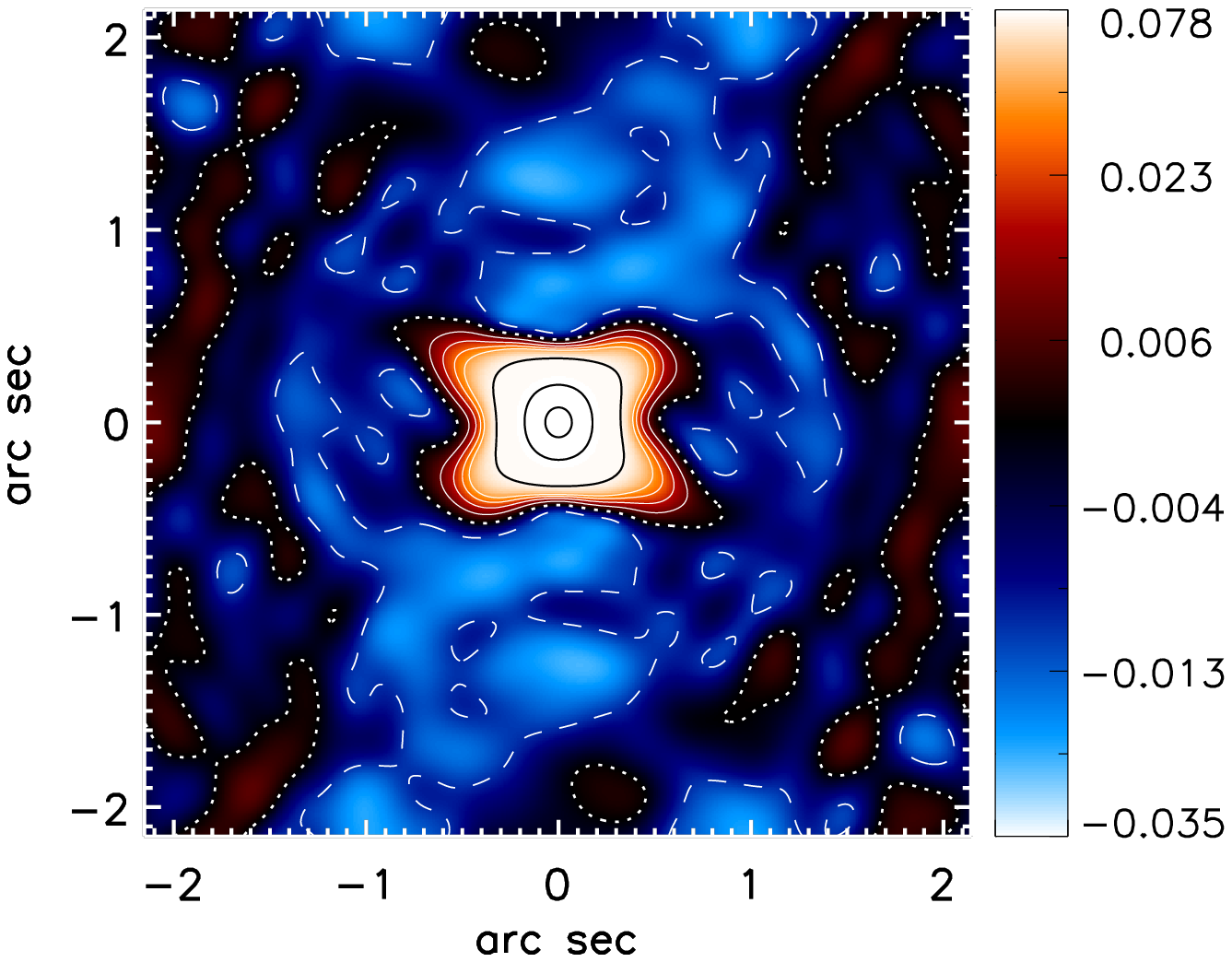} 
\includegraphics[trim=0.10in 0.3in 0.8in 0.2in, clip, width=2.75in]{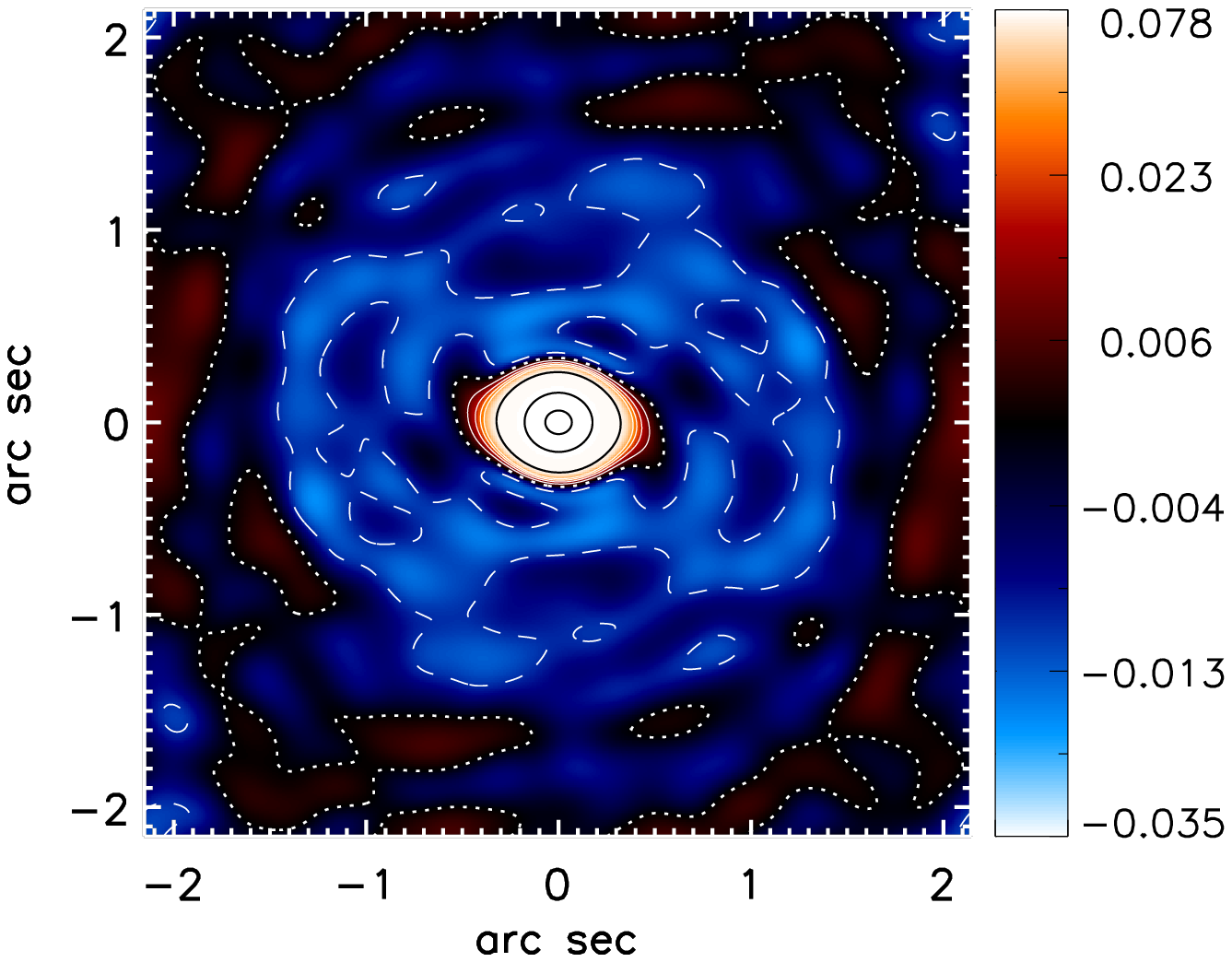} \\
\end{array}
$
\caption{
Beam patterns for 3 H-arrays tracking targets to the 
north ({\it right}) and south ({\it left}).
Top and bottom show the beam patterns for 
the 36 antenna configuration  s6p6 of figure \ref{fig:ska36fix}
and the 54 antenna configuration  s3p3d6 of figure \ref{fig:ska54-n}, respectively.
}
\label{fig:declinations}
\end{figure}

The previous examples show the beam patterns for a source that transits 
through zenith. Figure \ref{fig:declinations}
shows the beam patterns for two low side lobe designs, the 36-antenna s6p6, 
and the 54-antenna s3p3d6
configurations, tracking sources at declinations 
of $-23^\circ$ and $+46^\circ$ assuming that
the array is at a latitude of $+23^\circ$. 
The FWHM changes with declination as the baselines
are shortened by projection, but the arrays generally maintain
their good characteristics.


\section{Numerical optimization}

These examples show how easy 
it is to build H-arrays with excellent performance without
numerical optimization, but the arrays
could still be improved.
The optimization problem is much
simpler with H-arrays because the number of
parameters that describe the arrays,
the relative scaling of the levels and the subarrays, and the rotation 
of the subarrays, is fewer
than the number of antennas. For example, the description of a three-level
array with 216 antennas requires 14 numbers: two scalings
between the levels, and six scalings and six rotations for the subarrays.
An H-spiral can be described by as few two numbers.
In contrast, the description of the antenna locations individually requires 432 numbers
for the 216 X,Y pairs.
Optimizing the antenna positions with respect to each other is
a combinatorially explosive problem with the number of possible
configurations increasing exponentially with the number of antennas.
It is difficult for algorithms that optimize
the antenna positions to demonstrate convergence to a global minimum. 
One difficulty is that quasi-random configurations with 
quasi-Gaussian distributions represent particularly seductive local minima.
Another problem is that ordered patterns such as the Reuleaux triangle
require a degree of coherence that is difficult to obtain for algorithms that move
one antenna at a time.

\section{Comparison to other designs}\label{alma}

\begin{figure}[t]
$
\begin{array}{cc}
\includegraphics[trim=0.10in 0.3in 0.8in 0.2in, clip, width=2.75in]{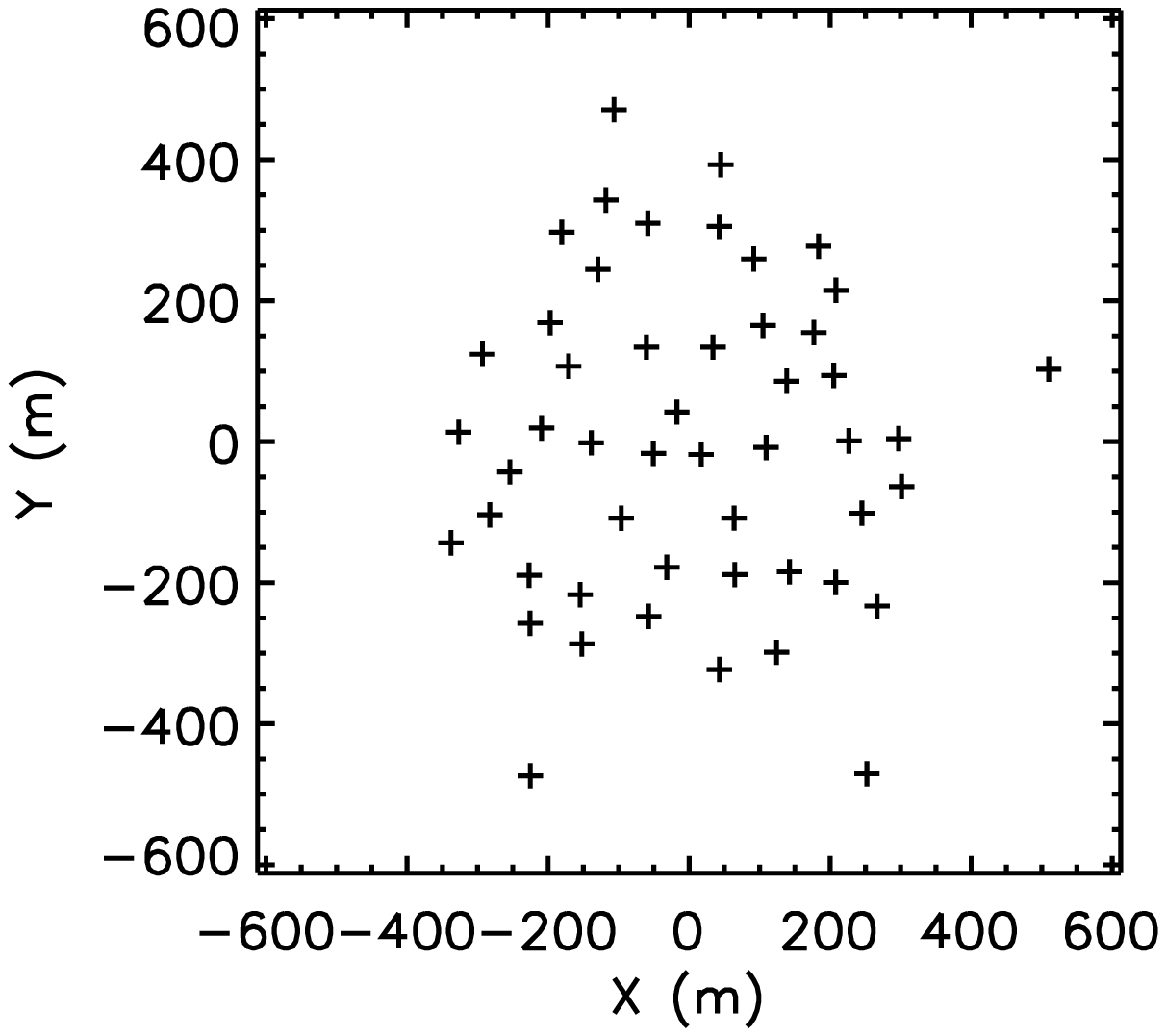} 
\includegraphics[trim=0.10in 0.3in 0.8in 0.2in, clip, width=2.75in]{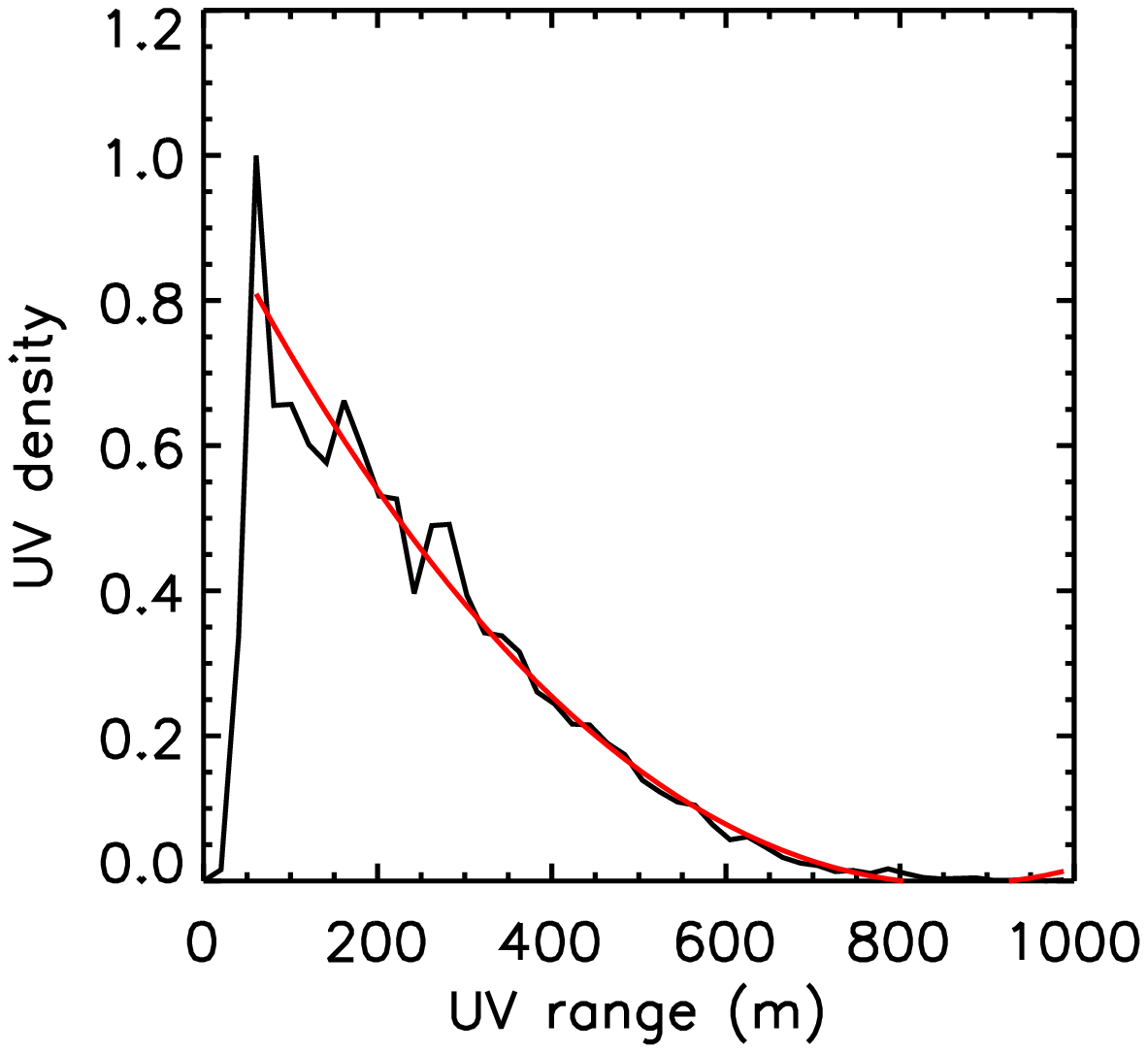} \\
\includegraphics[trim=0.10in 0.3in 0.8in 0.2in, clip, width=2.75in]{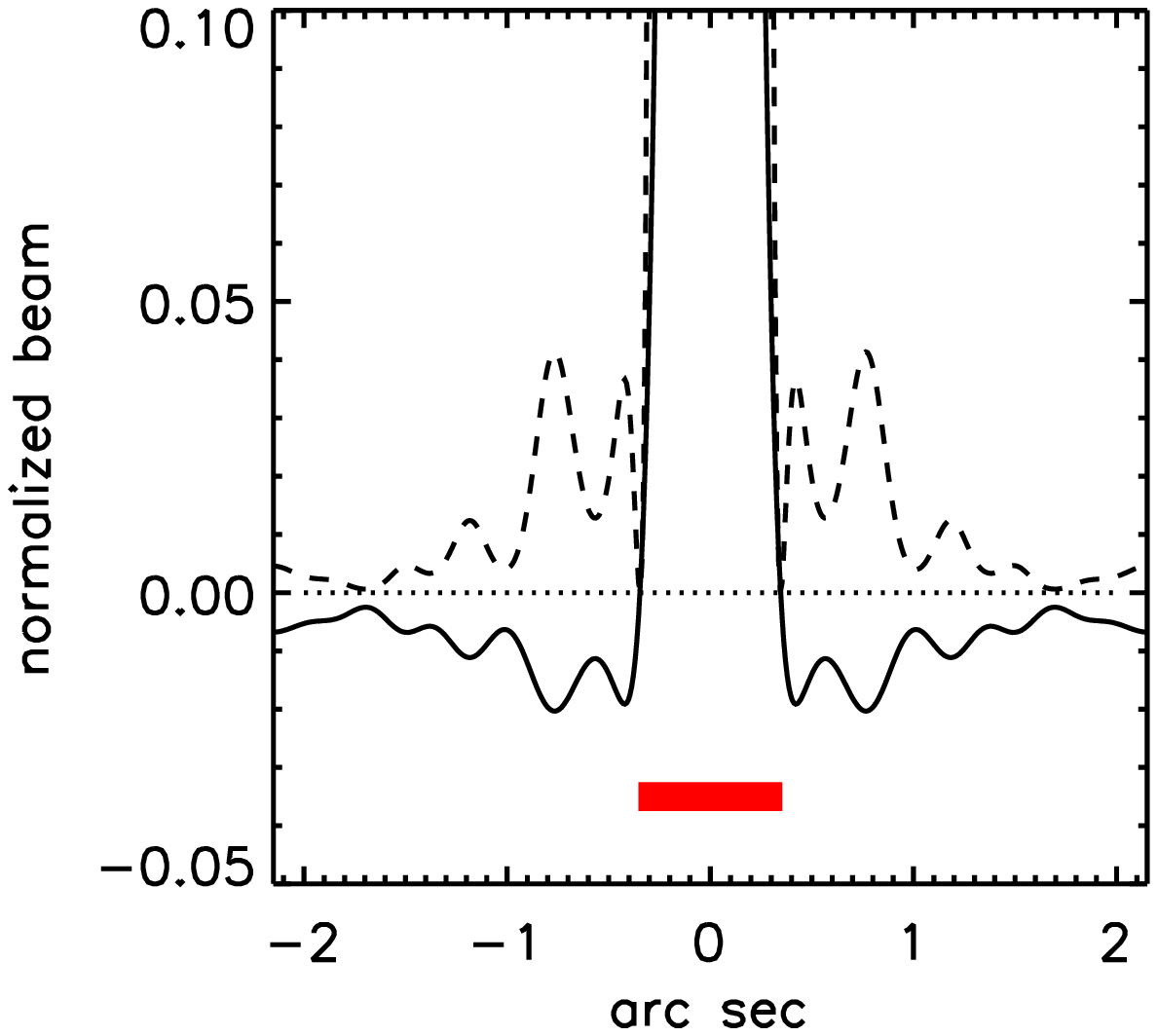} 
\includegraphics[trim=0.10in 0.3in 0.8in 0.2in, clip, width=2.75in]{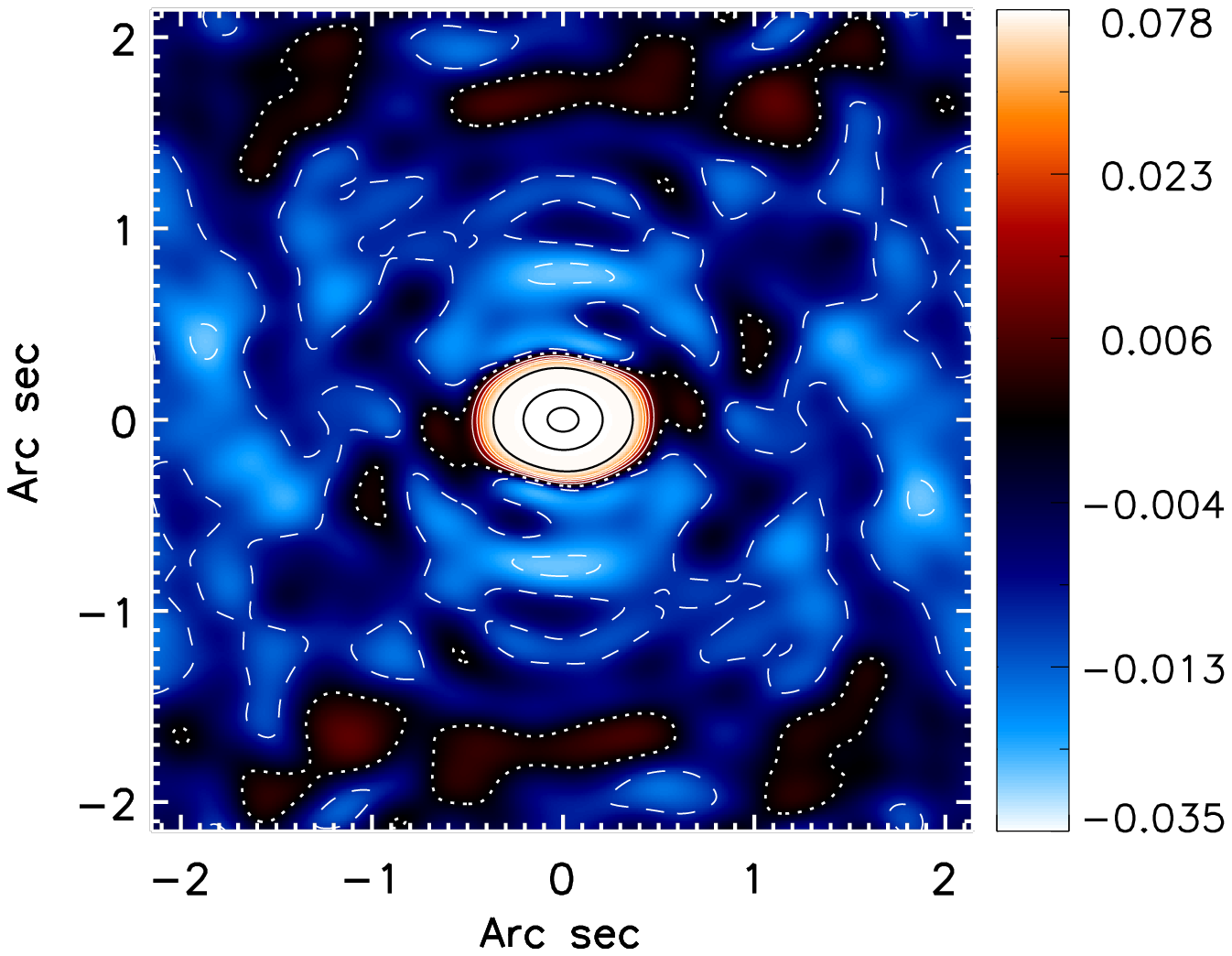} \\
\end{array}
$
\caption{{\it Top left:} Antenna locations, UV-distribution, encircled energy,
and beam pattern for the ALMA-02 configuration.
Same format as figure \ref{fig:ispiral6a}
The figures of merit are listed in table \ref{table:merit}.
}
\label{fig:alma02}
\end{figure}

\begin{figure}[t]
$
\begin{array}{cc}
\includegraphics[trim=0.10in 0.3in 0.8in 0.2in, clip, width=2.75in]{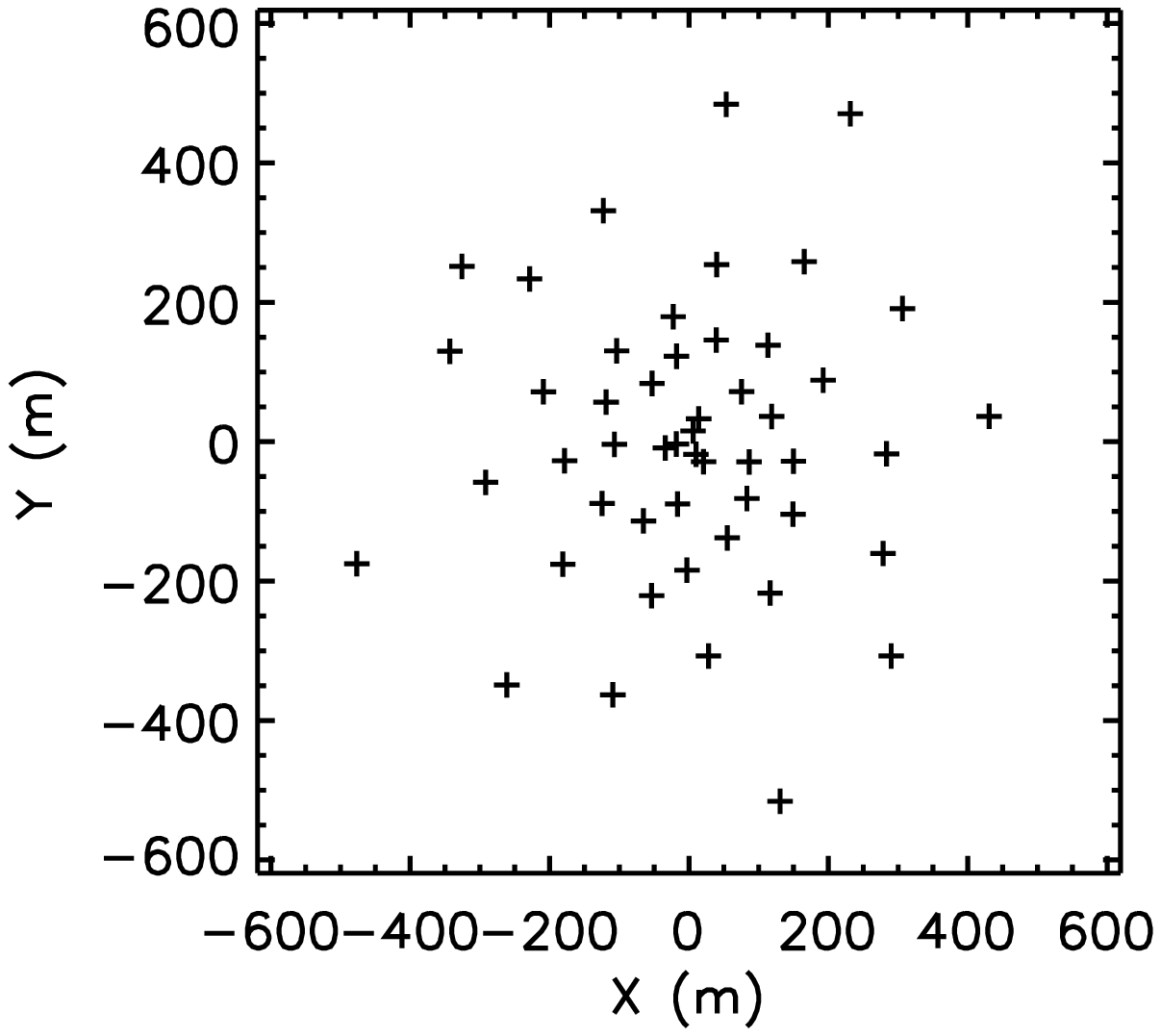} 
\includegraphics[trim=0.10in 0.3in 0.8in 0.2in, clip, width=2.75in]{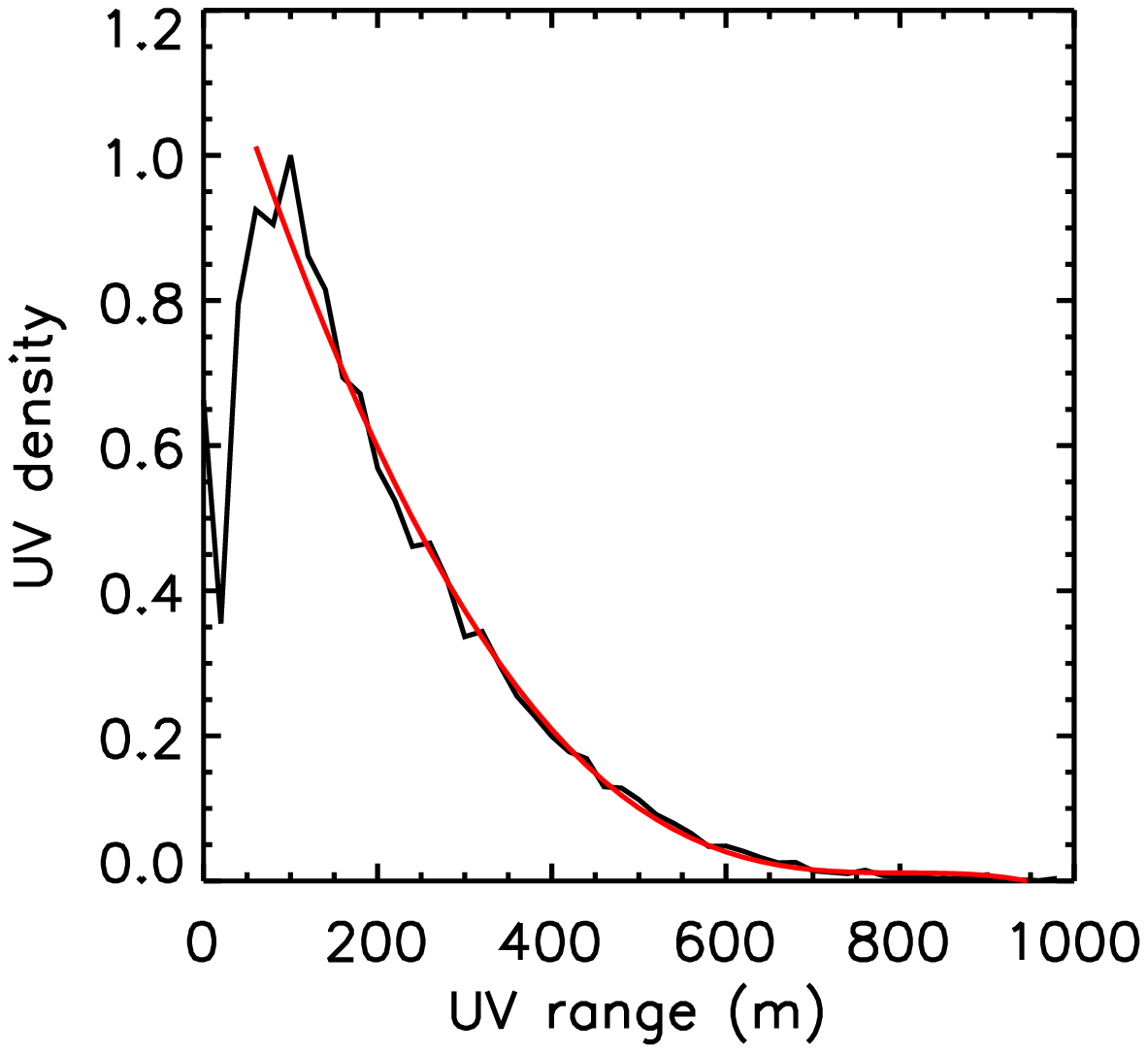} \\
\includegraphics[trim=0.10in 0.3in 0.8in 0.2in, clip, width=2.75in]{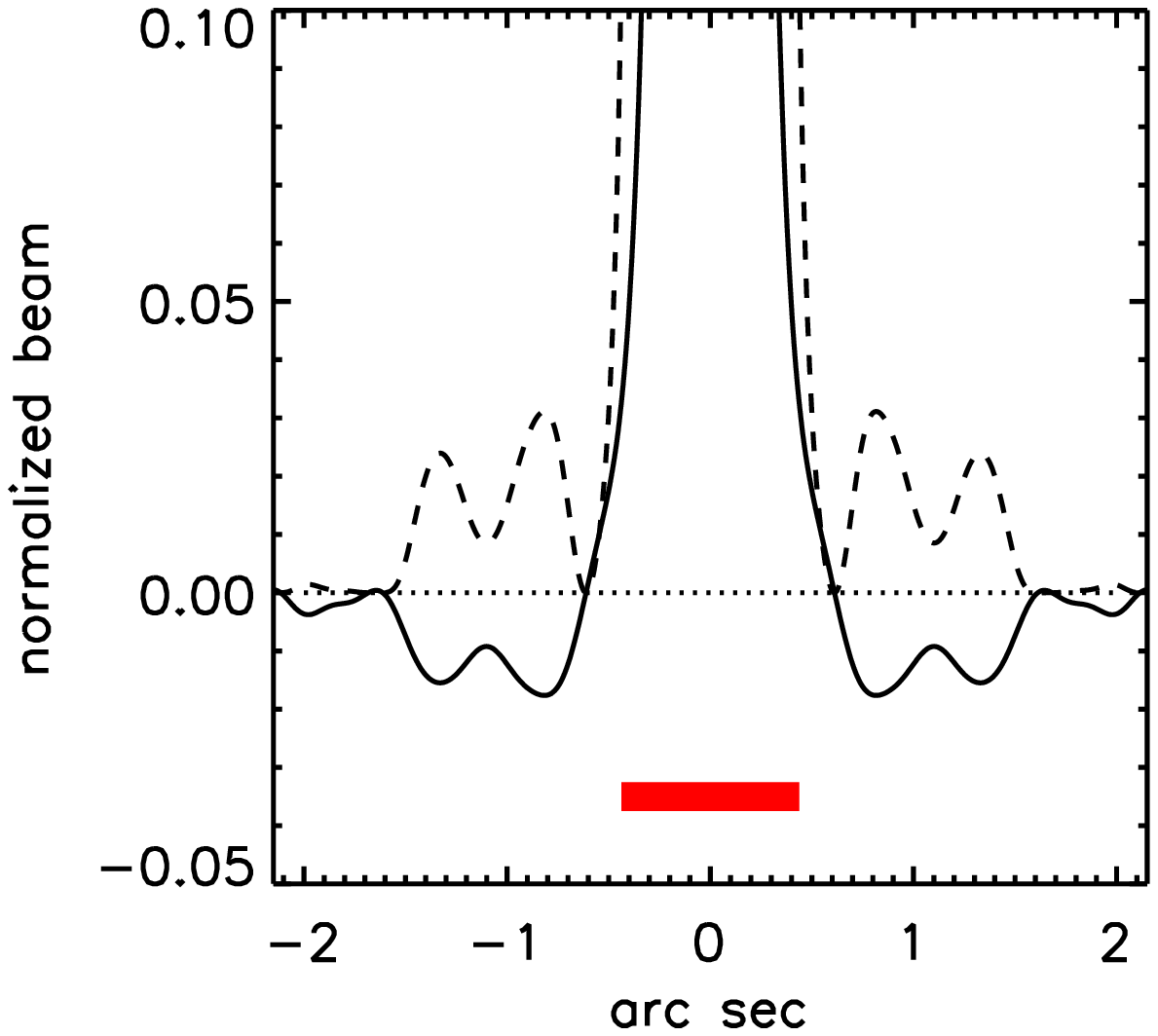} 
\includegraphics[trim=0.10in 0.3in 0.8in 0.2in, clip, width=2.75in]{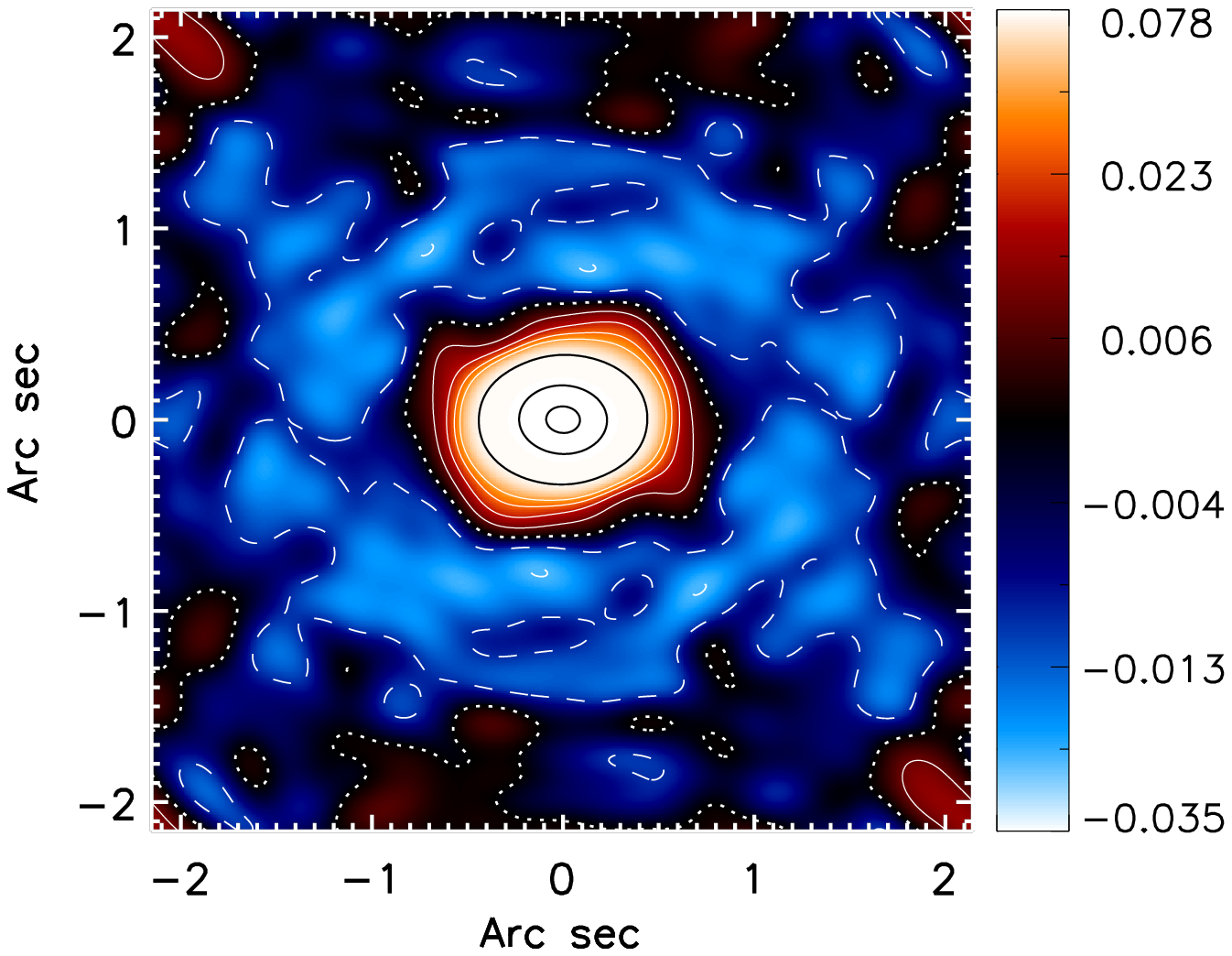} \\
\end{array}
$
\caption{{\it Top left:} Antenna locations, UV-distribution, encircled energy,
and beam pattern for the ALMA-14 configuration.
Same format as figure \ref{fig:ispiral6a}
The figures of merit are listed in table \ref{table:merit}.
}
\label{fig:alma14}
\end{figure}

\begin{figure}[t]
$
\begin{array}{cc}
\includegraphics[trim=0.10in 0.3in 0.8in 0.2in, clip, width=2.75in]{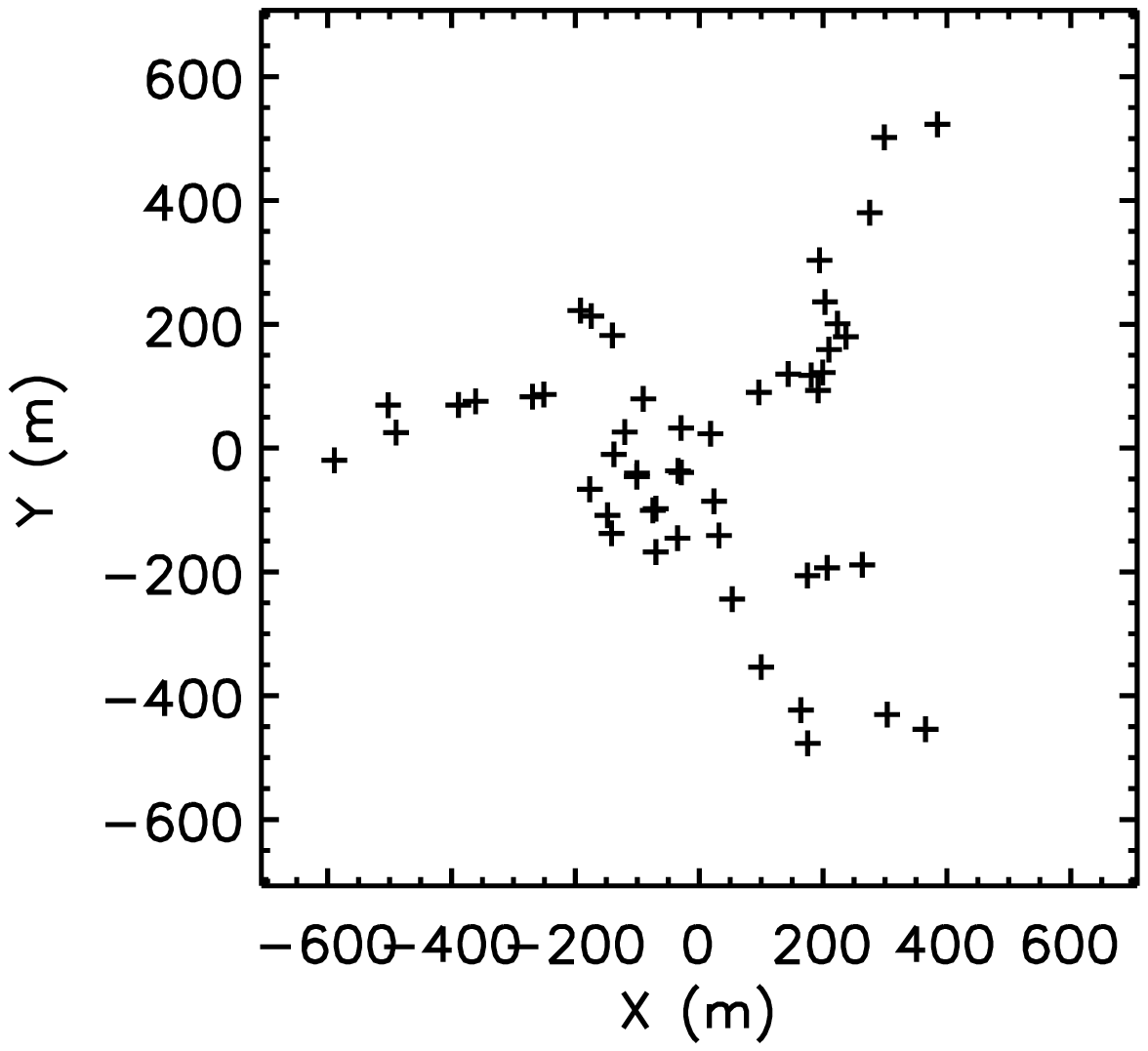} 
\includegraphics[trim=0.10in 0.3in 0.8in 0.2in, clip, width=2.75in]{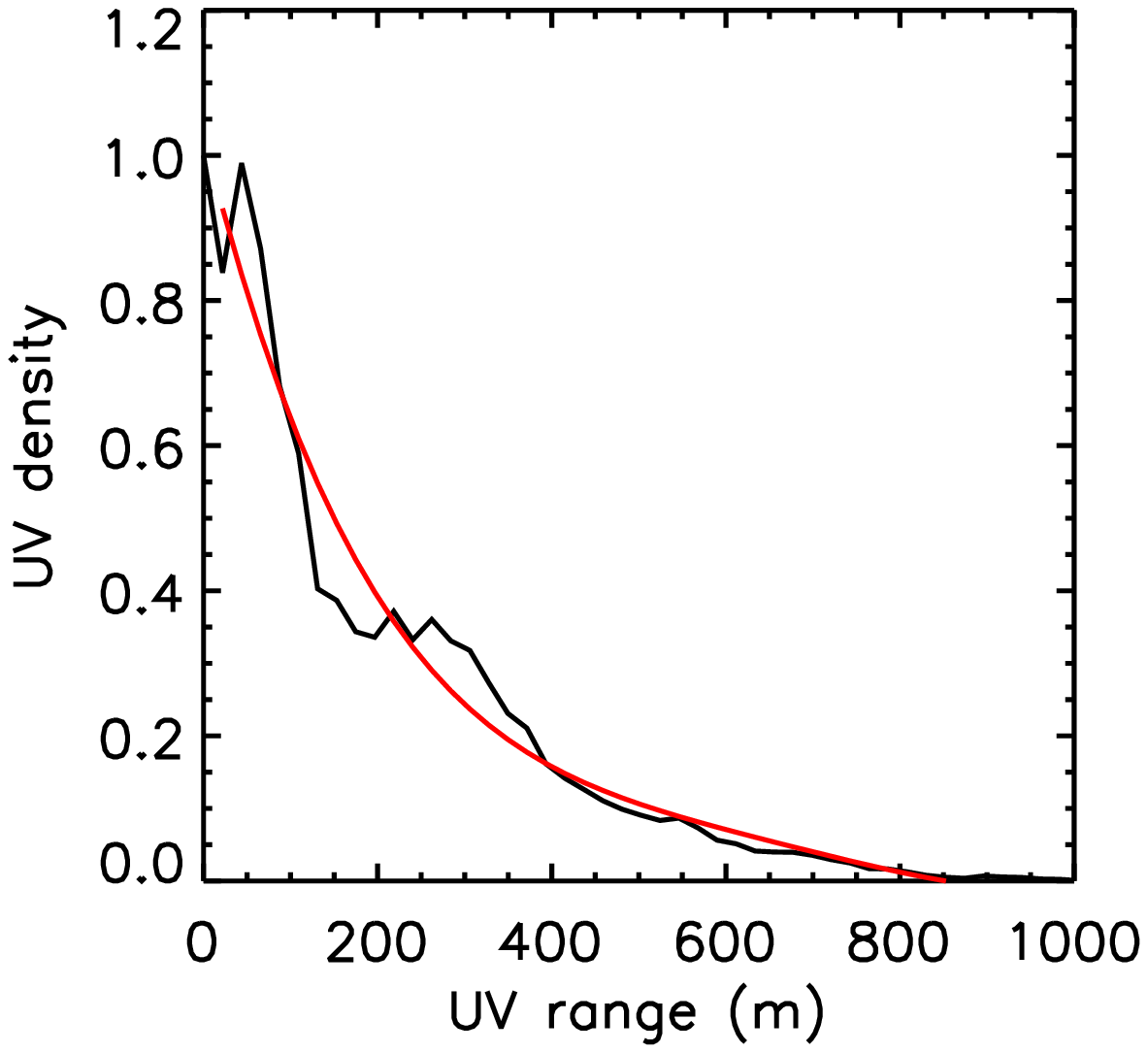} \\
\includegraphics[trim=0.10in 0.3in 0.8in 0.2in, clip, width=2.75in]{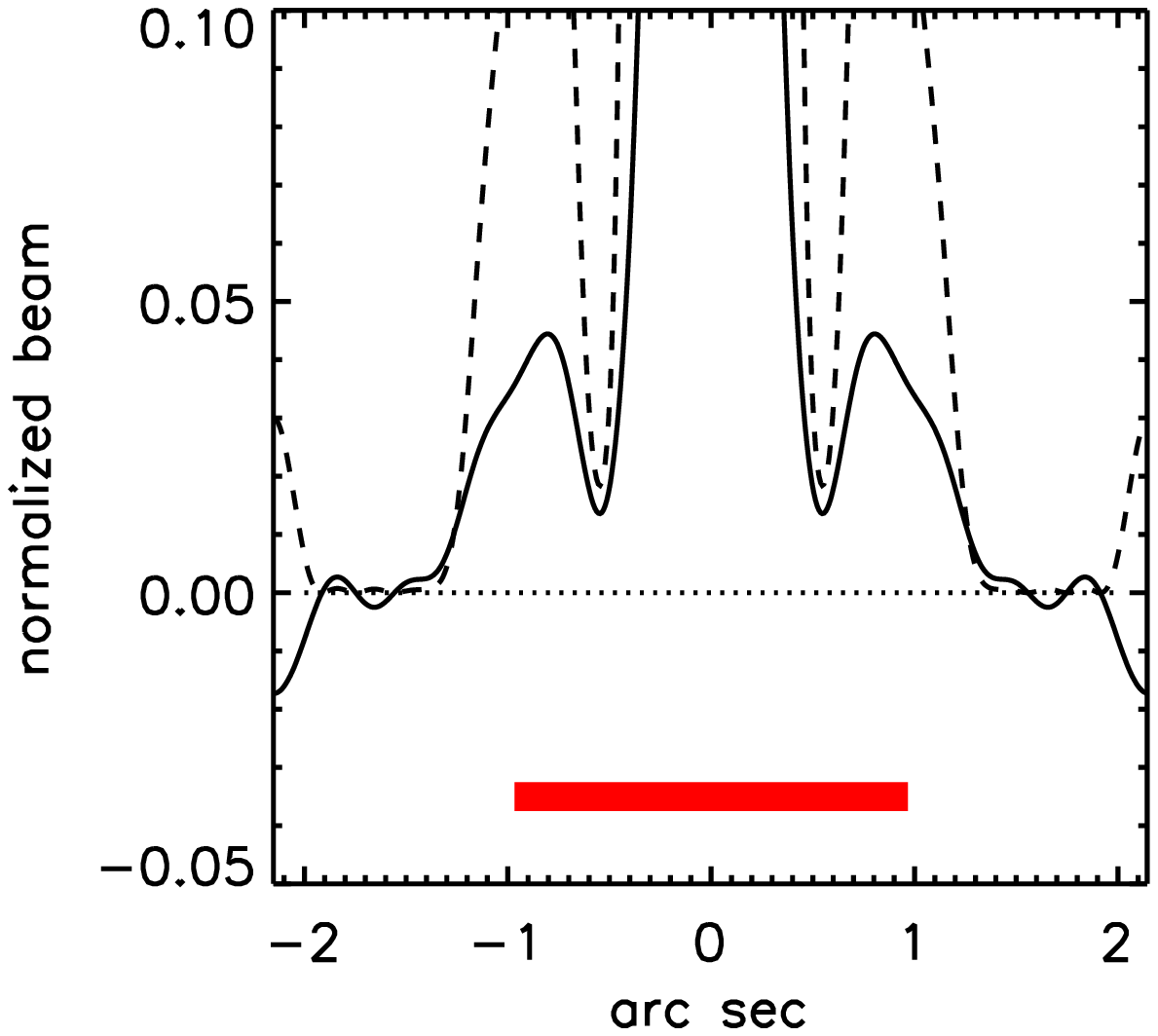} 
\includegraphics[trim=0.10in 0.3in 0.8in 0.2in, clip, width=2.75in]{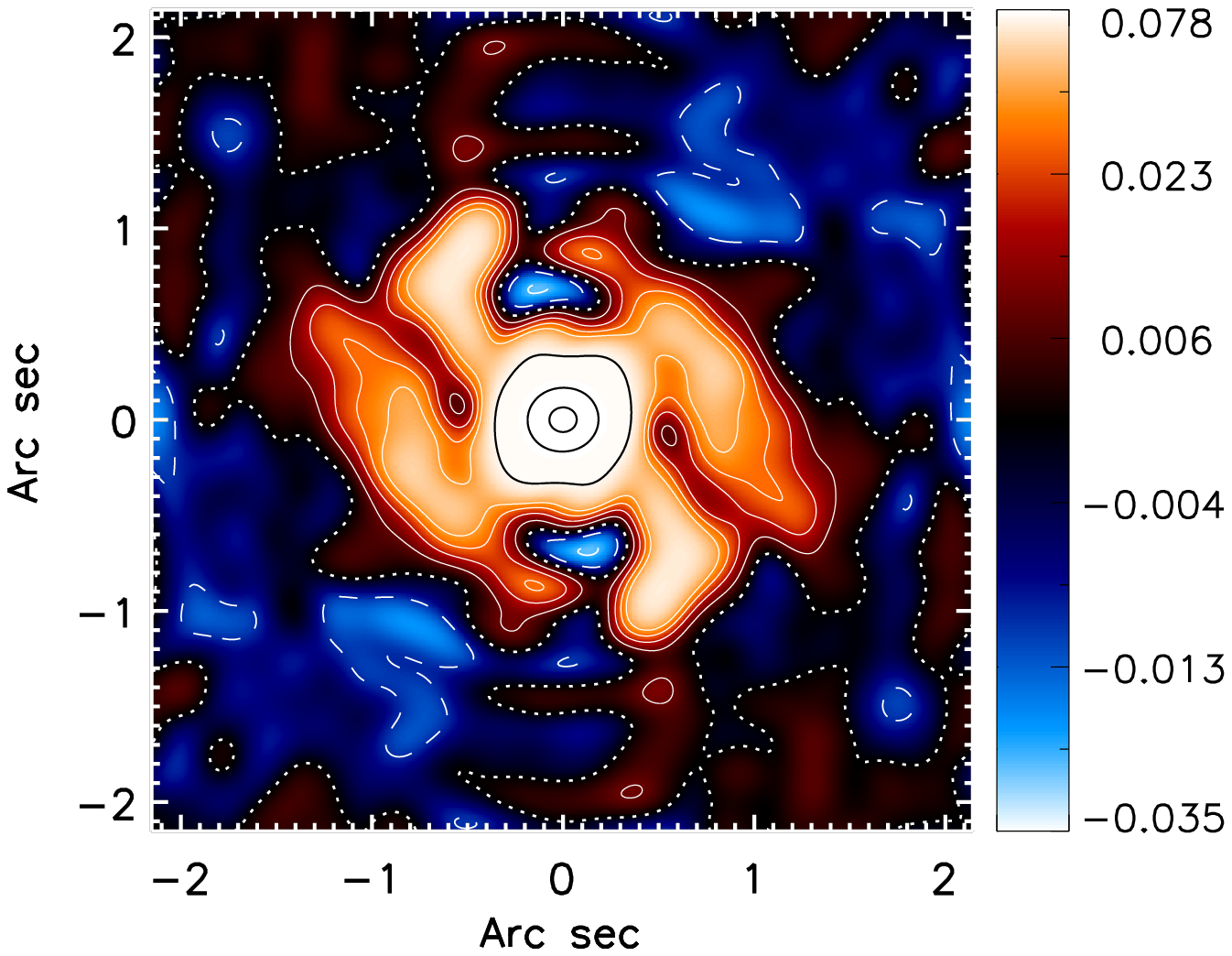} \\
\end{array}
$
\caption{{\it Top left:} Antenna locations, UV-distribution, encircled energy,
and beam pattern for the ALMA-28 configuration.
Same format as figure \ref{fig:ispiral6a}
The figures of merit are listed in table \ref{table:merit}.
}
\label{fig:alma28}
\end{figure}

It is worth comparing these example H-arrays to other designs. The recently
constructed Atacama Large Millimeter Array (ALMA) uses a 
variety of different configurations
listed in their CASA simulation software. Three of them, configurations ALMA-02, ALMA-14, and ALMA-28,
are shown in figures \ref{fig:alma02} to \ref{fig:alma28}. 
Aside from their different diameters, here normalized to 1000 m,
these three configurations are quite different from one another 
reflecting their different design goals as mentioned in the \S\ref{previous}. 
Configuration 
ALMA-02 is designed to minimize side lobe levels.
ALMA-14 is a spiral modified to approximate a Gaussian beam. 
ALMA-28 is a Y-pattern similar to the VLA. 
Both ALMA-02 and ALMA-14 have side lobe levels
comparable to the 36 or 54 antenna H-arrays, 
but the H-arrays, the appropriately scaled H-spirals, as well as the
Gaussian random array, all have better angular
resolution and a tighter concentration of beam energy 
encircled in a smaller radius. 
The figures of merit
for the three ALMA configurations are listed in table \ref{table:merit}


\section{Conclusions}

\begin{enumerate}
\item Different designs for cross-correlation interferometer arrays
exist in a continuous space of trade-offs between competing goals
of high angular resolution, concentrated beam power, compact array size,
and the smoothness of the distribution of the measured Fourier components.
\item Three figures of merit are useful in assessing the imaging performance of an array.
\begin{enumerate}
\item The $K_{nn}$ product of the array size and the radius which encompasses $nn$
of the total beam energy.
\item The angular resolution as measured by the beam FWHM.
\item The smoothness of the distribution of antenna separations.
\end{enumerate}
\item A technique of building a large array by scaling and rotating one simple
pattern of a small number of antenna locations can be used to construct
hierarchical arrays (H-array) or spiral arrays (H-spirals) with a large number
of antennas.
\item Hierarchical arrays (H-arrays) provide an excellent framework for 
locating the antennas of a cross-correlation imaging interferometer to
provide optimal imaging qualities.
\item By changing the scaling between the subarrays, both the H-arrays 
and the H-spirals can produce beams at a particular point
along the trade-off between angular resolution and concentrated beam power.
\item Sparsely populated H-arrays also have excellent performance and are useful for
interferometers designed with more antenna locations than antennas. This allows
the angular resolution to be changed by populating different subsets of
the antenna locations.
\item The construction procedure of scaling and rotating a basic subarray can also
be used to design high performance, intelligent spiral arrays, H-spirals. Spirals
are good for array designs emphasizing low side lobe designs rather than 
high angular resolution.
\item The H-arrays are easy to optimize because they can be described by a few
parameters, much fewer than the number of antennas themselves.
\item H-arrays are useful for future multi-element interferometers.
\end{enumerate}

\begin{sidewaystable}[ht] 
\caption{Figures of merit for example arrays in earth rotation synthesis}
\centering
\begin{tabular}{l l c c c c c }
\hline\hline
& \\
Array name & number of &shown as& FWHM& radius of 98\%   & $K_{98}$ product	& Normalized $\chi^2$ \\
           & antennas  &figure  &     & encircled energy \\
           &           &        & (arc sec) & (arc sec) & (m arc sec) & ($\times 1000$)\\ [0.5ex]
\hline
$J_1(r)/r$ & ... & \ref{fig:FTpair} {\it left}    	& 0.27 & 1.53 & 1530 & ...	\\
Gauss      & ... & \ref{fig:FTpair} {\it right} 	& 0.40 & 0.35 & 353  & ...	\\
s6p6       & 36 & \ref{fig:ska36_sinc_revB}         	& 0.17 & 1.38 & 1379 & 1.70  	\\
s6p6       & 36 & \ref{fig:ska36fix}	         	& 0.21 & 0.29 & 285  & 41.5  	\\ 
s3p3d6	   & 54 & \ref{fig:ska54_flat} 			& 0.17 & 0.89 & 891  & 0.20  	\\
s3p3d6	   & 54 & \ref{fig:ska54-n} 			& 0.23 & 0.29 & 294  & 1.79  	\\
H-spiral galaxy & 54 & \ref{fig:ispiral164}		& 0.23 & 0.29 & 285  & 1.47	\\
H-spiral sea star & 54 & \ref{fig:ispiral113}		& 0.23 & 0.29 & 285  & 0.81	\\
Gauss random   & 50 & \ref{fig:random}			& 0.26 & 0.33 & 327  & 7.80   \\
unif. random   & 50 & \ref{fig:random}			& 0.20 & 0.37 & 369  & 6.11   \\
ALMA-02    & 50 & \ref{fig:alma02}               	& 0.26 & 0.35 & 353  & 1.74  	\\
ALMA-14    & 50 & \ref{fig:alma14}               	& 0.29 & 0.44 & 437  & 0.79  	\\
ALMA-28    & 50 &\ref{fig:alma28}                	& 0.25 & 0.97 & 967  & 2.36	\\ [0.5ex]
\hline
\end{tabular}
\label{table:merit}
\end{sidewaystable}

\begin{sidewaystable}[ht] 
\caption{Figures of merit for spirals based on a 6-element subarray with different scaling}
\centering
\begin{tabular}{l l c c c c c }
\hline\hline
& \\
Scaling    & number of &shown as& FWHM& radius of 98\%   & $K_{98}$ product	& Normalized $\chi^2$ \\
factor     & antennas  &figure  &     & encircled energy \\
           &           &        & (arc sec) & (arc sec) & (m arc sec) & ($\times 1000$)\\ [0.5ex]
\hline
1.05       & 54 & \ref{fig:ispiral6a} {\it left} 	& 0.18 & 0.52 & 521  & 5.20	\\
1.15       & 54 & \ref{fig:ispiral6a} {\it right}      	& 0.23 & 0.26 & 260  & 0.88  	\\
1.25       & 54 & \ref{fig:ispiral6b} {\it left}	& 0.27 & 0.37 & 370  & 0.84  	\\ 
1.35	   & 54 & \ref{fig:ispiral6b} {\it right}	& 0.32 & 0.66 & 664  & 0.43  	\\
\hline
\end{tabular}
\label{table:meritcont}
\end{sidewaystable}



\end{document}